\pdfoutput=1 
\RequirePackage{silence} %
\documentclass[ twoside,openright,titlepage,numbers=noenddot,%
                headinclude,footinclude,cleardoublepage=empty,abstract=on,
                BCOR=5mm,paper=a4,fontsize=11pt,table,usenames,dvipsnames
                ]{scrreprt}

\PassOptionsToPackage{utf8}{inputenc}
  \usepackage{inputenc}

\PassOptionsToPackage{T1}{fontenc} %
  \usepackage{fontenc}

\PassOptionsToPackage{
  drafting=false,    %
  tocaligned=false, %
  dottedtoc=true,  %
  eulerchapternumbers=true, %
  linedheaders=false,       %
  floatperchapter=true,     %
  eulermath=true,  %
  beramono=true,    %
  palatino=true,    %
  style=classicthesis %
}{classicthesis}

\newcommand{\myTitle}{Around-Body Interaction\xspace}
\newcommand{\mySubtitle}{Leveraging Limb Movements for Interacting in a Digitally Augmented Physical World\xspace}
\newcommand{\myDegree}{Doctor rerum naturalium (Dr. rer. nat.)\xspace}
\newcommand{\myName}{Florian Benjamin Müller, M.Sc.\xspace}
\newcommand{\myNameNoTitle}{Florian Benjamin Müller\xspace}
\newcommand{\myProf}{Prof. Max Mühlhäuser (Technische Universität Darmstadt)\xspace}
\newcommand{\mySecondAdvisor}{Prof. Albrecht Schmidt (Ludwig-Maximilians-Universität München)\xspace}

\newcommand{\myFaculty}{Fachbereich Informatik\xspace}
\newcommand{\myDepartment}{Telecooperation Lab\xspace}
\newcommand{\myUni}{Technische Universität Darmstadt\xspace}
\newcommand{\myLocation}{Darmstadt\xspace}
\newcommand{\myTime}{November 2019\xspace}
\newcommand{\myVersion}{v1.0}
\newcommand{\myKennziffer}{Hochschulkennziffer D 17}

\newcommand{\submissionDate}{04. November 2019\xspace}
\newcommand{\disputationDate}{19. Dezember 2019\xspace}

\providecommand{\mLyX}{L\kern-.1667em\lower.25em\hbox{Y}\kern-.125emX\@}

\PassOptionsToPackage{ngerman,american}{babel} %
\usepackage{babel}

\usepackage{csquotes}
\PassOptionsToPackage{%
  backend=biber,bibencoding=utf8, %
  language=auto,%
  style=authoryear-comp,
  url=false,
  doi=true,
  sorting=nyt, %
  maxbibnames=99, %
  maxcitenames=3,
  dashed=false,   
  natbib=true %
}{biblatex}
    \usepackage{biblatex}

\PassOptionsToPackage{fleqn}{amsmath}       %
  \usepackage{amsmath}

\usepackage{graphicx} %
\usepackage{scrhack} %
\usepackage{xspace} %
\PassOptionsToPackage{printonlyused,smaller}{acronym}
  \usepackage{acronym} %

\usepackage{tabularx} %
\usepackage{subfig}

\usepackage{listings}
\usepackage{todonotes}
\usepackage{siunitx}
\usepackage{pdfpages}
\usepackage{booktabs,amssymb,makecell,rotating}
\usepackage{fontawesome}
\usepackage{xcolor}
\usepackage[framemethod=tikz]{mdframed}
\usepackage[tight]{minitoc}
\usepackage{nicefrac}
\usepackage[export]{adjustbox}
\usepackage{bookmark}
\usepackage{placeins}

\usepackage{tikz}
\usetikzlibrary{tikzmark}

\usepackage{classicthesis}

\definecolor{CTurl}{named}{Black}
\definecolor{CTlink}{named}{Black}
\definecolor{CTcitation}{named}{Black}

\hypersetup{%
  colorlinks=true, linktocpage=true, pdfstartpage=1, pdfstartview=FitV,%
  breaklinks=true, pageanchor=true,%
  pdfpagemode=UseNone, %
  plainpages=false, bookmarksnumbered, bookmarksopen=true, bookmarksopenlevel=1,%
  hypertexnames=true, pdfhighlight=/O,%
  urlcolor=CTurl, linkcolor=CTlink, citecolor=CTcitation, %
  pdftitle={\myTitle},%
  pdfauthor={\textcopyright\ \myName, \myUni, \myFaculty},%
  pdfsubject={},%
  pdfkeywords={},%
  pdfcreator={pdfLaTeX},%
  pdfproducer={LaTeX with hyperref and classicthesis}%
}

\makeatletter
\@ifpackageloaded{babel}%
  {%
    \addto\extrasamerican{%
    }%
    \addto\extrasngerman{%
    }%
    }{\relax}
\makeatother

\DeclareBibliographyCategory{ignore}
\addtocategory{ignore}{Muller2015b}
\addtocategory{ignore}{Muller2020}
\addtocategory{ignore}{Muller2019}
\addtocategory{ignore}{muller2017cloudbits}
\addtocategory{ignore}{muller2016proxiwatch}
\addtocategory{ignore}{muller2015a}
\addtocategory{ignore}{muller2018personalized}
\addtocategory{ignore}{muller2018smartobjects}
\addtocategory{ignore}{SmartObjects2018}

\def\iconYes{\faCheck}

\def\iconPartially{\faCircleThin}

\def\iconAward{\faTrophy}
\def\iconQuote{\faQuoteLeft}
\def\iconInteraction{\faStreetView}

\definecolor{thesisGreen}{RGB}{46, 204, 64}
\definecolor{thesisRed}{RGB}{255, 65, 54}

\DeclareFieldFormat{citehyperref}{%
  \DeclareFieldAlias{bibhyperref}{noformat}%
  \bibhyperref{#1}}

\DeclareFieldFormat{textcitehyperref}{%
  \DeclareFieldAlias{bibhyperref}{noformat}%
  \bibhyperref{%
    #1%
    \ifbool{cbx:parens}
      {\bibcloseparen\global\boolfalse{cbx:parens}}
      {}}}

\savebibmacro{cite}
\savebibmacro{textcite}

\renewbibmacro*{cite}{%
  \printtext[citehyperref]{%
    \restorebibmacro{cite}%
    \usebibmacro{cite}}}

\renewbibmacro*{textcite}{%
  \ifboolexpr{
    ( not test {\iffieldundef{prenote}} and
      test {\ifnumequal{\value{citecount}}{1}} )
    or
    ( not test {\iffieldundef{postnote}} and
      test {\ifnumequal{\value{citecount}}{\value{citetotal}}} )
  }
    {\DeclareFieldAlias{textcitehyperref}{noformat}}
    {}%
  \printtext[textcitehyperref]{%
    \restorebibmacro{textcite}%
    \usebibmacro{textcite}}}

\newcommand{\anovaCor}[6]{$F_{#1, #2}=#3$, $p#4$, $\epsilon=#5$\subEta{#4}{#6}}
\newcommand{\anova}[5]{$F_{#1, #2}=#3$, $p#4$\subEta{#4}{#5}}
\newcommand{\anovaCorWithoutEffect}[5]{$F_{#1, #2}=#3$, $p#4$, $\epsilon=#5$}
\newcommand{\anovaWithoutEffect}[4]{$F_{#1, #2}=#3$, $p#4$}

\newcommand{\subEta}[2]{%
	\ifthenelse{\equal{#1}{\string >.05}}
	{}
	{, $\eta^{2}=#2$}%
}

\newcommand{\valSi}[3]{$\mu = \SI{#1}{#3}$, $\sigma = \SI{#2}{#3}$}
\newcommand{\val}[2]{$\mu = #1$, $\sigma = #2$}
\newcommand{\se}[1]{$\sigma_{\overline{x}} = #1$}

\newcommand{\emm}[2]{\ac{EMM} $\mu = #1$, \se{#2}}
\newcommand{\emmSi}[3]{\ac{EMM} $\mu = \SI{#1}{#3}$, \se{\SI{#2}{#3}}}
\newcommand{\emmSiCI}[5]{\ac{EMM} $\mu = \SI{#1}{#5}$, \se{\SI{#2}{#5}}}

\newcommand{\emmP}[2]{\emm{#1\%}{#2\%}}

\newcommand{\pearson}[2]{$r = #1$, $p#2$}

\def\gge{$\epsilon$}

\newcommand{\ncite}[1]{[\cite{#1}]}
\newcommand{\typcite}[1]{\citeauthor{#1}~[\citeyear{#1}]}

\newcommand{\pquote}[2]{\enquote{\emph{#1}} (P#2)}
\newcommand{\mpquote}[2]{\enquote{\emph{#1}} (#2)}

\newcommand{\reffig}[1]{see fig. \ref{#1}}
\newcommand*{\fullref}[1]{\hyperref[{#1}]{\autoref*{#1} (\nameref*{#1})}} %

\makeatletter
\newcommand{\tempmaxup}[1]{\def\blx@maxcitenames{\blx@maxbibnames}#1}
\makeatother

\DeclareCiteCommand{\longfullcite}[\tempmaxup]
  {\usebibmacro{prenote}}
  {\usedriver
     {\DeclareNameAlias{sortname}{default}}
     {\thefield{entrytype}}}
  {\multicitedelim}
  {\usebibmacro{postnote}}

\DeclareCiteCommand{\citeyear}
{\usebibmacro{prenote}}
{\bibhyperref{\printfield{year}}\bibhyperref{\printfield{extrayear}}}
{\multicitedelim}
{\usebibmacro{postnote}}

\DeclareCiteCommand{\citeyearpar}[\mkbibparens]
{\usebibmacro{prenote}}
{\bibhyperref{\printfield{year}}\bibhyperref{\printfield{extrayear}}}
{\multicitedelim}
{\usebibmacro{postnote}}

\DeclareCiteCommand{\citeauthorfirstlast}
{\boolfalse{citetracker}%
	\boolfalse{pagetracker}%
	\DeclareNameAlias{labelname}{first-last}%
	\usebibmacro{prenote}}
{\ifciteindex
	{\indexnames{labelname}}
	{}%
	\printnames{labelname}}
{\multicitedelim}
{\usebibmacro{postnote}}
\newcommand{\colfig}[3][1]{

 \begin{figure}
  \centering
  \includegraphics[width=#1\linewidth]{Content/Figures/#2}
  \caption{#3}
  \label{fig:#2}
 \end{figure}

}

\newcommand{\colfigH}[3][1]{
	
	\begin{figure}[h!]
		\centering
		\includegraphics[width=#1\linewidth]{Content/Figures/#2}
		\caption{#3}
		\label{fig:#2}
	\end{figure}
	
}

\newcommand{\colfigVspace}[4][1]{

 \begin{figure}
  \centering
  \includegraphics[width=#1\linewidth]{Content/Figures/#2}
  \caption{#3}
  \label{fig:#2}
  \vspace{#4}
 \end{figure}

}

\newcommand{\textfig}[3][1]{

 \begin{figure*}
  \centering
  \includegraphics[width=#1\textwidth]{Content/Figures/#2}
  \caption{#3}
  \label{fig:#2}
 \end{figure*}

}

\newcommand{\textfigH}[3][1]{
	
	\begin{figure*}[h!]
		\centering
		\includegraphics[width=#1\textwidth]{Content/Figures/#2}
		\caption{#3}
		\label{fig:#2}
	\end{figure*}
	
}

\newcommand{\textfigStudybox}[1]{
	
	\begin{figure*}[h!]
		\centering
		\includegraphics[width=\textwidth]{Content/Figures/#1}
	\end{figure*}
	
}

\newlength{\twosubht}
\newsavebox{\twosubbox}

\newcommand{\sameheightpic}[5]{

\begin{figure*}[ht!]
	
	\sbox\twosubbox{%
		\resizebox{\dimexpr.99\textwidth-1em}{!}{%
			\includegraphics[height=3cm]{Content/Figures/#1}%
			\includegraphics[height=3cm]{Content/Figures/#3}%
		}%
	}
	\setlength{\twosubht}{\ht\twosubbox}

	\subfloat[#2\label{fig:#1}]
	{\includegraphics[height=\twosubht]{Content/Figures/#1}}\hfill
	\subfloat[#4\label{fig:#3}]
	{\includegraphics[height=\twosubht]{Content/Figures/#3}}
	\caption{#5}
\end{figure*}

}

\newcommand{\cptteaser}[2]{
\begin{figure}[h!]
	\begin{center}
		\includegraphics[width=\linewidth]{Content/Figures/#1}
		\caption{#2}
		\label{fig:#1}
	\end{center}
\end{figure}
} 

\def\reqYes{\iconYes}
\def\reqPartially{\iconPartially}
\def\reqNo{}
\mdfdefinestyle{infoboxstyle}{%
	rightline=true,
	innerleftmargin=10,
	innerrightmargin=10,
	linecolor=gray,
	outerlinewidth=1.0mm,
	topline=false,
	rightline=false,
	bottomline=false,
	skipabove=5pt,
	skipbelow=5pt
}

\noptcrule 
\DeclareSourcemap{
  \maps[datatype=bibtex]{
    \map[overwrite]{
      \step[fieldsource=month, match=\regexp{\A(j|J)an\Z}, replace=1]
      \step[fieldsource=month, match=\regexp{\A(f|F)eb\Z}, replace=2]
      \step[fieldsource=month, match=\regexp{\A(m|M)ar\Z}, replace=3]
      \step[fieldsource=month, match=\regexp{\A(a|A)pr\Z}, replace=4]
      \step[fieldsource=month, match=\regexp{\A(m|M)ay\Z}, replace=5]
      \step[fieldsource=month, match=\regexp{\A(j|J)un\Z}, replace=6]
      \step[fieldsource=month, match=\regexp{\A(j|J)ul\Z}, replace=7]
      \step[fieldsource=month, match=\regexp{\A(a|A)ug\Z}, replace=8]
      \step[fieldsource=month, match=\regexp{\A(s|S)ep\Z}, replace=9]
      \step[fieldsource=month, match=\regexp{\A(o|O)ct\Z}, replace=10]
      \step[fieldsource=month, match=\regexp{\A(n|n)ov\Z}, replace=11]
      \step[fieldsource=month, match=\regexp{\A(d|D)ec\Z}, replace=12]
    }
  }
}
\newcommand\circledmark[1][red]{%
 
}

\newcommand{\circledtext}[2]{
 \ooalign{%
	\hidewidth
	\kern0.65ex\raisebox{-0.9ex}{\scalebox{3}{\textcolor{#2}{#1}}}
	\hidewidth\cr
	$\checkmark$\cr
}
}
\usepackage{thmtools}

\definecolor{colorBackground}{HTML}{FFFFFF} %
\definecolor{colorDefinition}{HTML}{4CAF50} %
\definecolor{colorNotiz}{HTML}{81D4FA}      %
\definecolor{colorWarnung}{HTML}{FF9800}    %
\definecolor{colorBeispiel}{HTML}{0288D1}   %
\definecolor{colorSatz}{HTML}{009688}       %

\definecolor{colorRequirement}{HTML}{FF9800}
\definecolor{colorDefinition}{HTML}{4CAF50} 
\definecolor{colorQuote}{HTML}{0288D1}
\definecolor{colorInteractionSituation}{HTML}{8e24aa} %

\def\iconReq{\faExclamation}
\def\iconDef{\faBook}
\def\iconQuote{\faQuoteLeft}

\newcommand*\circled[2]{\tikz[baseline=(char.base)]{
		\node[shape=circle,fill=#1,draw,inner sep=1pt] (char) {#2};}}

\mdfdefinestyle{coloredBoxStyle}{
    topline=false,
    bottomline=false,
    rightline=false,
    linewidth=4pt,
    backgroundcolor=colorBackground,
    frametitlefont=\sffamily\bfseries\color{black},
    splittopskip=.5cm,
    frametitlebelowskip=.3cm,
}

\mdtheorem[
style=coloredBoxStyle,
linecolor=colorRequirement,
nobreak=true,
]{reqTheorem}{\hangindent=5em\hangafter=1\raisebox{-.4ex}{\Large \color{colorBackground}\circled{colorRequirement}{\iconReq}} Requirement}

\mdtheorem[
style=coloredBoxStyle,
linecolor=colorDefinition,
nobreak=true,
]{defTheorem}{\raisebox{-0ex}{\small \color{colorBackground}\circled{colorDefinition}{\iconDef}} Definition}

\mdtheorem[
style=coloredBoxStyle,
linecolor=colorQuote,
nobreak=true,
]{quoteTheorem}{\raisebox{-.1ex}{\small \color{colorBackground}\circled{colorQuote}{\iconQuote}} Quote}

\mdtheorem[
style=coloredBoxStyle,
linecolor=colorInteractionSituation,
nobreak=true,
]{interactionTheorem}{\hangindent=5em\hangafter=1\raisebox{-.4ex}{\Large \color{colorBackground}\circled{colorInteractionSituation}{\iconInteraction}} Alice's Day in the City}

\newcommand{\defbox}[3]{
	\begin{defTheorem*}[#2]
		\label{def:#1}
		#3
	\end{defTheorem*}
}

\newcommand{\citebox}[2]{
	\begin{quoteTheorem*}[\citeauthorfirstlast{#2}, \citeyear{#2}]
		\enquote{#1}
	\end{quoteTheorem*}
}

\newcommand{\interactionbox}[3]{
	\begin{interactionTheorem*}[#2]
		\label{int:#1}
		#3
	\end{interactionTheorem*}
}

\usepackage{keyval}

\newcommand{\reqbox}[3]{
	\begin{reqTheorem}[#2]
		\label{req:#1}

		#3
	\end{reqTheorem}
}

\newcommand{\refreq}[1]{
	R\ref{#1}
}

\numberwithin{reqTheorem}{section}

\newcommand{\refproject}[1]{
	\spacedlowsmallcaps{#1}
}

\def\projProximity{\refproject{Proximity-Based One-Handed Interaction}}
\def\projCheesyfoot{\refproject{Mind The Tap: Direct and Indirect Interaction using Foot-Taps}}
\def\projCheesyfootToGo{\refproject{Walk The Line: Lateral Shifts of the Walking Path as an Input Modality}}
\def\projCloudbits{\refproject{CloudBits: Spatially-Aware Interaction with Proactively Retrieved Information}}

\def\projProximityS{\refproject{Proximity-Based Interaction}}
\def\projCheesyfootS{\refproject{Mind The Tap}}
\def\projCheesyfootToGoS{\refproject{Walk The Line}}
\def\projCloudbitsS{\refproject{CloudBits}}

\def\mobilityLower{\protect\inlinegraphicsame{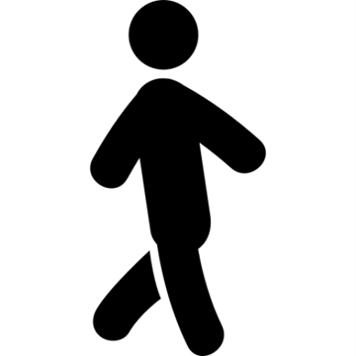}~mobility}
\def\inplaceLower{\protect\inlinegraphicsame{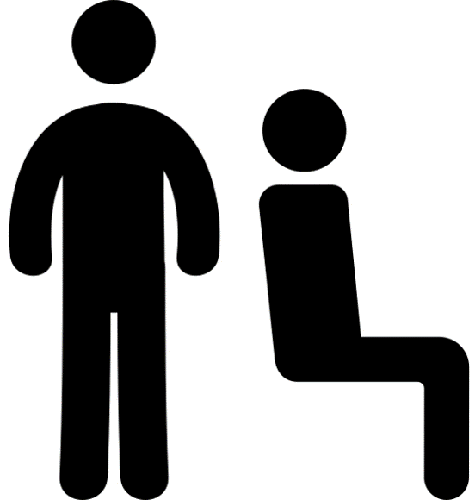}~in-place}
\def\singleuserLower{\protect\inlinegraphicsame{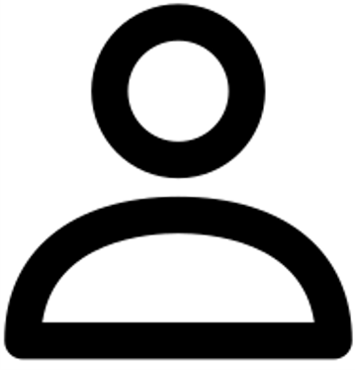}~single-user}
\def\multiuserLower{\protect\inlinegraphicsame{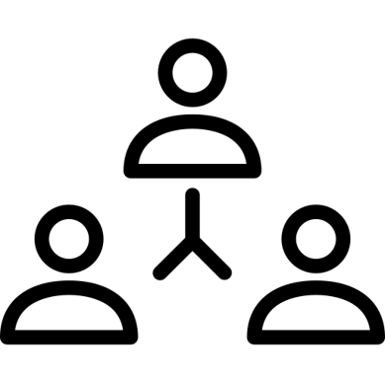}~multi-user}
\def\continuousLower{\protect\inlinegraphicsame{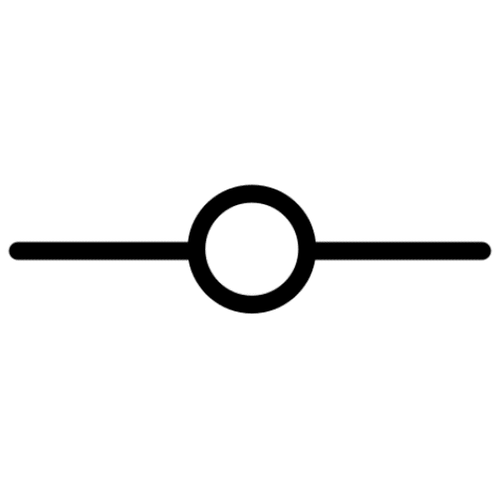}~continuous interaction}
\def\discreteLower{\protect\inlinegraphicsame{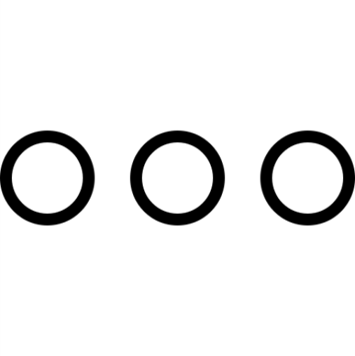}~discrete interaction}

\def\mobilityUpper{\protect\inlinegraphicsame{Content/Figures/General/mobility}~Mobility}
\def\inplaceUpper{\protect\inlinegraphicsame{Content/Figures/General/in-place}~In-Place}
\def\singleuserUpper{\protect\inlinegraphicsame{Content/Figures/General/single-user}~Single-User}
\def\multiuserUpper{\protect\inlinegraphicsame{Content/Figures/General/multi-user}~Multi-User}
\def\continuousUpper{\protect\inlinegraphicsame{Content/Figures/General/continuous}~Continuous Interaction}
\def\discreteUpper{\protect\inlinegraphicsame{Content/Figures/General/discrete}~Discrete Interaction}

\def\aroundbodyinteraction{\emph{around-body interaction}}
\def\Aroundbodyinteraction{\emph{Around-body interaction}}
\newcolumntype{Y}{>{\centering\arraybackslash}X}

\newcommand*\mybackmatter{%
\startcontents
\phantomsection
\addcontentsline{toc}{part}{}%
\@endpart}
\newlength\myheight
\newlength\mydepth
\settototalheight\myheight{Xygp}
\newcommand*\inlinegraphics[1]{%
  \settototalheight\myheight{Xygp}%
  \settodepth\mydepth{Xygp}%
  \raisebox{-1.8\mydepth}{\includegraphics[height=1.8\myheight]{#1}}%
}

\newcommand*\inlinegraphicsame[1]{%
	\settototalheight\myheight{Xygp}%
	\settodepth\mydepth{Xygp}%
	\raisebox{-1.5\mydepth}{\includegraphics[height=1.5\myheight]{#1}}%
}

\newcommand*\attributegraphic[2]{%
	\settototalheight\myheight{Xygp}%
	\settodepth\mydepth{Xygp}%
	\raisebox{-1.8\mydepth}{\includegraphics[height=1.8\myheight]{#1}}~by #2.
	\smallskip
}

\newcommand{\cptIntroBox}[2]{
\bigskip

\begin{mdframed}[style=infoboxstyle]
	\textbf{Publication:} This chapter is based on the following publication:
	
	\begin{small}
		\longfullcite{#1}
	\end{small}
	
	\textbf{Contribution Statement:} I led the idea generation, implementation and performed the data evaluation. #2
\end{mdframed}
}

\newcommand{\cptIntroBoxAward}[2]{
	\bigskip
	
	\begin{mdframed}[style=infoboxstyle]
		\textbf{Publication:} This chapter is based on the following publication:
		
		\begin{small}
			\longfullcite{#1}
			
			\iconAward~This paper received an honorable mention award.
		\end{small}
		
		\textbf{Contribution Statement:} I led the idea generation, implementation and performed the data evaluation. #2
	\end{mdframed}
}
\newcommand{\manumarkboth}[1]{
	\manualmark
	\markboth{\spacedlowsmallcaps{#1}}{\spacedlowsmallcaps{#1}}%
}

\begin{document}
\frenchspacing
\raggedbottom
\selectlanguage{american} %
\pagenumbering{roman}
\pagestyle{plain}
\begin{titlepage}
    \begin{addmargin}[-1cm]{-3cm}
    \begin{center}
        \large

        \hfill

        \vfill

        \begingroup
            \color{CTtitle}\spacedallcaps{\myTitle} \\ \medskip
        \endgroup

		\spacedlowsmallcaps{Leveraging Limb Movements for Interacting} \\
		
		\spacedlowsmallcaps{in a Digitally Augmented Physical World}

\end{center}

        \vfill

        Vom Fachbereich Informatik der Technischen Universität Darmstadt genehmigte
        
        \textbf{Dissertation}
        
        \mbox{zur Erlangung des akademischen Grades eines \myDegree{}.}

        Eingereicht von\\
        \textbf{\myName{}}

\vspace{1em} 

	\begin{tabular}{@{}ll@{}}
		
		Erstreferent: & \myProf\\
		Korreferent: & \mySecondAdvisor \\[1em]
		
		Tag der Einreichung: & \submissionDate{} \\ 
		Tag der Disputation: & \disputationDate{} \\
		
	\end{tabular}
	
	\vspace{1em} 
	\noindent
	
	\myFaculty\\
	\myDepartment\\
	\myUni\\
	\myKennziffer\\[1em]

        \vfill

\begin{center}
        \includegraphics[width=6cm]{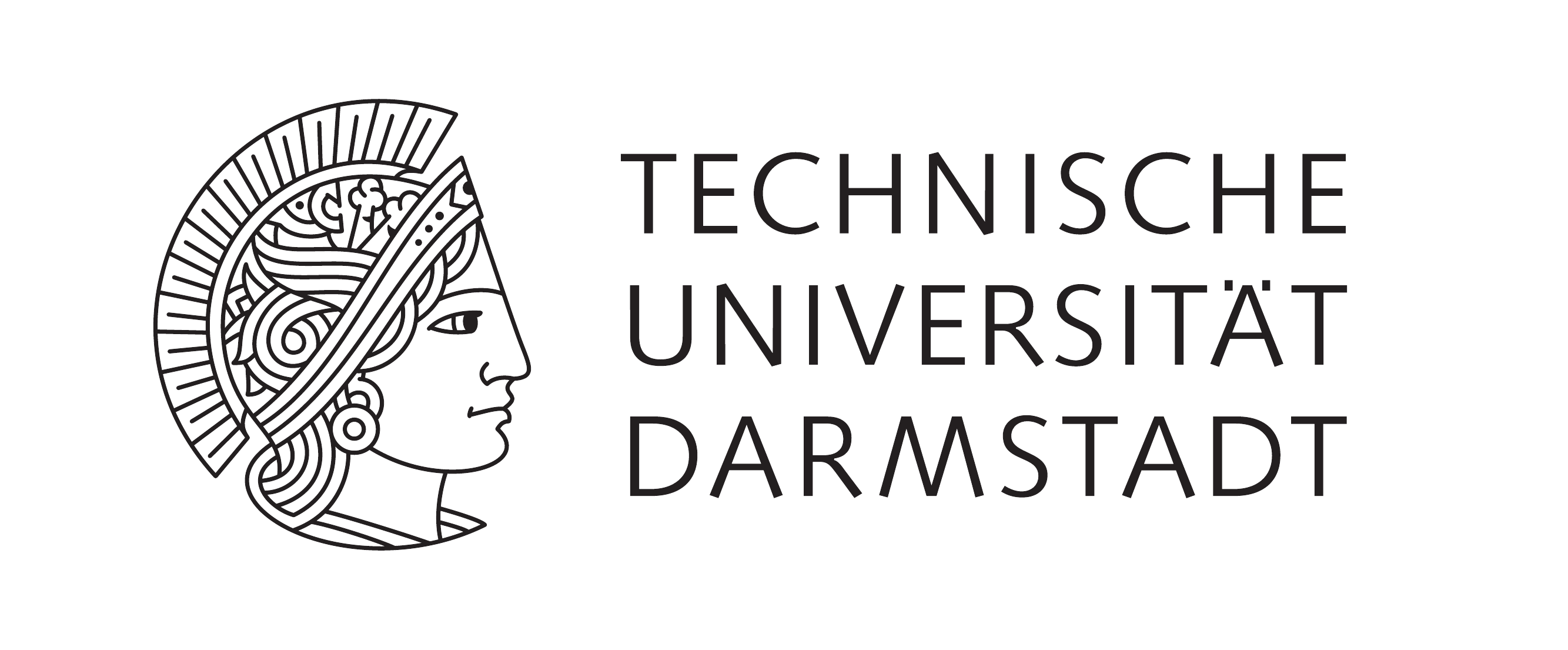} \\\medskip

   \end{center}
  \end{addmargin}
\end{titlepage}

\thispagestyle{empty}

\hfill

\vfill

\noindent\myNameNoTitle: \textit{\myTitle:} \mySubtitle\\[1em]
Darmstadt, Technische Universität Darmstadt\\[1em]
Tag der Einreichung: \submissionDate{}\\
Tag der Disputation: \disputationDate{}\\[1em]
URN: urn:nbn:de:tuda-tuprints-113886\\
DOI: \url{https://doi.org/10.25534/tuprints-00011388}\\
URI: \url{https://tuprints.ulb.tu-darmstadt.de/id/eprint/11388}\\
Jahr der Veröffentlichung der Dissertation auf TUprints: 2020\\[1em]
Texte: \textcopyright{} Copyright 2020 by Florian Benjamin Müller\\
Umschlag: \textcopyright{} Copyright 2020 by Florian Benjamin Müller\\[1em]
Verlag:\\
Florian Benjamin Müller\\
Rhönring 64\\
64289 Darmstadt\\
f\_m@outlook.com\\[1em]
Druck: epubli – ein Service der neopubli GmbH, Berlin

\begin{center}
	
	Published under CC BY-NC-ND 4.0 International.\\
	\url{https://creativecommons.org/licenses/by-nc-nd/4.0}
	
	\begin{figure}[h]
		\centering
		\includegraphics[width=0.3\textwidth]{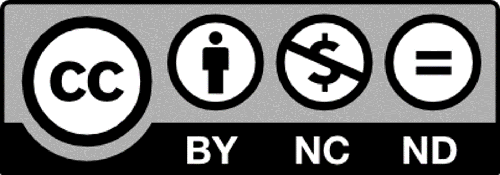}
		\label{fig:by-nd}
	\end{figure}
\end{center}

\cleardoublepage\pdfbookmark[1]{Abstract}{Abstract}
\begingroup
\let\clearpage\relax
\let\cleardoublepage\relax
\let\cleardoublepage\relax

\chapter*{Abstract}
Recent technological advances have made head-mounted displays (HMDs) smaller and untethered, fostering the vision of ubiquitous interaction with information in a digitally augmented physical world. For interacting with such devices, three main types of input - besides not very intuitive finger gestures - have emerged so far: 1) Touch input on the frame of the devices or 2) on accessories (controller) as well as 3) voice input. While these techniques have both advantages and disadvantages depending on the current situation of the user, they largely ignore the skills and dexterity that we show when interacting with the real world: Throughout our lives, we have trained extensively to use our limbs to interact with and manipulate the physical world around us.

This thesis explores how the skills and dexterity of our upper and lower limbs, acquired and trained in interacting with the real world, can be transferred to the interaction with HMDs. Thus, this thesis develops the vision of \aroundbodyinteraction{}, in which we use the space around our body, defined by the reach of our limbs, for fast, accurate, and enjoyable interaction with such devices. This work contributes four interaction techniques, two for the upper limbs and two for the lower limbs: The first contribution shows how the proximity between our head and hand can be used to interact with HMDs. The second contribution extends the interaction with the upper limbs to multiple users and illustrates how the registration of augmented information in the real world can support cooperative use cases. The third contribution shifts the focus to the lower limbs and discusses how foot taps can be leveraged as an input modality for HMDs. The fourth contribution presents how lateral shifts of the walking path can be exploited for mobile and hands-free interaction with HMDs while walking.

\vfill

\begin{otherlanguage}{ngerman}
\pdfbookmark[1]{Zusammenfassung}{Zusammenfassung}
\chapter*{Zusammenfassung}
Die jüngsten technologischen Fortschritte haben Head Mounted Displays (HMDs) kleiner und kabellos gemacht und fördern so die Vision von allgegenwärtiger Interaktion mit Informationen in einer digital erweiterten physikalischen Welt. Zur Interaktion mit solchen Geräten wird bislang Eingabe-seitig – neben wenig intuitiven Fingergesten – vor allem dreierlei verwendet: 1) Touch-Eingabe auf dem Gehäuse der Geräte oder 2) auf Zubehör (Controller) sowie 3) Spracheingabe. Während diese Techniken, abhängig von der aktuellen Situation des Benutzers, sowohl Vor- als auch Nachteile haben, so ignorieren sie weitgehend die Fähigkeiten und Geschicklichkeit, die wir im Umgang mit der realen Welt zeigen: Während unseres ganzen Lebens haben wir ausgiebig trainiert unsere Gliedmaßen zu benutzen, um mit der physischen Welt um uns herum zu interagieren und sie zu manipulieren.

Diese Arbeit untersucht, wie sich diese Fertigkeiten und Geschicklichkeit unserer oberen und unteren Gliedmaßen, die in der Interaktion mit der realen Welt erworben und trainiert wurden, auf die Interaktion mit HMDs übertragen lassen. So entwickelt diese Arbeit die Vision der \emph{Around-Body Interaction}, in der wir den Raum um unseren Körper, definiert durch die Reichweite unserer Gliedmaßen, für eine schnelle, genaue und angenehme Interaktion mit solchen Geräten nutzen. Diese Arbeit trägt vier Interaktionstechniken bei, jeweils zwei für die oberen und zwei für die unteren Gliedmaßen: Der erste Beitrag zeigt, wie der räumliche Abstand zwischen Kopf und Hand genutzt werden kann, um mit HMDs zu interagieren. Der zweite Beitrag erweitert die Interaktion mit den oberen Gliedmaßen auf mehrere Benutzer und veranschaulicht, wie die Registrierung von augmentierten Informationen in der realen Welt kooperative Anwendungsfälle unterstützen kann. Der dritte Beitrag verlagert den Fokus auf die unteren Gliedmaßen und diskutiert, wie Fußberührungen als Eingabemodalität für HMDs genutzt werden können. Der vierte Beitrag stellt vor, wie seitliche Verschiebungen des Gehweges für die mobile und freihändige Interaktion mit HMDs während des Gehens genutzt werden können.
\end{otherlanguage}

\endgroup

\vfill

\cleardoublepage\pdfbookmark[1]{Acknowledgments}{acknowledgments}

\begingroup
\let\clearpage\relax
\let\cleardoublepage\relax
\let\cleardoublepage\relax
\chapter*{Acknowledgments}

They say that writing is a lonely job. After the experience of writing this dissertation, I would like to object to that. Over all these years (three or so\ldots), I have had wonderful support from so many people without whom this work would not have been possible. This is to each and every one of you.

First of all, I would like to thank Max Mühlhäuser (Technische Universität Darmstadt, Germany), my doctoral supervisor and mentor, who over the years has always provided outstanding support, even for the craziest ideas. His undiminished interest in developing new ideas and his childlike joy in doing so are a great inspiration and source of motivation. I am also very grateful to Prof. Albrecht Schmidt (Ludwig-Maximilians-University of Munich, Germany) for acting as a co-referee for my work and contributing many exciting ideas.

Furthermore, I had the pleasure to spend the last years in a fantastic research group. Thanks to all current and former staff of the Telecooperation Lab -- may the TK spirit be with all of you! This applies, of course, especially to \inlinegraphicsame{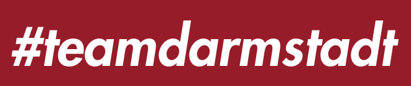} (formally known as the TI or HCI Group). Thank you to Sepp, Martin, Jan, Hesham, Karo, Julius, Markus, Roman, Mo, and Niloo for all the support and your contributions to my work. You've been wonderful, folks.

Likewise, my deepest gratitude goes to all my students who over the years have struggled with the technical shortcomings of several generations of AR glasses from META 1 to Hololens. I would like to especially mention Domi, Daniel as well as Alex\&Daniel. It was wonderful to throw problems at you and to see, how you were able to meet impossible deadlines when solving them. I owe you guys.

Finally, I would like to express my sincere gratitude to my girlfriend Isabell, my parents, Gertrud and Peter, as well as my friends. Without your support and the patience to endure my changing moods in times of insane work, none of this would have been possible. 

Thank You.

\endgroup

\cleardoublepage
\pagestyle{scrheadings}
\pdfbookmark[1]{\contentsname}{tableofcontents}
\setcounter{tocdepth}{2} %
\setcounter{secnumdepth}{3} %
\manualmark
\markboth{\spacedlowsmallcaps{\contentsname}}{\spacedlowsmallcaps{\contentsname}}
\tableofcontents
\automark[section]{chapter}
\renewcommand{\chaptermark}[1]{\markboth{\spacedlowsmallcaps{#1}}{\spacedlowsmallcaps{#1}}}
\renewcommand{\sectionmark}[1]{\markright{\textsc{\thesection}\enspace\spacedlowsmallcaps{#1}}}

\cleardoublepage
\pagestyle{scrheadings}
\pagenumbering{arabic}
\cleardoublepage
\setcounter{ptc}{1}
\ctparttext{\parttoc}
\part{Introduction and Background}\label{pt:introduction}
\setcounter{page}{1}
\chapter[Introduction]{Around-Body Interaction with Head-Mounted Displays}\label{ch:introduction}
{

\citebox{Imagine a time far into the future, when all knowledge about our civilization has been lost. Imagine further, that in the course of planting a garden, a fully stocked computer store from the 1980s was unearthed, and that all of the equipment and software was in working order. Now, based on this find, consider what a physical anthropologist might conclude about the physiology of the humans of our era?}{buxton1986more}

\section{Motivation}

\acresetall

In his 1986 essay, \typcite{buxton1986more} wondered what conclusions a future anthropologist would draw from the current computer technology on the physiology of its users. With a humorous tone, he concluded that \enquote{My best guess is that we would be pictured as having a well-developed eye, a long right arm, uniform-length fingers, and a \enquote{low-fi} ear. But the dominating characteristic would be the prevalence of our visual system over our poorly developed manual dexterity.}

Thirty years have passed since Buxton’s description of the State-Of-The-Art in interacting with information. During the time of his writing, indirect interaction through mediator devices such as mouse and keyboard dominated interaction with computing systems. Since then, more direct multi-touch interaction~\ncite{Shneiderman1982} has found its way into stationary devices such as desktop PCs and has taken the market for mobile devices by storm. 

However, the utilization of our body for interacting with information might be even more limited today than it was when Buxton’s essay was written: While we use all of our fingers to interact with a keyboard, interaction with today’s touch devices is often limited to just the thumb tapping and sliding on the device, while the remaining fingers are degraded to hold and stabilize~\ncite{Xiong2014}. Therefore, such interfaces even further ignore the highly developed skills and dexterity of our body that we show in our daily interaction with the real world, increasing the mismatch between human physical abilities and the design of the computer systems we surround ourselves with. This style of interaction is largely based on the inherent limitations of such devices: Information is visualized as \emph{Pictures Under Glass}~\ncite{victor_2011}, bound to a small 2D surface.%

As a possible solution, the advent of see-through \acp{HMD} and \ac{AR} technologies allows information to \emph{break the glass} and spread into the real world. Such devices consist of a head-mounted combination of two semi-transparent displays (one per eye) to enable stereoscopic output, as well as sensors that allow tracking of the position and orientation of the user’s head~\ncite{Shibata2002}. To allow for natural movement in a (partially) virtual environment~\ncite{Sutherland1968}, such devices use the tracking data to calculate and display a perspectively accurate image. This allows digital information to appear as homogeneous members of reality, thus, enabling and affording a more physical interaction with information.

However, looking at how we interact with today’s \acp{HMD}, we mainly find touch-input 1) on the frame of the \ac{HMD}~\ncite{Islam2018} or on 2) accessories~\ncite{Ashbrook2011} or 3) voice-based input~\ncite{He2018}. Although these interaction styles are often practical and useful, they still largely ignore the degrees of freedom of our body and have various other disadvantages: Touch input on \acp{HMD} and accessories such as the \emph{Hololens Clicker} do not support direct manipulation of content. Voice input is difficult to use in noisy environments and imposes privacy and social acceptability concerns in public areas~\ncite{Starner2002}. Therefore, despite many recent advances that turned \acp{HMD} into \emph{versatile output devices}, there is still a lack of appropriate interaction techniques to transform such devices into \emph{versatile input devices}.

In recent years, more physical styles of interaction with \acp{HMD} and in general have emerged in the field of \emph{body-based} or \emph{body-centric} interaction. As the most prominent example, \emph{on-body interfaces}~\ncite{Harrison2012} gained wide-spread attention. Such on-body interfaces allow (multi-) touch input -- often combined with visual output -- on the surface of our body (e.g., \ncite{Harrison2010}). In most of these systems, the user’s non-dominant hand or arm acts as a two-dimensional interactive surface on which the opposing hand interacts with the content, mimicking the interaction with a hand-held or body-worn touchscreen device. Besides all the advantages, the interaction space is bound to the two-dimensional surface of the body. Moreover, this style of interaction requires both hands and, therefore, hardly supports situations where users’ hands are busy.

\section{Around-Body Interaction}
\label{sec:introduction:aroundbodyinteraction}

In order to overcome the limitations connected with the interaction \emph{on} the body, this thesis focuses on interaction in the space \emph{around} our body using our limbs, leveraging the learned skills and the dexterity that we show while interacting with the real world for interaction with \acp{HMD}.

\typcite{Chen2014} shaped the term of \aroundbodyinteraction{} for interactions that expand \enquote{the input space beyond the device’s screen, [situating] interaction in the space within arm’s reach around the body}. The authors proposed to sense the distance and orientation of a hand-held device relative to the user’s body and presented example applications to use this information for around body interaction techniques. However, the authors only focused on the area around the upper body, ignoring the lower limbs, and mainly proposed interaction techniques tailored explicitly to smartphones. 

This work in the area of \aroundbodyinteraction{} shaped this thesis to a great extent. The central idea of expanding the interaction area to the space \emph{around} the user, leaving behind the limited interaction areas of devices and the surface of the user’s body is a fundamental building block of the work presented here. However, this thesis goes beyond the State-of-the-art and the prior definition of around-body interaction by 1) also exploring the lower limbs for interaction, 2) considering the effects of different forms of visualization, and 3) tailoring the interaction techniques to the requirements of \acp{HMD}.

Therefore, this thesis proposes the following extended definition of \aroundbodyinteraction{}:

\defbox{aroundbody}{Around-Body Interaction}{Around-body interaction leverages the movement of our upper and lower limbs to interact with information in the space around our body, defined by the reach of our limbs.}

Such \aroundbodyinteraction{} techniques stand in a long tradition of works that leverage the degrees of freedom of our limbs for interaction. As the most prominent example for the upper limbs, mid-air gestures~\ncite{Colaco2013} received considerable attention in the field of body-based interaction with \acp{HMD}. Such gestures cover a far greater amount of the possible movements of our limbs, making it possible to use not only the surface of our body but also the area \emph{around} our body for interaction. 

However, despite a variety of examples of such interaction techniques around the body, challenges remain: Most of the presented approaches focus only on the upper limbs neglecting the degrees of freedom of our legs and feet. Thus, such approaches cannot support situations where the user’s hands are not available. In addition, most of these approaches only provide a fixed gesture set, limiting the expressiveness of the interaction. Further, these interaction techniques have always been considered independently, limiting their applicability to specific use cases. 

This thesis, in contrast, contributes to the vision of a concept that supports a variety of interaction techniques that leverage the degrees of freedom of our \emph{upper and lower} limbs for fast and natural interaction with information in the space around our body. Therefore, this work proposes to conceptualize these interactions in a common framework as \aroundbodyinteraction{}s to build a comprehensive concept for interaction with \acp{HMD}.

\subsection{Research Challenges}
\label{chp:introduction:challenges}

To illustrate the vision of ubiquitous interaction with information in a digitally augmented physical world, we consider a day in Alice’s life. In the following chapters, this thesis will revisit Alice’s day to demonstrate the suitability of the presented interaction techniques. 

\interactionbox{lbl}{Introduction}{Alice spends a day in the city. She will go shopping, meet friends, and roam the streets. Throughout the day, she is time and again faced with situations where she needs support from technology in various areas, from communication and navigation to information retrieval and entertainment. The interaction with the information takes place in various situations: While sitting, standing or walking, alone or during (local or remote) conversations with other people and while being hampered in her interaction because she carries her shopping.}

Based on the vision of ubiquitous \aroundbodyinteraction{} in a digitally augmented world, a number of research challenges arise that lead to the contributions of this thesis.

\begin{description}
	\item[Interaction Techniques for Upper and Lower Limbs] Depending on the context of use, one or more body parts and, thus, interaction techniques tailored to these body parts may not be available because of \emph{situational hindrances}. For example, when we carry things in our hands, we cannot use our hands for interaction. Or when we walk, the feet are not available for interaction. Therefore, a single interaction technique with a fixed set of limbs is not capable of supporting interaction in every situation. As a consequence of these situational hindrances, it is necessary to support interaction situations in which upper or lower limbs are not available for interaction. Therefore, interaction techniques for both limb-groups are necessary.
	
	\item[Support for Different Visualizations] Suitable interaction techniques are not only dependent on the limbs used for input, but must also be adapted to the output of the system. Different tracking technologies of \acp{HMD} allow different types of visualization that 1) move as the user moves (\emph{body-stabilized}) or 2) are anchored in the real world (\emph{world-stabilized}) and, thus, allow the user to move independently of the visualization. These different visualization techniques impose different requirements on the interaction with the system and, thus, require specific interaction techniques. 
	
	\item[Support for Interaction Situations] In addition to the general challenges of input and output, there are also special challenges that arise from specific interaction situations:
	
	\begin{description}
		\textfigH{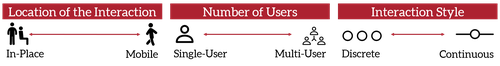}{A design space for \aroundbodyinteraction{} based on the 1) location of the interaction (\inplaceUpper{} or \mobilityUpper{}), 2) number of user (\singleuserUpper{} or \multiuserUpper{}), and 3) interaction style (\discreteUpper{} and \continuousUpper{}).}
		
		\item[\protect\inlinegraphicsame{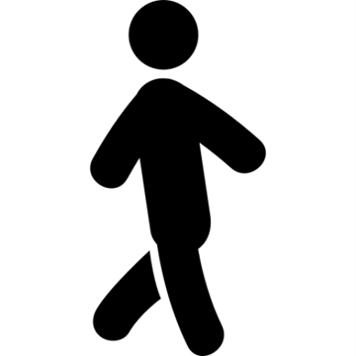} Mobility] \acp{HMD} are inherently mobile devices. Due to their unique placement on the user’s head, such devices will be ultimately - once today’s technical limitations are overcome - always available and can support many situations without having to reach for a device. However, mobile use also poses particular challenges, especially during locomotion, which require interaction techniques to support such situations optimally.
		\item[\protect\inlinegraphicsame{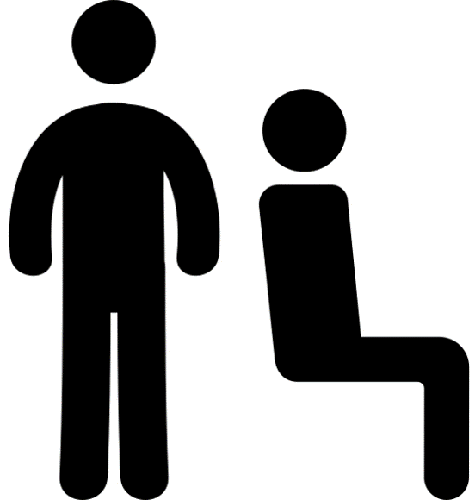} In-Place] 
		Despite the inherent mobility of \acp{HMD}, interaction during stationary periods (e.g., while sitting or standing) will continue to take place in the future. Therefore, interaction techniques are required that support users in such situations.
		\item[\protect\inlinegraphicsame{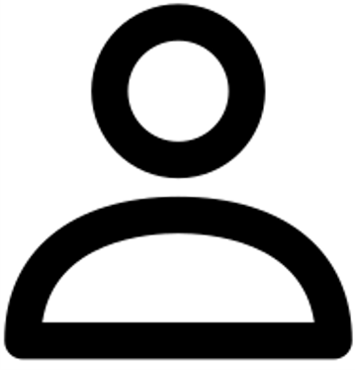} Single-User] \acp{HMD} are radically private devices: Visual output and interaction with information is only available to the wearing user. Therefore, interaction techniques for single users are required.
		\item[\protect\inlinegraphicsame{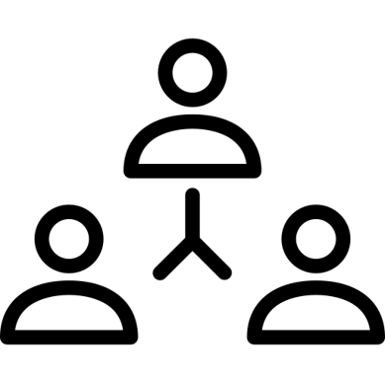} Multi-User] 
		While the focus on individual users given by the shape of \acp{HMD} offers a multitude of advantages (e.g., increased privacy), this also deprives \acp{HMD} of inherent opportunities for collaboration which, due to their physical design, are naturally available on other device classes. Therefore, interaction techniques are necessary that support multi-user collaboration.
		\item[\protect\inlinegraphicsame{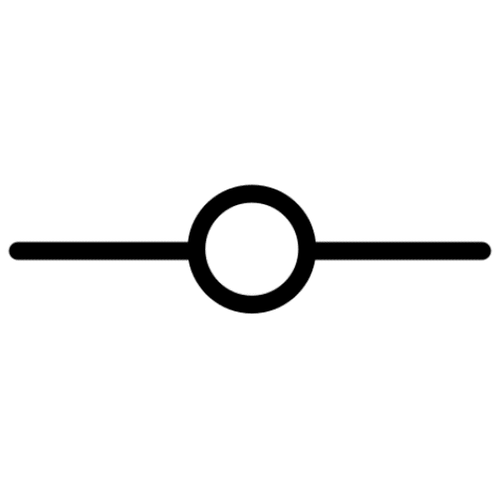} Continuous Interaction] Continuous control like the gradual adjustment of a slider is important for fine-grained manipulation, especially in complex interaction situations. Therefore, interaction techniques supporting continuous interaction are necessary.
		\item[\protect\inlinegraphicsame{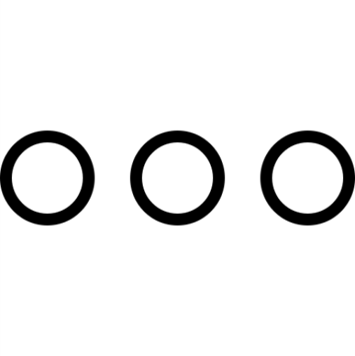} Discrete Interaction] In addition to fine-grained and precise continuous value changes, discrete interactions can offer the possibility for fast and immediate interactions. Such discrete interactions can, for example, support shortcuts to trigger frequently used interactions or allow to traverse cascading menus.
	\end{description}
	
\end{description}

\subsection{Contributions}

\textfigH{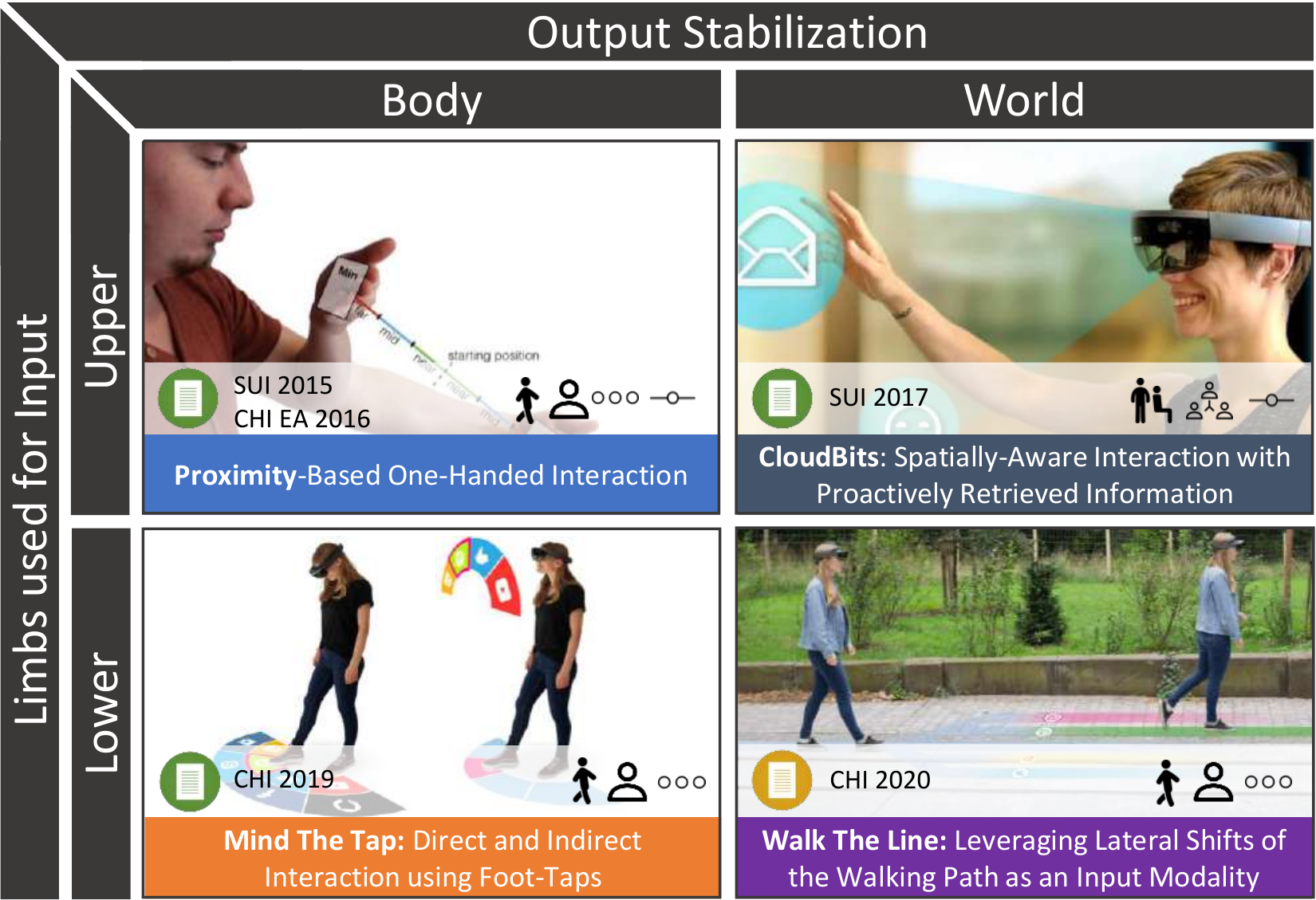}{The main contributions of this thesis, structured by \emph{limb used for input} and \emph{output stabilization}. Further, each contribution focuses on specific interaction situations (\inplaceUpper{}, \mobilityUpper{}, \singleuserUpper{}, \multiuserUpper{}, \continuousUpper{}, \discreteUpper{}). \protect\inlinegraphics{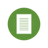} depicts published papers, \protect\inlinegraphics{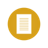} depicts papers currently under submission.}%

Based on the research challenges, this thesis contributes four interaction techniques for \aroundbodyinteraction{}, each focusing on a unique combination of the \emph{limbs used for input} and the \emph{output stabilization} (see section \ref{chp:introduction:challenges}). Within the respective quadrant of the design space, each contribution focuses on a subset of the identified interaction situations (see figure \ref{fig:relatedwork/interaction_situations}). Further, each contribution discusses possible extension points and directions for future work to support further interaction situations. Based on this design space (see figure \ref{fig:introduction/contributions}), the contributions of this thesis are:

The first contribution is \projProximity{} and focuses on interaction using the \emph{upper limbs} with \emph{body-stabilized} interfaces that are augmented to the user’s body. This contribution focuses on how the degree of freedom offered by the elbow joint, i.e., \emph{flexion} by moving the hand towards and \emph{extension} by moving the hand away from the body on the user’s line of sight can be leveraged for interacting with \acp{HMD}. For this, the interaction space in front of the user is divided into multiple parallel planes where each plane corresponds to a layer with visual content. When moving the hand through the interaction space, the visual content is augmented to the user’s palm, allowing the user to browse through successive layers.

The second contribution is \projCloudbits{} and explores interacting using the \emph{upper limbs} with \emph{world-stabilized} interfaces. The contribution focuses on how the world-stabilization of information can be leveraged for new use cases that give meaning to the spatial location of information: Like documents in the real world, users can sort and group information and use the spatial layout to add meta-information to the actual information implicitly. Further, the chapter explores how the context-aware and proactive retrieval of information can support users.

The third contribution, \projCheesyfoot{}, focuses on using the \emph{lower limbs} with \emph{body-stabilized} interfaces, proposing foot-tapping as an input modality for interaction with \acp{HMD}. More precisely, the contribution explores the interaction with a semi-circular grid in the reachability of the user’s feet while standing. Part of this contribution is also the comparison of two different visualization techniques and their influence on the performance of the users: 1) direct interaction with interfaces that are displayed on the floor and require the user to look down to interact and 2) indirect interaction with interfaces that, although operated by the user’s feet, are displayed as a floating window in front of the user.

The fourth contribution, \projCheesyfootToGo{}, focuses on interacting using the \emph{lower limbs} with \emph{world-stabilized} interfaces. The focus of this contribution is on how users can interact with \acp{HMD} while walking without losing the connection to reality and, thus, getting themselves into potentially dangerous situations. Therefore, this thesis investigates the idea of using minimal shifts of the user’s walking path to interact with a visualization augmented to the ground.

\subsection{Publications}

All main contributions of this thesis have been published at international peer-reviewed conferences. This thesis uses parts of the content of the respective publications verbatim.

\projProximity is based on the publications 

\begin{small}
	\begin{addmargin}[25pt]{0pt} 
		\longfullcite{muller2015a}
		
		\longfullcite{muller2016proxiwatch}
	\end{addmargin}
\end{small}

\projCloudbits is based on the publication

\begin{small}
	\begin{addmargin}[25pt]{0pt} 
		\longfullcite{muller2017cloudbits}
	\end{addmargin}
\end{small}

\projCheesyfoot is based on the publication

\begin{small}
	\begin{addmargin}[25pt]{0pt} 
		\longfullcite{Muller2019}
		
	\end{addmargin}
\end{small}

\projCheesyfootToGo is based on the publication

\begin{small}
	\begin{addmargin}[25pt]{0pt} 
		\longfullcite{Muller2020}
	\end{addmargin}
\end{small}

\subsection{Research Methodology}
\label{sec:rw:methodology}

The contributions of this thesis are situated in the field of \acp{HCI}, a field \enquote{concerned with the design, evaluation and implementation of interactive computing systems for human use and with the study of major phenomena surrounding them}~\ncite{Hewett2014}. Each of the main contributions presented in this thesis is substantiated in one or more observational studies or controlled experiments. All of the studies and experiments were designed, conducted, and analyzed according to widely accepted standards of the \ac{HCI} community~\ncite{Lazar2010}. This section describes the general approach to the design and analysis, which applies to all studies presented. Any exceptions are mentioned and justified in the respective chapters.

\subsubsection{Study Design}

In this thesis, two general types of study designs are applied, controlled experiments and observational studies.

\paragraph{Controlled Experiment}

Controlled experiments are used to gain quantitative insights into a domain while excluding external factors as far as possible. These experiments, depending on the experiment design, vary one or more independent variables to assess their influence on measured dependent variables. In order to avoid first order carryover effects (e.g., learning effects between the conditions), the experiments vary the sequence of conditions between participants according to a balanced Latin square as proposed by \typcite{Williams1949}. As dependent variables, performance metrics (e.g., \acf{TCT}, accuracy) accepted and widely used by the \ac{HCI} community were recorded. The respective methodology sections name and define them with regards to the presented experiment. In addition to the performance metrics recorded by tracking the users, the experiment designs use standardized and accepted questionnaires to collect further data:

\begin{description}
	\item[NASA TLX] as proposed by \typcite{Hart1988} to quantify the perceived mental load of participants. For the analysis of the NASA TLX questionnaires, the raw method indicating an overall workload as described by \typcite{Hart2006} is used.
	\item[AttrakDiff] as proposed by~\typcite{Hassenzahl2003} to quantify the participants’ opinions about the user experience of concepts.
	\item[Custom Questionnaires in Likert Scales] as proposed by~\typcite{likert1932technique} to assess participants’ attitudes towards certain aspects of the proposed concepts.
\end{description}

Besides the resulting quantitative results, additional qualitative feedback was collected between or after the respective experiments in semi-structured interviews or focus groups~\ncite{longhurst2003semi} to gain further insights into the user experience of the participants.

\paragraph{Observational Study}

In addition to controlled experiments, this work also uses observational studies where users are observed interacting with systems without actively intervening. This type of study is used in this thesis to collect qualitative findings of the interaction of users with a system. As for the controlled experiments, the observational studies were concluded with semi-structured interviews to collect further data.

\subsubsection{Analysis}

The collected data were carefully analyzed using widely accepted quantitative and qualitative methods. This section describes the methods used for analysis.

\paragraph{Parametric Analysis}

The recorded continuous dependent variables were analyzed using (multi-way) repeated measures analysis of variance (RM ANOVA) to unveil significant effects of the influence of the respective factors. This is an accepted approach to hypothesis testing in frequentist statistics~\ncite{girden1992anova}. Before analyzing, the data were tested for the fulfillment of the assumptions of RM ANOVA using the standard tests~\ncite{Field2003}: First, the data were tested for normality using Shapiro-Wilk’s test. If the test indicated a violation of the assumption of normality, the data were treated as non-parametric. Second, the data were tested for sphericity using Mauchly’s test. If Mauchly’s test indicated a violation of the assumption of sphericity, the degrees of freedom of the RM ANOVA were corrected using the Greenhouse-Geisser method, and this thesis reports the respective \gge. 

When RM ANOVA revealed significant effects, the analysis used paired-samples t-tests for pairwise post-hoc comparisons. To assure the reliability of the results of the t-tests, the results were corrected using the conservative Bonferroni method. 

\paragraph{Non-Parametric Analysis}

For the non-parametric hypothesis testing, the data was analyzed using the test as proposed by \typcite{Friedman1937} (for single-factor designs) or using an Aligned Rank Transformation (ART) followed by a RM ANOVA as proposed by \typcite{Wobbrock2011} (for multi-factorial designs).

When significant effects were revealed, pairwise signed-rank tests as proposed by \typcite{Wilcoxon1945} were performed for post-hoc analysis and, as for the parametric results, corrected using Bonferroni’s method. Wilcoxon's pairwise signed rank test is a non-parametric alternative to the t-test that does not assume a normal distribution.

\paragraph{Qualitative Analysis}

The qualitative feedback in semi-structured interviews and the conversations of participants in observational studies were recorded and transcribed. Afterward, the data were analyzed using an open coding~\ncite{Strauss1998} approach to, in the next step, identify common themes across participants.

\subsubsection{Reporting of the Results}
This thesis reports the eta-squared $\eta^{2}$ as an estimate of the effect size and uses Cohen’s suggestions to classify the effect size~\ncite{Cohen1988}. As an estimate of the influence of the individual factors, the thesis reports the \ac{EMM} as proposed by \typcite{Searle1980}. For the measured raw values as well as for the \acp{EMM}, the thesis reports the mean value $\mu$, the standard deviation $\sigma$ and the standard error $\sigma_{\overline{x}}$.

\section{Structure}

This thesis is structured as follows:

\begin{description}
	\item[Chapter \ref{ch:relatedwork}] discusses definitions and approaches from related works for interaction with \acp{HMD} (section \ref{sec:relatedwork:hmds}), body-based interaction (section \ref{sec:relatedwork:bodybased}) and \aroundbodyinteraction{} (section \ref{sec:rw:aroundbody}) and, further, establishes requirements.
	\item[Chapter \ref{ch:proximity:merged}] presents \projProximity, exploiting the proximity-dimension between the user's hand and head as an input modality for interacting with \acp{HMD}.
	\item[Chapter \ref{ch:cloudbits}] proposes \projCloudbits, leveraging the spatial dimension of world-stabilized output to support collaborative interaction with information in a shared information space.
	\item[Chapter \ref{ch:cheesyfoot}] contributes \projCheesyfoot, exploring direct and indirect interaction with augmented information using foot-taps.
	\item[Chapter \ref{ch:walktheline}] introduces \projCheesyfootToGo, leveraging lateral shifts of the walking path as an input modality for \acp{HMD}.
	\item[Chapter \ref{ch:conclusion}] integrates the previous contributions of this thesis (section \ref{sec:conclusion:summary}) and presents an outlook to future directions for research (section \ref{sec:conclusion:futurework}). 
\end{description}
}
\chapter[Background and Related Work]{Background and Related Work}\label{ch:relatedwork}
{
This chapter presents definitions and background information for the related research areas of 1) \nameref{sec:relatedwork:hmds} (section \ref{sec:relatedwork:hmds}),  2) \nameref{sec:relatedwork:bodybased} (section \ref{sec:relatedwork:bodybased}) and 3) \nameref{sec:rw:aroundbody} (section \ref{sec:rw:aroundbody}). Further, this chapter establishes requirement for the interaction with \acp{HMD} which inform the contributions of this thesis. 

\section{Interacting with Head-Mounted Displays}
\label{sec:relatedwork:hmds}

\acresetall

In the following, this thesis presents the related work in interacting with \acp{HMD}. First, the section gives an overview of the definitions of \acp{HMD}, \ac{AR} and \ac{VR} (section \ref{sec:relatedwork:hmds:def}) and presents a brief historical outline of the technical realization of such devices (section \ref{sec:relatedwork:hmds:implementation}). Second, the section introduces a set of requirements for the interaction with \acp{HMD} with respect to the vision of ubiquitous interaction in a digitally augmented physical world (section \ref{sec:relatedwork:hmds:requirements}). Finally, the section discusses four different streams of research that enable interaction with \acp{HMD} using 1) \nameref{sec:rw:hmd:accessory}, 2) \nameref{sec:rw:hmd:device}, 3) \nameref{sec:rw:hmd:voice} and 4) \nameref{sec:rw:hmd:body} and assesses these research streams with regard to the established requirements.

\subsection{Definitions and Background}
\label{sec:relatedwork:hmds:def}

\citebox{The ultimate display would, of course, be a room within which the computer can control the existence of matter. [...] With appropriate programming such a display could literally be the Wonderland into which Alice walked}{Sutherland1965}

With these words, \typcite{Sutherland1965} described his vision of the \enquote{Ultimate Display}. In his famous article, Sutherland made two fundamental demands to such a display: First, such a display should exert forces that can be perceived and utilized by the user so that \enquote{A chair displayed in such a room would be good enough to sit in}. Second, the display should be a \enquote{kinesthetic display} that can \enquote{make the display presentation depend on where we look}.

Three years later, as a very early step towards such an ultimate display, Sutherland presented what later became known as the \emph{Sword of Damocles}, the first ever \acl{HMD}. \typcite{Sutherland1968} described his \enquote{three-dimensional display} as a device \enquote{to present the user with a perspective image which changes as he moves. [...] The image presented [...] must change in exactly the way that the image of a real object would change for similar motions of the user’s head.}

Much has changed in the fifty years since Sutherland’s \emph{Sword of Damocles}: \acp{HMD} have become smaller and lighter~\ncite{Billinghurst2015}, we have seen the first wave of \ac{VR} applications come and go~\ncite{Cruz-Neira2015}, and with Google Glass and Microsoft Hololens, the first see-through \acp{HMD} have set out to bring mobile \ac{AR} to the end user market. 

Still, many of the concepts present in Sutherland’s glasses can be found in today’s \acp{HMD}: Such devices are \enquote{image display units that are mounted on the head. A unit consists of a helmet and small CRTs or liquid-crystal displays (LCDs) in a pair of goggles.}~\ncite{Shibata2002}. 

\acp{HMD} can be used to add virtual information to the experience of the real world perceived by the user in different levels: Various gradations on the continuum between true reality and complete virtuality are conceivable~\ncite{Milgram1995}. This range between reality and virtuality is referred to as \emph{Mixed Reality}. Yet, no clear definition of mixed reality exists until today~\ncite{Speicher2019}. 

\subsubsection{Virtual and Augmented Reality}

For the scope of this work, we now consider the two extremes of this continuum, \acf{VR}, and \acf{AR}. \typcite{Sherman2002} defined \emph{\ac{VR}} as \enquote{a medium composed of interactive computer simulations that sense the participant’s position and actions and replace or augment the feedback to one or more senses, giving the feeling of being mentally immersed or present in the simulation (a virtual world)}. \typcite{Azuma1997a} defined \ac{AR} as a variation of \ac{VR} that, while \ac{VR} \enquote{completely immerse a user inside a synthetic environment [...] allows the user to see the real world, with virtual objects superimposed upon or composited with the real world. Therefore, AR supplements reality, rather than completely replacing it}. This allows users to see, access, and modify digital information right in front of their eyes anytime, anywhere without losing the connection to the real world. Therefore, \ac{AR} allows the seamless connection of bits and atoms~\ncite{Ishii1997}, representing the digital and physical world.

On the technical side, \ac{VR} glasses use fully opaque displays. Thereby, the user is entirely surrounded by the virtual world; the (visual) reality is suppressed. \ac{AR} systems, in contrast, use partially transparent see-through displays, allowing the user to keep engaging with the surrounding reality. Additional sensing capabilities enable these devices to understand and interpret the environment and merge digital content with the physical world~\ncite{Kress2017}.

Just as \ac{VR} and \ac{AR} share many characteristics, so do the \acp{HMD} built for such applications. Therefore, the requirements for interaction with such devices often show overlaps. Nevertheless, both types have strengths and weaknesses: Due to the increased detachment from reality, \ac{VR} systems show a higher level of immersion~\ncite{Speicher2019}. While the accompanying loss of the (visual) connection to the surroundings is acceptable or even desired in a familiar and static environment like the living room (e.g., for games~\ncite{Pausch1997}), it becomes a problem in unfamiliar or rapidly changing environments like in mobile situations. 

Due to the advantages of \ac{AR} in mobile scenarios and the associated increased practicability, the focus of this work is on \ac{AR} \acp{HMD}. For easier readability, the remainder of this thesis will refer to \ac{AR} \acp{HMD} as \acp{HMD}.

\subsubsection{Output Stabilization, Tracking and Devices}
\label{sec:relatedwork:hmds:implementation}

The visual output of \acp{HMD} can be presented in different ways, depending on which parts of the world they are anchored to. \typcite{Billinghurst1998} defined the possible display methods through stabilization points: 

\begin{description}
	\item[Head-stabilized] interfaces are fixed to the user’s viewpoint, i.e., they provide an interface that moves together with the user and are always displayed at the same position (e.g., the \ac{HUD} interface of Google Glass was intrinsically restricted to this stabilization category).
	\item[Body-stabilized] interfaces are fixed relative to the user’s body position, i.e., they provide an interface that moves together with the user and allows the user to see different parts of the interface through rotating the head.
	\item[World-stabilized] interfaces are fixed to real-world locations, i.e., they provide an interface that stays always at the same physical location with the same orientation.
\end{description}

The possible display methods, namely different stabilization points for information, have increasing technical requirements for the underlying tracking system of the \ac{HMD}: While \emph{head-stabilized} systems do not require any head tracking, \emph{body-stabilized} systems require head-orientation tracking and \emph{world-stabilized} systems require full orientation and position tracking of the user’s head~\ncite{Billinghurst1998}.

Various approaches exist to provide this tracking information of the user's head position and orientation with different advantages and disadvantages depending on the use case. These approaches can be roughly grouped by the hardware used~\ncite{Zhou2008}:

\begin{description}
	\item[Sensor-based approaches] using e.g., gyroscopes, magnetic or mechanical sensors.
	\item[Vision-based approaches] using computer vision techniques on (depth) images.
	\item[Hybrid approaches] using a combination of sensor-based and vision-based approaches.
\end{description}

The tracking technologies can be further grouped by the location of the tracking hardware:

\begin{description}
	\item[Outside-In] systems use tracking integrated into the environment. The system recognizes the headset in the room and calculates estimations for the position and orientation of the device, which is then used to render appropriate visual output~\ncite{Dorfmuller1998}.
	\item[Inside-Out] systems, in contrast, have the necessary sensing hardware integrated into the headset and recognize features in the environment that are, in turn, used to estimate the position and orientation of the device~\ncite{Dorfmuller1999}. Research proposed multiple approaches for inside-out tracking based on (infrared) markers~\ncite{Rekimoto1998} or (color, grayscale or depth) image features~\ncite{Wuest2005}. 
\end{description}

Newer approaches further combine the process of inside-out tracking with the simultaneous creation of a map of the environment, a process known as \ac{SLAM}~\ncite{Izadi2011, Kerl2013, Henry2014}. 

In spite of the great progress achieved in recent years in the field of tracking and registering information in the real world, today's devices still face many problems. While tracking in closed spaces works reasonably well, there is still no conclusion on tracking in large outside and urban areas~\ncite{Pascoal2018}. By design, outside-in tracking systems are unsuitable for such large areas and inside-out systems are (at least today) still overloaded with the size of the area to be tracked. However, this tracking information is essential for a proper registration of information in the real world as discussed above. Furthermore, today's devices still suffer from a low resolution and field of view, as well as large and bulky form factors. However, the development of such devices indicates that these limitations may eventually vanish over time.

\subsection{Requirements}
\label{sec:relatedwork:hmds:requirements}

\def\reqHmdNoDevice{Minimize the Dependence on the Physical Appearance}
\def\reqHmdDirect{Maximize Directness of Interaction}
\def\reqHmdAcceptability{Maximize Social Acceptability}
\def\reqHmdSecondary{Minimize the Dependence on Secondary Control Devices}

In the following, this thesis establishes requirements for the interaction with \acp{HMD} that are later used to classify existing research streams and the interaction concepts used in today’s commercially available \acp{HMD}.

\subsubsection{The Disappearing Tangibility}
\label{sec:relatedwork:hmds:requirements:physicality}

A central difference between \acp{HMD} and other device classes is the increasing disappearance of tangibility. The course of development teaches us that \acp{HMD} are getting smaller and smaller, potentially to the point where they will have largely disappeared as physical devices, existing, for example, as smart contact lenses~\ncite{Conrad2014, TaehoKim2015}. Yet, the physical appearance of devices already gives us an indication of how we can communicate with them. With the disappearance of \acp{HMD} as physical and tangible devices, these communication cues (known as affordances~\ncite{Norman1999}) of \acp{HMD} as physical things are also disappearing and, thus, can no longer be used to encourage interaction (e.g., sliding with a finger alongside the frame of the device for Google Glass). Further, as these devices become smaller and smaller, on-device input might no longer be possible at all.

Therefore, interaction techniques should not rely on the physical appearance of \acp{HMD}. 

\reqbox{hmd:nodevice}{\reqHmdNoDevice}{With \acp{HMD} disappearing as physical devices, on-device input becomes infeasible. Interaction techniques should, therefore, not depend on the physical appearance and tangibility of such devices.}

Not only is the tangibility of the \ac{HMD} as a physical device disappearing, but the displayed information also has no tangibility: Information is represented as images, created in the head of the user, manifested as pixels on displays in front of the eyes. This problem of disembodiment of information is, of course, also present in other device classes: For example, in recent years, touch-based interaction has prevailed in many areas, especially for mobile use, as a central interaction concept. Such devices also display non-physical information. However, there are differences in the types of interaction called for by the visualization: The user interfaces of touch-devices are flat and exist behind a layer of glass, so that touching the display can be perceived as touching the visualization. Using \acp{HMD}, on the other hand, information \emph{breaks the glass} and can be spatially distributed in the space, merged with the real world and appearing as equal members of reality. 

Therefore, information leaves the limited space offered by today’s widespread device classes such as desktop PCs or smartphones, spreading into the physical world. Despite \emph{looking} like an equal member of reality, however, augmented information itself has no physical properties. This prevents the application of the types of interaction we learned from the real world: We mainly touch things to manipulate them. However, since virtual things have no tangibility, these interaction techniques fail with \acp{HMD}. 

As a result, the possibilities of direct interaction with information vanish. This becomes a challenge for interaction with \acp{HMD}, as more direct types of interaction are seen as more \enquote{natural} and \enquote{compelling}~\ncite{Forlines2007}: One of the reasons for the rapid and breakthrough success of smartphones was the radical focus on direct interaction: Such devices omitted physical keyboards and trackballs and relied purely on direct interaction via the touchscreen~\ncite{west2010browsing}.

As a possible solution, interfaces can register virtual information to physical objects in the real world, giving them a (proxy-) body to allow users to directly interact with the information. By design, this solution to the missing tangibility of digital information is only available for \emph{world-stabilized} interfaces and requires additional tracking of the (proxy-) objects. In addition, this solution restricts the mobility of users, since a suitable object must be available at the place of interaction.

As this solution is only feasible for a limited number of interaction situations and, further, only with \emph{world-stabilized} interfaces, other approaches are needed. Therefore, this thesis focuses on other approaches to compensate for the disappearing tangibility of the presented information and the resulting decrease of direct interaction possibilities.

\reqbox{hmd:direct}{\reqHmdDirect}{Interaction techniques for \acp{HMD} should maximize the directness of the interaction (e.g., by providing means for direct interaction) to compensate for the missing tangibility of the displayed information.}

\subsubsection{Head-mounted Displays in a mobile context}

Since Sutherland’s \emph{Sword of Damocles} in 1968, \acp{HMD} have become smaller~\ncite{Billinghurst2015}, untethered~\ncite{Feiner1997} and, thus, more mobile. This increased mobility poses further challenges for the interaction with HMDs to support mobile situations.

Throughout the day, we find ourselves in different spatial and social contexts: We are standing in crowded subways or wandering alone across a wide field, participating in large meetings or intimate conversations. These changes in context lead to requirements for interaction techniques to be suitable for such situations:

First, depending on the context, interaction techniques can be perceived as socially inappropriate from the outside or as embarrassing by the user himself because it might affect the social image, especially for new and unknown classes of devices~\ncite{Koelle2017}. For example, speaking to technology in public environments~\ncite{Efthymiou2016} or touching the groin area~\ncite{Harrison2014} of the body are perceived as inappropriate.

Therefore, interaction techniques should focus on socially acceptable ways of interaction.

\reqbox{hmd:appropriate}{\reqHmdAcceptability}{Interaction techniques for \acp{HMD} should be socially accepted in a variety of contexts.}

Second, mobility leads to frequent changes in location. Each of these changes of location entails the chance to forget things, leave things behind, or lose them. Alongside the general problems that arise as a result, this is particularly critical for interaction with \acp{HMD} when the interaction is bound to a secondary device. 

Therefore, interaction techniques should work autonomously and not depend on secondary devices.

\reqbox{hmd:nosecondary}{\reqHmdSecondary}{Interaction techniques for \acp{HMD} should not depend on secondary devices as these might be misplaced or lost.}

\subsection{Related Work}

In the following, this section discusses the four major streams of research in the field of interacting with \acp{HMD} with regard the the established requirements. This includes work in the fields of 1) \nameref{sec:rw:hmd:accessory}, 2) \nameref{sec:rw:hmd:device}, 3) \nameref{sec:rw:hmd:voice}, and 4) \nameref{sec:rw:hmd:body}.

\subsubsection{Accessory-Based Interaction}
\label{sec:rw:hmd:accessory}

As discussed in the context of \refreq{req:hmd:direct}, the design of head-mounted displays renders traditional touch-based interaction techniques unusable as touch on the display itself is not feasible. Research, as well as manufacturers of \acp{HMD}, tried to transfer touch-based interaction to this new class of devices in different ways. Interaction techniques for Sony’s and Epson’s \acp{HMD}, SmartEyeglasses and Moverio, are built around a wired handset for indirect pointer-based interaction, imitating the mouse interaction known from desktop computers. Research has also provided further possibilities to interact with such devices using physical accessories: As prominent examples, \typcite{Ashbrook2011} presented an interactive ring and \typcite{Dobbelstein2015} proposed an interactive belt for unobtrusive touch input. Other proposed accessories include augmentations to the user’s pocket~\ncite{Dobbelstein2017} or sleeves~\ncite{Schneegass2016}. 

Such accessory interfaces are not bound to the physical appearance of one specific device and are considered socially acceptable~\ncite{Tung2015}. However, such interfaces do not provide means for direct interaction~\ncite{Hsieh2016} and, further, can be misplaced and lost.

\subsubsection{On-Device Interaction}
\label{sec:rw:hmd:device}

As another approach, Google used \emph{on-device} input on a one-dimensional touch-pad for interacting with its \ac{HMD}, Google Glass. Research added use cases for this style of interaction. For example, \typcite{Islam2018} proposed tapping gestures for authentication on the frame of the device. Other examples for on-device tapping and sliding techniques include games~\ncite{Hsu2014} or text entry~\ncite{Grossman2015}.

Such on-device interaction techniques show a high social acceptability~\ncite{Alallah2018} and do not require secondary devices. However, on-device input is based on specific physical properties of the devices and does not provide means for direct interaction. Further, user tests showed little enthusiasm for this type of input for \acp{HMD}~\ncite{Tung2015}.

\subsubsection{Voice-based Interaction}
\label{sec:rw:hmd:voice}

Speech is a central component of human-to-human interaction~\ncite{Kohler2017}: It is always available and provides a natural way of transporting information. Recent advances in speech recognition~\ncite{Hinton2012} and natural language processing~\ncite{manning2014stanford} have led to systems that are suitable for everyday use and can be deployed to mobile devices with limited resources, even for offline use~\ncite{He2018}. Taking together the naturalness of input and the technical feasibility of such interfaces, speech-based input appears to be a compelling way to interact with \acp{HMD}. The industry seems to agree: From Google Glass and Microsoft Hololens to Magic Leap: Voice input is one of the fundamental interaction concepts of these devices. This widespread use of voice-based interaction techniques is based on a number of strong advantages of such interfaces: Naturally, language has no inherent connection to a physical device and representation.

However, voice-based interfaces show problems in many areas: Users might have problems to build a mental model of the system, resulting in systems failing to \enquote{bridge the gap between user expectation and system operation}~\ncite{Luger2016}. Further, language can only describe, not directly manipulate and, thus, excludes the possibility of direct interaction with content~\ncite{Frohlich1993}. \typcite{Shneiderman2000} depicts further problems of voice-based interfaces: \enquote{Speech is slow for presenting information, is transient and therefore difficult to review or edit, and interferes significantly with other cognitive tasks.}

In particular, voice-based interfaces impose problems for mobile use with \acp{HMD}: In the mobile context, other people may be nearby while interacting with the device, raising the question of social acceptance and privacy concerns~\ncite{EaswaraMoorthy2015}. Referring to these situations, \typcite{Alallah2018} compared different input modalities for \acp{HMD} regarding their social acceptability and found the lowest approval rates for voice input compared to other input modalities. In addition, the recognition quality of speech input depends on background noise, rendering such techniques difficult to use in urban environments~\ncite{Starner2002}.

\subsubsection{Body-based Interaction}
\label{sec:rw:hmd:body}

With the increasing proliferation of (low-cost) sensor hardware and advances in computer vision, research began to incorporate the human body as an input modality for interacting with \acp{HMD}. Such \emph{body-based} interaction techniques leverage movements of (parts of) our body as an input modality. The possibilities of such body-based interactions are manifold, reaching from touch input on the surface of our body~\ncite{Harrison2010} to gesture-based interfaces~\ncite{Colaco2013} and eye-based gaze interaction~\ncite{Piumsomboon2017}.

While there are great differences depending on the interaction technique, these techniques nevertheless share vast similarities: The interaction does not rely on the physical properties of the \acp{HMD} nor a secondary device. Further, research showed the social acceptability of such interfaces~\ncite{Hsieh2016} and provided examples for direct interaction~\ncite{Harrison2011a}.

\subsection{Conclusion}

\renewcommand*\theadfont{\bfseries}
\settowidth\rotheadsize{\theadfont Interaction Situation 1: Interaction}
\renewcommand\theadgape{}
\renewcommand\theadalign{lc}
\renewcommand\rotheadgape{}
\begin{table}[h!]
	\centering
	\begin{tabular}{m{5cm}ccccc}
		& \rothead{\refreq{req:hmd:nodevice}: \reqHmdNoDevice} & \rothead{\refreq{req:hmd:direct}: \reqHmdDirect} & \rothead{\refreq{req:hmd:appropriate}: \reqHmdAcceptability} & \rothead{\refreq{req:hmd:nosecondary}: \reqHmdSecondary} \\
		\midrule
		On-Device Interaction & \reqNo & \reqNo & \reqYes & \reqYes \\
		Accessories Interaction & \reqYes & \reqNo & \reqYes & \reqNo \\
		Voice-Based Interaction & \reqYes & \reqNo & \reqNo & \reqYes \\
		Body-Based Interaction & \reqYes & \reqYes & \reqYes & \reqYes \\
		\bottomrule
	\end{tabular}
	\caption{Comparison of the different approaches to interaction with \acp{HMD} with respect to the identified requirements. \reqYes indicates whether interaction techniques in this group can potentially fulfill this requirement.}
	\label{tab:req:hmd}
\end{table}

Comparing the basic interaction concepts, body-based techniques show the strongest suitability with respect to the previously established requirements (see table \ref{tab:req:hmd}). In the following, this thesis will, therefore, focus on body-based interaction techniques for \acp{HMD}.

\section{Body-based Interaction}
\label{sec:relatedwork:bodybased}

This section presents the related work in body-based interaction techniques. While many of the presented approaches are not directly tailored to the interaction with \acp{HMD}, the concepts can be transferred to this class of devices as \acp{HMD} can mimic the visual output of other systems (e.g., by virtually projecting visual output). Therefore, this section does not distinguish between interaction techniques for \acp{HMD} and other device classes.

First, this section gives an overview of the definition and history of body-based interaction (section \ref{sec:relatedwork:body:def}). Second, the section introduces a set of requirements for the body-based interaction with \acp{HMD} with respect to the vision of ubiquitous interaction in a digitally augmented physical world (section \ref{sec:relatedwork:body:requirements}). Finally, the section discusses three different streams of research on body-based interaction with \acp{HMD}: 1) \nameref{sec:rw:body:onbody}, 2) \nameref{sec:rw:body:headgaze}, and 3) \nameref{sec:rw:body:whole} (section \ref{sec:rw:body}).

\subsection{Definitions and Background}
\label{sec:relatedwork:body:def}

With the increasing proliferation of (low-cost) sensor hardware and advances in computer vision, research began to incorporate the human body as an input modality. These works belong to the area of \emph{body-based} (also \emph{body-centric}) interaction. Although not a new phenomenon, body-based interaction was first conceptualized and used as a term in a work in the field of input control for games in 2009. \typcite{Silva2009} described an approach which \enquote{we call body-based interaction [...]. The use of real-world movements to control virtual actions can make use of players’ physical skills from previous experiences with the world (e.g., muscle memory). Body movements allow players to utilize input modalities beyond the hand and fingers. [..] In VR, body-based interaction has promoted benefits such as better spatial understanding, and higher sense of presence.}

During the first wave of \ac{VR} in the mid-90s, such systems typically used expensive and complicated hardware setups for human motion tracking, e.g., the A/C magnetic sensors used by the CAVE project~\ncite{Cruz-Neira1992}. In the following years, research moved towards human motion recognition using only visual information using color and edge feature detection on 2D images~\ncite{Wang2018}. With the availability of inexpensive integrated color and depth cameras (RGB-D cameras, e.g., Microsoft Kinect), the focus of research has shifted to the utilization of such devices~\ncite{Zhang2012}. The progress in computer vision algorithms led to the possibility of body joint estimation~\ncite{Shotton2011} and was further improved by fitting it into a model of human kinesthetics, excluding unlikely positions and movements~\ncite{Corazza2006}, to allow for novel input techniques as discussed below. In recent years, progress in the field of computer vision and machine learning techniques~\ncite{Omran2018} has led to a decline in interest to the analysis of depth images. Instead, the focus again moved to the mere analysis of RGB images.

Many body-based interfaces are built upon the sense of proprioception: This proprioception, the \enquote{sensation of body position and movement}~\ncite{Tuthill2018}, gives us an innate understanding of the relative position and orientation of our body parts to each other. We can move our limbs without visual attention. This sense allows building user interfaces that minimize the interference with the user's interaction with the real world, as the (visual) attention is not entirely captured by the interaction. 

This sense allows for interactions without explicit feedback: \typcite{Lopes2015} leveraged this sense of proprioception and presented eyes-free interaction techniques, shaping the term \emph{Proprioceptive Interaction}. Subsequent research showed different use cases for such interactions, from controlling a music player~\ncite{Lissermann2013} or a TV set~\ncite{Dezfuli2012} to operating a phone~\ncite{Gustafson2011a}.

\subsection{Requirements}
\label{sec:relatedwork:body:requirements}

The following subsection establishes requirements for body-based interaction with \acp{HMD} with respect to the vision of ubiquitous interaction in a digitally augmented physical world, which are then used to classify existing research streams and the interaction concepts used in today’s commercially available \acp{HMD}.

\def\reqBlocking{Minimize Blocking of Body Parts}
\def\reqNatural{Minimize unnatural movements}
\def\reqInputSpace{Support for 3D-Interaction}
\def\reqExpressiveness{Maximize the expressiveness of Input}

\subsubsection{The Tension Between Interaction and Reality}

We use our body and especially our limbs to interact with the world: Our hands grasp and carry things, we walk and pedal with our feet, and we use our head to look around and point at things. Adding the ubiquitous interaction with an computing system imposes additional tasks on our body, entailing additional burdens. 

This becomes a particular challenge when an interaction with the \acp{HMD} coincides with an interaction with the real world. Any part of the body involved in an interaction with the \acp{HMD} is no longer (or, at least, no longer entirely) available for interaction with the real world. 

Therefore, interaction techniques should minimize the blocking of individual body parts.

\reqbox{body:blocking}{\reqBlocking}{Interaction techniques should require as few body parts as possible at the same time in order not to complicate regular interaction with the real world.}

We perform the above interactions with the world in very specific ways that respect the characteristics of our body. This is also required for interaction with HMDs: For natural interaction, it is essential not to force users to make uncomfortable and unnatural movements.

Therefore, interaction techniques should accept the natural boundaries of the human locomotor system.

\reqbox{body:natural}{\reqNatural}{Interaction techniques should not require unnatural and uncomfortable movements.}

\subsubsection{Interaction after Breaking the Glass}

As discussed in section \ref{sec:relatedwork:hmds:requirements:physicality}, \acp{HMD} allow information to leave the limited display area of today’s devices and spread throughout the real world, adding a third dimension. This dispersion of the output in the world also requires appropriate interaction techniques. 

Therefore, in particular, interaction techniques are required which support three-dimensional information spaces.

\reqbox{body:inputspace}{\reqInputSpace}{Interaction techniques should leverage the entire interaction space available, which is created by the output in the three-dimensional space.}

Today’s smartphones have developed into mobile all-rounders: We use them - just as examples - for communication, navigation, entertainment, and information retrieval. \acp{HMD} meet the technical requirements to perform all of these tasks in possibly exciting new ways due to the additional output capabilities. 

However, communicating these tasks as well as the necessary information to complete the tasks to the system can require a variety of necessary interaction steps. 

Therefore, expressive interaction techniques are required.

\reqbox{body:expressiveness}{\reqExpressiveness}{Interaction techniques should provide a high level of expressiveness to transfer a maximum of information in a minimum of time.}

\subsection{Related Work}
\label{sec:rw:body}

In the following, this section discusses three different streams of research for body-based interaction techniques for \acp{HMD} using 1) \nameref{sec:rw:body:onbody}, 2) \nameref{sec:rw:body:headgaze}, and 3) \nameref{sec:rw:body:whole}.

\subsubsection{On-Body Interaction}
\label{sec:rw:body:onbody}

Recent advances in input and output technology have led to the emergence of so-called \emph{on-body}~\ncite{Harrison2012} interfaces, leveraging the human body as an interactive surface for both, input and output. In these systems the input is performed by touch-based interactions on projected~\ncite{Winkler2014, Mistry2009, Wilson2010}, augmented~\ncite{Ha2014} or imaginary~\ncite{Dezfuli2012, Gustafson2013, Oh2014} user interfaces.

Harrison et al. paved the way for such on-body interfaces in their two groundbreaking works, Skinput~\ncite{Harrison2010}, and Omnitouch~\ncite{Harrison2011a}. The papers presented approaches for the acoustic and, respectively, optical localization of touch events on the arm and hand of the user together with projected visual output on these body parts. Focussing again on the topic of on-body interfaces, \typcite{Harrison2014} further explored the implications of different locations for input and output on the human body. Building on this work, \typcite{Weigel2014} increased the input space for on-body interactions by analyzing how additional input modalities, such as pulling, pressing, and squeezing, can be used for more expressive interactions. \typcite{Mehta2016} added itching and scratching to the input space. \typcite{Bostan2017} contributed a user-elicitation study, collecting and comparing on-body gestures.

The vast majority of the work concentrated on the hands and forearms of the users due to the anatomically easy accessibility and because they are often unclothed and socially acceptable to touch~\ncite{Wagner2013}. Yet, other parts of the body, such as the abdomen~\ncite{Vo2014} or the ear~\ncite{Lissermann2013}, were also examined for input.

\paragraph{Further Sensing Techniques}

Beyond the approaches presented by Harrison et al. and adopted by the other works cited, research proposed other sensing techniques. For example, \typcite{Saponas2009b} proposed to sense touch through sensing human muscle activity. Further, \typcite{Wang2016} showed how to reconstruct finger movement on the body using the skin stretch, and \typcite{Matthies2015} analyzed how unique electric signatures of different body parts can be used to localized touch events on any part of the body.

\paragraph{Accessories}

With the packaging of sensor techniques into body-worn accessories, research developed several approaches for supporting on-body interaction on the move while, at the same time, realizing Mark Weisser’s demand for disappearing technologies~\ncite{weiser1991computer}. These accessories range from attachable skin buttons~\ncite{Laput2014} and clothing~\ncite{Heller2014, Ueda2018}, to belts~\ncite{Dobbelstein2015}, wristbands~\ncite{Dobbelstein2018} and rings~\ncite{Ashbrook2011}.

\paragraph{Body Parts as Public Displays}

Visual output on body parts can also be interpreted as a visualization on a public display~\ncite{Olberding2013}. \typcite{Hoang2018} used such on-body visual output to use the human body as a canvas to communicate information. Such visual output can also reflect the use of tattoos as a means of communicating information intended for the public. This resulted in a stream of research focusing on integrating sensing technology into tattoo-styled skin-worn accessories ~ \ncite{Weigel2015, Lo2016, Kao2016}. \typcite{Strohmeier2016} analyzed the tension between the input on and with one’s own body, perceived as something very personal, and the perception of the skin as a public display that is also (passively) accessible to other people which must be considered in the design of such systems. 

\paragraph{Conclusion}

Such on-body interfaces have shown considerable advantages over other approaches for the mobile use of computing systems (e.g., during sports~\ncite{Hamdan2017, Vechev2018}): No second device is necessary for interaction, which can be forgotten or lost. In addition, the limited interaction surface of today’s mobile devices - tablets, smartphones, or smartwatches - is replaced with the much larger surface of the user’s body. Leveraging the sense of proprioception, such interfaces can be even operated without visual attention (see section \ref{sec:relatedwork:body:def}).

However, a closer look reveals challenges which remain unsolved, especially in the context of mobile use with \acp{HMD}. In most of these systems, the user’s non-dominant hand acts as a two-dimensional interactive surface on which the opposing hand interacts with content through (multi)-touch gestures. While useful and practical, the interaction space is bound to the two-dimensional surface of the hand, imitating the interaction with a hand-held mobile device and ignoring the benefits of three-dimensional output and input. Moreover, this style of interaction requires both hands and, therefore, hardly supports situations, where users are encumbered.

\subsubsection{Head- and Gaze-based Interaction}
\label{sec:rw:body:headgaze}

As another approach to body-based interaction, research proposed the usage of gaze-input as a hands-free input modality for \acp{HMD}~\ncite{Lukander2013}. Such systems face the challenge of distinguishing between intentional input and regular body movements, known as the \emph{Midas Touch Problem}~\ncite{Jacob1995}. To overcome this challenge, research proposed to exploit dwell time~\ncite{Wobbrock2008}, active pupil size manipulation~\ncite{Ekman2008} or gaze gestures~\ncite{Rantala2015}. 

However, introducing dwell times increases interaction times and gaze gestures reduce the general applicability of the approaches. In addition, research showed that gaze interactions feel unnatural since the natural task of the eyes is to capture sensory information, and it is fatiguing to use the gaze for continuous manipulation tasks~\ncite{Zhai1999}, decreasing the expressiveness of gaze~\ncite{Chatterjee2015}. Also, today’s sensors require time-consuming calibration steps and suffer from decreasing accuracy because of movements of the glasses ~\ncite{Kyto2018}.

Recently, also head-pointing~\ncite{Morris2000} has emerged as a promising interaction techniques for \acp{HMD} in both industry (e.g., Microsoft Hololens) and research~\ncite{Clifford2017}. 

However, research showed that, while being fast and precise~\ncite{Qian2017}, head-pointing can require unnatural movements and, further, cause fatigue~\ncite{Kyto2018}.

\subsubsection{Whole-body and Embodied Interaction}
\label{sec:rw:body:whole}

As another approach, research proposed to harness all available information about the current state of the user for \emph{Whole-Body Interaction}. \typcite{England2009} defined \emph{Whole-Body Interaction} as \enquote{The integrated capture and processing of human signals from physical, physiological, cognitive and emotional sources to generate feedback to those sources for interaction in a digital environment}~\ncite{England2009}. In a later publication, \typcite{England2011} further added that \enquote{The key word here is integration [...] to use two or more [...] input categories in combination so that we can [...] get a richer picture of what the user intends when they move their body [...]}.

The central idea behind \emph{Whole-Body} interaction is to include as much information as possible about the user in the interpretation of actions. This does not only involve physical movements but also - with references to the field of affective computing~\ncite{Picard1997} - the emotional state of the user.

As a highly related field of research, \emph{Embodied Interaction} has attracted much attention in recent years. In embodied interaction, the concept of embodiment has not only references to the involvement of the body but stands in the philosophical tradition of phenomenology as proposed by \typcite{Husserl1913} and \typcite{heidegger1977sein}. \typcite{Dourish2004} defined such embodied interactions as \enquote{the creation, manipulation, and sharing of meaning through engaged interaction with artifacts} and further explained that \enquote{When I talk of \emph{embodied interaction}, I mean that interaction is an embodied phenomenon. It happens in the world, and that world (a physical world and a social world) lends form, substance, and meaning to the interaction}~\ncite{Dourish1999}. %

Having all this physical, physiological, and cognitive information at hand, research applied the concept of whole-body and embodied interaction to a plethora of application areas. \typcite{Maes1997} showed how whole-body interaction could be used to interact with agents in a virtual world. \typcite{Walter2015} proposed techniques for interacting with public displays. \typcite{Price2016} and \typcite{Freeman2013} explored whole-body interaction to engage visitors in museums. \typcite{Cafaro2012} proposed allegories to encourage data exploration using whole-body interaction. \typcite{Dezfuli2012} used implicit and pose-based embodied interactions to control a TV set. Further examples of whole-body and embodied interaction include virtual training~\ncite{Reidsma2011}, assisted living~\ncite{Altakrouri2014}, support for elderly~\ncite{Ferron2019}, interaction with virtual environments~\ncite{Hasenfratz2004}, supporting collaboration~\ncite{Malinverni2015}, interaction with smartphones~\ncite{Khalilbeigi2011} or human-robot interaction~\ncite{Peternel2013}.  More abstracted from specific application domains, \typcite{Fogtmann2008} described how whole-body interaction could serve as a foundation for designing interactive systems.

Further, whole-body interfaces received particular attention in the area of games and gamified~\ncite{Deterding2011} systems. \typcite{Gerling2012} used whole-body and motion-based game controls for the elderly and found that such controls \enquote{can accommodate a variety of user abilities, have a positive effect on mood and, by extension, the emotional well-being of older adults}. Following this thread of research, \typcite{schonauer2015full} showed how whole-body interaction with games can help patients in motor rehabilitation following, e.g., strokes or traumatic brain injuries. \typcite{Bianchi-Berthouze2007} analyzed the effect of whole-body interaction on the engagement levels of players of games and found a significant increase in engagement compared to controller-based input. %

By including such a large amount of information about the user and his context in the evaluation and interpretation of interactions, both positive and negative effects emerge. Such interactions can communicate a large amount of information with small interactions and, thus, show a high expressiveness. Further, the interactions are not limited to a specific area and fully support interaction in the 3D space.

However, monitoring such a multitude of information (not only physiological but also cognitive) also requires a large amount of sensors and, thus, poses technical challenges that are not yet been finally solved. Further, the existing solutions are always tailored to a specific application, a general interaction framework for use in all situations such as WIMP for the desktop~\ncite{VanDam1997} is not yet in sight. This lack of a general basis of interaction primitives can lead to overstraining users, who have to remember a different style of interaction for each device or application. Additionally, since whole-body interaction considers the entire body, such interaction techniques can hardly support situations in which body-parts are not available, for example, because of carrying something in hand.

\subsection{Conclusion}

\renewcommand*\theadfont{\bfseries}
\settowidth\rotheadsize{\theadfont Interaction Situation 1: Interaction}
\renewcommand\theadgape{}
\renewcommand\theadalign{lc}
\renewcommand\rotheadgape{}
\begin{table}[h!]
	\centering
	\begin{tabular}{m{5cm}ccccc}
		& \rothead{\refreq{req:body:blocking}: \reqBlocking} & \rothead{\refreq{req:body:natural}: \reqNatural} & \rothead{\refreq{req:body:inputspace}: \reqInputSpace} & \rothead{\refreq{req:body:expressiveness}: \reqExpressiveness} \\
		\midrule
		On-Body Interaction & \reqNo & \reqYes & \reqNo & \reqYes \\
		Head/Gaze-based Interaction & \reqYes & \reqNo & \reqNo & \reqNo \\
		Whole-Body Interaction & \reqNo & \reqYes & \reqYes & \reqYes \\
		\bottomrule
	\end{tabular}
	\caption{Comparison of the different approaches to body-based interaction with \acp{HMD} with respect to the identified requirements. \reqYes indicates whether interaction techniques in this group can potentially fulfill this requirement.}
	\label{tab:req:body}
\end{table}

Comparing the basic concepts for body-based interaction with \acp{HMD} to the established requirements, none of the presented approaches can fulfill all of the requirements \refreq{req:body:blocking} - \refreq{req:body:expressiveness} (see table \ref{tab:req:body}). The following section introduces \aroundbodyinteraction{} as an alternative interaction concept for \acp{HMD} and demonstrates that this type of interaction can fulfill all of the requirements.

\section{Around-Body Interaction}
\label{sec:rw:aroundbody}

This section presents definitions and background (section \ref{sec:rw:aroundbody:definitions}) relevant to the topic of \aroundbodyinteraction{} and proposes a classification of such interfaces based on the limb used for input and the stabilization of the output (section \ref{sec:rw:aroundbody:classification}).

\subsection{Definitions and Background}
\label{sec:rw:aroundbody:definitions}

The use of the degrees of freedom offered by our upper and lower limbs for interacting with computer systems and \acp{HMD} is not new; there exists already a variety of interaction techniques for \acp{HMD} that leverage the space \emph{around} the user for interaction. However, despite a multitude of examples of such \aroundbodyinteraction{} techniques, these interaction techniques were only considered separately and not as a common concept. Therefore, this work proposes to conceptualize these interactions in the common framework of \aroundbodyinteraction{} to build a comprehensive concept for interaction with \acp{HMD}.

The following section as well as the remainder of this thesis uses the extended definition of \aroundbodyinteraction{} introduced in section \ref{sec:introduction:aroundbodyinteraction}, which goes beyond the initial definition proposed by~\typcite{Chen2014}. However, it is not possible to separate \aroundbodyinteraction{} sharply from other body-based interaction techniques presented in section \ref{sec:rw:body:onbody}, there are overlaps in these areas. For example, a touch of a body part, depending on the point of view, can be considered as on-body interaction as well as \aroundbodyinteraction{}. The difference lies in the focus of the view: While in on-body interaction the touch itself is the focus of the interpretation, in \aroundbodyinteraction{} it is the movement itself. 

\subsubsection{Limb Movement}

\textfigH{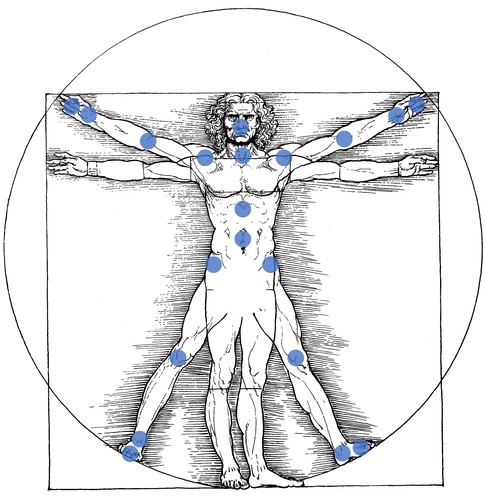}{The skeleton joint model used by Microsoft Kinect.}

The possible movements of the human body result from the degrees of freedom of the connecting points of the bones, called joints~\ncite{Whiting2018}. Not all of these joints are movable, and some are only movable to a very limited extent~\ncite{Archer2003}. Therefore, to assess the possible movements of the human body, a simplified model of the human skeleton is typically used in the body of related works, considering the 15-25 joints most important to human motion.~\ncite{Lee1985, Rocha2015}. 

Looking at these models, the human limbs stand out in particular. For example, the skeleton tracking by the Microsoft Kinect uses a total of 20 body joints (see figure \ref{fig:introduction/skeleton}) with 16 of these joints belonging to the users' limbs. This is also reflected in our daily experience: The interactions with the world in everyday life are mainly driven by changes in the degrees of freedom of the joints of the limbs.

\subsubsection{Gesture-based Interaction}

\Aroundbodyinteraction{} relies on leveraging limb movements to communicate information to the \acp{HMD}. The information is, thus, encoded in the movement of the limbs, a principle known \emph{gestures}. More precisely, \typcite{Kurtenbach1990} defined gestures as \enquote{motion of the body that conveys information. Waving goodbye is a gesture. Pressing a key on a keyboard is not a gesture because the motion of a finger on its way to hitting a key is neither observed nor significant. All that matters is which key was pressed.} \typcite{Billinghurst2018} added that \enquote{this is true regardless of the gesture that was used to push the key. It could have been pushed lovingly or in anger. Either could be easily sensed by an observer. But both are irrelevant to the computer, which only cares about what key was pushed when.}

There is a wide range of different classifications of gestures depending on the context of usage. Therefore, this section cannot be a conclusive discussion of gestures, but merely presents the definitions relevant to this thesis.

\typcite{cadoz2000gesture} proposed a classification of such gestures based on their relationship with the environment as \emph{ergotic} (modify the environment), \emph{epistemic} (receive knowledge from the environment) and \emph{semiotic} (transmit information to the environment). For this thesis, the classification of gestures as a means of transmitting information to the environment (semiotic) is most relevant: If in this classification, the \ac{HMD} is defined as the relevant environment, then this classification quite precisely describes the core idea of \aroundbodyinteraction{}: Using movements of the limbs to transmit information to the \ac{HMD}. 

The temporal aspect can also be included as a further level of classification. However, the distinction is not concerned with the total duration of the interaction, but whether information a) is transmitted by the execution and completion (discrete gesture) or b) is transmitted continuously during the execution (continuous gesture)~\ncite{buxton2007multi}.

\subsection{Classification of Around-Body Interfaces}
\label{sec:rw:aroundbody:classification}

\textfigH{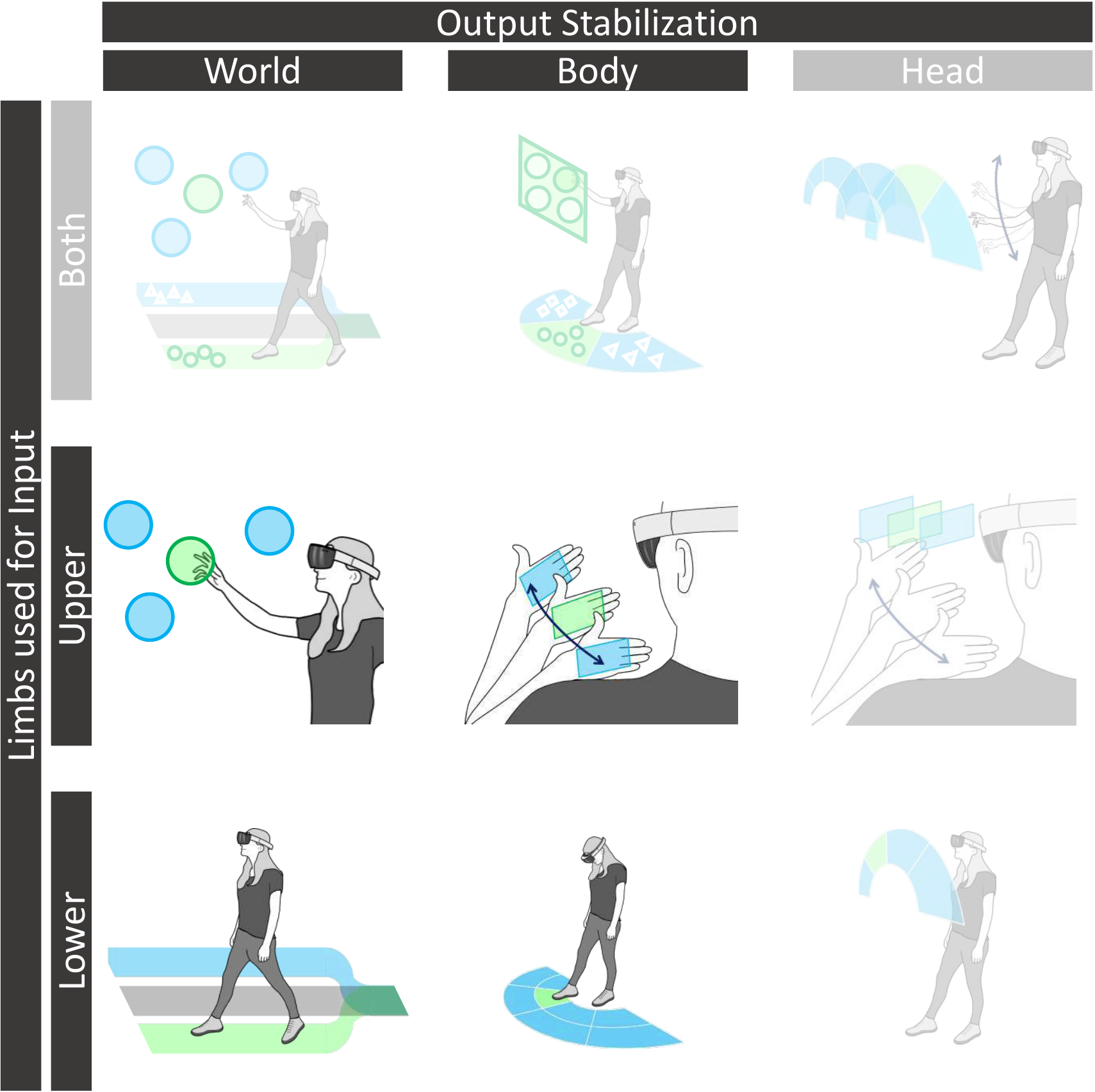}{A classification of \aroundbodyinteraction{} techniques based on the 1) \emph{limbs used for input} and the 2) \emph{stabilization method of the output}. As discussed in section \ref{sec:rw:aroundbody:classification}, this thesis focuses on \emph{upper} and \emph{lower} limbs and \emph{body-stabilized} and \emph{world-stabilized} interfaces.}

In this section we propose a classification of \aroundbodyinteraction{} techniques based on the 1) limbs used for input and the 2) stabilization method of the output.  

It is difficult to distinguish the body parts involved in movements since individual interaction steps typically comprise the seamless interaction of the movement of several joints of the body at the same time~\ncite{Lu2012}. When we use our hand to open a door, we first use the degrees of freedom of the shoulder and arm to position the hand before the degrees of freedom of the hand are used to perform the interaction. In this work, the affected degrees of freedom of the individual joints are, therefore, not studied independently of each other, but as a collective movement of the entire limb.

\subsubsection{Input - Upper and Lower Limbs}
\label{sec:rw:around:upper}

As discussed before, the limbs allow us to perform interactions in the space around our body. The input can be classified by the group of limbs involved: upper, lower, or both.

\paragraph{Input using Upper Limbs}

The \emph{upper limbs} can support interactions in front of the upper part of the body. This can be achieved by using the degrees of freedom of our shoulders, elbows, and wrists, as well as the many further degrees of freedom offered by our hands and fingers.

Figure \ref{fig:introduction/designspace_reduced} shows examples of interaction techniques using the upper limbs together with the different output stabilization methods. These examples are discussed in more detail in chapters \ref{ch:proximity:merged} and \ref{ch:cloudbits}.

\paragraph{Input using Lower Limbs}
\label{sec:rw:around:lower}

The \emph{lower limbs} can support interactions in front of the lower part of the body by leveraging the degrees of freedom offered by the hip, knee, and ankle joints. Such interactions are particularly suitable for situations where the user's hands are not available.

Figure \ref{fig:introduction/designspace_reduced} shows examples of interaction techniques using the upper limbs together with the different output stabilization methods. These examples are discussed in more detail in chapters \ref{ch:cheesyfoot} and \ref{ch:walktheline}.

\paragraph{Input using Both Groups of Limbs}

As a result of requirement \refreq{req:body:blocking} (\reqBlocking) introduced in section \ref{sec:relatedwork:body:requirements}, this thesis focuses on interaction techniques that use as few body parts as possible at the same time. Interaction techniques using upper and lower limbs at the same time complicate the usage in the real world, as they might collide with the regular interaction of the user with the real world.

Therefore, this thesis will not cover \aroundbodyinteraction{} using a combination of upper and lower limbs. However, the increased expressiveness by using multiple limbs simultaneously can also be beneficial in certain interaction situations. Therefore, chapter \ref{sec:conclusion:futurework} presents directions for future work on such interfaces.

\subsubsection{Output - Head, Body and World Stabilization}

As discussed in section \ref{sec:relatedwork:hmds:def}, \acp{HMD} support visual output on a scale from head-stabilization to world-stabilization. These different types of output influence the techniques suitable for interaction.

\paragraph{Head-Stabilized Output}

By design, \emph{head-stabilized} interfaces do not merge information with the real world, but only offer a static overlay. In recent years, such \ac{HUD} interfaces have diminished in importance because of the wide availability of low-cost and robust methods for tracking the orientation of \acp{HMD}.

Based on this decline in interest, this thesis will not cover \aroundbodyinteraction{} with such interfaces in the main contributions. However, chapter \ref{sec:conclusion:futurework} gives an outlook on possible interaction techniques for this kind of visual output.

\paragraph{Body-Stabilized Output}

\emph{Body-stabilized} output can display interfaces that move as the user moves. Together with hand and foot tracking, this allows for interfaces that are registered to parts of the body, regardless of where the user is and whether he is currently moving.

Figure \ref{fig:introduction/designspace_reduced} shows examples for such interaction techniques using the upper and lower limbs. These examples are discussed in more detail in chapters \ref{ch:proximity:merged} and \ref{ch:cheesyfoot}.

\paragraph{World-Stabilized Output}

\emph{World-stabilized} output can display interfaces that are not registered to the user (or parts of the user), but the real world. Such interfaces do not move along with the user. As a consequence, the user can move around or along such interfaces and thus change the relative position and orientation to the interface. This relative position and orientation can further be used as an additional input dimension.

Figure \ref{fig:introduction/designspace_reduced} shows examples for such interaction techniques using the upper and lower limbs. These examples are discussed in more detail in chapters \ref{ch:cloudbits} and \ref{ch:walktheline}.

\subsubsection{Interaction Situations}

In addition to the two-dimensional classification according to the \emph{limbs used for input} and the \emph{stabilization of the output}, this thesis proposes to classify \aroundbodyinteraction{} techniques based on their support for common interaction situations as introduced in section \ref{chp:introduction:challenges}: \inplaceUpper{}, \mobilityUpper{}, \singleuserUpper{}, \multiuserUpper{}, \continuousUpper{} and \discreteUpper{}.

\subsection{Conclusion}

The two-dimensional classification of \aroundbodyinteraction{}s by the limbs used for input and the stabilization of output shows four areas for such interaction techniques. This work will focus more closely on these four areas and offer a corresponding interaction technique for each of these combinations. More precisely, 

\begin{description}
	\item[Chapter \ref{ch:proximity:merged}] presents \projProximity{} and focuses on interaction with the upper limbs for body-stabilized interfaces, contributing solutions for \singleuserLower{} \mobilityLower{} situations, providing support for \continuousLower{} and \discreteLower{}.
	\item[Chapter \ref{ch:cloudbits}] presents \projCloudbits{} and focuses on interaction using the upper limbs and world-stabilized interfaces, contributing solutions for \continuousLower{} in \multiuserLower{} \inplaceLower{} situations.
	\item[Chapter \ref{ch:cheesyfoot}] presents \projCheesyfoot{} and focuses on interaction using the lower limbs and body-stabilized interfaces, contributing solutions for \discreteLower{} for \singleuserLower{} and \mobilityLower{} situations.
	\item[Chapter \ref{ch:walktheline}] presents \projCheesyfootToGo{} and focuses on interaction using the lower limbs and world-stabilized interfaces, contributing solutions for \discreteLower{} for \mobilityLower{} and \singleuserLower{} situations.
\end{description}

As each combination of these input and output possibilities entails specific requirements for interaction, each chapter will present the requirements and a review of the related works for the respective areas.

}

\setcounter{ptc}{2}
\ctparttext{\parttoc}
\part{Interaction using the Upper Limbs}\label{pt:hand}

\chapter[Proximity-based Interaction]{Proximity-based Hand Input for One-handed Mobile Interaction}\label{ch:proximity:merged}
{

\cptteaser{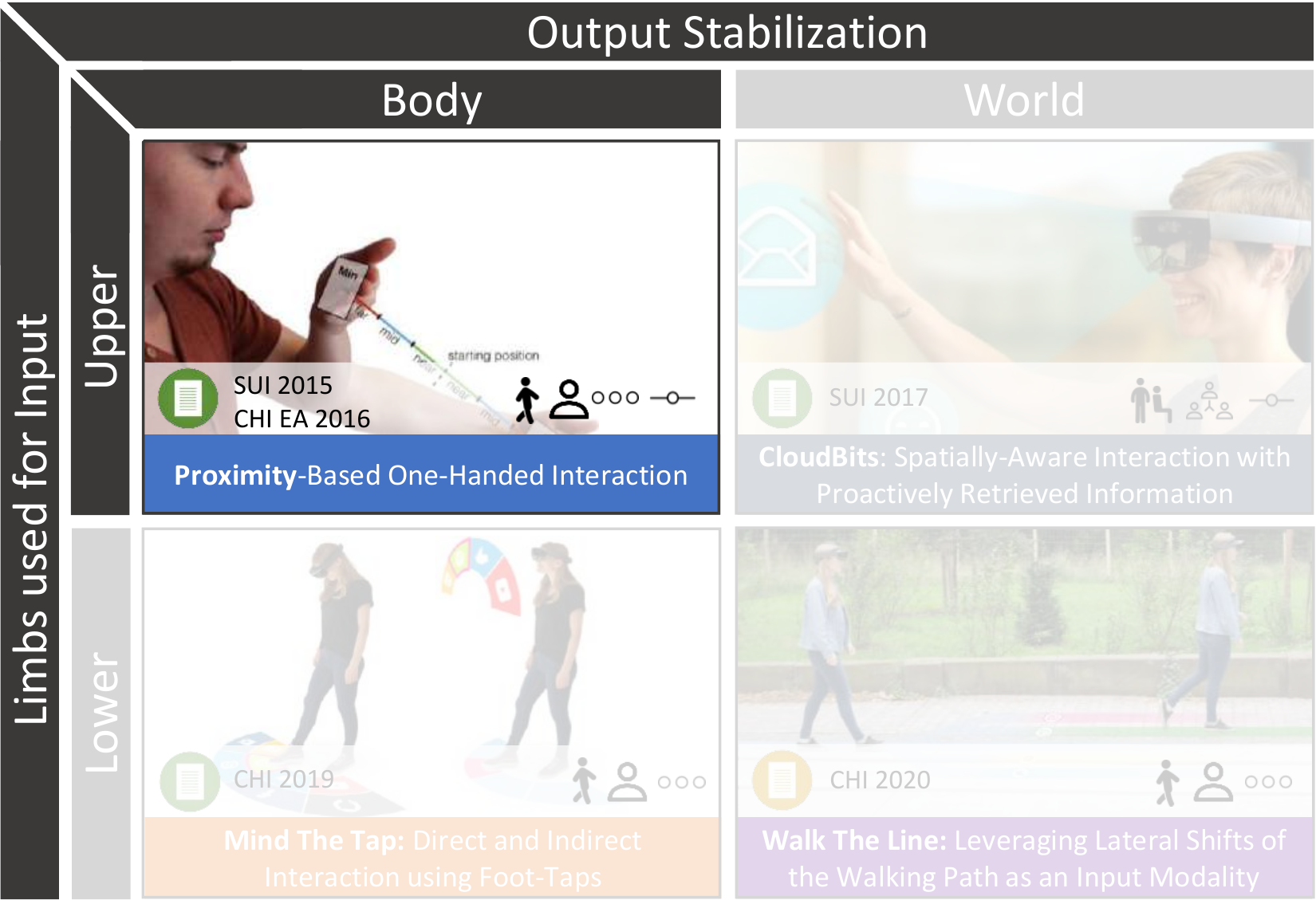}{This chapter presents an interaction technique for body-stabilized interfaces leveraging the upper limbs for input.}
	
The previous chapter introduced the emergence of a new class of body-based interfaces for mobile interaction with \acp{HMD} that extend the interaction from the surface \emph{on} our body into the space \emph{around} our body defined by the reachable range of our limbs. The discussion of such interfaces led to a classification by limbs used for input and stabilization point of the output. This chapter discusses \emph{upper limb} interfaces for use with \emph{body-stabilized} interfaces (see section \ref{sec:relatedwork:hmds:implementation}) with an emphasis on support for the interaction situations \mobilityLower{}, \singleuserLower{}, \discreteLower{} and \continuousLower{}.

As outlined in section \ref{sec:rw:around:upper}, it is with our hands and arms that we show the greatest dexterity in interacting with the real world. Further, the body stabilization of the visualization allows the user to \emph{carry along} an interface, thus affording mobile interaction. Therefore, interaction techniques in this quadrant of the design space (see figure \ref{fig:proximity/overview_proximity}) are particularly suitable for precise interactions in the mobile context.

The contribution of this chapter is three-fold. First, the chapter presents the results of two controlled experiments investigating a one-handed hand input modality for 1) continuous and 2) discrete interaction with \emph{body-stabilized} interfaces. Second, based on the findings of the experiments, this chapter presents a set of guidelines for designing such \aroundbodyinteraction{} techniques in this quadrant of the design space. Third, building on the design guidelines, this chapter presents use cases for such interfaces and shows the applicability of the presented interaction technique beyond \acp{HMD} through a prototype implementation for smartwatches.

The remainder of this chapter is structured as follows: After reviewing the related works (section \ref{sec:proximity:rw}) and based on the established requirements, the chapter presents the concept for an interaction technique (section \ref{sec:proximity:intro}). Afterward, section \ref{sec:proximity:exp1} and \ref{sec:proximity:exp2} present the methodology and results of two controlled experiments investigating the interaction technique presented. Based on the results, section \ref{sec:proximity:discussion} presents guidelines for the future use of such interfaces. Section \ref{sec:proximity:applicability} provides hints for the applicability of the presented interaction technique for \acp{HMD} and other wearable device classes. The chapter concludes with limitations and guidelines for future work (section \ref{sec:proximity:limitations}).

\bigskip

\begin{mdframed}[style=infoboxstyle]
	\textbf{Publication:} This chapter is based on the following publications:
	
	\begin{small}
		\longfullcite{muller2015a}
		
		\longfullcite{muller2016proxiwatch}
	\end{small}
	
	\textbf{Contribution Statement:} I led the idea generation, implementation, and performed the data evaluation. The student \emph{Sebastian Günther} implemented the study client application. \emph{Mohammadreza Khalilbeigi}, \emph{Niloofar Dezfuli}, \emph{Alireza Sahami Shirazi} and \emph{Max Mühlhäuser} supported the conceptual design and contributed to the writing process.
\end{mdframed}
\section{Related Work}
\label{sec:proximity:rw}

Chapter \ref{ch:relatedwork} discusses the related works on interaction techniques for \acp{HMD}. The following section presents a set of requirements for interacting with \acp{HMD} using the upper limbs and, in the following, categorizes relevant research with regards to the requirements. 

\subsection{Requirements}

The following section presents a set of requirements for one-handed interactions with \acp{HMD} derived from the related works. The requirements are then used to compare the most relevant related work (see tables \ref{tab:req:proximity_twohanded}, \ref{tab:req:proximity_discrete} and \ref{tab:req:proximity_continuous}).

\begin{description}
	\item[R3.1: Independent Usage] Particularly in a mobile context, it is necessary to burden as few body parts as possible with the interaction so that they are available for the normal interaction with the world. Therefore, interaction techniques should focus on one-handed interaction and not require any additional body parts for interaction.
	\item[R3.2: Direct Interaction with Content] A direct spatial connection of input and output (i.e., input and output happen at the same physical location) allows for more \enquote{natural} and \emph{compelling} interactions~\ncite{Forlines2007}. Therefore, systems should provide such a direct spatial connection.
	\item[R3.3: Support for Peripheral or Proxemic Interactions] In many situations, it can be a hindrance to completely draw the user’s (visual) attention to the interface. Therefore, systems should allow interactions without complete visual focus.
	\item[R3.4: Discrete Interaction] Discrete interaction allows fast and short interactions through shortcuts. Therefore, systems should provide means for discrete interaction with content.
	\item[R3.5: Continuous Interaction] Continuous interaction allows fine granular interaction with information. Therefore, systems should provide means for continuous interaction with information.
\end{description}

\subsection{Interaction with the Upper Limbs}

There is a large body of related works on interaction with the upper limbs, most of which relate to the field of gesture-based interaction.

The first approaches to support such interactions started appearing towards the end of the 1980s. \typcite{Zimmerman1987} presented the \emph{DataGlove}, a device supporting real-time position, orientation, and posture tracking of the user’s hand. \typcite{Quam1990} demonstrated how such a device could be used for gesture recognition. These devices typically consisted of a glove, worn by the user, and augmented with a set of sensors or a mechanical construction to track the position and orientation~\ncite{Sturman1994a}.

To overcome the limitations associated with the need to wear a glove, research proposed optical systems that use computer vision to track the user’s hands without any hardware attached to the body ~ \ncite{Maqueda2015}. While the first systems were only able to recognize static postures of the user’s hand~\ncite{Pavlovic1997}, research brought forth various approaches to also understand the temporal dimension of gestures~\ncite{Wu1999}.

Research proposed multiple techniques for interacting with the hands in front of the upper body of the user without a secondary device. These techniques fall into the group of mid-air gestures~\ncite{aigner2012understanding}, also referred to as \emph{Free-Hand}~\ncite{Ren2013} or \emph{Bare-Hand}~\ncite{VonHardenberg2001}. 

\subsubsection{Two-Handed Gestures}

As a prominent example, \typcite{Mistry2009} presented a wearable interface supporting natural gesture interaction. Building on this, \typcite{Datcu2013} proposed two-handed mid-air interaction with \acp{HMD} through hand and finger gestures. Further, \typcite{Benko2012} proposed a combination with tangibles and a system to manipulate virtual objects the \enquote{same way [as] users manipulate real world objects}. 

Such two-handed mid-air gestures have also been proposed for interacting with various types of computing systems, from stationary systems such as tabletops~\ncite{Hilliges2009}, public displays~\ncite{Muller2014} or television sets~\ncite{Sang-HeonLee2013} to highly mobile systems such as smartphones~\ncite{Aslan2014} and smartwatches~\ncite{ArefinShimon2016}.

\renewcommand*\theadfont{\bfseries}
\settowidth\rotheadsize{\theadfont Interaction Without Full}
\renewcommand\theadgape{}
\renewcommand\theadalign{lc}
\renewcommand\rotheadgape{}
\begin{table}[h!]
	\centering
	\begin{tabular}{m{4cm}ccccc}
		& \rothead{R1: Independent Usage} & \rothead{R2: Direct Interaction with Content} & \rothead{R3: Support for Peripheral or Proxemic Interactions} & \rothead{R4: Discrete Interaction} & \rothead{R5: Continuous Interaction} \\
		\midrule
		\multicolumn{6}{c}{\emph{Two-Handed Interaction}} \\
		\cite{Benko2012} & \reqNo & \reqYes & \reqNo & \reqNo & \reqYes \\
		\cite{Datcu2013} & \reqNo & \reqPartially & \reqNo & \reqYes & \reqPartially \\
		\cite{Hilliges2009} & \reqNo & \reqPartially & \reqNo & \reqNo & \reqYes \\
		\cite{Muller2014} & \reqNo & \reqPartially & \reqPartially & \reqYes & \reqYes \\
		\cite{Sang-HeonLee2013} & \reqNo & \reqNo & \reqNo & \reqYes & \reqNo \\
		\cite{Aslan2014} & \reqNo & \reqNo & \reqYes & \reqYes & \reqNo \\
		\cite{ArefinShimon2016} & \reqNo & \reqPartially & \reqYes & \reqYes & \reqNo \\
		\cite{Mistry2009} & \reqNo & \reqPartially & \reqNo & \reqPartially & \reqYes \\
		\cite{Whitmire2017} & \reqNo & \reqNo & \reqYes & \reqYes & \reqYes \\
		\bottomrule
	\end{tabular}
	\caption{Fulfillment of requirements of the related works. \reqYes~ indicates that a requirement is fulfilled, \reqPartially~ indicates a partial fulfillment.}
	\label{tab:req:proximity_twohanded}
\end{table}

While useful and practical, two-handed interaction limits the suitability of interaction techniques during daily use as both hands need to be free for interaction. Table \ref{tab:req:proximity_twohanded} compares the related work in two-handed gestures to the requirements presented above.

\subsubsection{One-Handed Gestures}

To overcome the problems of two-handed interfaces, research proposed one-handed interaction techniques. In the following, this thesis presents these approaches, grouped as discrete and continuous interaction techniques.

\paragraph{Discrete Gestures}

Research presented a variety of discrete one-handed interaction techniques. As two prominent examples of this group of interfaces prominent example, \typcite{Colaco2013} showed how to capture and interpret fine-grained single-handed gestures, and \typcite{Akkil2016} compared the accuracy of different pointing gestures to communicate locations. Beyond sole interacting with the \ac{HMD} itself, \typcite{Kollee2014} proposed a set of interaction techniques for interacting with surrounding devices in a smart space.

Further examples include interaction techniques in the industrial context~\ncite{Witt2006} and the combination of hand gestures with finger gestures~\ncite{Hsieh2016}, head-pointing~\ncite{Kyto2018} or other body parts~\ncite{Heo2017}.

Highly related, \typcite{Xu2018} proposed a body-stabilized mid-air interface that is bound to the user’s wrist. Users can select items in this interface by pointing to items using the index finger. However, this approach requires an outstretched arm and is, therefore, prone to fatigue. In addition, the approach only displays a selection menu - actual content appears outside the interface. Therefore there are no possibilities for direct interaction with content. Further, all of the presented approaches are missing means for continuous interaction with information. Table \ref{tab:req:proximity_discrete} compares the related work in discrete one-handed gestures to the requirements presented above.

\renewcommand*\theadfont{\bfseries}
\settowidth\rotheadsize{\theadfont Interaction Without Full}
\renewcommand\theadgape{}
\renewcommand\theadalign{lc}
\renewcommand\rotheadgape{}
\begin{table}[t!]
	\centering
	\begin{tabular}{m{4cm}ccccc}
		& \rothead{R1: Independent Usage} & \rothead{R2: Direct Interaction with Content} & \rothead{R3: Support for Peripheral or Proxemic Interactions} & \rothead{R4: Discrete Interaction} & \rothead{R5: Continuous Interaction} \\
		\midrule
		\multicolumn{6}{c}{\emph{One-Handed Discrete Interaction}} \\
		\cite{Xu2018} & \reqYes & \reqPartially & \reqYes & \reqYes & \reqNo \\
		\cite{Witt2006} & \reqYes & \reqNo & \reqYes & \reqYes & \reqNo \\
		\cite{Akkil2016} & \reqYes & \reqNo & \reqNo & \reqNo & \reqYes \\
		\cite{Kollee2014} & \reqYes & \reqNo & \reqYes & \reqYes & \reqNo \\
		\cite{Hsieh2016} & \reqYes & \reqNo & \reqPartially & \reqPartially & \reqYes \\
		\cite{Heo2017} & \reqPartially & \reqNo & \reqYes & \reqYes & \reqPartially \\
		\cite{Heo2010} & \reqYes & \reqYes & \reqNo & \reqYes & \reqNo \\
		\cite{Kyto2018} & \reqNo & \reqNo & \reqYes & \reqYes & \reqPartially \\
		\bottomrule
	\end{tabular}
	\caption{Fulfillment of requirements of the related works. \reqYes~ indicates that a requirement is fulfilled, \reqPartially~ indicates a partial fulfillment.}
	\label{tab:req:proximity_discrete}
\end{table}

\paragraph{Continuous Gestures}

Beyond discrete gestures, research also presented approaches for continuous one-handed interaction with information. For example, \typcite{Whitmire2017} presented one-handed finger gestures and \typcite{Buchmann2004} presented a system for direct manipulation of virtual objects in \ac{AR}. Further, \typcite{Khademi2014} presented continuous one-handed mid-air gestures for stroke rehabilitation. Highly related, \typcite{Chen2014} proposed a mid-air \aroundbodyinteraction{} technique leveraging the degrees of freedom of the elbow joint. 

These gesture-based approaches do not relay on the physical appearance of devices, nor on secondary devices and can, further, help to provide more direct interaction while preserving social acceptability of devices. 

However, all the presented approaches only consider continuous interaction and, thus, cannot support shortcuts as a fast technique for interacting with information. Table \ref{tab:req:proximity_continuous} compares the related work in discrete one-handed gestures to the requirements presented above.

\renewcommand*\theadfont{\bfseries}
\settowidth\rotheadsize{\theadfont Interaction Without Full}
\renewcommand\theadgape{}
\renewcommand\theadalign{lc}
\renewcommand\rotheadgape{}
\begin{table}[h!]
	\centering
	\begin{tabular}{m{4cm}ccccc}
		& \rothead{R3.1: Independent Usage} & \rothead{R3.2: Direct Interaction with Content} & \rothead{R3.3: Support for Peripheral or Proxemic Interactions} & \rothead{R3.4: Discrete Interaction} & \rothead{R3.5: Continuous Interaction} \\
		\midrule
		\multicolumn{6}{c}{\emph{One-Handed Continuous Interaction}} \\
		\cite{Colaco2013} & \reqYes & \reqYes & \reqNo & \reqYes & \reqPartially \\
		\cite{Chen2014} & \reqYes & \reqYes & \reqYes & \reqNo & \reqYes \\
		\cite{Khademi2014} & \reqYes & \reqNo & \reqNo & \reqNo & \reqYes \\
		\cite{Buchmann2004} & \reqYes & \reqYes & \reqNo & \reqNo & \reqYes \\
		\bottomrule
	\end{tabular}
	\caption{Fulfillment of requirements of the related works. \reqYes~ indicates that a requirement is fulfilled, \reqPartially~ indicates a partial fulfillment.}
	\label{tab:req:proximity_continuous}
\end{table}

\newpage
\section{Concept}
\label{sec:proximity:intro}

\cptteaser{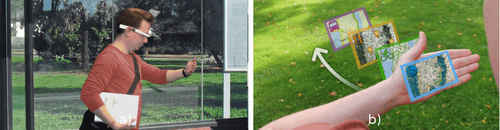}{A map application as an example of one-handed (a) proximity-based interaction with a linear layered information space. The user can browse map layers by moving his hand through the space (b).}

Especially in mobile situations, one hand of users is often busy interacting with the world, rendering the large body of related work on two-handed interaction techniques unsuitable for such situations. One-handed interaction techniques, in contrast, can reduce the interference with regular interactions with the real world and, thus, cover a larger amount of interaction situations.

Therefore, this chapter focuses on how the large number of degrees of freedom offered by our hands and arms can support one-handed interactions. 

The degrees of freedom of movement of the arm and hand are defined by the degrees of freedom of the joints involved. In particular, this includes the shoulder joint, the elbow joint, the wrist as well as the countless joints of the individual fingers. Since movements of the wrist and fingers are relatively close to (on-body) touch interactions already explored in the body of related work, and movements controlled by the shoulder joint are known to cause fatigue~\ncite{Hincapie-Ramos2014}, this work focuses as a first step on the degrees of freedom of the elbow joint. The elbow joint is movable by \emph{flexion} (i.e., moving the hand towards the body) and \emph{extension} (i.e., moving the hand away from the body). 

This chapter, therefore, explores the proximity dimension defined by the elbow joint as an additional input modality for one-handed mobile interaction: The interaction space alongside the user's line of sight can be divided into multiple parallel planes. Similar to \ncite{Subramanian2006}, each plane corresponds to a layer with visual content that can be displayed on the user's hand. The user can move the hands to browse through successive layers (see figure \ref{fig:proximity/teaser_small}). Further, through the sense of proprioception, users can perform these actions unconsciously, reducing the mental load of interaction and allowing peripheral or completely eyes-free interactions.

This chapter explores two different interaction techniques for \emph{continuous} and \emph{discrete} interaction in this one-dimensional interaction space: For \emph{continuous} interaction, the user can move his hands to browse through successive layers (see figure \ref{fig:proximity/teaser_small}). This movement can, for example, represent scrolling through a list. For discrete interaction, the user can raise his arm at a specific distance and, thus, directly select a layer. Each layer can be mapped to a shortcut action, allowing for fast and immediate interaction.

\section{Experiment I: Continuous Interaction}
\label{sec:proximity:exp1}

The following section presents the methodology (see section \ref{sec:proximity:methodology}) and the results (see section \ref{sec:proximity:results}) of a controlled experiment investigating the human capabilities for a proximity-based one-hand input modality in multi-layer information spaces.

\subsection{Methodology}
\label{sec:proximity:methodology}

This section presents the methodology of a controlled experiment focusing on proximity-based continuous interaction. More specifically, the controlled experiment addressed three main research questions:

\begin{enumerate}
	\item[RQ1] How accurate and efficient users can interact with the layered information space in a search task scenario?
	\item[RQ2] How does the direction of interaction and the side of the hand affect the efficiency and accuracy of the interaction?
	\item[RQ3] How to design the interaction space in terms of layer thickness, number of layers, and convenient boundaries of the physical interaction volume?
\end{enumerate}

For this, 14 participants (P1-P14: 4 female, 1 left-handed), aged between 24 and 29 years ($\mu=26$, $\sigma=1.6$), were recruited. The average height was 177cm ($\sigma=9.5$cm) with an average arm length (measured from armpit to carpus) of 59cm ($\sigma=3.6cm$). No compensation was provided.

\subsubsection{Design and Task}

\textfigH{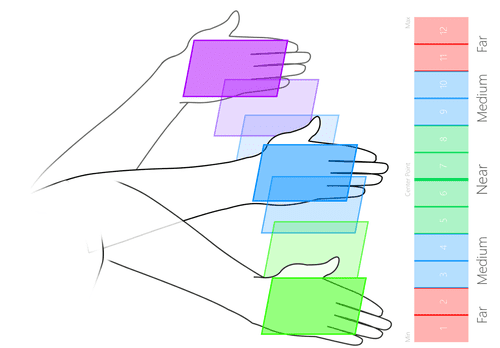}{The information space alongside the participants' line of sigh used in the experiment, grouped by the distance to the starting point.}

To answer the research questions presented above, the design of the task was based on a basic multi-layer information space alongside the participants' line of sight (see figure \ref{fig:proximity/conditions}) consisting of randomized integer numbers (each layer displayed one number) similar to~\ncite{Spindler2012}. The conditions varied the \emph{number of layers} in the available interaction space (which directly correlates with the layers' thickness) as an independent variable with the values of \emph{12, 24, 36, 48, 60} and \emph{72}. In addition, the conditions also varied the \emph{direction of interaction} between \emph{flexion} and \emph{extension} as a second as well as the \emph{side of the hand} as (\emph{palm} or \emph{backside}) as a third independent variable. The subsequent analysis examined the influence of the individual factors on the participants' performance in terms of accuracy and efficiency. 

The experiment varied the independent variables with 6 levels for \emph{numbers of layers}, 2 different \emph{hand sides}, and 2 \emph{directions of interaction} with 6 repetitions (two from each zone) for each combination in a repeated measure design, resulting in $6\times2\times2\times6 = 144$ trials per participant. Informal pre-tests suggested that these levels would provide the highest accuracy. The order of the conditions was counterbalanced using a Balanced Latin Square design for the number of layers and the direction of the interaction. For practicality reasons, the \emph{side of the hand} condition was excluded from the Latin Square design because remounting the trackable marker resulted in also recalibrating the system. However, half of the participants performed all palm-side trials first, while the other half started with the backside trials.

\begin{figure}[t!]
	\begin{center}
		\includegraphics[width=\linewidth]{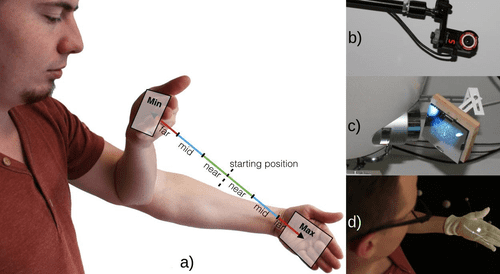}
		\caption{Traveling distance zones (a) and setup of the experiment (b-d).}
		\label{fig:proxi:concept}
	\end{center}
\end{figure}

The participants' first task was to search for the one red colored number in the stack of white colored numbers (see fig. \ref{fig:study}). Once found, participants confirmed the discovery by pressing a button with their non-interacting hand. Directly afterwards, as the second task, participants had to hold the hand steady at the respective position for 3 seconds to measure the accuracy while trying to hold on a layer.

The system defined the maximum boundary of the interaction space with the participant's individual arm-length and the minimum boundary as the near point of the human's eye of young adults (not closer than 12.5cm to the user's face~\ncite{Kulp1999}). Furthermore, the starting point of all trials was defined as half of the distance between the minimum and the maximum interaction distance, resulting in an elbow joint deflection of around 100 degrees. Informal pre-tests showed this to be a natural and relaxed holding position for the hand. To systematically analyze influences of the traveling distance of the user's hand, the total available interaction space in each direction was divided into three equal-sized zones for later analysis: near, medium, and far as shown in Figure \ref{fig:proximity/conditions}).

\subsubsection{Experiment Setup and Apparatus}

The setup used an optical tracking system (OptiTrack, see fig. \ref{fig:proxi:concept} b) to precisely measure the linear distance between the participant's hand and eyes alongside the participant's line of sight. To reliably track the position and orientation of the participant's head and hand, the participants wore two trackable apparatuses: A glasses frame and a glove, each augmented with a number of small retro-reflective markers (see fig. \ref{fig:proxi:concept} d). The system further used the real-time tracking information to fit the projected feedback to the participant's hand (see fig. \ref{fig:study}) to simulate an \ac{AR} system. For each trial, we measured:

\begin{enumerate}
	\item the \emph{task completion time (TCT)} as the timespan between starting the trial and confirming the discovery of the target.
	\item the \emph{overshooting error} as the maximum deviation in the distance (in mm) between the center of the target layer and the participant's hand after first reaching the target layer before confirming the discovery.
	\item the \emph{holding error} as the maximum distance (in mm) from the starting point of the holding task.
\end{enumerate}

\subsubsection{Procedure}

\begin{figure}[t!]
	\begin{center}
		\includegraphics[width=\linewidth]{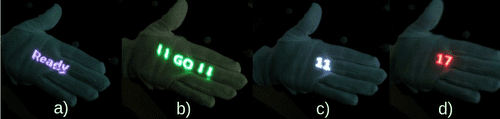}
		\caption{Visual feedback in the experiment: After reaching the starting position (a), the system showed the direction of interaction (b). The participants task was to browse through a stack of white colored numbers (c) to find the one red colored number (d).}
		\label{fig:study}
	\end{center}
\end{figure}

The investigator introduced the participants to the concept and experiment setup and asked the participants to put on the two trackable apparatuses before calibrating the system to adapt it to the respective arm size. Before starting each trial, the system guided the user to the starting position through visual feedback displayed on the user's hand. Once in the starting position, the system displayed the direction of the interaction. Each trial started by pressing the button. Once the target was found, the participant confirmed the discovery through another click. After that, the system informed participants to hold their current position for three seconds. Participants did not receive any feedback during the holding task and were not informed on the current layer thickness.

After each condition, participants took a 30 seconds break. The experiment concluded with a semi-structured interview focusing on the participants' overall opinion about the concept, preferred interaction boundaries (minimum/maximum distance), and differences between the tested conditions. The experiment took about 60 minutes per participant.

\textfigStudybox{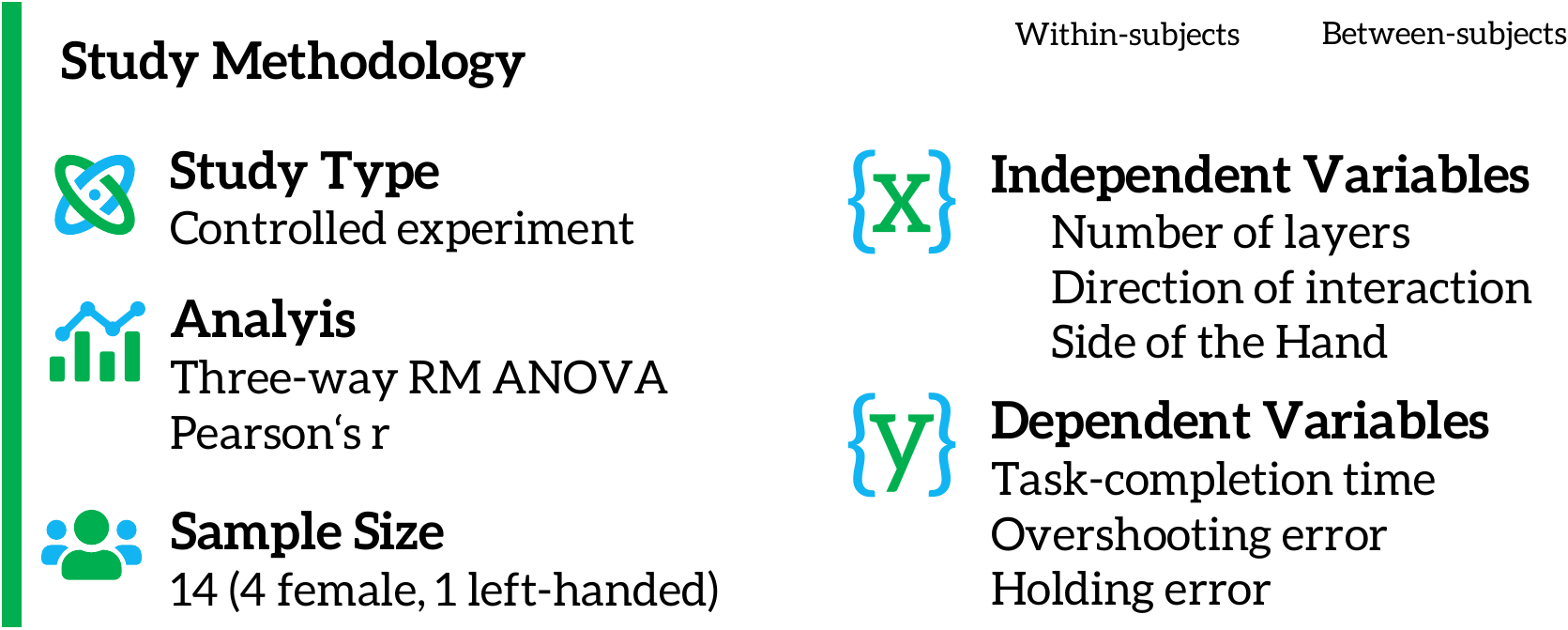}
\subsection{Results}
\label{sec:proximity:results}

This section reports the results of the controlled experiment investigating the research questions RQ1, RQ2, and RQ3 as described in section \ref{sec:proximity:methodology}. The analysis of the data was performed as described in section \ref{sec:rw:methodology}. 

\subsubsection{Task Completion Time}

\begin{figure}[t!]
	\begin{center}
		\includegraphics[width=\linewidth]{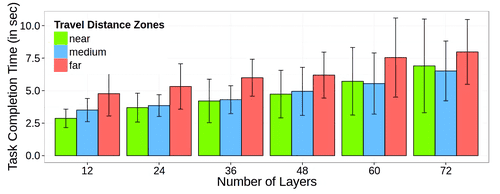}
		\caption{Mean TCT and SD for different numbers of layers.}
		\label{fig:taskcompletiontime}
	\end{center}
	\vspace{-.5cm}
\end{figure}

The analysis unveiled that the traveling distance (measured in the three groups \emph{near}, \emph{medium} and \emph{far}) of the hand had a significant effect on the \ac{TCT} (\anovaWithoutEffect{2}{62}{23.27}{<.001}). Post-hoc tests confirmed that the \ac{TCT} for near (\valSi{4.7}{3.6}{s}) and medium zone (\valSi{4.8}{2.7}{s}) targets were significantly smaller ($p<.001$) than for those in the far zone (\valSi{6.4}{3.4}{s}). Post-hoc test did not indicate significantly different \acp{TCT} between medium and near zone targets. Table \ref{tab:proximity/tct_distance} lists the \acp{TCT} for all zones.

Further, the number of layers and, thus, the size of the individual layers had a significant effect on the \ac{TCT} (\anovaCorWithoutEffect{2.45}{31.36}{45.68}{<.001}{.49}). Post-hoc tests confirmed a significantly ($p<.01$) larger \ac{TCT} for higher numbers of layers between all groups. The mean \ac{TCT} increased from  \valSi{3.7}{1.8}{s} for 12 layers to \valSi{7.2}{4.7}{s} for 72 layers. While the mean \ac{TCT} was faster for extension (\valSi{5.5}{3.5}{s}) than flexion (\valSi{5.1}{3.1}{s}), the analysis could not proof any significant effects (\anovaWithoutEffect{1}{13}{2.8}{>.05}). Also, no significant effect of the hand orientation on \ac{TCT} was found (\anovaWithoutEffect{1}{13}{.15}{>.05}, Palm: \valSi{5.2}{3.2}{s}, Back: \valSi{5.3}{3.4}{s}). Also, the analysis could not find interaction effects between the conditions. Figure \ref{fig:taskcompletiontime} shows the \ac{TCT} for the explored numbers of layers and target zones.

\begin{table}[t!]
	\centering

\begin{tabularx}{\linewidth}{YYYYY}
	&   &   & \multicolumn{2}{c}{\textbf{95\% Confidence Interval}}\\
  	\cmidrule(lr){4-5}
	\textbf{Zone} & $\pmb{\mu}$ & $\pmb{\sigma_{\overline{x}}}$ & \textbf{Lower} & \textbf{Upper}\\
	\midrule
	Near   & 4.7s  & .9s      & 2.81s                     & 6.56s                     \\
	Medium & 4.8s  & .7s      & 3.39s                     & 6.21s                     \\
	Far    & 6.4s  & .9s      & 4.62s                     & 8.18s                     \\
	\bottomrule
	\end{tabularx}

	\caption{The Task Completion Time (TCT) per distance zone (in seconds). The table reports the mean value $\mu$, the standard error $\sigma_{\overline{x}}$ and the 95\% confidence interval.}
	\label{tab:proximity/tct_distance}
\end{table}

\subsubsection{Overshooting Error}

\begin{figure}[b!!]
	\begin{center}
		\includegraphics[width=\linewidth]{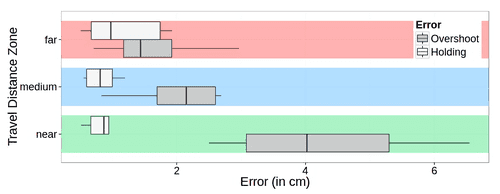}
		\caption{Error measurements for the three traveling distance zones.}
		\label{fig:errorperinteractionzone}
	\end{center}
\end{figure}

The traveling distance had a significant effect on the overshooting error (\anovaCorWithoutEffect{1.26}{16.38}{39.44}{<.001}{.63}). Post-hoc tests confirmed significant differences between all zones (all $p<.05$). The observations during the experiment showed that participants initially started with fast movements and slowed down towards their physical boundaries in the far zones, resulting in higher overshooting errors in the near (\valSi{4.4}{1.7}{cm}) and medium ($\mu$=2.1cm, $\sigma$=1.0cm) zones compared to the far ($\mu$=1.6cm, $\sigma$=0.7cm) zone. Table \ref{tab:proximity/overshoot_distance} lists the mean overshoot and corresponding standard errors per target zone, figure \ref{fig:errorperinteractionzone} compares the values graphically.

The analysis showed neither any significant influence of the direction of interaction on the overshooting error (\anovaWithoutEffect{1}{13}{.0008}{>.05}, flexion: \valSi{2.5}{3.0}{cm}, extension: \valSi{2.6}{3.2}{cm}) nor the hand orientation (\anovaWithoutEffect{1}{13}{.11}{>.05}, palm: \valSi{2.6}{3.1}{cm}, back: \valSi{2.6}{3.0}{cm}). Furthermore, the analysis could not show any significant influence (\anovaWithoutEffect{5}{64}{.64}{>.05}) of the number of layers (Min: \valSi{2.2}{3.1}{cm} for 12 layers, Max: \valSi{2.8}{3.4}{cm} for 36 layers). Also, the analysis did not show any significant correlation between the participants' arm-length and their accuracy ($r(166) = -0.8376, p>.05$) in the recorded data.

\begin{table}[h!]
	\centering

\begin{tabularx}{\linewidth}{YYYYY}
	&   &   & \multicolumn{2}{c}{\textbf{95\% Confidence Interval}}\\
  	\cmidrule(lr){4-5}
	\textbf{Zone} & $\pmb{\mu}$ & $\pmb{\sigma_{\overline{x}}}$ & \textbf{Lower} & \textbf{Upper}\\
	\midrule
	Near   & 4.4cm & .45cm    & 3.51cm                    & 5.29cm                    \\
	Medium & 2.1cm & .27cm    & 1.58cm                    & 2.62cm                    \\
	Far    & 1.6cm & .19cm    & 1.23cm                    & 1.97cm                    \\
	\bottomrule
	\end{tabularx}

	\caption{The overshooting error per distance zone (in cm). The table reports the mean value $\mu$, the standard error $\sigma_{\overline{x}}$ and the 95\% confidence interval CI.}
	\label{tab:proximity/holding_distance}
\end{table}

\subsubsection{Holding Error}

The analysis showed significant effects of the distance between the starting point and the holding point on the holding error (\anovaCorWithoutEffect{1.12}{14.56}{5.53}{<.05}{.56}). Post-hoc tests confirmed significant effects (all $p<.05$) between targets in the far (\valSi{1.6}{1.8}{cm}) and the medium (\valSi{1.0}{.9}{cm}) zone as well as between the far and near (\valSi{1.1}{1.1}{cm}) zone. The difference between near and medium zones, however, was not significant ($p>.05$). Table \ref{tab:proximity/holding_distance} lists the mean overshoot and corresponding standard errors per target zone, figure \ref{fig:errorperinteractionzone} compares the values graphically.

The analysis showed neither any significant influences of the direction of interaction (\anovaWithoutEffect{1}{13}{1.65}{>.05}, flexion: \valSi{.7}{.8}{cm}, extension: \valSi{.8}{.8}{cm}) nor of the hand orientation (\anovaWithoutEffect{1}{13}{1.37}{>.05}, palm: \valSi{.8}{.8}{cm}, back: \valSi{.7}{.7}{cm}) on the holding error.

However, the analysis could show a significant (\anovaCorWithoutEffect{1.45}{18.85}{7.21}{<.001}{.29}) influence on the number of layers. Post-hoc tests confirmed a significant ($p<.01$) bigger holding error for 12 layers (\valSi{1.2}{.9}{cm}) compared to all higher numbers of layers. The mean holding error further decreased for increasing numbers of layers (min: \valSi{.6}{.4}{cm} for 72 layers), but was not significant.

\begin{table}[t!]
	\centering
	
\begin{tabularx}{\linewidth}{YYYYY}
	&   &   & \multicolumn{2}{c}{\textbf{95\% Confidence Interval}}\\
		\cmidrule(lr){4-5}
		\textbf{Zone} & $\pmb{\mu}$ & $\pmb{\sigma_{\overline{x}}}$ & \textbf{Lower} & \textbf{Upper}\\
		\midrule
		Near   & 1.1cm & .29cm    & .52cm                     & 1.68cm                    \\
		Medium & 1.0cm & .24cm    & .53cm                     & 1.47                      \\
		Far    & 1.6cm & .48cm    & .66cm                     & 2.54cm                    \\
		\bottomrule
	\end{tabularx}
	
	\caption{The holding error per distance zone (in cm). The table reports the mean value $\mu$, the standard error $\sigma_{\overline{x}}$ and the 95\% confidence interval CI.}
	\label{tab:proximity/overshoot_distance}
\end{table}

\subsubsection{Qualitative Results}

In general, all participants appreciated the idea of being able to interact with multi-layer information spaces through movements of their hand. There was a strong consensus among participants (11 out of 14) that this input modality is suitable for immediate and short-term interactions, such as the serendipitous discovery of contents, fast peeking into information or executing a shortcut. The participants' comments suggested that the convenient boundaries for interaction are approximately the near and middle zones in each direction. Far zones turned out to cause more fatigue on the arm and upper arm muscles.

Regarding the hand side, participants mentioned mixed opinions. Six participants preferred the back side of the hand for interactions. P9 commented, e.g., \enquote{I know this movement, that is like looking at my watch}. The remaining eight participants preferred interactions with the palm side of the hand. Participants did not feel an influence of the direction of the interaction. P13 commented on this: \enquote{Both directions are okay for me, as long as the target is not too far away or too close to my head}.

\section{Experiment II: Discrete Interaction}
\label{sec:proximity:exp2}

The following section presents the methodology (see section \ref{sec:proximitywatch:methodology}) and the results (see section \ref{sec:proximitywatch:results}) of a controlled experiment investigating discrete interactions in a multi-layered interaction space.

\subsection{Methodology}
\label{sec:proximitywatch:methodology}

\textfigH{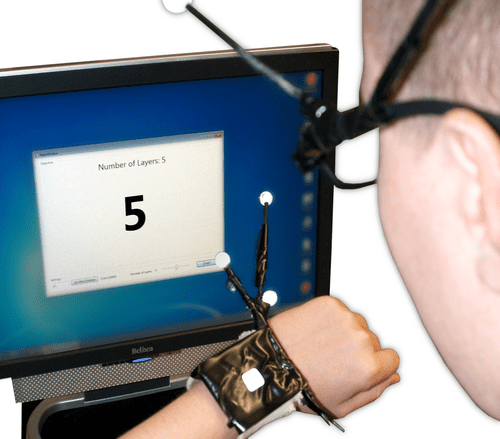}{The setup of the controlled experiment with two retro-reflective apparatuses mounted on the participant’s head and wrist, and the display showing the current task.}

This section presents the methodology of the second controlled experiment focusing on discrete interactions for proprioceptive interactions without visual feedback. More specifically, the second experiment addressed the following research questions:

\begin{enumerate}
	\item[RQ4] How accurate and efficient users can raise the hand to a given target position in the space in front of them without any visual feedback on their performance?
	\item[RQ5] Where are the targets located in the participants’ mental model?
\end{enumerate}

For this, 15 participants (5 female, 2 left-handed), aged between 19 and 30 years, were recruited. No compensation was provided.

\subsubsection{Design and Task}
The design of the experiment was similar to the design used in the first experiment. Again, the experiment defined a basic information space alongside the participants’ line of sight, evenly split into multiple layers and numbered in ascending order. The participants’ task was to raise their arm at a specified target layer without any visual feedback. The conditions varied the number of layers as an independent variable with integer values from 2 to 8. As the results from the first experiment did not show an effect of the hand side on any of the dependent variables, the second experiment disregarded the hand side as an independent variable. As in the first experiment, the system defined the maximum boundary of the interaction space as the participant’s individual arm-length and the minimum boundary as the near point of the human’s eye (not closer than 12.5cm to the user’s face). However, the investigator told the participants to use the space that is most comfortable for them as an interaction space.

The experiment used a repeated measure design with 7 levels for the number of layers (2, 3, \ldots, 7, and 8). For each level, the participants targeted each layer with 5 repetitions. This resulted in a total of $(2+3+4+5+6+7+8) * 5 = 175$ trials per participant. The order of conditions, as well as the order of targets within each condition, was counterbalanced using a Balanced Latin Square design (see section \ref{sec:rw:methodology}).

\subsubsection{Experiment Setup}

The system used an optical tracking system (OptiTrack) to measure the distance alongside the line of sight between the participant’s wrist and eyes. Participants wore a wristband on their non-dominant hand and a pair of glasses, each augmented with a set of retro-reflective markers, during the experiment (\reffig{fig:proxiwatch/setup}). A display in front of the participants showed the current task (layer subdivision and target layer within this subdivision). Additionally, a button was mounted within reach of the participant’s dominant hand. For each trial, the system recorded 

\begin{enumerate}
	\item the \emph{distance} between wrist and eyes after completing the task.
	\item the \emph{task completion time (TCT)} as the timespan between starting the trial until pressing the confirmation button.
	\item the \emph{target layer} of the condition.
	\item the \emph{total number of layers} in the current condition.
\end{enumerate}

\subsubsection{Procedure}

After welcoming the participants, the investigator introduced them to the concept and the setup of the experiment and asked them to put on the two trackable apparatuses. Then, the system was calibrated to adapt it to the respective arm length. Before each condition, the system informed the participants about the layer subdivision for this task. Each trial was started by asking the participant to stand relaxed and lower the non-dominant arm. Once ready, the participant pressed the button to start the trial.

After that, the system showed the target layer as a number from 1 (nearest layer to the body) to the highest layer of the current condition (2-8). Then, the participants raised their hand at the position where they imagined the respective layer. The investigator told the participants to look at the center of the trackable apparatus on their wrist in order to to keep the measured distances comparable. After raising their hand, participants had to confirm their action by pressing the nearby mounted button with their non-interacting hand. In the following, the system asked the user to take their hand down and enforced a 5-second break before starting the next trial.

The investigator told the participants to focus on the accuracy instead of the speed. Participants did not receive any feedback during the experiment. After each condition, participants took a 30-second break. The complete experiment took about 30 minutes for each participant.

\textfigStudybox{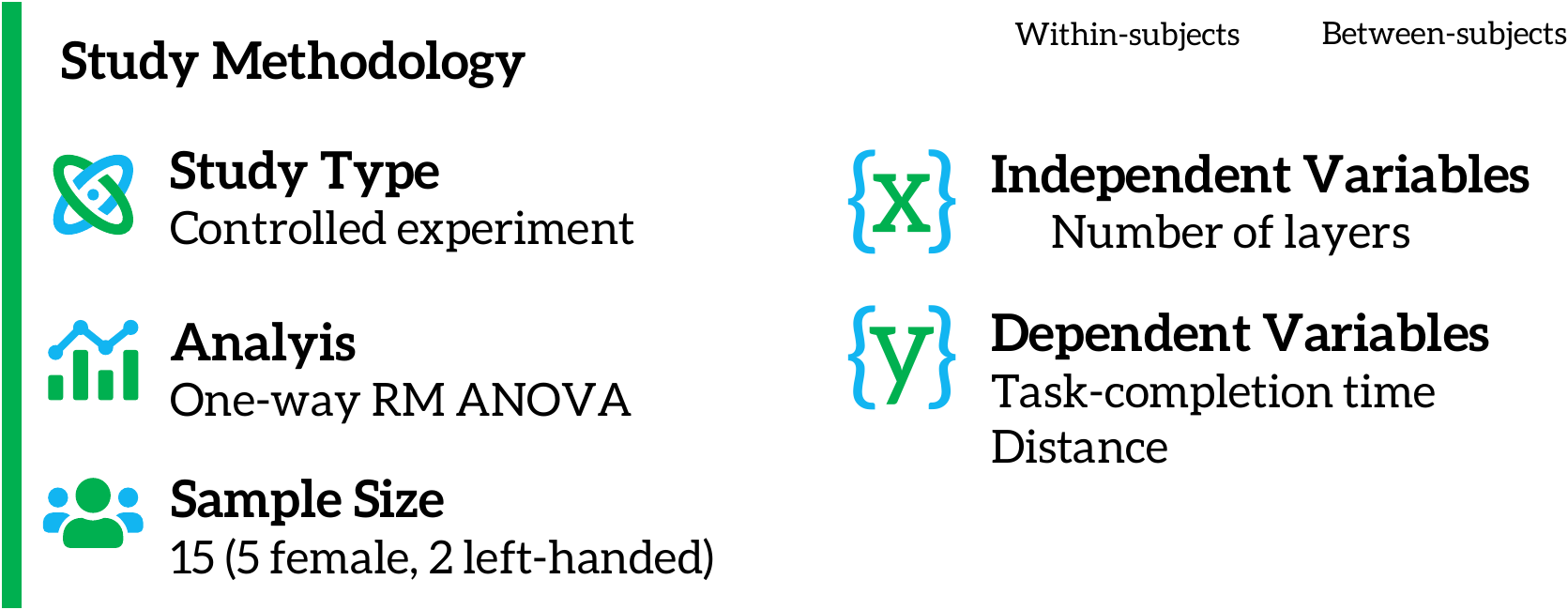}
\subsection{Results}
\label{sec:proximitywatch:results}

For the analysis, the recorded data was normalized to the respective arm length of the participant into a scale from $0\ldots1$. In this scale, $0$ refers to the near point of the human eye (\SI{12.6}{cm}) and $1$ to the arm length of the participant from shoulder to wrist. This maximum arm reach was measured in the calibration process of the system with the same optical tracking system.  The analysis of the data was performed as described in section \ref{sec:rw:methodology}. 

\subsubsection{Task Completion Time (TCT)}

The analysis showed a significant (\anovaCorWithoutEffect{2.92}{40.85}{5.17}{<.01}{.486}) influence of the number of layers on the \ac{TCT}. Post-hoc tests confirmed significantly rising \acp{TCT} for higher numbers of layers when comparing 2 to 4 and 6 (both $p<.05$), 7 ($p<.01$) and 8 ($p<.001$) layers. Further, the analysis showed a significant effect between 3 and 8 layers ($p<.05$). Table \ref{tab:proxiwatch/tct_post_hoc} lists the mean differences for all post-hoc comparisons together with the significant conditions. The analysis showed generally small \acp{TCT}, ranging from \valSi{1.7}{.14}{s} for 2 target layers to \valSi{2.37}{.14}{s} for 8 target layers. Table \ref{tab:proxiwatch/tct_mean} lists all mean \acp{TCT} for the conditions.

\begin{table}[t!]
	\centering
	
	\begin{tabularx}{\linewidth}{YYYY}
		
		\multicolumn{2}{c}{\textbf{Comparison}} & & \\
		\cmidrule(lr){1-2}
		\textbf{Layers} & \textbf{Layers} & $\pmb{\Delta	\mu}$ & \textbf{sig} \\
		\midrule
		2 & 4 & -.42s & *\\
		2 & 6 & -.46s & *\\
		2 & 7 & -.59s & **\\
		2 & 8 & -.64s & ***\\
		3 & 8 & -.44s & *\\
		\bottomrule
	\end{tabularx}
	
	\caption{The post-hoc tests for the mean \acp{TCT} between the tested numbers of layers. Only the significant comparisons are listed. \textbf{sig} denotes the significance level: * $p<.05$, ** $p<.01$, *** $p<.001$}
	\label{tab:proxiwatch/tct_post_hoc}
\end{table}

\begin{table}[b!]
	\centering

\begin{tabularx}{\linewidth}{YYYYY}
	&   &   & \multicolumn{2}{c}{\textbf{95\% Confidence Interval}}\\
	\cmidrule(lr){4-5}
	\textbf{Nr of Layers} & $\pmb{\mu}$ & $\pmb{\sigma_{\overline{x}}}$ & \textbf{Lower} & \textbf{Upper}\\
	\midrule
	2    & 1.73s & .14s     & 1.44s                     & 2.02s        				\\
	3    & 1.93s & .14s     & 1.64s                     & 2.22s                     \\
	4    & 2.15s & .14s     & 1.86s                     & 2.44s         			\\
	5    & 2.07s & .14s     & 1.78s                     & 2.36s                     \\
	6    & 2.20s & .14s     & 1.91s                     & 2.48s                     \\
	7    & 2.32s & .14s     & 2.03s                     & 2.61s                     \\
	8    & 2.37s & .14s     & 2.08s                     & 2.66s						\\ 
	\bottomrule
\end{tabularx}

	\caption{The Task Completion Times (TCT, in seconds) for different numbers of layers. The table reports the mean value $\mu$, the standard error $\sigma_{\overline{x}}$ and the 95\% confidence interval.}
	\label{tab:proxiwatch/tct_mean}
\end{table}

\subsubsection{Personal Interaction Space}

\textfigH{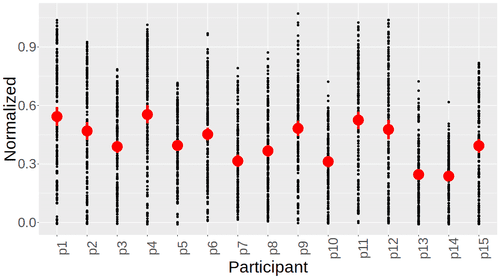}{The interaction space of the participants. Black dots show the recorded distances normalized to the arm size, and red dots show the center point of the interaction space. The used space differs significantly between participants.}

The investigator told the participants to use the area and to separate the interaction space into layers in a way that is convenient for them. The analysis showed that the interaction space used by the participants as well as the center point of all interactions differs significantly (\anovaWithoutEffect{14}{84}{8.6}{<.001}) between participants (Min: P14, $.0-.51$ with one outlier, center point of interaction $\mu=.24$, Max: P4, $.0-.98$, center point of interaction $\mu=.54$). Figure~\ref{fig:proxiwatch/interactionSpace} shows the interaction space of all participants normalized to their arm length. This personal interaction space for each participant remained constant for different layer subdivisions. For all further evaluations, the data was scaled based on the personal interaction space of each participant with 0 as the closest and 1 the most distant data point.

\subsubsection{Directly Accessible Layers}

\textfigH{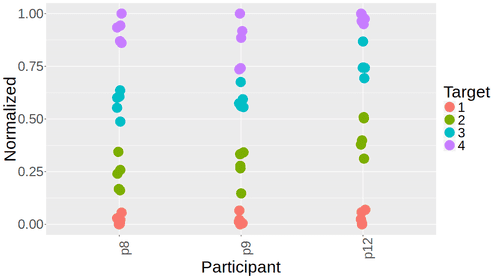}{The distances for three exemplary participants for layer subdivision $n=4$, scaled to their personal interaction space. Finding: A global model to classify points to a target layer is not possible. However, individual models per user to classify point with regards to the target layer seems feasible.}

The analysis showed that the size and the location of the center points of the layers differ significantly (Size: \anovaWithoutEffect{14}{98}{2.6}{<.01}, Location: \anovaWithoutEffect{14}{98}{6.9}{<.001}) between participants even after scaling the data to the personal interaction space of each participant. A generalized model that is able to map points from every user into the respective target layer is, therefore, not feasible for layer subdivisions $>2$, as the layers largely overlap between participants. Within the data of individual participants, however, a more fine-grained differentiation between the layers can be archived with no overlapping layers for a subdivision of at least 4 for all participants. As an example of this finding, figure~\ref{fig:proxiwatch/targetZones4} shows the data points for three participants for a subdivision of four layers. The analysis further showed smaller layers for the outer regions (i.e.,\ close to and far away from the body) compared to the inner regions (Inner: $\mu=.16,~\sigma=.06$, Outer: $\mu=.12,~\sigma=.07$). This is not influenced by the personal interaction spaces of the participants.

\section{Discussion and Guidelines}
\label{sec:proximity:discussion}

The quantitative, as well as the qualitative results of both experiments, indicated that one-handed proximity-based interactions could be a viable interaction technique for \acp{HMD} that can support convenient and fun interactions for both, discrete and continuous interactions. Based on the results of the experiments and the related works, this section proposes three guidelines for the future design of such interfaces.

\subsection{Partition the Space by Layer Thickness}
\label{sec:proximity:discussion:guideline1}

\begin{figure}[h!]
	\centering
	\includegraphics[width=\textwidth]{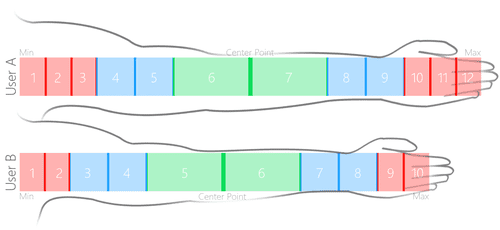}
	\caption{\emph{Design Guideline: Partition the space by layer thickness} The interaction space should be designed based on the absolute layer thickness, not on a desired number of layers to account for different arm sizes.}
	\label{fig:proximity:guideline:1}
\end{figure}

The results indicate that various sizes of participants' arms do not influence the accuracy - measured as an error of absolute distance - of hand movement. Therefore, for users with smaller arms, too many and, thus, thin layers would decrease the accuracy. On the other hand, for taller users with greater arm length, insufficient numbers of layers would result in greater traveling distances and, therefore, decreased efficiency. 

Hence, as a result of the experiment, the interaction space should be designed based on the layers specific thickness. This way, the design results in different numbers of layers for different arm sizes, allowing the user to interact within the borders of their physical abilities (see figure \ref{fig:proximity:guideline:1}).

\subsection{Use an Uneven Layer Thickness}
\label{sec:proximity:discussion:guideline2}

\begin{figure}[h!]
	\centering
	\includegraphics[width=\textwidth]{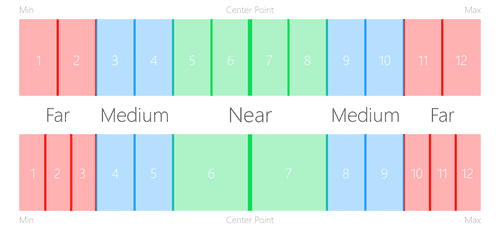}
	\caption{\emph{Design Guideline: Use an uneven layer thickness} The interaction space should be designed with a decreasing layer size towards the outer layers to account for the higher overshooting errors found in the experiment.}
	\label{fig:proximity:guideline:2}
\end{figure}

The traveling distance of the hand proved to be the most critical factor to influence the depended values in both experiments. The first experiment focusing on continuous interactions showed that the typical overshooting error decreases towards outer regions. Furthermore, the second experiment indicated that the outer layers of the mental model of participants were smaller compared to the inner layers. Therefore, this chapter proposes to use uneven and descending layer thicknesses towards outer regions (see figure \ref{fig:proximity:guideline:2}).

This layer subdivision allows for smaller layers in outer regions without increasing the interaction time that is introduced due to overshooting the target. Based on the quantitative results, a layer thickness of 7.8 cm for near, 4.2 cm for medium and 3.0 cm for far targets (the respective mean overshoot plus the double standard deviation) would result in $>$95\% accuracy for all traveling distances.

\subsection{Respect the Personal Interaction Space}
\label{sec:proximity:discussion:guideline3}

\begin{figure}[h!]
	\centering
	\includegraphics[width=\textwidth]{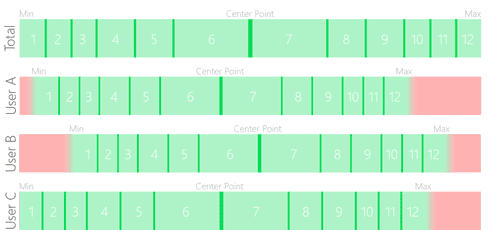}
	\caption{\emph{Design Guideline: Respect the Personal Interaction Space} The experiment showed differing convenient interaction spaces for each participant. This should be reflected by a system implementation.}
	\label{fig:proximity:guideline:3}
\end{figure}

The experiment showed that different users have different personal convenient interaction spaces. These convenient interaction spaces are not generalizable over multiple users. Thus, a system should not force the user into a fixed set of layers that spans larger or smaller than the user's personal interaction space (see figure \ref{fig:proximity:guideline:3}).

Therefore, a general model over all users is not feasible and, thus, a personal model is necessary to achieve high recognition rates for higher subdivisions than two.

\subsection{Focus on the Convenient Range} 
\label{sec:proximity:discussion:guideline4}

\begin{figure}[h!]
	\centering
	\includegraphics[width=\textwidth]{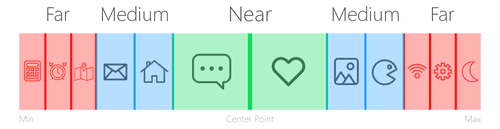}
	\caption{\emph{Design Guideline: Focus on the Convenient Range} Use the near and medium layers for frequent and common interactions, use the far layers for irreversible interactions.}
	\label{fig:proximity:guideline:4}
\end{figure}

The qualitative feedback from participants showed that interactions in the far zones are less convenient compared to the closer regions. Therefore, we propose to focus on the near and medium zones for frequent and common interactions. (see figure \ref{fig:proximity:guideline:4}) 

As proposed by \typcite{Benford2012}, the slightly uncomfortable hand position in the far zones can be leveraged for important and not reversible actions such as deleting a file or sending an e-mail.

\section{Use Cases and Applicability}
\label{sec:proximity:applicability}

This section presents use cases for the usage of one-handed proximity-based interaction for \acp{HMD} and, as an example for the applicability beyond \acp{HMD}, for interacting with a smartwatch.

\subsection{Use Cases for Interacting with Head-Mounted Displays}

A real-world system supporting proximity-based one-handed interaction can leverage the space in front of the user to present a layered information space (virtually augmented on the user's palm). Similar to \ncite{Ha2014}, the hand's 3D features can be extracted from an RGB-D attached to the head-mounted display. This section presents four possible use cases to demonstrate the design space of such proximity-based interaction.

\textfig{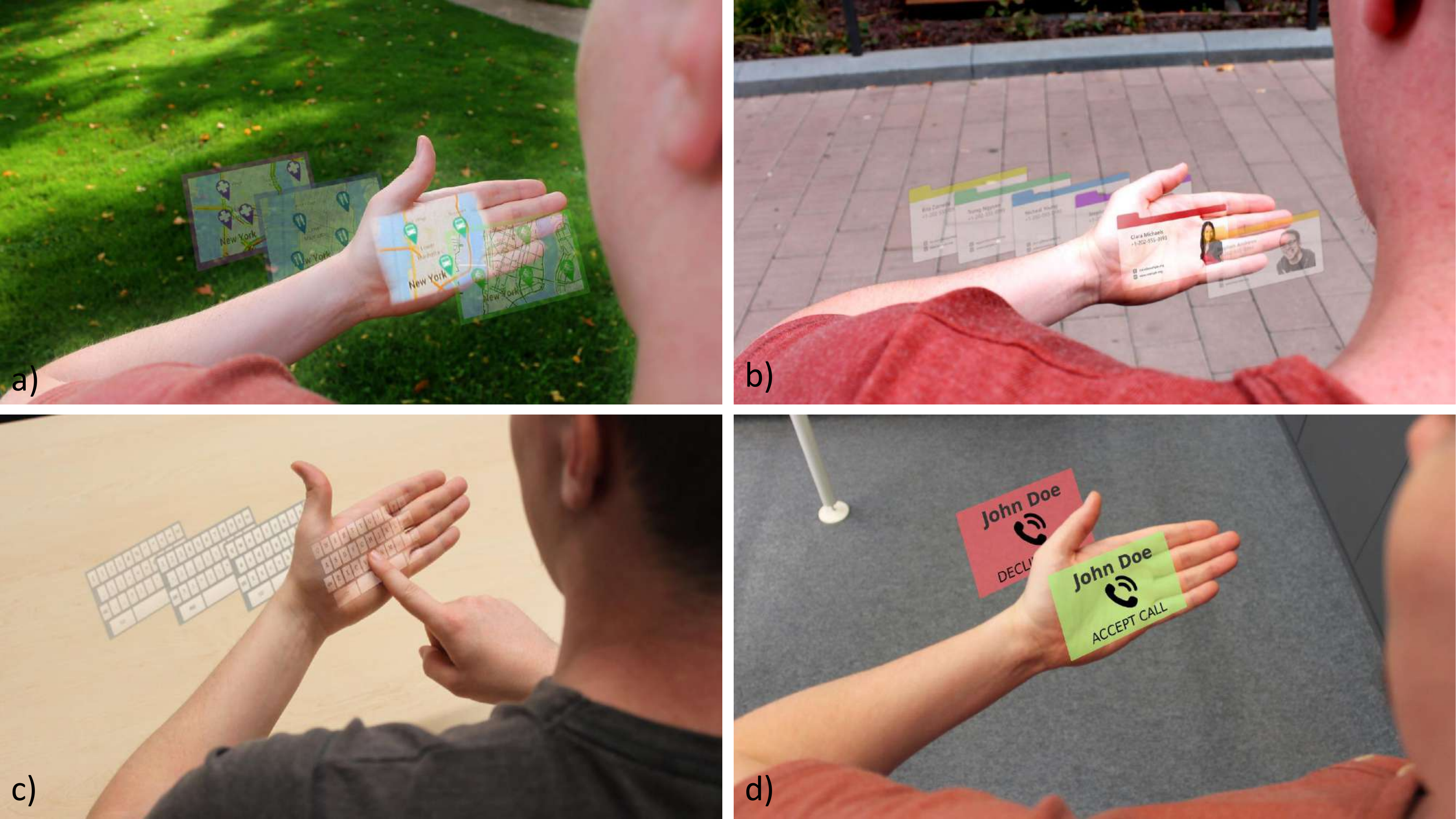}{Example useases for proximity-based interaction with \acp{HMD}. Proximity-based interaction can be used to browse through different views and visualizations of a data set (a) or to scroll through a list of items (b). Additional touch input can be used to interact with different visualizations (c). Further, shortcuts can be used to perform actions by raising the arm at certain distances from the body (d).}

\subsubsection{Exploring Information Layers}

Proximity-based interaction can be used to explore multidimensional information structures in which different information layers refer to the same underlying data set. 

A map application can serve as an example of this type of interaction. While the underlying data set to be displayed does not change, a different view of the data can be selected by moving the hand. This process allows the user to switch through different information layers quickly - e.g., elevation map, topographic map, traffic intensity - without needing a second hand (see figure \ref{fig:proximity/usecases} a). According to guideline 2 and 4 (see section \ref{sec:proximity:discussion:guideline2} and \ref{sec:proximity:discussion:guideline4}), the most frequently used information layers are presented in the near and medium zones and occupy a larger part of the interaction space then the other layers.

\subsubsection{Browsing Lists}

As a second example, proximity-based interaction can be used to scroll through list structures by moving the hand, following the metaphor of scrolling through a file cabinet. 

Two implementations are possible, depending on the size of the list:  For short lists, list elements can be bound directly to layers of the information space, allowing direct access of elements.  For longer lists, a mapping can be used, where the inner (near) layers are used to select a list element, while the outer (medium and far) layers are used for slow or fast scrolling in the respective direction (see figure \ref{fig:proximity/usecases} b). According to guideline 1 (see section \ref{sec:proximity:discussion:guideline1}), the number of layers displayed depends on the arm length of the user to ensure fast and accurate interactions.

\subsubsection{Combination with Touch-Input}

In addition to the selection of elements through proximity-based interaction, methods of on-body interaction can be used to realize additional touch-input on the respective layers.

As an example, imagine a proximity-enabled keyboard. The proximity-dimension can be used to select different layers (e.g., small letters, capital letters, numbers) of the keyboard. On each layer, the opposing hand can provide touch input to select keys (see figure \ref{fig:proximity/usecases} c).

\subsubsection{Discrete Interaction for Shortcuts}

In addition to the continuous interaction by the movement of the hand through the interaction space, this thesis explored discrete interactions by raising the hand at specific distances from the user's body. Such interactions can be performed directly and are based only on raising the hand; no additional interaction is necessary. Therefore, this type of interaction is suitable for fast shortcuts. According to guideline 3 (see section \ref{sec:proximity:discussion:guideline3}), the position of the shortcut layer is tailored to the user in order to adapt to their personal interaction area.

As an example, such discrete shortcuts can be used to accept or decline incoming calls without the need of a second hand to interact (see figure \ref{fig:proximity/usecases} d).

\subsubsection{Conclusion}

This section presented four use cases to demonstrate the applicability of proximity-based interaction with \acp{HMD} in different situations. In combination with other input modalities such as touch or further proximity dimensions, further styles of interaction ar possible. Section \ref{sec:proximity:limitations} discusses these possible extensions.

\subsection{Use Cases for Smartwatches}

The two styles of interaction introduced - discrete and continuous - can also support other device classes beyond \acp{HMD} for immediate and joyful interactions. As an example, this section presents a proof of concept for the applicability of the concepts for interaction with smartwatches.

A closer look reveals that there are great similarities between the challenges found in interacting with smartwatches with the challenges of interacting with \acp{HMD}: Interface elements cannot be touched (or only with difficulty), the interaction takes place in a mobile context with a possibly occupied hand. Therefore, this section explores how the presented proximity-based interaction techniques can be transferred from the field of \acp{HMD} to Smartwatches. Based on the findings presented earlier in this chapter, this section introduces ProxiWatch: A one-handed proximity-based hand input modality for smartwatches along with two main interaction techniques. Further, this section presents the design of two example applications to show the usefulness for varying scenarios.

\subsubsection{Motivation and Background}

Highly capable smartwatches have become an emerging class of wearable devices that allow ubiquitous and mobile interaction with digital contents. Such devices usually consist of small multi-touch displays, bundled together with computing and sensing hardware of a smartphone, worn on the user's wrist. Therefore, smartwatches allow users to see, access, and modify information right at their wrist, anytime, and anywhere. The screen size of such devices is a trade-off between wearing comfort and interaction space. On the one hand, the display should be big enough for meaningful touch interaction. On the other hand, bigger display sizes result in bulky devices and reduced wearing comfort. The evolution of wearable devices shows a trend towards small and elegant devices with small interaction space~\ncite{Ni2009}.

Therefore, traditional interaction techniques are not directly applicable to smartwatches. Current consumer devices (e.g., Apple Watch, Android Wear, Pebble Watch) mainly focus on 1) touch-based interfaces, 2) physical input controls on the frame of the device (e.g., a digital crown, buttons) and 3) off-device input modalities such as voice input. While practical and useful, those styles of interaction have various drawbacks in the context of smartwatches. Traditional touch-based interaction techniques suffer from small screen sizes as the user's interacting finger occludes a big part of the screen. %
Physical input controls such as a digital crown allow interacting with the content without occluding the screen; however, these approaches do not support direct targeting and selecting of UI elements. In addition, touch interfaces, as well as physical input controls, require both hands of the user and, thus, may diminish the user experience in situations where the user is encumbered~\ncite{Ng2014}. Voice input lacks direct manipulation and is difficult to use in noisy environments~\ncite{Starner2002} (see also section \ref{sec:rw:hmd:voice}).

In recent years, work is emerging that addresses these interaction challenges: Research presented on-device input modalities beyond traditional touch-based interfaces using a finger-mounted stylus~\ncite{Xia2015} or tapping gestures~\ncite{Oakley2015} on the device. As another approach, off-device input modalities have been proposed that increase the interaction space of such devices by leveraging the space around the device: Using infrared, acoustic or magnetic sensors embedded into the frame, the device can track the location of the fingertip of the user's dominant hand. This tracking allows users to perform off-screen (multi-)touch~\ncite{Butler2008, Harrison2010} or air gestures~\ncite{Harrison2009, Kim2007, Knibbe2014} around the device or on surrounding surfaces~\ncite{VanVlaenderen2015} and, thus, without occluding the content on the screen. Despite the advantages, the presented approaches still require both hands of the user. As another approach, one-handed interfaces have been proposed that allow users to trigger a set of actions by performing gestures with their finger or hand~\ncite{Kerber2015, Knibbe2014, Rekimoto2001, Xu2015a} of the arm wearing the watch. While such interfaces can be operated with one hand, they do not support continuous interactions.

\subsubsection{Proxiwatch: Concept}
\label{sec:proxiwatch:concept}

When interacting with smartwatches, proximity-based interaction can support the user in overcoming the problems of the small touch area and speech interaction discussed above. This section presents two interaction styles, that can serve as an additional input modality for smartwatches.

\textfigH{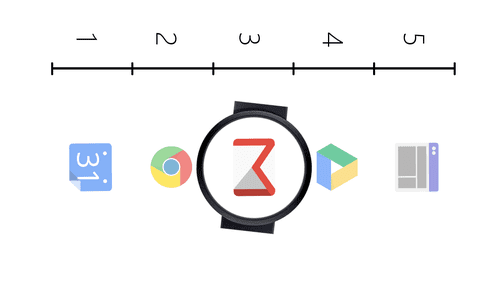}{Example application for discrete interaction: An application launcher that allows users to launch applications by raising their arm.}

\paragraph{Discrete Interaction}

Discrete interaction is based on the directly accessible layers, as presented in section \ref{sec:proximity:exp2}. Within this set of layers, each layer can be mapped to a functionality in terms of a shortcut. Such a system can be used to directly select various items in a fixed set of objects (e.g., launcher, media player controls) by raising the arm within the boundaries of one of the target zones in front of the body. The analysis of the experiment presented above found that 4 layers are easily distinguishable for users. As an example for this style of interaction, this section focuses on an application launcher for a smartwatch. This launcher allows users to open up a favorite application by raising the arm (\reffig{fig:proxiwatch/apps}). Based on guidelines 2 and 4 (see section \ref{sec:proximity:discussion:guideline2} and \ref{sec:proximity:discussion:guideline4}), the most frequently used applications are presented as larger layers and close to the center of the interaction space. Further, based on guideline 3 (see section \ref{sec:proximity:discussion:guideline3}), the position of the shortcut positions is tailored to the user in order to adapt to their personal interaction area.

\textfigH{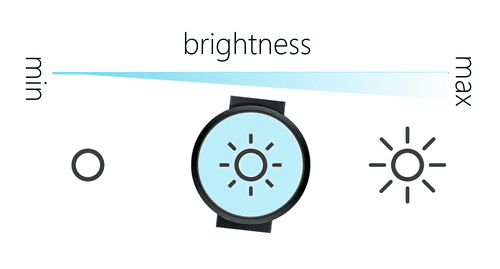}{Example application for continuous interaction: A brightness control that lets the user modify the value by moving the arm alongside the line of sight.}

\paragraph{Continuous Interaction}

Continuous interaction allows the user to adjust a continuous value by moving the hand within the bounds of the personal interaction space, as presented in section \ref{sec:proximity:exp1}. Through visual feedback on the smartwatch, the user is able to quickly and efficiently adjust a value (e.g.,\ slider) just by moving the arm. For smartwatches, this style of interaction  can be used as an extension to discrete interaction for lists with more elements. ProxiWatch illustrates this technique with a brightness control for the smartwatch that allows users to adjust the value by moving the hand. The user can select a value by lowering the hand. Based on guideline 3 (see section \ref{sec:proximity:discussion:guideline3}), the part of the interaction area used for adjusting values is adopted to the user's personal interaction space. 

Both interaction techniques presented can be combined (e.g., discrete interaction to launch an application, continuous interaction to adjust a value in this application). In addition, traditional touch-based input is possible on each layer.

\paragraph{Implementation and Prototype}
\label{sec:proxiwatch:implementation}

This subsection illustrates the building process of a stand-alone, wireless prototype to enable proximity-based interaction concepts on a consumer smartwatch (Motorola Moto 360) in an iterative design process (\reffig{fig:proxiwatch/prototypes}).

The implementation of the prototype was based on a battery powered Arduino Nano with two infrared distance sensors (Sharp GP2Y0A21YK0F). Pre-tests indicated that a single distance sensor cannot cover the needed field of view to use the complete degree of freedom provided by the elbow joint. Therefore, the final design contained two distance sensors, slightly tilted relative to each other (\reffig{fig:proxiwatch/prototypes} f). When holding the hand in a rotation that allows to read the display of the watch, both sensors were directed towards the body of the user. The system reported the value of the first sensor as long as this sensor has the body of the user within its field of view. Otherwise, the value of sensor two was reported (\reffig{fig:proxiwatch/distance}). Depending on the body structure of the user, the handover point was found to be around \SI{20}{cm} distance. Because of the limited processing capabilities of the Arduino, the system transmitted the raw sensor data to a processing application on a mobile phone (Samsung Galaxy S4) via Bluetooth. The phone handled the incoming raw data and selected the appropriate sensor. In addition, the system used a Kalman filter to reduce the statistical noise of the returned sensor values. After processing, the phone sent the estimated distance to the smartwatch. To detect if a user raised the arm, the system used the acceleration sensor of the smartwatch. This allowed the system to support the discrete interaction technique presented above. Furthermore, the acceleration sensor was used to register a shake-wrist gesture, which can be used for secondary actions (i.e., select item).

\textfig{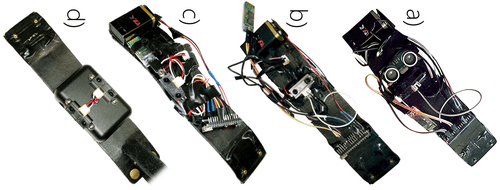}{Iterative design of the prototype: a) First version using a generic ultrasonic and infrared sensor b) second version using two infrared sensors c) third version with tilted infrared sensor alignment d) current version with boxed design.}

To confirm the viability of the prototype, the estimated distance values of the prototype were compared to the real distance (measured using an optical tracking system). The recorded data showed that the prototype robustly recognizes the distance to the user for the complete interaction space (average deviation \valSi{1.9}{1.19}{cm}, \reffig{fig:proxiwatch/distance}). Future work is needed to improve the quality of the measurements and to further miniaturize the prototype.

\textfigH{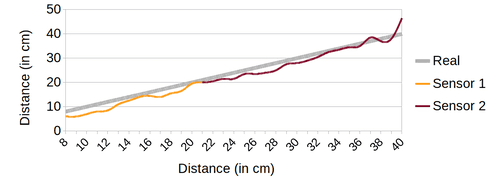}{The estimated distance values reported by the two sensors compared to the real distance. The handover between both sensors is at $\sim$21cm.}

\section{Limitations and Future Work}
\label{sec:proximity:limitations}

The study design, as well as the results of the controlled experiment, hint at some limitations and directions for future work. This section lists these points and provides possible solutions.

\subsection{Beyond the One-Dimensional Interaction Space}

This chapter focused on rectilinear interactions alongside the user's line of sight. This limitation was deliberately chosen in order to provide a sound and rigorous assessment of human capabilities for such types of interaction. However, the other degrees of freedom provided by the shoulder, elbow and wrist joints allow for a multitude of other movements and, thus, other proximity-based interactions in front of the user, that may be beneficial for future use of \acp{HMD}. The design space opened here offers many starting points for future contributions in this area.

\subsection{Combination with other Input Modalities}

This chapter mainly focused on the \emph{selection} of information. Future systems will, therefore, also require means to \emph{manipulate} information. There are several approaches to address this: First, the second hand could be used for on-body touch input on the currently active layer, similar to~\ncite{Dezfuli2012}. Second, proximity-based interaction could be supplemented, similar to~\ncite{Whitmire2017}, with finger gestures of the same hand: Users can use their thumbs to provide discrete and continuous input on the remaining fingers or the palm. Furthermore, a combination with further input modalities would be conceivable, e.g., eye tracking.

\subsection{Anchor Point of Proximity}

This chapter focused on interpreting the proximity between the user's hand and head as an input modality. There are also other possible anchor points of proximity conceivable, such as the abdomen or chest of the user. The relocation of the anchor point could alleviate the problem of unpleasant interactions directly in front of the user's face. In addition, multiple anchor points could be active simultaneously to provide the user with multiple dimensions for interaction. For future use of such interfaces, it is, therefore, necessary to establish an understanding of the influence of the anchor point on the accuracy, efficiency and overall user experience of such interfaces.

\section{Conclusion}

This chapter explored an interaction design for one-handed and proximity-based gestures for interacting with \acp{HMD}. More precisely, this chapter investigated the efficiency and accuracy of the interaction in a layered information space for 1) continuous interaction and 2) discrete interaction. The results confirmed the viability and feasibility of this input modality. The traveling distance to the target layer proved to be the primary influence for accuracy and efficiency. 

This chapter added to the body of research in multiple areas:

\begin{enumerate}
	\item This chapter considered the spatial position of the hand in relation to the body as an input dimension for \acp{HMD}. This style of interaction has not been considered by previous work on interacting with \acp{HMD}. Thus, the work presented in this chapter opens up a new field of research for future body-based and one-hand interfaces for \acp{HMD}.
	\item This chapter contributed the design and results of two controlled experiments assessing discrete and continuous proximity-based interaction in a layered information space in front of the body. The results confirmed the viability of this interaction style for accurate and fast interactions. The results, as well as the guidelines derived from the results, provide a reliable basis for the future use and further refinement of such interaction techniques.
	\item  Furthermore, this chapter presented four use cases for proximity-based interactions with \acp{HMD}. Additionally, the chapter demonstrated how the presented interaction techniques could be implemented for the interaction with a smartwatch. Thereby, the chapter highlighted that the presented body-based interaction technique could also be utilized in other areas beyond \acp{HMD}. 
	
\end{enumerate}

\subsection{Integration}

\textfigH{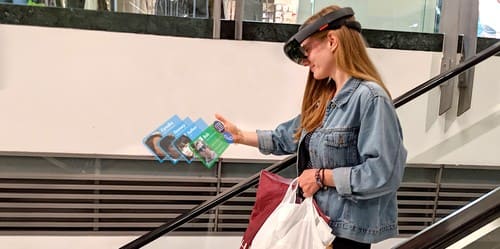}{Alice uses proximity-based interaction to select a phone contact.}

\interactionbox{alice_proximity}{At the Mall}{
	Alice is at the mall carrying a shopping bag. She wants to take a break and thinks of her friend Bob, who lives nearby. She decides to call Bob. For this, she raises her free hand and scrolls through her contact list by moving her arm back and forth (see figure \ref{fig:proximity/alice}). After she has found the contact, she confirms the call by clicking with her thumb. Bob wants to have some coffee. Alice uses her free hand to scroll through the list of applications and selects the map application. In the following, she uses proximity-based interaction to set filters on coffee shops and set the radius of the card. The two agree on a coffee shop and set off.
	
	The entire interaction takes place one-handed and allows Alice to interact quickly and concurrently with information while still holding the shopping bag in her hand.
}

The interaction technique presented in this chapter allows for fast and direct interaction with \emph{body-stabilized} interfaces, leveraging our \emph{upper limbs} for one-handed input. In particular, the one-handed operation allows mobile interaction with information, while the second hand is still available for interaction with the real world (supporting \mobilityLower{}). The two interaction styles presented allow the operation of different common types of interfaces, such as cascading menus (\discreteLower{}) or sliders (\continuousLower{}). Therefore, the chapter contributes to the vision of ubiquitous \aroundbodyinteraction{} with \acp{HMD} as introduced in section \ref{sec:introduction:aroundbodyinteraction}.

As described in the section \ref{sec:proximity:limitations}, the design space of interactions with \emph{body-stabilized} interfaces using the \emph{upper limbs} exhibits much more degrees of freedom than could be covered in the scope of this work. The extension of the concept by further input dimensions (i.e., 2D or 3D movements) as well as the connection with on-body touch input (carried out by the fingers of the same hand) could enhance the expressiveness while potentially maintaining the presented advantages of proximity-based interactions. While these ideas were outside the scope of this work, the ideas and quantitative results presented in this chapter can serve as a reliable baseline for further explorations of the design space.

\subsection{Outlook}

Interfaces registered to the body of the user have an inherent focus on \singleuserLower{} as they move together with the user by design, rendering such interfaces unsuitable for multiple users. Chapter \ref{ch:cloudbits} discuss how \emph{world-stabilized} interfaces can be used to support such \multiuserLower{} situations.

Furthermore, interaction with the \emph{upper limbs} cannot support situations in which both hands are occupied, e.g., when something is carried in both hands. Chapters \ref{ch:cheesyfoot} and \ref{ch:walktheline} present solutions for such situations leveraging the \emph{lower limbs}.
}

\chapter[CloudBits]{CloudBits: Supporting Conversations Through Augmented Zero-query Search Result Visualization}\label{ch:cloudbits}
{

\cptteaser{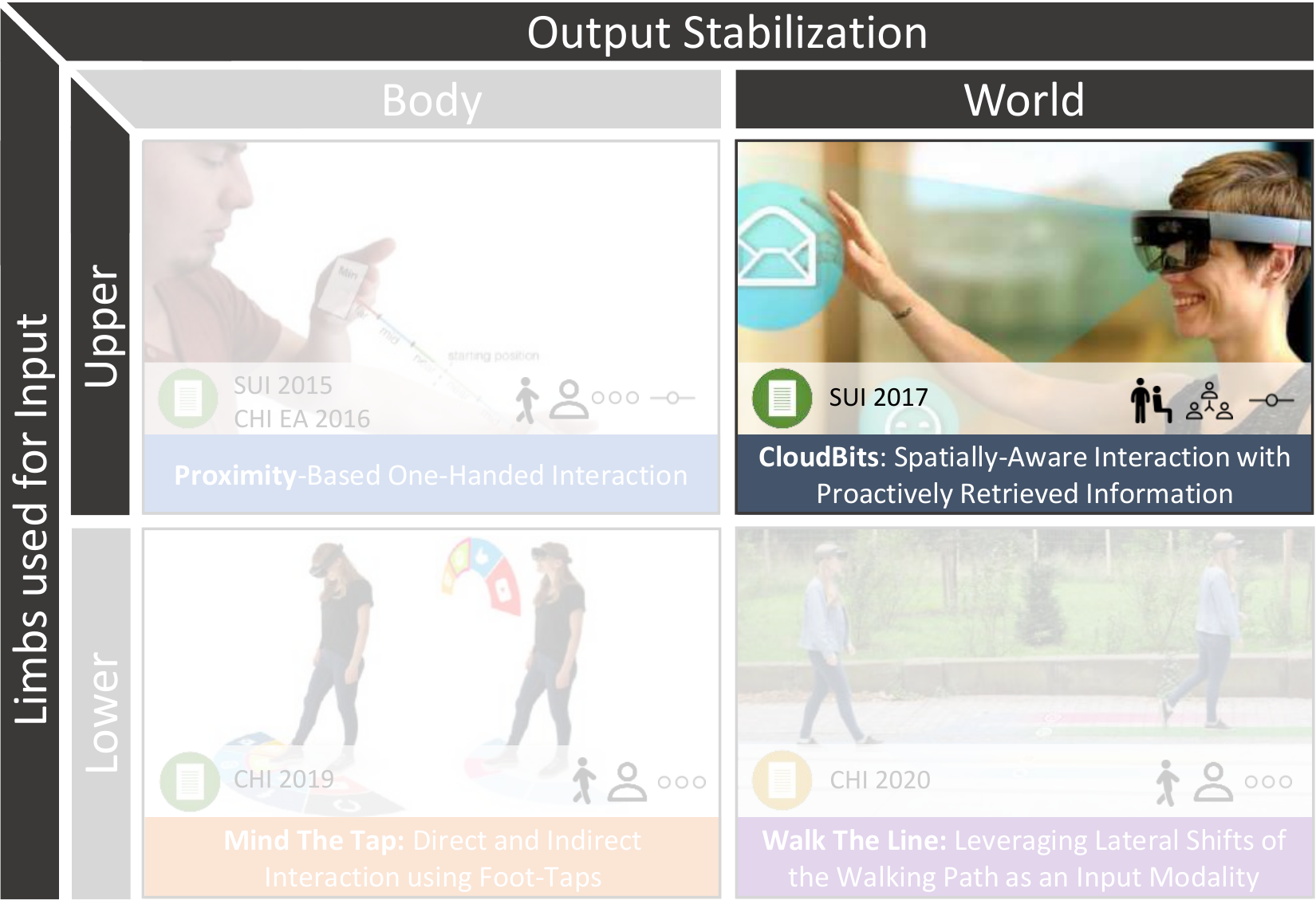}{This chapter presents an interaction technique for world-stabilized interfaces leveraging the upper limbs for input.}

The previous chapter explored how the degrees of freedom offered by the \emph{upper limbs} and, in particular, the elbow joint can be used for interactions with a \emph{body-stabilized} interface. The presented interaction technique registers the visualization to the body (in this specific case, the palm) of the user and, thus, enables the interface to be available anytime and anywhere for fast and immediate interactions.

This chapter shifts the stabilization point of the visualization from the user's body to the real world, focusing on interaction techniques for \emph{world-stabilized} interfaces (see section \ref{sec:relatedwork:hmds:implementation}) which are still operated using the \emph{upper limbs} (see figure \ref{fig:cloudbits/overview_cloudbits}). Since information is no longer bound to the position and movements of the user (or specific body-parts), the information can be freely positioned in space, and the user can view it from different perspectives by moving around. This positioning enables new use cases for \inplaceLower{} scenarios that give meaning to the spatial location of information: Like documents in the real world, users can sort and group information and use the spatial layout to add meta-information to the actual information implicitly. The \emph{world-stabilization} is particularly relevant for \multiuserLower{} scenarios if the information is anchored to the same real-world position for all participating users, allowing collaborative interaction with information. To gain insights into this problem domain, this chapter focuses on co-located conversation and meeting scenarios as an exemplary application domain, where users frequently interact with public and private information. Despite the focus on this application domain, the resulting interaction technique can as well as the insights gained can also be transferred to other domains, as highlighted in section \ref{sec:cloudbits:limitations:usecase}.

This chapter 1) contributes the results of an exploratory study investigating the requirements for the design of a user interface to support the interaction with information during conversations. Based on the results of the study, this chapter 2) presents an interaction technique for collaborative interaction with \emph{world-stabilized} interfaces using the \emph{upper limbs}, along with a prototype implementation. Finally, this chapter 3) reports the findings of a qualitative evaluation of the interaction techniques and concludes with guidelines for the design of such user interfaces.

The remainder of this chapter is structured as follows: First, this section presents a review of the related works in the area of \acp{HMD} in social interactions (section \ref{sec:cloudbits:relatedwork:hmd}). Second, based on a focus-group study to establish requirements and the review of related works, the chapter introduces an interaction technique to support collaborative interactions with information (section \ref{sec:cloudbits:introduction}). In the following, section \ref{sec:cloudbits:study2} presents the methodology and section \ref{sec:cloudbits:study2results} reports the results of the lab study on the behavior of users when interacting in a shared information space. Based on the results, section \ref{sec:cloudbits:guidelines} provides guidelines for the future use of such interfaces. The chapter concludes with a discussion of limitations and with guidelines for the future use of such interfaces (section \ref{sec:cloudbits:limitations}).

\newpage

\cptIntroBox{muller2017cloudbits}{The student \emph{Azita Hosseini Nejad} implemented the study client application and supported the conduct of the study. \emph{Sebastian Günther}, \emph{Niloofar Dezfuli}, \emph{Mohammadreza Khalilbeigi} and \emph{Max Mühlhäuser} supported the conceptual design and contributed to the writing process.}
\section{Related Work}
\label{sec:cloudbits:relatedwork:hmd}

The related works regarding interaction with \acp{HMD} were covered in chapter \ref{ch:relatedwork} already. The related works concerning interaction using the upper limbs were addressed in section \ref{sec:proximity:rw}. This section further adds related work in the area of \acp{HMD} in social interaction.

\subsection{Head-Mounted Displays in Social Interaction}

As discussed in section \ref{sec:relatedwork:hmds}, \acp{HMD} are a promising technology for immediate and direct interaction with information. Despite all benefits, research showed that the use of such interfaces introduces problems in social interactions: The form factor of \acp{HMD}, as well as the visibility of the \ac{HMD}’s output restricted to the wearer only, can have a negative impact on attentiveness, concentration, and eye-contact, and, thus, lead to less natural conversations~\ncite{McAtamney2006, Due2015}.

While some of the presented problems can be solved through technological advances (e.g., better eye contact through less bulky devices), other problems (e.g., the private experience of information) are inherently connected to the use of \acp{HMD}. As a possible way to increase social acceptance and to mitigate for these effects, \typcite{Koelle2015} showed that the offering of awareness to \enquote{communicate the intention of use} helps to build interfaces that overcome problems presented above. Following this stream of research, \typcite{Gugenheimer2017} proposed interaction techniques between \acp{HMD} persons wearing \acp{HMD} and external persons and \typcite{Chan2017} showed how an additional display on the \acp{HMD} could help to involve other persons.

As another approach to overcome the effects of the inherently private display, several \acp{HMD} can synchronously access a shared information space (i.e., Interface elements are displayed for multiple users at a synchronized real-world position) as proposed by Billinghurst et al~\ncite{Billinghurst1998a, Billinghurst2000}. Based on this inspirational work, such shared information spaces have been used in various contexts such as gaming~\ncite{Szalavari1998} or learning~\ncite{Bacca2015}. Recent advances in computer vision allow remote attendees to seamlessly integrate into a conversation as a lifelike 3D model, overcoming spatial boundaries ~ \ncite{Orts-Escolano2016}.

However, to the best of our knowledge, no prior work has been concerned with the retrieval of content in such shared information spaces. Retrieving information and the associated involvement in the interaction with technology can, however, cause participants to lose contact with the conversation~\ncite{Su2015}. Therefore, we have chosen the area of information retrieval as the use case for this chapter and will discuss this in more detail in the next section.

\section{Concept}
\label{sec:cloudbits:introduction}

The following section presents an interaction technique for the \emph{upper limbs} to interact with \emph{world-stabilized} interfaces. Since the focus of this contribution is on the utilization of such world-stabilized interfaces to support \multiuserLower{} situations, this chapter is build around a use case for collaborative \ac{AR} systems. As the review of related work revealed a gap in the field of information retrieval support for such collaborative \ac{AR} concepts, this chapter centers around this use case. However, the interaction techniques presented are not tied to this particular use case and, thus, can support other situations as outlined in section \ref{sec:cloudbits:limitations:usecase}.

The remainder of this section is structured as follows: After a more detailed introduction to the use case (section \ref{sec:cloudbits:introduction:introduction}), the section presents the methodology (section \ref{sec:cloudbits:study1}) and results (section \ref{sec:cloudbits:study1:results}) of an exploratory study to establish requirements for a system to support this use case. Next, the section presents related work in the area of conversation support systems and classifies them according to the requirements (section \ref{sec:cloudbits:relatedwork}). Last, the section presents the concept for \acp{HMD}-based conversation support leveraging \emph{world-stabilized} interfaces that are operated using the \emph{upper limbs} (section \ref{sec:cloudbits:cloudbits}).

\subsection{Introduction and Background}
\label{sec:cloudbits:introduction:introduction}

Today, the retrieval of digital information during conversations and meetings (for both, private and shared use) happens using personal (smart) devices such as smartphones or laptops. However, the interaction with the smart device requires the user to shift the (visual) attention to the device and, thus, away from the conversation and other tasks. This cognitive focus switching between conversation and smart device can hamper the flow of the conversation: Users can lose the connection to the conversation~\ncite{Przybylski2013}, or even favor the interaction with the smart device over the actual conversation, a phenomenon known as \emph{phubbing}~\ncite{Coehoorn2014}. This can decrease mutual awareness of user's activities or otherwise hamper the joint experience. Furthermore, sharing retrieved information with other participants of the conversation can be cumbersome: Users need to connect their device to a public display or pass round the device which imposes privacy issues~\ncite{Karlson2009}.) 

Prior work proposed ambient voice search~\ncite{Radeck-Arneth} as a first step towards supporting conversation scenarios through proactive information retrieval. Such systems automatically retrieve relevant auxiliary information through voice recognition and topic extraction and present it on a shared (large-scale) public display to all users. This can help to diminish the need for individual information retrieval and, thus, to mitigate the challenges set out above. However, interaction with the presented information is limited to touch-based interaction on the public display itself. Furthermore, such a system cannot provide individual per-user output and, therefore, support private information.

In contrast to prior systems that present auxiliary information on screens (including mobile phones and public displays), we argue in this chapter that the representation of information using \acp{HMD} in the periphery of users has a great potential to unobtrusively support the interaction with information. Therefore, this chapter presents a novel approach to visualize and interact with public and private information in the  user's periphery to support conversations. 

In order to establish the requirements for such a system, a focus group-based study was conducted, which will be presented in the next section (section \ref{sec:cloudbits:study1}). Based on the results of the study, section \ref{sec:cloudbits:relatedwork} presents a classification of the related work in this area. Building up on the restults on the study and the review of the related works, section \ref{sec:cloudbits:cloudbits} presents the final design of CloudBits.
\subsection{Exploratory Study}
\label{sec:cloudbits:study1}

This section presents the design and results of an exploratory study to gain insights into the design of a user interface to support conversations and to establish requirements for such a system. More specifically, the study investigated the following research questions:

\begin{description}
	\item[RQ1] How can a system effectively support information retrieval in co-located conversations?
	\item[RQ2] What are the requirements for the user interface of such a system?
\end{description}

For the study, 7 participants (4 male, 3 female, 30 years on average) were invited for individual semi-structured interview sessions. No compensation was provided.

\subsubsection{Design and Procedure}

As a starting point for the brainstorming sessions, five different conversation scenarios (S1-S5) were defined. Based on~\ncite{Pask1976, Dubberly2009}, the five scenarios were designed to include a wide variety of circumstances of conversations in terms of (1) \emph{location}, (2) \emph{objectives} and (3) \emph{mood} of the participants as well as different (4) \emph{relationships}.

\begin{description}
	\item[S1: Consultation] A conversation between persons with different levels of information and understanding of a problem space, e.g., a medical consultation.
	\item[S2: Meeting] A conversation between peers with the same level of information, e.g., a meeting between coworkers.
	\item[S3: Authority Gradient] A conversation between persons with different levels of information and an authority gradient, e.g., a trainer teaching a trainee.
	\item[S4: Informal Talk] A conversation between peers in an informal setting, e.g., friends at a bar.
	\item[S5: Different Intentions] A conversation between persons with different intentions, e.g., a sales meeting with an estate agent.
\end{description}

Remembering special experiences (both positive and negative) is easier than remembering ordinary experiences.~\ncite{Sharp2007}. Therefore, the investigator asked participants about their positive and negative experiences with information retrieval in the respective scenarios, focusing on problems with the current systems. If participants did not have specific experiences in the respective scenario, the scenario was skipped. The study lasted around two hours per single-user session. For data gathering, the sessions were recorded on video.

\subsection{Results and Requirements}
\label{sec:cloudbits:study1:results}

The recorded sessions were analyzed using an open coding approach. The coders selected salient quotes for further analysis. The following section presents the results of our study with respect to the research questions.

In general, all participants stated that they currently use mobile information retrieval in conversation scenarios. When asked about the kind of retrieved information, participants stated that they primarily looked up unknown terms or abbreviations, factual information from public sources and personal information such as appointments or e-mails. The comments of the participants showed clusters in three areas, which led to the identification of three main requirements for the design of a user interface to support information retrieval in conversations.

\subsubsection{R4.1: Unsolicited and Real-Time Service}
In the study, participants stated that the shame of nescience is one of the major reasons for information retrieval using personal devices in all of the discussed situations. This includes not only formal situations but also informal talk with friends. P4 said: \enquote{If I think that it’s too easy or I don’t listen to something, I won’t ask anybody because it’s embarrassing}, P7 added: \enquote{I don’t ask other people because of shyness}. As another reason, participants remembered multiple situations in which fast and immediate retrieval of relevant information was necessary for the continuation of the conversation. Participants stated that breaks during the conversation, caused by the necessity for information retrieval, were \enquote{really upsetting} (P4). Additionally, the interviews showed that information should stay available for immediate re-retrieval as the same information might be needed again within short time frames. 

To support the presented situations, a system should provide direct and unsolicited service to all participants without the need to explicitly ask for information. The information should be available in real-time (i.e., available at the right moment) and time-varying (i.e., available as long as needed) fashion.

\subsubsection{R4.2: Supporting Fluid Transition and Re-Engagement} 

We found that participants have the feeling that they spend a significant amount of time for information retrieval in conversations which \enquote{leads to missing other parts} (P2) of the conversation. This even led participants to refrain from searching (P2, P5) in multiple situations. Participants felt that the time spent on the mobile device caused them to \enquote{lose connection} (P3) to the actual conversation because their focus shifted towards the interaction with the device and the retrieved information. Even more, participants felt \enquote{let down} (P5) when the other person in a conversation focused on their mobile device. Participants named other instances (such as having to leave to room) that caused them to lose the connection to the topics of the conversation and, thus, forced an immediate re-engagement process after returning to the conversation. 

Therefore, a system should provide a means for a fast and smooth transition between information retrieval and the actual conversation to prevent users from losing the connection. In the case of inevitable disruptions, the system should support the user in the re-engagement process. As a further consequence, systems should avoid being a source of distraction from conversation through their visualization.

\subsubsection{R4.3: Selective Sharing from the Public-Private Information Spectrum}

In the analysis, we found the sharing of the retrieved information with other participants of the conversation to be cumbersome. The retrieved information is only available on the personal device of the retrieving user and, thus, shared through sharing the complete device by handing the mobile phone to someone. Participants felt \enquote{uncomfortable} (P3) doing this, not only in formal but also in more intimate situations. Besides privacy issues, participants recalled multiple situations (particular regarding S1 and S5) where this turned out to be frustrating for users because of the limited screen space.

Thus, a system should support 1) selective sharing of specific contents and 2) collaborative interaction with information in a large shared information space. 
\subsection{Related Work}
\label{sec:cloudbits:relatedwork}

The following section discusses the relevant prior work regarding information retrieval in conversations with regard to both, prevalent problems and approaches to overcome them. In the following, this section compares the approaches to the requirements established through the exploratory user study (see table \ref{tab:req:cloudbits}).

\subsubsection{Information Retrieval in Conversations}

Many studies investigated the influence of information retrieval using mobile devices on the quality of conversations. \typcite{Su2015} found that smartphones help to enhance conversations through additional information but can also cause disruptions to the ad-hoc and informal nature of conversations. The use of such devices \enquote{force[s] people to isolate themselves rather than engage in their immediate surroundings}. Continuing on this, \typcite{Porcheron2016} found that, while additional information retrieval may help to solve open issues during conversations, the process of information retrieval also causes people to get distracted from the actual conversation. After the transient focus on the mobile device, people also showed problems to re-engage with the discussion. \typcite{Brown2015} found that information retrieval can be a vivid part of a conversation and \enquote{rather than search being solely about getting correct information, conversations around search may be just as important.}

The perseverative interaction with mobile devices can lead to encapsulation in a mobile bubble, a phenomenon defined as \emph{phubbing}~\ncite{Coehoorn2014}. Emphasizing the influence on the quality of conversations, \typcite{Przybylski2013} found that the interaction with mobile devices reduces closeness and trust as well as interpersonal understanding and empathy between the participants. Regarding family meal situations, \typcite{Moser2016} found that \enquote{attitudes about mobile phone use at meals differ depending on the particular phone activity and on who at the meal is engaged in that activity, children versus adults.}%

Regarding meeting scenarios, \typcite{Bohmer2013} found that phone usage interferes with and decreases productivity and collaboration. Individuals have the feeling that they make productive use of their smart devices but perceive the usage of others as unrelated.

\subsubsection{Approaches for Conversation Support Systems}

Various approaches have been presented to overcome the presented problems and to support information retrieval in conversations.

\typcite{SusLundgren8601} proposed to use a tablet as a public display to provide awareness for the activity of persons working on their smartphones. \typcite{Ferdous2016} proposed to use personal devices as a combined shared display to support interactions and conversations at the family dinner table. To support conversations between strangers, \typcite{Nguyen2015} proposed to display potential conversation topics of mutual interest through \acp{HMD}.

Further approaches focus on managing the time users focus on their mobile devices. \typcite{Lopez-Tovar2015} propose to assess the importance of notifications and whether the user needs to be interrupted. As another approach, \typcite{Eddie2015} presented a solution that proactively interrupts users to discourage excessive mobile phone usage during conversations.

While all of the presented approaches are practical and helpful in several ways, none of the approaches offers comprehensive support of the requirements for conversation support systems established in the explorative study (see table \ref{tab:req:cloudbits}).

Highly related, \typcite{Suh2007} evaluated such a shared information space for collaboration, sharing, and interaction with contents. However, the authors a) did not focus on the area of content retrieval and b) focused on mobile \ac{AR} using smartphones. In this work, we argue that \acp{HMD} are a better fit for this use case as they leave the user's hands-free for interaction.

\subsubsection{Zero-Query Search}

To reduce the time needed to retrieve data, zero-query search has been proposed as a proactive means to retrieve necessary information~\ncite{Rhodes2000}. Such systems use contextual cues such as location, time, or usage history to retrieve and proactively present information to the user. In recent years, zero-query search-based systems such as Google Assistant or Microsoft Cortana were broadly implemented in consumer devices. This was accompanied by a stream of research focusing on how contextual cues can be used to derive search queries and when they should be presented to the user~\ncite{Yang2016, Shokouhi2015}.

\renewcommand*\theadfont{\bfseries}
\settowidth\rotheadsize{\theadfont Unsolicited and Real-Time}
\renewcommand\theadgape{}
\renewcommand\theadalign{lc}
\renewcommand\rotheadgape{}
\begin{table}[t!]
	\centering
	\begin{tabular}{m{5cm}cccc}
		& \rothead{R4.1: Unsolicited and Real-Time Service} & \rothead{R4.2: Supporting Fluid Transition and Re-Engagement} & \rothead{R4.3: Selective Sharing from the Public-Private Information Spectrum} \\
		\midrule
		\cite{SusLundgren8601} & \reqYes & \reqPartially & \reqNo  \\
		\cite{Ferdous2016} & \reqNo & \reqNo & \reqYes \\
		\cite{Nguyen2015} & \reqYes & \reqNo & \reqYes \\
		\cite{Lopez-Tovar2015} & \reqPartially & \reqPartially & \reqNo \\
		\cite{Eddie2015} & \reqPartially & \reqPartially & \reqNo \\
		\cite{Radeck-Arneth} & \reqYes & \reqPartially & \reqNo \\
		\cite{Andolina2015} & \reqYes & \reqPartially & \reqNo \\
		\cite{Suh2007} & \reqNo & \reqYes & \reqYes \\
		\bottomrule
	\end{tabular}
	\caption{Fulfillment of requirements of the related works. \reqYes~ indicates that a requirement is fulfilled, \reqPartially~indicates partial fulfillment.}
	\label{tab:req:cloudbits}
\end{table} 

Building on the concept of zero-query search, work on ambient voice search~\ncite{Radeck-Arneth} supports users in a conversation scenario by providing relevant information to all participants of the conversation on a public display. This allows users to interact with the information through direct (touch) interaction on the display. Focusing on collaborative idea generation, \typcite{Andolina2015} presented a similar system to support users through displaying related keywords based on the topics of their conversation. However, the presented concepts do not provide complete support for fluid transition and, further, do not support the sharing of private information.
\subsection{CloudBits: Interaction Techniques and Prototype}
\label{sec:cloudbits:cloudbits}

\cptteaser{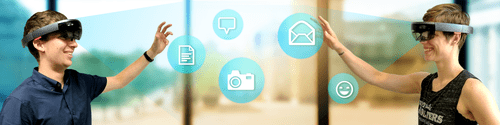}{CloudBits provides users with auxiliary information based on the topics of their conversation, proactively retrieved by means of zero-query search. The information is visualized as augmented \emph{information bits} falling from the \emph{cloud}.}

Based on the findings from the exploratory study and the related work, this section presents the CloudBits concepts and prototype implementation.

CloudBits leverages the metaphor of \emph{cloud} and \emph{drops}, where retrieved units of information are visualized as small drops of information, gracefully falling from an imaginary cloud above the users, in sync with the flow of the conversation (see figure \ref{fig:cloudbits/teaser}). The usage of \acp{HMD} allows to present both, 1) a public shared information space visualized as jointly visible information drops, and 2) individual private information visualized as only privately visible information drops, and supports direct and immediate interaction right in front of the eyes.

The \emph{world-stabilized} nature of the visualization enables spatial interaction with information: Users can sort and group information in the complete 3D space provided by the physical location, leveraging affordances of the real world (e.g. placing something on the table). The spatial dimension of public information is shared between all users so that information appears at the same physical location, allowing for collaborative interaction with information.

The remainder of this section discusses the CloudBits concepts with regards to the established requirements (see section \ref{sec:cloudbits:study1:results}).

\subsubsection{Augmented Zero-Query Information Drops for Unsolicited and Real-Time Service}
\label{sec:cloudbits:cloudbits:focuspluscontext}

CloudBits is an augmented reality system for \acp{HMD} that supports users through small units of information. To fulfill requirement R1, these information units appear in real-time and time-varying to support the current context of the conversation. The information units are visualized as information drops, depicting a preview image, dropping slowly from the metaphorical \emph{cloud} above the users (see fig. \ref{fig:cloudbits/cloudbits} b). Information drops exist in a shared information space, i.e., position and movement of information bits are synchronized between the users. Thus, the information drops appear at the same real-world coordinates but rotated towards each user, allowing users to naturally refer to individual information drops (\emph{Look, there!}, see figure \ref{fig:cloudbits/cloudbits} a). If interested, users can interact with the information drops or, if not, just let them drop slowly to the ground. Once an information bit hits the ground, which means that its lifespan is over, it disappears without further interaction from the user. 

CloudBits unobtrusively transcribes conversations in the background through several microphones and a voice recognition system. Based on the transcribed text, the ambient voice search engine deduces the topics of the conversation. CloudBits uses those topics as zero-query search terms to proactively retrieve information for the users from public (e.g., map data, websites) and private (e.g., e-mail, calendar) information sources. The individual spawn position of the information drops is calculated to be in the peripheral vision of the users in order to lessen the visual clutter and the imposed distraction~\ncite{Kruijff2010}.

\subsubsection{Supporting Fluid Transition between Focus and Context}

To support the fluid transition between the conversation and the process of information retrieval (R3), CloudBits proposes a focus+context~\ncite{Card1999} approach for interaction with information in a conversation setting. 

While the conversation is the \emph{focus} of the user, CloudBits provides \emph{context} through small information drops visualizing the course of the conversation. Vice versa, when interacting with information, CloudBits becomes the \emph{focus} of the user. In contrast to information retrieval using a mobile device, which restricts the participation in the conversation to the auditory channel, the augmented reality nature of CloudBits still allows audio-visual participation as \emph{context}, as the other persons of the conversation are still in the peripheral vision. The tight integration and synchronization of CloudBits with the conversation allows for a fast and smooth transition of the focus between the actual conversation and the information retrieval.

The presented focus+context nature of CloudBits supports users in re-engaging with the content of the conversation through the always-available context of the conversation. Furthermore, as we will introduce in the next section, the vital information drops can be pinned in the information space and always accessible just by a quick glance.

\subsubsection{Immediate Interaction with Information}

\textfig{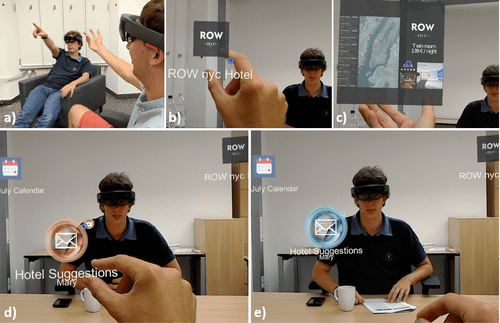}{The interaction techniques of CloudBits: Users can (b) grab\&move information drops freely in the space. To access information, users can grab an information bit and (c) open the hand with the palm facing upwards. Private information drops can (d,e) be shared with other users through grabbing and moving them towards another user. The color encodes if they are public or private.}

CloudBits provides a set of interaction techniques that allow for easy and immediate interaction with the information. All interactions with public information (see section \ref{sec:cloudbits:concept:sharing}) are shared between the users, i.e., if one user changes the position of an information bit or shows its content, this is visualized for all users. This provides mutual awareness as users can understand (1) other users' interactions with the system and (2) the context of their interactions as they can also see the information they are interacting with.

\begin{description}
	\item[Grab \& Move] Users can grab (see fig. \ref{fig:cloudbits/cloudbits}, b) information drops and drag them from the stream of falling drops. Bits can be freely moved around in the real world. \vspace{1em}
	\item[Grab \& Pin] Interesting information drops can be kept for future access through pinning them to a real-world position. Pinning is initiated by moving an information bit to the desired position and releasing the Grab \& Move gesture. Users can unpin an information bit through tapping. \vspace{1em}
	\item[Grab \& Throw] When no longer needed, users can discard information drops by grabbing and throwing them away.\vspace{1em}
	\item[Grab \& Show] To access the content of an information bit, users can unfold it through dragging the information bit into the center of their vision and opening the hand with the palm facing upwards (see fig. \ref{fig:cloudbits/cloudbits} b,c). Similar to the closed drops, the expanded information is presented at the same world coordinate but individually rotated towards each user. To close information drops, users can perform the reversed gesture. 
\end{description}

\subsubsection{Selective Information Sharing from the Public-Private Information Spectrum}
\label{sec:cloudbits:concept:sharing}

To fulfill requirement R3 and to overcome the privacy issues of information sharing on personal smart devices~\ncite{Karlson2009} and in traditional ambient voice search systems, CloudBits supports private information that is only visualized for the respective user. Users can distinguish private and public information through a color-coding (orange for private, blue for public, see figure \ref{fig:cloudbits/cloudbits} d,e). Private information drops provide the same interaction techniques, as outlined in the last section. Additionally, private information can be selectively shared with other users through the \emph{Grab \& Share} gesture.

\begin{description}
	\item[Grab \& Share] Users can share information by grabbing (see figure \ref{fig:cloudbits/cloudbits}, d) and moving it towards another user (see figure \ref{fig:cloudbits/cloudbits}, e), resembling the natural gesture of handing an object to another person. 
\end{description}

\subsubsection{Prototype Implementation}
\label{sec:cloudbits:prototype}

As mentioned earlier, the CloudBits prototype is based upon the implementation of an ambient voice search engine presented by~\typcite{Radeck-Arneth}. The system implementation is based on two main components: (1) a centralized server and (2) a client visualization application for the \acp{HMD}.

The centralized server receives the topics from the ambient voice search engine. The server then orchestrates the spawn positions of information drops in world coordinates and distributes those to the client applications. The server selects the spawn positions such that the direct line of sight between the users remains clear, and the drops appear in their peripheral area.

The CloudBits client application was implemented for the Microsoft HoloLens using Unity3D. All interactions from users are synchronized with the centralized server in real-time (delay <0.2s, 20 fps) and, in the case of public information drops, broadcasted to the other connected clients.

\section{Evaluation}
\label{sec:cloudbits:study2}

This section presents the methodology of a laboratory study to investigate if and how CloudBits supports co-located conversation scenarios. In particular, the experiment focused on if and how\ldots

\begin{enumerate}
	\item \ldots users leverage the surrounding space for acquiring and interacting with information,
	\item \ldots CloudBits provides mutual awareness of activities and eases the (re)-engagement into the conversation and
	\item \ldots CloudBits enables selective sharing from the private-public information spectrum.
\end{enumerate}

For this, 12 participants (P1-P12: seven female, aged between 25 and 35 years) were recruited in six groups of two persons each. The two-person pairs knew each other before in order to stimulate conversation and collaboration similar to real-life situations. The recruitment included pairs of persons in various relationships: work colleagues, close friends, and spouses. None of them had prior experience with augmented reality glasses. No compensation was provided.

\subsection{Design and Task}

Inspired by the study design presented by~\typcite{Lissermann2014}, the overarching goal for the participants in the study was to plan a vacation trip collaboratively. The design of the scenario was chosen to require participants to search, explore, and share both private and public information with their partners and individually. 

The study tested CloudBits and information retrieval via smartphones as two conditions in a within-subjects design. In both conditions, the participants received a destination name, the available budget, and a list of tasks. Participants received all tasks upfront, and they were free to choose the order of processing. The investigator asked the participants to note down their decisions on a provided paper. The four tasks were:

\begin{description}
	\item[Task 1] required participants to agree on the departure date and length of the trip by reviewing their personal calendars and finding possible time slots.
	\item[Task 2] required participants to agree on a flight for their trip based on the selected dates and the price. Therefore, participants had to check the offers they personally received from their travel agencies via e-mail.
	\item[Task 3] required participants to select a hotel based on their budget, the location of the hotel, and online reviews. Furthermore, both participants received personal e-mails with suggestions from friends who traveled to the respective destination before. 
	\item[Task 4] required participants to select a restaurant for the first evening based on location, reviews, and the type of food served.  
\end{description}

To understand if and how CloudBits supports (re)-engagement in the conversation, the investigator induced an (for the participants) unanticipated break in both condition to interrupt the discussions and distract the participants from their current tasks. These interruptions were realized through faked technical problems. After five minutes, the investigator pretended to have fixed the problems and asked the participants to continue from where they left off.

\subsection{Study Setup and Apparatus}

\textfig{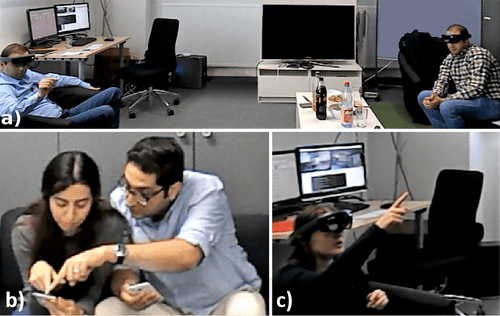}{The setting of the study. Participants were free to choose a spatial arrangement for both, the CloudBits (a) and the smartphone (b) condition.}

The study setup used the prototype application, as presented in section \ref{sec:cloudbits:prototype} deployed to two Microsoft HoloLens devices. For the smartphone condition, the study used two Google/LG Nexus 5X devices. 

The study was conducted in Wizard-of-Oz style to have full control on when and what the participants saw during the study and to eliminate system errors caused by problems in the speech recognition or language processing as these parts were not the focus of the evaluation. Therefore, a wizard application was implemented that allowed the investigator to prepare information drops upfront and to send them on demand to the individual participants (see figure \ref{fig:cloudbits/fig_card_study}).

Similarly, the smartphones used by the participants during the study were prepared with the content (such as e-mails and calendar entries) that they needed to complete the tasks. 

The sessions were videotaped with an external camera and, for the CloudBits condition, the personal views of the participants were recorded through the HoloLens \enquote{Mixed-Reality Capture}\footnote{\url{https://docs.microsoft.com/en-US/windows/mixed-reality/mixed-reality-capture}, [last downloaded 10.07.2019]}. This allowed recording the participants view into the real world together with the augmented information drops. The study concluded with a semi-structured interview for each participant pair. The recorded data from the study was analyzed using an open coding~\ncite{Strauss1998} approach. 

\subsection{Procedure}

\textfig[.7]{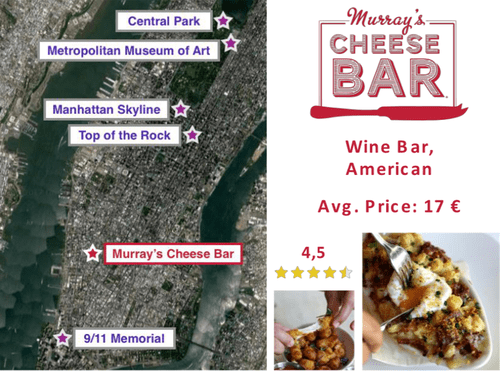}{Content of an information drop as used in the evaluation: A hotel in New York City.}

The order of the two conditions was counterbalanced by randomly assigning the starting condition to the participant pairs. Further, the destination and date of the task were changed between both conditions to avoid learning effects. 

After welcoming the participants, the investigator introduced the participants to the setup of the study and gave them 15 minutes to acclimatize to the Hololens and its general interaction and visualization techniques.

For the smartphone condition, the investigator handed them the two prepared smartphones and informed them about their task. For the CloudBits condition, the investigator observed the conversation of the two participants and sent them appropriate information drops during the study. 

After both conditions, participants took a five-minute break. The experiment concluded with a semi-structured interview focusing on the participants' overall opinion about the CloudBits concept and the differences between the tested conditions. The experiment took 180 minutes per participant pair.

\textfigStudybox{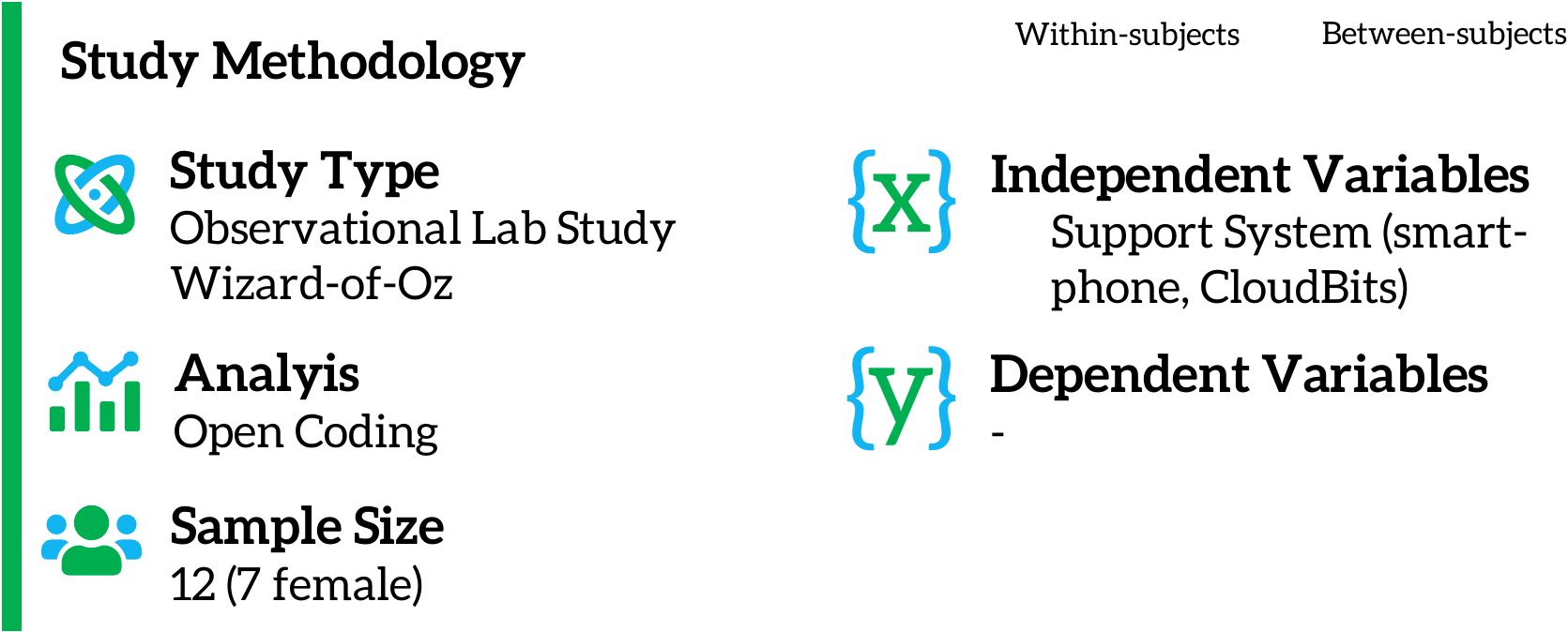}
\section{Results and Discussion}
\label{sec:cloudbits:study2results}

The following section reports on the results of the evaluation with respect to the research questions presented above. Therefore, the section reports the results grouped by 1) the spatial arrangement and use of the 3D space, 2) the participants' usage of working and storage zones, 3) the mutual awareness of participants, 4) CloudBits' support for (re-) engagement, and 5) the selective sharing from the public-private information spectrum. The analysis of the data was performed as described in section \ref{sec:rw:methodology}.

\subsection{Spatial Arrangement and Use of the 3D Space}
\label{sec:cloudbits:study2results:spatialarrangement}

The participants highly appreciated the general idea of in-situ support through augmented information drops. Notably, the possibility for an individual arrangement of information drops, visualized in a shared 3D space, was received enthusiastically. The analysis found that most participants used the complete available ($\sim$ 30 square meter) space to sort the information and that participants used all available dimensions (top/down, left/right and front/back). Nine participants expressed their satisfaction of using a wide space or even the whole room as an information space in the semi-structured interview. P6 commented: \enquote{Compared to the mobile scenario, where the information space is restricted to a very small screen, CloudBits big scene filled with information is extremely desired.} 

\textfig{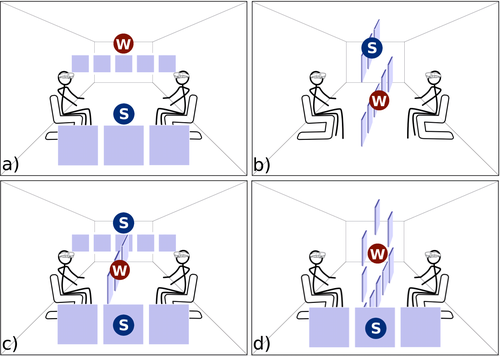}{Spatial arrangement of shared information in the study. Participants divided the information drops into working (W) and storage (S) zones based on their current task.}

The human capabilities allow creating a cognitive map that contains relative positions and orientations of objects~\ncite{Manns2009}. This so-called spatial memory allowed participants to place, arrange, and relocate information drops naturally. P8 compared this to pervasive work practices for knowledge workers on desks: \enquote{It is like... I could arrange my documents on the desktop and easily memorize where to find them. However, doing that in 3D is much more fun.} P10 added: \enquote{I can easily categorize the retrieved information in space. Since the arrangement is personally customized, I can immediately remember where is what.} The participants used this shared augmented information space to refer to information drops through natural gestures such as pointing and looking (see figure \ref{fig:cloudbits/study_setting}, c). P10 commented on this: \enquote{I just accidentally pointed at the information and said \emph{Look there!}. It was amazing that my [P9] could also see the same information on the same place and understood what I meant.}

Participants found that the smartphone condition required constant focus switches between different information sources on the smartphone and between the smartphone and the actual conversation and, thus, constant re-engagement. P3 commented: \enquote{When I need to search for information, using the mobile phone required me to constantly switch my focus from one application or piece of information to the other. So I will lose detail of one information when I switch to the next and need to repeatedly do the switching.}

In comparison, CloudBits allowed the participants to \enquote{see more information at a glance} (P7) while still being able to focus on the actual conversation, supporting the \emph{focus+context} (see section \ref{sec:cloudbits:cloudbits:focuspluscontext}) nature of CloudBits.

\subsection{Working and Storage Zones}
\label{sec:cloudbits:study2results:zones}

Participants used different spatial configurations to sort and categorize information drops. The spatial arrangements were created collaboratively in an on-demand manner. While the participant groups created individual categories for categorization, we found that participants across all groups divided information drops into 

\begin{description}
	\item[Working Zones] containing the information drops that participants were actively using for their current task.
	\item[Storage Zones] containing the information drops not used at that time, but that participants kept for later use.
	
\end{description}

While the spatial layout of these zones differed over all participant groups (see figure \ref{fig:cloudbits/working_zones_public}), the general usage of these zones proofed to be consistent over all participants, a behavior of users that was already found when interacting with tabletops~\ncite{Scott2003}.

The participants' tasks required them to make decisions and re-retrieve this information later on. Participants used the storage zones to pin relevant drops for later use while explicitly removing or letting fade out unused information drops. Five participants pointed out that they found CloudBits pin concept \enquote{very intuitive} (P2, P6) as \enquote{when I pin my to-do post-it on the kitchen board} (P2).

In the smartphone condition, participants reacted to the requirement to keep information for later access with different techniques: Participants wrote the information down on paper or created screenshots on the smartphones. P12 commented on the problems: \enquote{I need to browse and remember which snapshots are relevant as they look all similar and include a lot of text.}

\subsection{Awareness}
\label{sec:cloudbits:study2results:awareness}

The analysis showed that participants followed the actions of their partner through brief glances at their actions. In the following interview, all participants reported that they could gain insights about the current state of the work of their partner. P8 explained that \enquote{While using CloudBits, I was really happy that I could see what my partner is looking at and interacting with.}

The observations showed that the missing awareness in the smartphone condition caused a management overhead in the conversation. Participants were forced to give regular updates about their current actions and whether they were ready to continue the conversation with regard to the content. P3 described the problems: \enquote{We both wanted to search [...] each using our own mobile devices. [...] when I was still in the search process, she found her desired answer and started speaking about the next step we needed to do. But I was still engaged with the searching process of the last needed information piece and could not get what she was talking about.}

\subsection{Supporting (Re)-Engagement}
\label{sec:cloudbits:study2results:reengagement}

During both conditions, the investigator enforced a distraction through faked technical problems in the study setup. After five minutes, the investigator told the participants to continue from where they left off.

The participants' comments, as well as the observations, showed that CloudBits provided them with means for easy and fast re-engagement. After the break, participants started the re-engagement process in the CloudBits condition by looking around the room, using the information drops to get back to the conversation. The analysis showed that participants used both, 1) the falling information drops (covering the latest topics of the conversation) as well as the pinned information drops in their working zones to re-engage with the conversation and their individual tasks. During the interviews, seven participants explicitly appreciated that the necessary information to re-engage was directly available without the need to interact with the system actively. 

In contrast, the observations, as well as the comments from the participants, clearly showed that the smartphone condition did not provide sufficient support for re-engagement. P4 said that \enquote{If I lose my attention to the topic of conversation, I need to concentrate for a while in order to be able to switch back to the topic, using my mobile phone does not help at all and might be even more distracting}. P10 added \enquote{I usually have lots of open information tabs on my mobile device which needs to be browsed to skim them, but I am not able to immediately remember where I have stopped.}

\subsection{Selective Sharing from the Public-Private Information Spectrum}
\label{sec:cloudbits:study2results:sharing}

\textfig{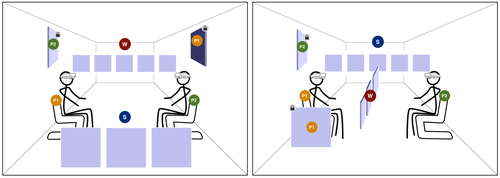}{Spatial arrangement of public (W,S) and private (P1, P2) information in the study.}

All participants showed enthusiasm regarding the possibility to access both public and private information in a shared workspace at the same time. When asked for the reasons, participants reported that this enabled them to selectively share information without the need to share the complete device. P9 explained: \enquote{CloudBits let me share a part of the information which needs to become public [...]. I always have concerns about other persons having access to all my data while sharing information with others through my mobile phone.} P10 further added: \enquote{I really did not want to share my personal device to my partner, but it was also kind of impolite to ask him to search for the same information himself. This meant I have no trust in you or I do not want to help you.}

Participants did not mix public and private information in the same zones (see figure \ref{fig:cloudbits/working_zones_private_back}). In particular, participants chose spatial arrangements to keep private information drops far apart from the shared work zones. We observed two basic patterns for the spatial arrangement of private information: Half of the participants positioned private information drops close to themselves (see figure \ref{fig:cloudbits/working_zones_private_back} P1), while the other half of the participants moved them to an unused space preferably far away (see figure \ref{fig:cloudbits/working_zones_private_back} P2).

While the study setup did not impose any restriction on the physical arrangement of the participants in the room, the observations showed that participants chose different arrangements for the conditions to support the process of information sharing. In the CloudBits condition, all participants arranged themselves in a face-to-face setting (see figure \ref{fig:cloudbits/study_setting} a). The interview revealed that this provided them with a comfortable position for the conversation and shared information access and, further, gave them the necessary space to perform mid-air gestures.

In the smartphone condition, participants showed two different approaches for the spatial arrangement to support information sharing: Three pairs constantly changed their position between face-to-face for individual work and side-by-side for sharing information through each other's smartphone screen (see figure \ref{fig:cloudbits/study_setting} b). The other three pairs stayed in a face-to-face arrangement during the whole session and tried to exchange the found information orally.

Comparing both conditions, participants commented that information sharing felt more immediate and efficient in the CloudBits condition. P5 explained: \enquote{Similar to the real world, sharing information using CloudBits occurs by just naturally changing the virtual position of the information to where my partner is. This experience reminds me exactly to when I pass a physical object to someone to share it.}  P8 added that \enquote{CloudBits information sharing saves the effort currently is needed [...] to share [...].}

\subsection{AttrakDiff}
\label{sec:cloudbits:study2results:attrakdiff}

\sameheightpic{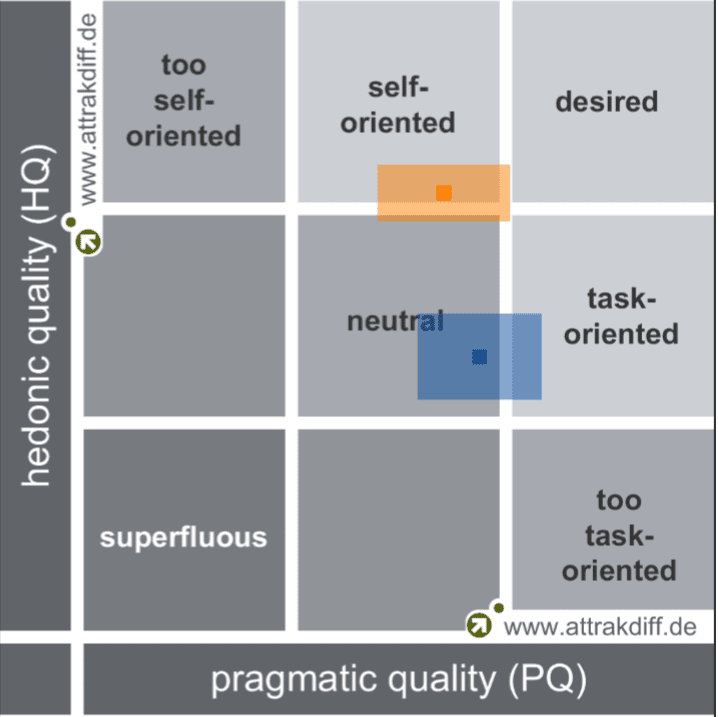}{AttrakDiff Comparison}
{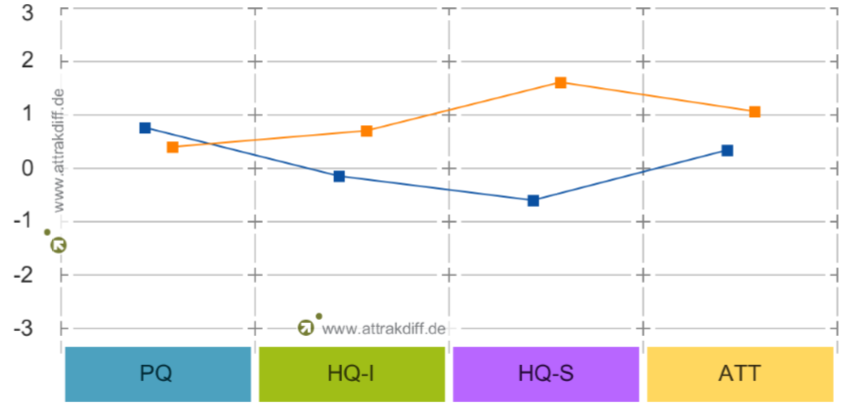}{Detailed Results}
{General (Smartphone: blue, CloudBits: Orange) (a) and detailed (b) results of the AttrakDiff comparison between the two conditions for the four AttrakDiff quality dimensions (left to right: pragmatic, hedonic-identity, hedonic-stimulation, attractiveness).}

The AttrakDiff~\ncite{Hassenzahl2003} questionnaire indicated higher qualities for CloudBits in both hedonic dimensions (HQ-I: Identity and HQ-S: Stimulation) compared to traditional information retrieval using smartphones. The pragmatic qualities were rated on a similar level with slight advantages for the more traditional smartphone condition. In total, CloudBits achieved a higher result for attractiveness (ATT, see figure \ref{fig:cloudbits/attrak_diff_details}).

Based on the feedback of the participants, the lower values for the practical quality dimensions could be based on the new and unfamiliar interaction with \acp{HMD}. Furthermore, the technical limitations of today's \acp{HMD} and of the study client implementation also influenced the results and might let to participants transferring those problems to the general concepts. Further work is needed in this area to identify the detailed reasons for these differences.

\section{Guidelines}
\label{sec:cloudbits:guidelines}

Based on the results of the study, the following section presents a set of guidelines for the design of user interfaces for conversation support systems:

\subsection{Leverage the Surrounding Physical Space}

The analysis of the results of the study presented showed a great interest of the participants in the possibility of arranging information in space, categorizing it and using it for cooperation. This interest manifested itself both in the observed interactions of the participants and in the qualitative feedback given by the participants in the semi-structured interviews (see section~\ref{sec:cloudbits:study2results:spatialarrangement}). The participants used the entire available space of the room in all three dimensions to work with information in \emph{working} and \emph{storage} areas (see section  \ref{sec:cloudbits:study2results:zones}). Therefore, conversation support systems should enable the usage of the entire available space for categorization of and interaction with the information.

\subsection{Provide Means for Fluid Transition and Re-Engagement}

The results of the study suggested that today's conversation support systems do not adequately support users in (re)-engaging in conversations. The analysis found that the lack of awareness of the activities of other conversation partners can lead to a feeling of exclusion, which in turn can manifest itself in immersing oneself in interaction with technology and losing connection to the conversation (see section \ref{sec:cloudbits:study2results:awareness}). This becomes a problem especially after interruptions of the conversation, whether caused by unwanted external influences or by the temporary focus on the reception of information (see section \ref{sec:cloudbits:study2results:reengagement}). It is, therefore, necessary for conversation support systems to provide the context to help users get back into the conversation after breaks. Furthermore, such systems should provide awareness of the actions of other participants during the conversation in order to prevent participants from drifting away.

\subsection{Support Selective Sharing from the Public-Private Information Spectrum}

The analysis of the related works and the results of the study presented in this chapter showed that today's systems do not offer sufficient possibilities for fast and easy sharing of information (see section \ref{sec:cloudbits:study2results:sharing}). In particular, the analysis showed privacy concerns when sharing information by handing around the private smartphone. Furthermore, the analysis also highlighted the importance of sharing information in collaboration. Therefore, conversation support systems should support selective sharing for privacy-preserving sharing of private information without the need to share the complete device.
\section{Limitations and Future Work}
\label{sec:cloudbits:limitations}

The study design, as well as the results of the controlled experiment, hint at some limitations and directions for future work. This section lists these points and provides possible solutions.

\subsection{Scope of the Study}

This chapter presented a Wizard-Of-Oz style evaluation of the concepts. This restriction was necessary in order to focus the study on evaluating the concepts presented in this chapter. Furthermore, this design decision eliminated the influences of uncontrollable variables (e.g. the success rate of speech recognition). However, due to the Wizard-Of-Oz style of the study, it was necessary to closely confine the scope of the study in terms of scenario and relationship. A large-scale study in the wild might therefore yield further insights into the problem area.

\subsection{Interaction Concepts for Larger Groups}

This chapter presented concepts and an evaluation of interaction techniques for groups of two conversation partners. While some of the concepts are directly transferable to larger groups (e.g, \emph{Grab\&Show}, some of the concepts are geared to the interaction between two persons. For example, the \emph{Grab\&Share} gesture employs the metaphor of handing over an object to a person. For larger groups, sharing must at least distinguish between sharing with one user and making information generally accessible to all users. Further, in the current concept, the \emph{Grab\&Show} gesture opens information drops for all participating users. For larger groups, this might worsen the overview. Therefore, an additional technique for privately opening information drops might be beneficial. In summary, future work is necessary to focus on concepts beyond the 1-on-1 collaboration scenarios addressed in this chapter.

\subsection{Interaction in the 3D Space}

The presented study mainly focused on the evaluation of the interaction techniques introduced earlier in this chapter. However, the analysis of the observations and the qualitative feedback of the participants found many interesting aspects regarding the use of a shared 3D space. Hereby, this chapter offers interesting starting points for a deeper and more focused look on how people interact in a shared 3D information space to arrange themselves and information spatially. In addition, further work is needed to understand the impact of such working practices on users' accuracy and efficiency.

\subsection{CloudBits Interaction Beyond the Specific Use Case}
\label{sec:cloudbits:limitations:usecase}

This section covered the presented interaction techniques in the context of the specific use case of information retrieval in co-located conversation scenarios. However, we are convinced that the CloudBits interaction techniques can also be used in other areas. For example, CloudBits can also be beneficial for individual users: Such a system could visualize notifications for the user. Depending on the urgency of the notification, color and size, as well as spawn position (between peripheral area and directly in the line of sight) could be varied. Future work is needed on how to adopt the interaction techniques to the requirements of individual users.

\newpage
\section{Conclusion}
\label{sec:cludbits:conclusion}

This chapter explored how multiple users interact with information in a shared information space and presented CloudBits together with its prototype implementation. We evaluated our concepts in a qualitative lab study and presented guidelines for the design of user interfaces for in-situ collaboration in a shared information space. The results of the controlled experiment confirmed the feasibility of the concept. 

To conclude, this chapter added to the body of research on interacting with \acp{HMD} in multiple areas:

\begin{enumerate}
	\item This chapter demonstrated how information could be visualized time-dependently based on the current information needs of users: Therefore, relevant information appears unsolicited in the peripheral view of the user. As time passes, the information drops fall down until they finally disappear, playfully visualizing the transience of information. While this visualization was used in this chapter to provide additional information during conversations, there are many more areas of application where such an unobtrusive time-dependent visualization of information could be of use (e.g., incoming e-mails or calendar appointments). Therefore, this work has paved the way for further research investigating the viability of such interfaces in other scenarios.
	\item In addition, this chapter contributed interaction techniques for collaborative interaction in a shared information space. Therefore, this work added to the body of research on interacting with \acp{HMD} by providing interaction techniques that break through the isolation of the inherently single-user focused device class. This work can thus serve as a foundation for further contributions in this area.
	\item Finally, this work contributes to the area of conversation support by unobtrusively providing additional information based on the topics of the conversation. The evaluation of the concept showed great advantages over state-of-the-art information retrieval during conversations. Thus this work opened a promising way to support future (co-located and remote) conversations.
	
\end{enumerate}

\subsection{Integration}

\textfigH{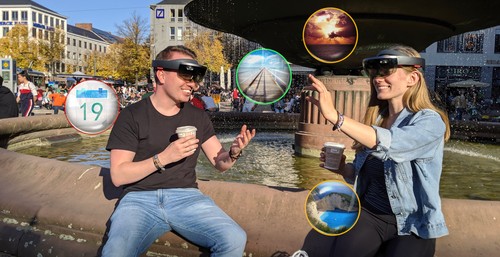}{Alice and Bob share memories during a conversation. Cloudbits provides them with retrieved digital information.}

\interactionbox{alice_cloudbits}{In the Coffee Shop}{Alice meets her friend Bob for a coffee. The two talk about their shared memories of the past year. During the conversation CloudBits automatically retrieves images of the two from their respective photo albums, which the two subsequently share with each other (see figure \ref{fig:cloudbits/alice}). The two of them come to the conclusion that you should once again do something together. During their discussion about different possibilities, CloudBits automatically displays appropriate information from the Internet and displays the respective calendars privately to help them schedule the appointment.
	
	Over the course of the conversation, Alice and Bob interact quickly and easily with the information, grouping and sharing it through natural hand movements. Through the shared information space, the other person understands what the partner is doing.}

The interaction technique presented in this chapter allows for interactions with \emph{world-stabilzed} interfaces leveraging our \emph{upper limbs} for input. The interaction technique uses the world-stabilization to provide a shared information space between users, supporting \continuousLower{} in co-located \inplaceLower{} \multiuserLower{} scenarios. This contribution is particularly important due to the natural focus of \acp{HMD} on single user interactions, which results from the limited visibility of displayed information restricted to the wearing user (see section \ref{sec:relatedwork:hmds}). Thus, the contribution extends the applicability of \aroundbodyinteraction{} (see section \ref{sec:introduction:aroundbodyinteraction}) to multi-user situations and contributes to the overall vision.

In this chapter, the presented interaction techniques were investigated based on the specific application area of information retrieval and interaction with information in a co-located two-person meeting. While this restriction to the specific use case was necessary to the scope of this work, other use cases may yield different results. As discussed in section \ref{sec:cloudbits:limitations}, further modifications to the design (e.g., by supporting more than two users) are necessary for future interaction with \acp{HMD}. The ideas and qualitative results presented in this chapter can serve as a reliable baseline for further explorations of the respective design spaces.

\subsection{Outlook}

This chapter, together with chapter \ref{ch:proximity:merged}, investigated interaction techniques with the \emph{upper limbs}. However, there are situations in which interaction with the upper limbs is not desired (e.g., while using the hands to interact with the real world) or even physically possible (e.g., while carrying something in both hands). Chapters \ref{ch:cheesyfoot} and \ref{ch:walktheline} therefore present hands-free interaction techniques that draw on our \emph{lower limbs}.
}

\setcounter{ptc}{3}
\ctparttext{\parttoc}
\part{Interaction using the Lower Limbs}\label{pt:foot}
\chapter[Mind the Tap]{Mind the Tap: Assessing Foot-Taps for Interacting with Head-Mounted Displays}\label{ch:cheesyfoot}
{
\def\upStyle{indirect}
\def\downStyle{direct}

\newcommand{\studyOneColor}{green}
\newcommand{\studyTwoColor}{red}

\def\plaintitle{\systemname: Assessing Foot-Taps for Interacting with Head-Mounted Displays}

\def\plainauthor{First Author, Second Author, Third Author,
Fourth Author, Fifth Author, Sixth Author}
\def\emptyauthor{}
\def\plainkeywords{Human Factors; Foot Interaction; HMD; User Study}

\def\mlAlgorithm{Support Vector Machine}
\newcommand\naive{na{\"\i}ve}

\newcommand{\row}[1]{#1-row}
\newcommand{\col}[1]{#1-column}

\newcommand{\direct}[0]{\emph{direct}}
\newcommand{\indirect}[0]{\emph{indirect}}

\cptteaser{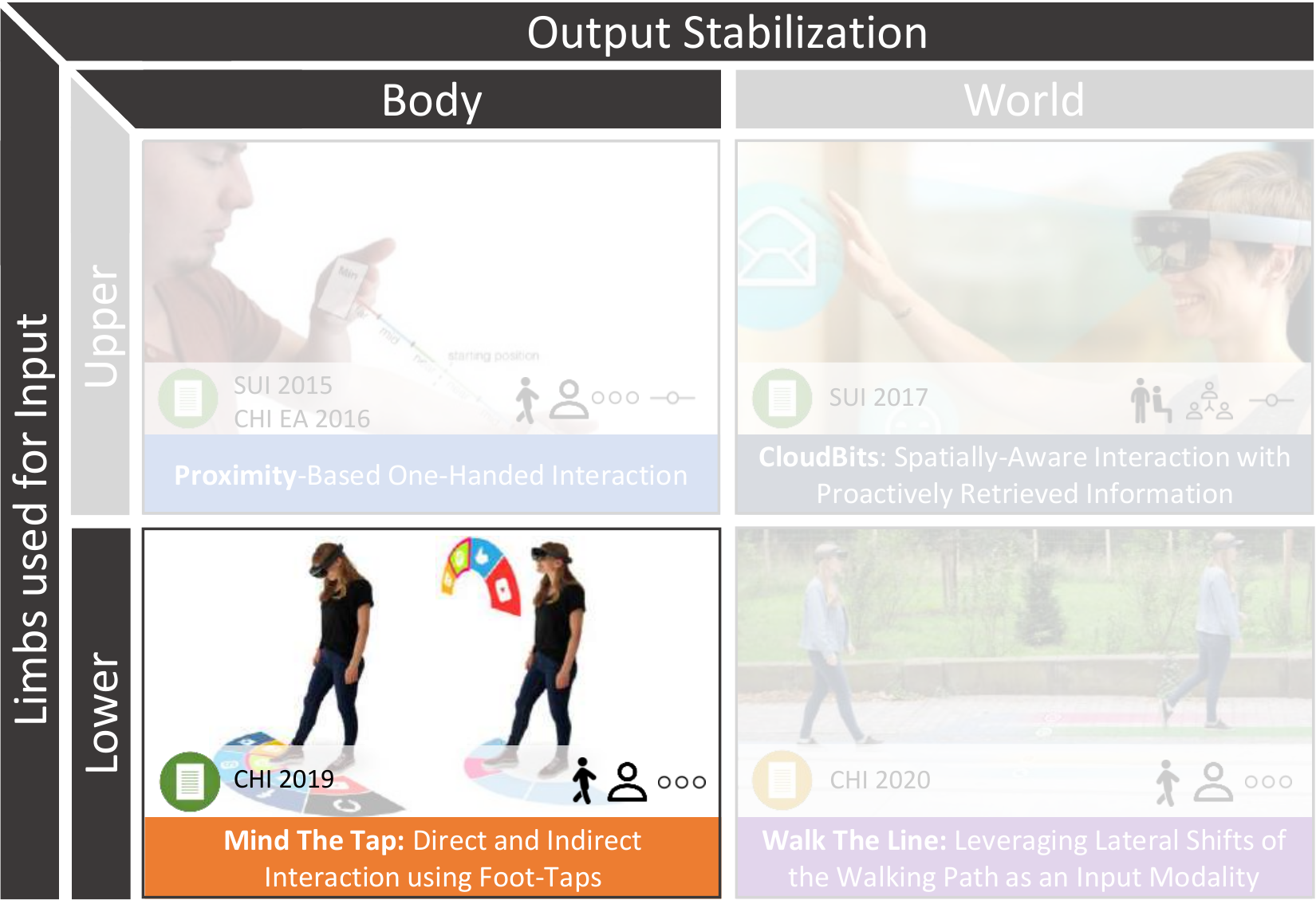}{This chapter presents an interaction technique for body-stabilized interfaces leveraging the lower limbs for input.}

The previous chapters of this thesis investigated the use of our \emph{upper limbs} for interacting with \acp{HMD} for \emph{body-stabilized} and \emph{world-stabilized} visualizations. As introduced in chapter \ref{ch:introduction} and \ref{ch:relatedwork}, interaction techniques utilizing the upper limbs are not sufficient to use \acp{HMD} in all everyday situations: Such interaction techniques cannot support a wide range of situations where both of the user's hands are busy interacting with the real world (e.g., due to carrying something in hand).

As a possible solution for such situations, this chapter proposes the usage of the \emph{lower limbs} to interact with \emph{body-stabilized} interfaces (see section \ref{sec:relatedwork:hmds:implementation}) with an emphasis on support for the interaction situations \mobilityLower{}, \singleuserLower{}, and \discreteLower{}. Interaction techniques in this quadrant of the design space allow users to keep their hands free to interact with the real world, while their feet can provide input for a system. By stabilizing the interface to the body of the user, it is always available for interaction, further increasing the mobility of interaction techniques in this quadrant of the design space (see figure \ref{fig:cheesyfoot/overview_mindthetap}). While there is a long history of using the lower limbs for foot-based interfaces in various areas of computing systems, to the best of our knowledge, foot-based interactions have not yet been systematically evaluated for interaction with \acp{HMD}.

This chapter aims to close this gap and to add to the body of research on interacting with \acp{HMD} by exploring a foot-based input modality for \acp{HMD}. The contribution of this chapter is twofold: First, the chapter presents the results of two controlled experiments, assessing the benefits and drawbacks of two styles of interaction leveraging the \emph{lower limbs} to interact with \emph{body-stabilized} interfaces. Second, based on the results of the two experiments, this chapter provides a set of guidelines for designing such user interfaces for both types of interaction.

The remainder of this chapter is structured as follows: After the review of related works (section \ref{sec:cheesyfoot:rw}), section \ref{sec:cheesyfoot:concept} presents two interaction styles for foot-based interaction with \acp{HMD}. In the following, \ref{sec:cheesyfoot:methodology} describes the methodology and research questions of two controlled experiments focusing on \direct{} and \indirect{} interaction with content, respectively. After that, sections \ref{sec:cheesyfoot:evaluation1_results} and \ref{sec:cheesyfoot:evaluation2_results} present the results of the two controlled experiments. Based on a comparison of both styles of interaction (section \ref{sec:cheesyfoot:comparison}), section \ref{sec:cheesyfoot:discussion} provides design recommendations for \direct{} and \indirect{} foot-based user interfaces for interacting with \acp{HMD}. The chapter concludes with the limitations of the approach and directions for future work (section \ref{sec:cheesyfoot:limitations}).

\cptIntroBoxAward{Muller2019}{The student \emph{Joshua McManus} implemented the study client application. \emph{Sebastian Günther}, \emph{Martin Schmitz}, \emph{Markus Funk} and \emph{Max Mühlhäuser} supported the conceptual design and contributed to the writing process.}
\section{Related Work}
\label{sec:cheesyfoot:rw}

Chapter \ref{ch:relatedwork} discusses the related works on interaction techniques with \acp{HMD}. The following section presents a set of requirements for foot-based interactions with \acp{HMD} and, in the following, categorizes relevant research with regards to the requirements. Further, this work was strongly inspired by proprioceptive and imaginary user interfaces, to be discussed at the end of the section.

\subsection{Requirements}

The following section presents a set of requirements for foot-based interactions with \acp{HMD} derived from the related works. The requirements are then used to compare the most relevant related work (see table \ref{tab:req:mindthetap}).

\begin{description}
	\item[R5.1: Independent Usage] In many mobile situations, the user's hands are encumbered (e.g., by carrying something). To support such situations, it is necessary that the interaction with the system is possible without the additional use of hands.
	\item[R5.2: Visual Output] To close the feedback loop between user and system, means for output are required in addition to the foot-based input.
	\item[R5.3: Interaction Without Full Visual Attention] The lack of visual attention can quickly lead to dangerous situations in mobile scenarios. Therefore, the user's visual attention should not be completely absorbed by the interface, allowing the user to keep a connection to the real world.
	\item[R5.4: Direct Interaction] A direct spatial connection of input and output (i.e., input and output happen at the same physical location) allows for more \enquote{natural} and \emph{compelling} interactions~\ncite{Forlines2007}. Therefore, systems should provide such a direct spatial connection.
\end{description}

\subsection{Foot-based Interaction}

Foot-based input has a long history in operating industry machines~\ncite{barnes1942pedal,Corlett1975,Keoemer1971,Pearson1986, Barnett2009} and also appeared early as a possible solution in the field of \ac{HCI}: In 1986 Douglas Engelbart gave a famous presentation on the basic concepts of \acp{GUI} and the mouse as an input modality, which would later go down in history as \emph{The Mother of All Demos}~\ncite{Engelbart1968}. One year before, Engelbart's team also worked on other input modalities: \typcite{english1967display} presented a system using knee and foot control for text selection. Since then, foot input frequently emerges as a potential input modality for novel computing systems and has been explored for seated~\ncite{Velloso2015a}, standing~\ncite{Saunders2016}, and walking~\ncite{Yamamoto2008} users in different scenarios.

For example, foot controls have been used to increase the input space for desktop~\ncite{Silva2009} or mobile~\ncite{Lv2014a} games or to operate a smartphone in the pocket of the user~\ncite{Barnett2009, Fan2017, Han2011}. Besides the sole use as an input modality, foot interaction has been used in conjunction with hand-gestures~\ncite{Lv2013, Lv2014, Lv2015} or gaze-input~\ncite{Rajanna2016, Gobel2013}. \typcite{Pakkanen2004} investigated foot-based interaction as a second input channel for non-accurate spatial tasks and found that foot interaction is appropriate, \enquote{maintaining adequate accuracy and execution time}. Highly related, \typcite{Saunders2016} explored indirect interaction with ring-shaped foot interfaces. However, the exploration by Saunders et al. was limited to 1) indirect interfaces and 2) two different layouts.

Further, research proposed multiple use cases for such foot-based input modalities. \typcite{Yin2003} presented an interactive animation system, controlled using foot gestures. \typcite{Simeone2014} used foot-based input for 3D~interaction tasks, \typcite{Schoning2009} presented support for navigating spatial data. Further examples include support for the interaction with large displays~\ncite{Jota2014, Felberbaum2016}, interactive floors~\ncite{Augsten2010a} and other public interfaces~\ncite{Fischer2014}. More general, \typcite{Alexander2012} and \typcite{Felberbaum2018} proposed user-defined foot-gestures for typical GUI tasks in different domains.

In recent years, research also focused on the applicability of foot-based interfaces for \acp{HMD}. \typcite{Matthies2013} presented a technical prototype to provide hands-free interaction for VR applications. \typcite{Fukahori2015} used the shifting of the user's weight on their foot for subtle gestures to control \acp{HMD} interfaces. Furthermore, \typcite{Fan2016} focused on foot-based interaction techniques for exploring a VR representation of a planet. Highly related, \typcite{Lv2015} used foot-based interaction techniques for controlling an \ac{AR} game. However, to the best of our knowledge, there is no systematic investigation of the human ability to interact with HMDs through foot-taps.

\renewcommand*\theadfont{\bfseries}
\settowidth\rotheadsize{\theadfont Interaction Without Full}
\renewcommand\theadgape{}
\renewcommand\theadalign{lc}
\renewcommand\rotheadgape{}
\begin{table}
	\centering
	\begin{tabular}{m{5cm}ccccc}
		& \rothead{R5.1: Independent Usage} & \rothead{R5.2: Visual Output} & \rothead{R5.3: Interaction Without Full Visual Attention} & \rothead{R5.4: Direct Interaction} \\
		\midrule
		\multicolumn{5}{c}{\emph{Foot-Based Interfaces}} \\
		\cite{Silva2009} & \reqNo & \reqYes & \reqYes & \reqNo  \\
		\cite{Lv2014a} & \reqNo & \reqYes & \reqNo & \reqYes \\
		\cite{Velloso2015a} & \reqYes & \reqYes & \reqYes & \reqNo  \\
		\cite{Saunders2016} & \reqYes & \reqYes & \reqPartially & \reqNo  \\
		\cite{Yamamoto2008} & \reqYes & \reqNo & \reqYes & \reqNo  \\
		\cite{Fan2017} & \reqYes & \reqNo & \reqYes & \reqNo  \\
		\cite{Han2011} & \reqNo & \reqYes & \reqNo & \reqPartially  \\
		\cite{Lv2015} & \reqPartially & \reqYes & \reqNo & \reqPartially  \\
		\cite{Rajanna2016} & \reqNo & \reqYes & \reqYes & \reqNo  \\
		\cite{Gobel2013} & \reqNo & \reqYes & \reqNo & \reqNo  \\
		\cite{Pakkanen2004} & \reqYes & \reqYes & \reqNo & \reqNo  \\
		\cite{Simeone2014} & \reqPartially & \reqYes & \reqNo & \reqNo  \\
		\cite{Schoning2009} & \reqNo & \reqYes & \reqNo & \reqNo  \\
		\cite{Augsten2010a} & \reqYes & \reqYes & \reqNo & \reqYes  \\
		\cite{Felberbaum2016} & \reqYes & \reqYes & \reqNo & \reqNo  \\
		\cite{Felberbaum2018} & \reqYes & \reqYes & \reqNo & \reqPartially  \\
		\multicolumn{5}{c}{\emph{Interfaces for \acp{HMD}}} \\
		\cite{Matthies2013} & \reqYes & \reqNo & \reqYes & \reqNo  \\
		\cite{Fukahori2015} & \reqYes & \reqPartially & \reqYes & \reqNo  \\
		\cite{Lv2015} & \reqNo & \reqYes & \reqNo & \reqPartially  \\
		\bottomrule
	\end{tabular}
	\caption{Fulfillment of requirements of the related works. \reqYes~ indicates that a requirement is fulfilled, \reqPartially~ indicates a partial fulfillment.}
	\label{tab:req:mindthetap}
\end{table}

\subsection{Imaginary and Proprioceptive User Interfaces}

Gustafson et al.~\ncite{Gustafson2010, Gustafson2011a} introduced imaginary interfaces as a novel approach to interaction without any visual feedback, leveraging the human's ability to map the spatial memory to (physical) surfaces. \typcite{Dezfuli2012} extended this idea using proprioception~\ncite{Schmicking2010, Lopes2015}, the subconscious knowledge about the relative position and orientation of our body parts, showing that users were able to create a mental mapping between on-screen user interfaces and eyes-free touch on the hand. This work extends these ideas for foot-based interactions.
\FloatBarrier
\section{Concept}
\label{sec:cheesyfoot:introduction}
\label{sec:cheesyfoot:concept}

\cptteaser{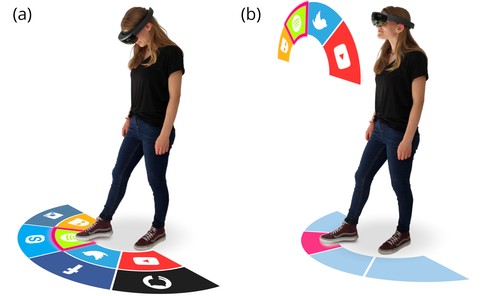}{This chapter presents two interaction techniques, leveraging foot-taps as a \direct{} (a) and \indirect{} (b) input modality for interacting with \acp{HMD}.}

As outlined above, foot-based interaction techniques frequently emerge as a potential input modality for novel computing systems. In recent years, research started to transfer the ideas of foot-based interaction to the field of interacting with \acp{HMD}~\ncite{Matthies2013, Fukahori2015}. This chapter contributes to this promising stream of research by exploring the feasibility of using foot-taps as an input modality for \acp{HMD}, assessing the benefits and drawbacks of 1) \emph{direct interaction} with interfaces that are displayed on the floor and require the user to look down to interact (see figure~\ref{fig:cheesyfoot/teaser_sepp} a) and 2) \emph{indirect interaction} with interfaces that, although operated by the user's feet, are displayed as a two-dimensional window floating in the space in front of the user (see figure~\ref{fig:cheesyfoot/teaser_sepp} b).

For this, we consider a semicircular interaction wheel that is anchored to the dominant foot of the user’s standing position. The interaction wheel is divided into a grid by multiple rows and columns. Each cell of the grid represents an option the user can select though a foot tap (see figure~\ref{fig:cheesyfoot/teaser_sepp}). This chapter explores two different styles of interaction with such an interactive grid that share the same style of input, but vary the visualization:

\begin{description}
	\item[Direct Interaction] The semicircular grid is visualized within leg reach on the floor in front of the participant. Therefore, there is a \direct{} connection between the location of input and output. The user can interact with the system by looking to the ground and tapping the location where the respective grid cell is visualized.
	\item[Indirect Interaction] The \indirect{} interface moves the visualization from the floor to the air in front of the user. However, despite the changed location of the visualization, users still can interact with the system using foot taps through the sense of proprioception. This sense allows users to move their feet without looking at them.
\end{description}

\section{Methodology}
\label{sec:cheesyfoot:methodology}

This chapter presents the methodology of two controlled experiments assessing the accuracy and efficiency of \direct{} and \indirect{} interfaces for foot-based interactions with \acp{HMD} as presented in section \ref{sec:cheesyfoot:concept}. More specifically, the experiments investigated the following research questions:

\begin{description}
	
	\item[RQ1] How does the layout of the semicircle in terms of number of columns and rows affect the accuracy, efficiency and user experience of \direct{} interfaces?
	\item[RQ2] How does the layout of the semicircle in terms of number of columns and rows affect the accuracy, efficiency and user experience of \indirect{} interfaces?
	\item[RQ3] How does the the visualization (as \direct{} or \indirect{}) affect the accuracy, efficiency and user experience of such an interface?
	
\end{description}

To avoid learning effects, RQ1 and RQ2 were addressed in two separate experiments, although the basic design and procedure show large overlaps. To keep the results of both experiments comparable, only the visualization technique was changed as \direct{} and \indirect{} between the experiments. No participant took part in both experiments. The analysis used the results of both experiments to address RQ3. The following description of the methodology applies to both experiments unless stated otherwise.

\colfigVspace{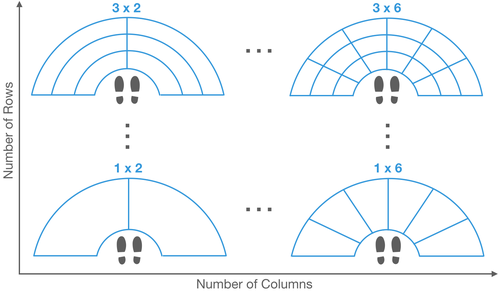}{The independent variables (number of rows and number of columns) tested in the two experiments.}{0em}

\subsection{Design and Task}

The conditions varied the number of \emph{rows} and \emph{columns} that divide the semicircular grid into several targets (see section \ref{sec:cheesyfoot:concept}) as independent variables in a repeated measures design. The independent variables were varied in three levels for the number of rows (1,2,3) and three levels for the number of columns (2,4,6). Therefore, the experiment tested grids from $1*2=2$ to $3*6=18$ targets (\reffig{fig:cheesyfoot/rowcolumns2}). We considered these variables to assess their impact on participants’ performance regarding accuracy and efficiency. The experiment required at least three repetitions of each target (i.e., based on the most complex condition \row{3}, \col{6}: $3*6*3 = 54$). To prevent the influence of fatigue, the experiment was designed with an equal number of trials in each condition. This design resulted in a total of $3*3*54=486$ trials per participant. The order of conditions was counterbalanced using a Balanced Latin Square design. For each condition, the series of targets were randomized while maintaining an equal number for each target.

\begin{figure*}[ht!]
\subfloat[Foot Tracking\label{fig:study:feet}]
  {\includegraphics[width=.49\linewidth]{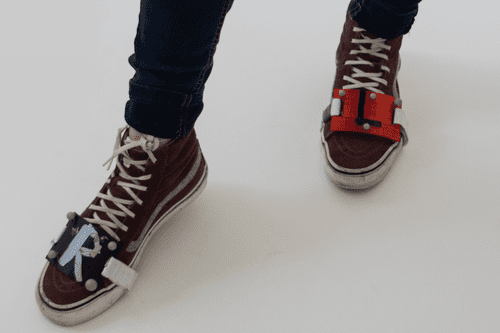}}\hfill
\subfloat[HoloLens Tracking\label{fig:study:head}]
  {\includegraphics[width=.49\linewidth]{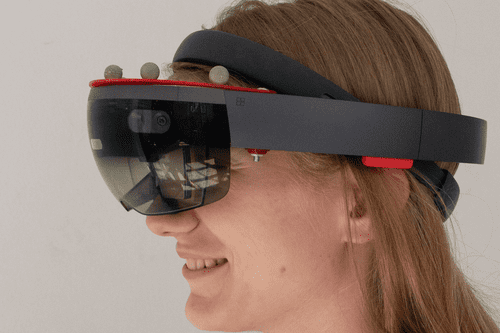}}\hfill
\subfloat[Direct Visualization\label{fig:study:direct}]
  {\includegraphics[width=.49\linewidth]{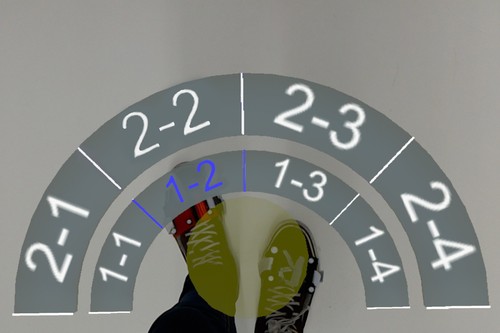}}\hfill
\subfloat[Indirect Visualization\label{fig:study:indirect}]
    {\includegraphics[width=.49\linewidth]{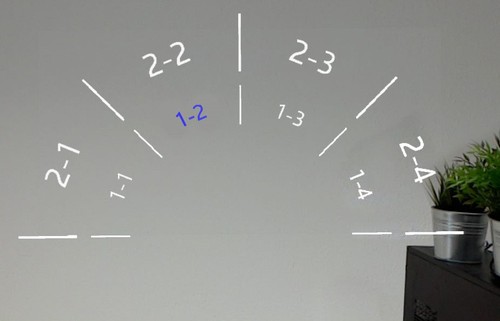}}
\caption{We tracked the position and orientation of the feet (a) and the hololens (b). During the first experiment, we used a \direct{} visualization on the floor (c). In the second experiment, we used an \indirect{} floating visualization (d).}
\end{figure*}

\subsubsection{Experiment I: Direct Visualization}
\label{sec:methodology:floor}

The system visualized the semicircular grid within leg reach on the floor in front of the participant. Depending on the condition, the semicircle was divided into a grid with 2-6 horizontal columns of equal size and 1-3 rows of equal size. The columns filled the complete semicircle (\reffig{fig:study:direct}). Based on the average human leg length \ncite{Eveleth1990}, the system used a fixed height of \SI{8.5}{cm} for each row. This size was chosen to allow all participants to reach the goals within the \row{3} conditions comfortably. The participants’ task was to look at the floor in front of them and to tap highlighted targets.

\subsubsection{Experiment II: Indirect Visualization}
\label{sec:methodology:heads-up}

The second experiment used an \indirect{} \ac{HUD} visualization, floating in front of the eyes of the user (\reffig{fig:study:indirect}). The ultimate goal of this experiment was to understand how the participants would naturally map the presented target areas to the ground in front of them. Therefore, the system did not give the participants feedback about the position of their feet. Such feedback would have given the participants an indication of the size of the target areas, thereby distorting the results. The participants’ task was to tap the floor at the position where they expected the targets highlighted in the floating visualization.

\subsection{Study Setup and Apparatus}

The system used an optical tracking system (OptiTrack) to measure the position of the participant’s feet. For this, the participants wore 3D-printed parts, each augmented with a set of retro-reflective markers, on both feet (\reffig{fig:study:feet}) and a Microsoft Hololens (also with retro-reflective markers, \reffig{fig:study:head}) which displayed the respective visualization.

A study client application was implemented that allowed the investigator to set the task from a desktop located next to the participant. For each trial, the system logged the trace of the participants’ \emph{feet movements} and \emph{head (HoloLens) movements} to establish a matching between the visual feedback and the foot-taps. Furthermore, the system measured the time between displaying the task and touching the floor with the foot as the \emph{\ac{TCT}} and logged it together with the \emph{foot used for interaction}, the \emph{tap position} (relative to the participant), the \emph{target} and the \emph{condition} for later analysis.

\subsection{Procedure}
\label{study1:procedure}

After welcoming the participants, the investigator introduced them to the concept and the setup of the study. During this, the method proposed by \typcite{Chapman1987} was used to measure the foot preference of the participants. For this, the investigator asked the participants to write their names with their feet, like they would in the sand on the beach. The investigator observed the participants and noted which foot they used. Further, the investigator measured the height and leg length of the participants to analyze their impact on the participants’ performance later. Then, the participants mounted the trackable apparatuses on their feet and were asked them to put the Hololens on their head. To avoid learning effects, the experiment started with five minutes warming phase without data recording to get accustomed to the hardware and the interfaces.

The system was calibrated with the participants standing relaxed and looking straight ahead. After starting the condition, the participants saw the respective visualization. Once ready and in starting position (both feet together), the investigator started the condition to be evaluated. The system then colored the respective target to be reached in blue (\reffig{fig:study:direct}, \ref{fig:study:indirect}) and informed the participant about the start of the trial with an additional audio signal. Then, the participant moved the foot and tapped the floor on the target position. The investigator did not enforce the usage of a specific foot but told the participants to use the foot that seemed most comfortable for each trial. After tapping the target, the system changed the target color to green to inform the participant that the measurement was recorded and that the participant should move the foot back to the starting position. Once reached, the system waited 2 seconds before proceeding to the next target.

Participants were instructed to focus on the accuracy (tapping the center of the target) instead of the speed. Participants did not receive any feedback regarding their performance during the study. After each condition, participants completed a NASA TLX~\ncite{Hart2006} questionnaire and answered questions regarding their experiences on a 5-point Likert-scale (1: strongly disagree, 5: strongly agree). Further, the system enforced a 5-minute break between the conditions during which participants gave qualitative feedback in a semi-structured interview. Each experiment took about 60 minutes per participant.

\textfigStudybox{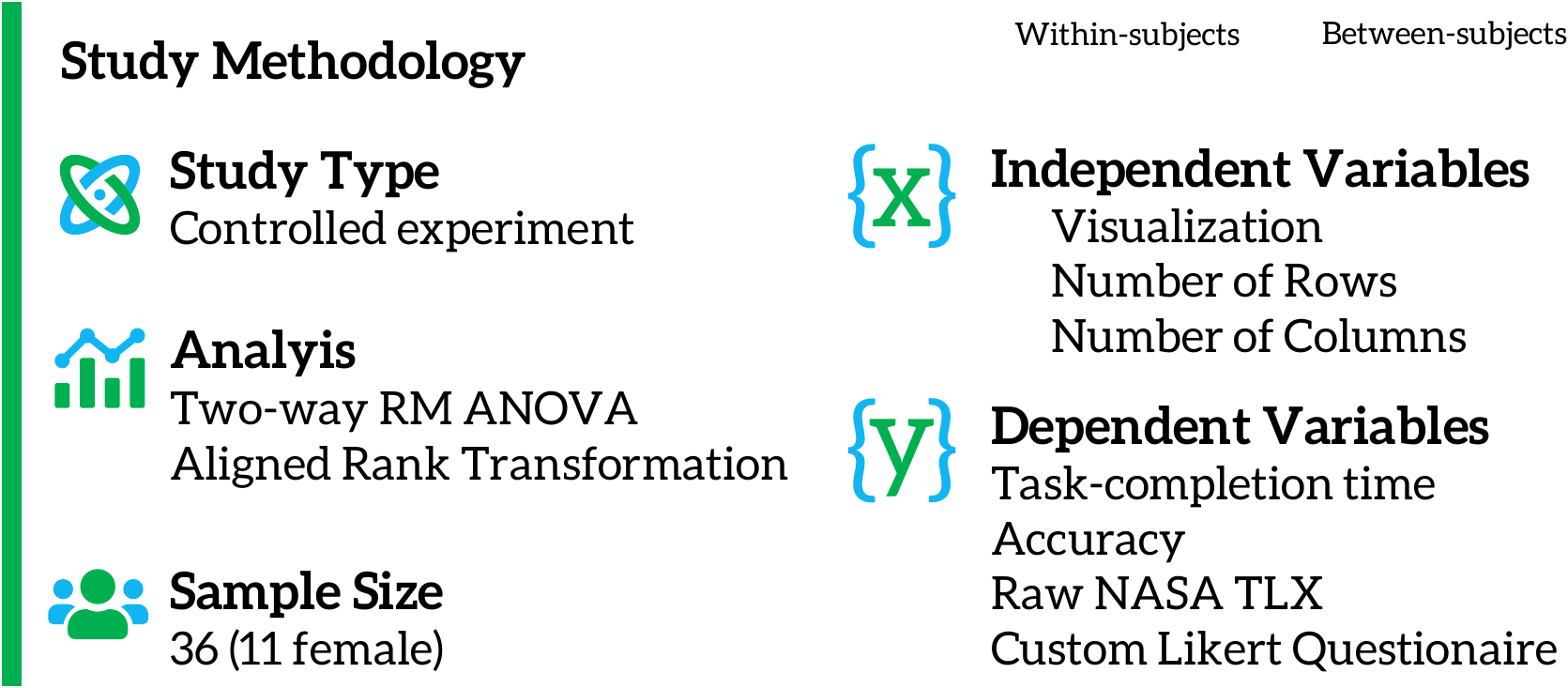}

\section{Experiment I: Direct Interaction}
\label{sec:cheesyfoot:evaluation1_results}

This section reports the results of a controlled experiment investigating RQ1 and, thus, focusing on \direct{} interfaces using the visualization on the floor as described in section \ref{sec:methodology:heads-up}. For this, 18 participants (6 female), aged between 21 and 30 years (\val{24.9}{3.0}), were recruited using our University’s mailing list. Three of them had prior experience with \ac{AR}. 7 out of 8748 trials were excluded in the analysis as outliers due to technical problems. The analysis of the data was performed as described in section \ref{sec:rw:methodology}. 

Section \ref{sec:cheesyfoot:comparison} compares the results of this experiment (focusing on \direct{} interfaces) to the results for \indirect{} interfaces. Section \ref{sec:cheesyfoot:discussion} discusses the results of both experiments with regards to the research questions.

\subsection{Accuracy}
\label{sec:cheesyfoot:experiment1:accuracy}

\textfigH{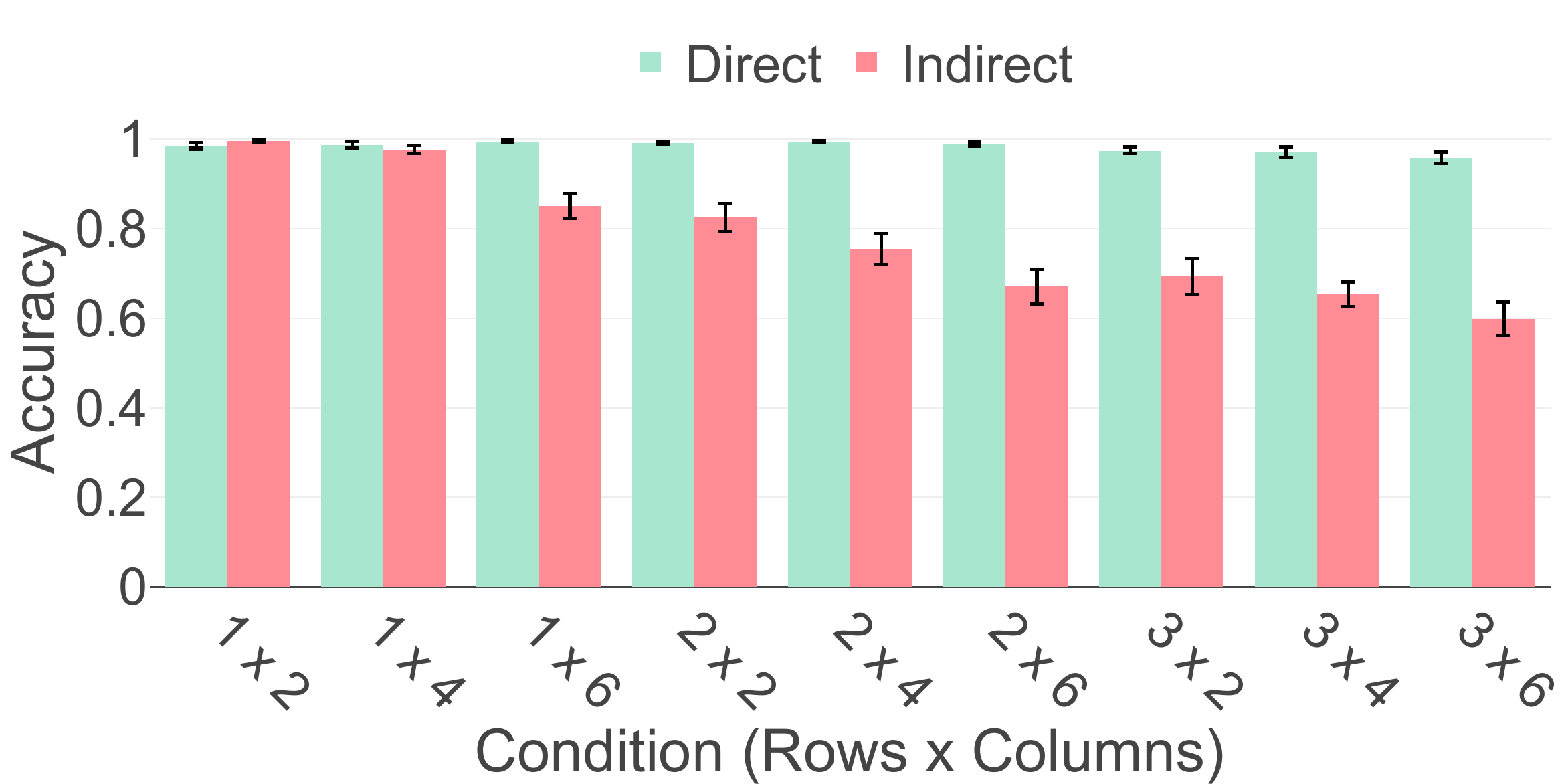}{The measured accuracy rates in both experiments. All error bars depict the standard error.}

The physical dimensions of the targets (visible on the floor through the \ac{HMD}) were used to classify the taps of the participants as hits and errors to obtain an accuracy rate. The analysis revealed that the number of rows had a significant (\anovaCor{1.32}{22.51}{4.068}{<.05}{.662}{.099}) influence on the accuracy with a small effect size. Post-hoc tests confirmed significantly higher accuracy rates for the \row{1} (\emmSiCI{98.9}{.6}{97.7}{100.2}{\%}) and \row{2} (\emmSiCI{99.1}{.6}{97.9}{100.4}{\%}) conditions compared to the \row{3} (\emmSiCI{96.8}{.6}{95.6}{98.1}{\%}) conditions.

The analysis did not show any significant influence of the number of columns (\anova{2}{34}{.515}{>.05}{.003}) or interaction effects between both factors (\anovaCor{2.63}{44.70}{1.699}{>.05}{.657}{.019}). Overall, the results showed high accuracy rates up to the highest (\row{3}, \col{6}) condition (\val{95.9 \%}{.5 \%}). Figure \ref{fig:cheesyfoot/accuracy} (\studyOneColor{}) depicts the measured accuracy rates for all conditions, table \ref{tab:cheesyfoot/tables/accuracy} lists the \acp{EMM} for the individual factors.

\begin{table}
	\centering
	
	\begin{tabularx}{\linewidth}{ccYYYY}
		& &   &   & \multicolumn{2}{c}{\textbf{95\% CI\textsuperscript{1}}}\\
		\cmidrule(lr){5-6}
		\textbf{Rows} & \textbf{Columns} & $\pmb{\mu}$ & $\pmb{\sigma}$ & \textbf{Lower} & \textbf{Upper}\\
		\midrule
		1 & 2 & 0.990 & 0.021 &  0.938 & 1.043  \\
		
		& 4 & 0.982 & 0.035 & 			0.930 & 1.035  \\
		
		& 6 & 0.925 & 0.112 & 			0.872 & 0.978  \\
		
		2 & 2 & 0.911 & 0.129 & 			0.858 & 0.963  \\
		
		& 4 & 0.878 & 0.161 & 			0.825 & 0.930  \\
		
		& 6 & 0.841 & 0.193 & 			0.789 & 0.894  \\
		
		3 & 2 & 0.834 & 0.193 & 			0.782 & 0.887  \\
		
		& 4 & 0.816 & 0.189 & 			0.763 & 0.868  \\
		
		& 6 & 0.783 & 0.217 & 			0.730 & 0.836  \\
		\bottomrule
	\end{tabularx}
	
	\caption{The accuracy rates of direct interfaces as measured in the first experiment. The table reports the recorded mean values $\mu$ together with the standard deviation $\sigma$. \textsuperscript{1} The confidence interval CI is based on the fitted \ac{EMM} model.}
	\label{tab:cheesyfoot/tables/accuracy}
\end{table}

\subsection{Task Completion Time}
\label{sec:cheesyfoot:experiment1:tct}
\textfigH{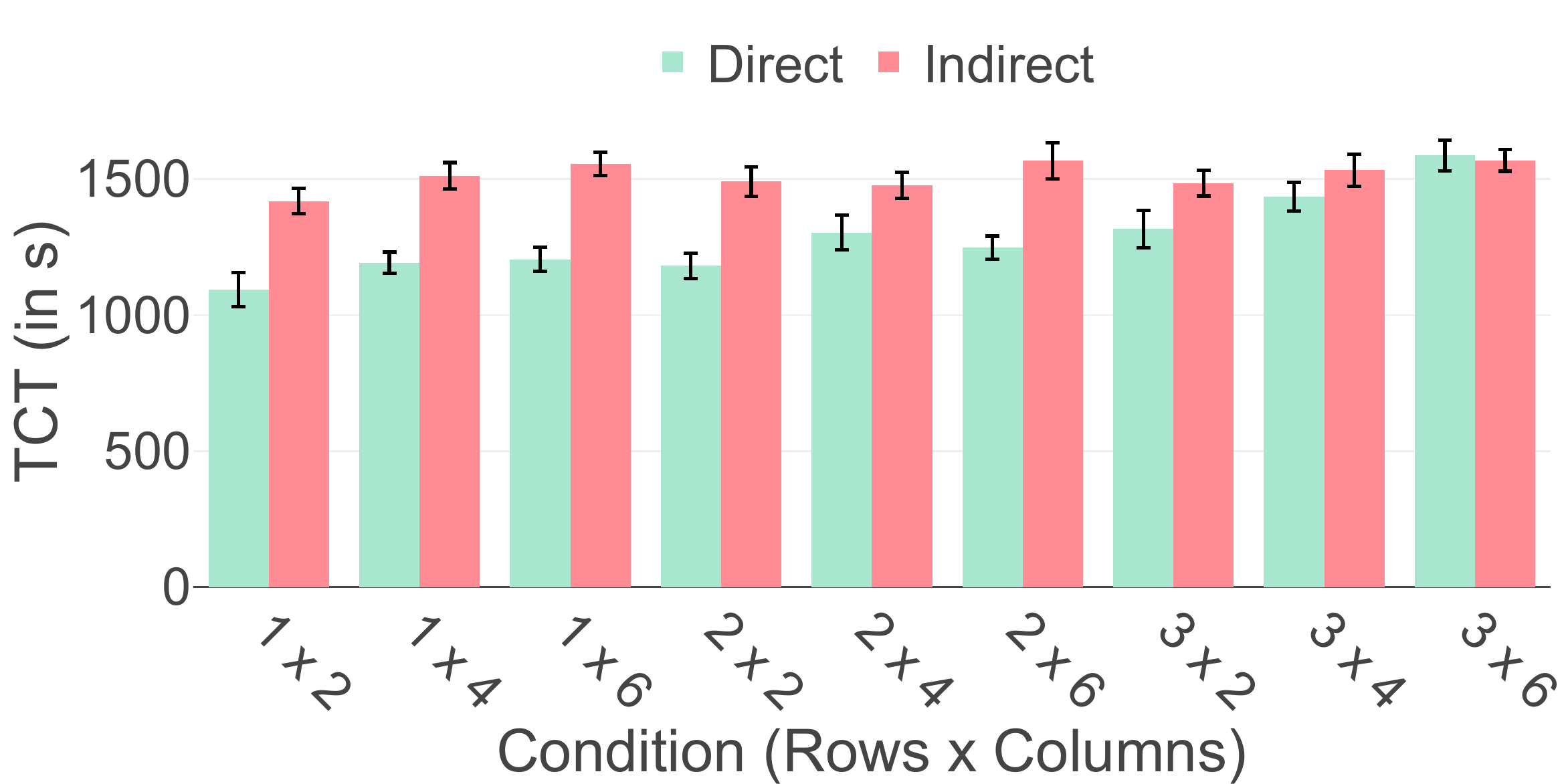}{The measured task-completion times in both experiments. All error bars depict the standard error.}

The analysis unveiled that both, the number of rows (\anova{2}{34}{14.47}{<.001}{.059}) and the number columns (\anova{2}{34}{43.39}{<.001}{.203}) had a significant influence on the \acl{TCT} with medium and large effect size, respectively. Further, the analysis found interaction effects between the number of rows and the number of columns (\anovaCor{2.16}{36.67}{3.22}{<.05}{.539}{.024}) with a medium effect size.

Post-hoc tests confirmed significantly rising \acp{TCT} for higher numbers of rows (\row{1}: \emmSiCI{1.163}{.046}{1.068}{1.258}{s}, \row{2}: \emmSiCI{1.243}{.046}{1.149}{1.338}{s}, \row{3}: \emmSiCI{1.445}{.046}{1.351}{1.540}{s}) between all levels ($p<.05$ between \row{1} and \row{2}, $p<.001$ otherwise). For the number of columns, post-hoc tests showed significant differences between the \col{2} (\emmSi{1.196}{.045}{1.102}{1.290}{s}) and \col{6} (\emmSiCI{1.346}{.045}{1.252}{1.440}{s}) conditions ($p<.001$) as well as between the \col{4} (\emmSiCI{1.310}{.45}{1.252}{1.440}{s}) and the \col{6} conditions. Figure \ref{fig:cheesyfoot/tct} (\studyOneColor{}) shows the \acp{TCT} for all conditions, table \ref{tab:cheesyfoot/tables/tct} lists the \acp{EMM} for the individual factors.

\begin{table}
	\centering
	
	\begin{tabularx}{\linewidth}{ccYYYY}
		& &   &   & \multicolumn{2}{c}{\textbf{95\% CI\textsuperscript{1}}}\\
		\cmidrule(lr){5-6}
		\textbf{Rows} & \textbf{Columns} & $\pmb{\mu}$ & $\pmb{\sigma}$ & \textbf{Lower} & \textbf{Upper}\\
		\midrule
			1 & 2 & 1.09s & .26s & 0.98s & 1.20s  \\
& 4   & 1.19s & .17s & 1.08s & 1.30s  \\
& 6   & 1.20s & .19s & 1.10s & 1.31s  \\
2 & 2 & 1.18s & .20s & 1.07s & 1.29s  \\
& 4   & 1.30s & .27s & 1.19s & 1.41s  \\
& 6   & 1.24s & .18s & 1.14s & 1.36s  \\
3 & 2 & 1.31s & .29s & 1.20s & 1.42s  \\
& 4   & 1.43s & .22s & 1.33s & 1.54s  \\
& 6   & 1.58s & .33s & 1.48s & 1.69s  \\
		\bottomrule
	\end{tabularx}
	
	\caption{The task-completion times of direct interfaces as measured in the first experiment (in seconds). The table reports the recorded mean values $\mu$ together with the standard deviation $\sigma$. \textsuperscript{1} The confidence interval CI is based on the fitted \ac{EMM} model.}
	\label{tab:cheesyfoot/tables/tct}
\end{table}

\subsection{Footedness and Foot Used for the Interaction}

The analysis could not find any influence of the footedness of the participants on the accuracy (\anova{1}{16}{.570}{>.05}{.007}) nor on the TCT (\anova{1}{16}{1.42}{>.05}{.042}). Interestingly, although the system left it up to the participants to decide which foot they wanted to use, virtually all targets to the left of the participants’ line of sight were performed with the left foot and vice versa ($\mu > 96\%$ for all conditions). Matching this, the results showed no significant influences of the number of rows (\anova{2}{32}{.408}{>.05}{.002}), the number of columns (\anovaCor{1.21}{19.28}{.292}{>.05}{.603}{.003}) or the footedness (\anova{1}{16}{.451}{>.05}{.014}) on the foot used for interaction.

\subsection{Size of the Target Areas}

\textfig{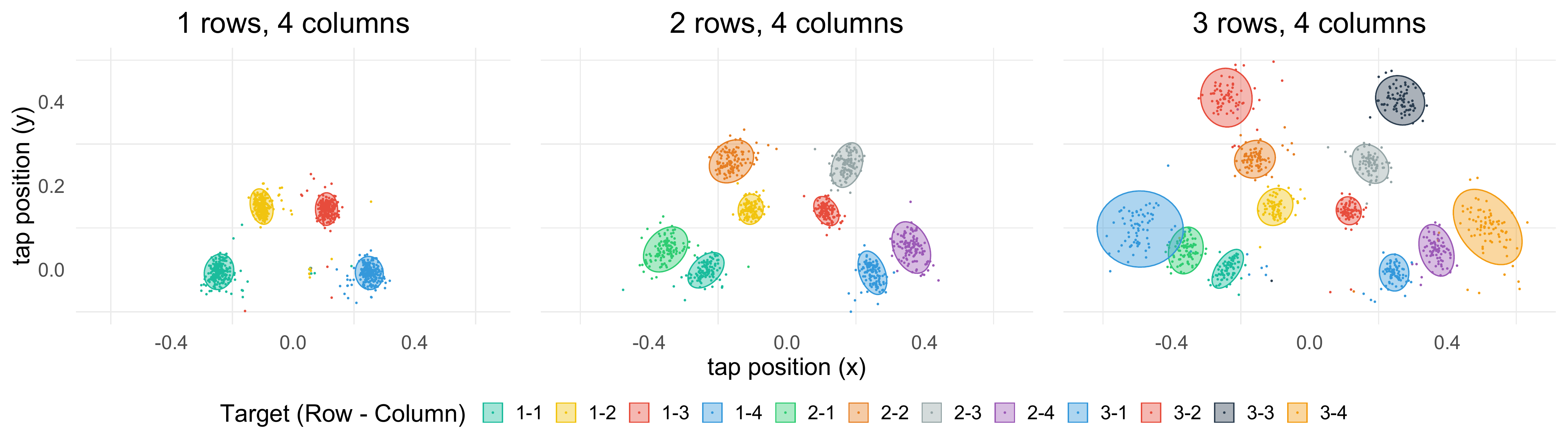}{Scatter plots with 95\% data probability ellipses for the \col{4} conditions with \direct{} interfaces in the first experiment. All target areas can be separated. The outer (\row{3}) target areas are larger than the nearer targets.}

This section analyzes the influence of the target position (as target row and target column) on the spread of the recorded tapping positions. As a measurement for the spread of data, individual 95\% data probability ellipses (i.e., ellipses containing 95\% of the recorded points for this target) were calculated per participant and target position and compared the other data ellipses.

The analysis showed a significant influence of the target row on the area of the targets (\anova{2}{34}{13.36}{<.001}{.04}) with a small effect size. Post-hoc tests confirmed significantly larger areas if the target was located in \row{3} (\emmSiCI{.0454}{.006}{.033}{.058}{m^2}) compared to \row{1} (\emmSiCI{.005}{.006}{.000}{.017}{m^2}) and \row{2} (\emmSiCI{.008}{.006}{.000}{.020}{m^2}), both $p<.001$. Despite rising means, the analysis could not show significant effects between targets in the \row{1} and \row{2} conditions.

The analysis could not find a significant influence of the target column (\anova{11}{187}{1.62}{>.05}{.026}) nor interaction effects between the number of rows and the number of columns (\anova{22}{374}{1.62}{>.05}{.0051}). The overlap between the target areas was not analyzed as the \direct{} visualization limited the size of the target areas. Figure \ref{fig:cheesyfoot/scatters_study1} depicts the 95\% data probability ellipses for the \col{4} conditions and illustrates the rising area sizes for targets in outer rows.

\subsection{TLX and Questionnaire}

\textfigH{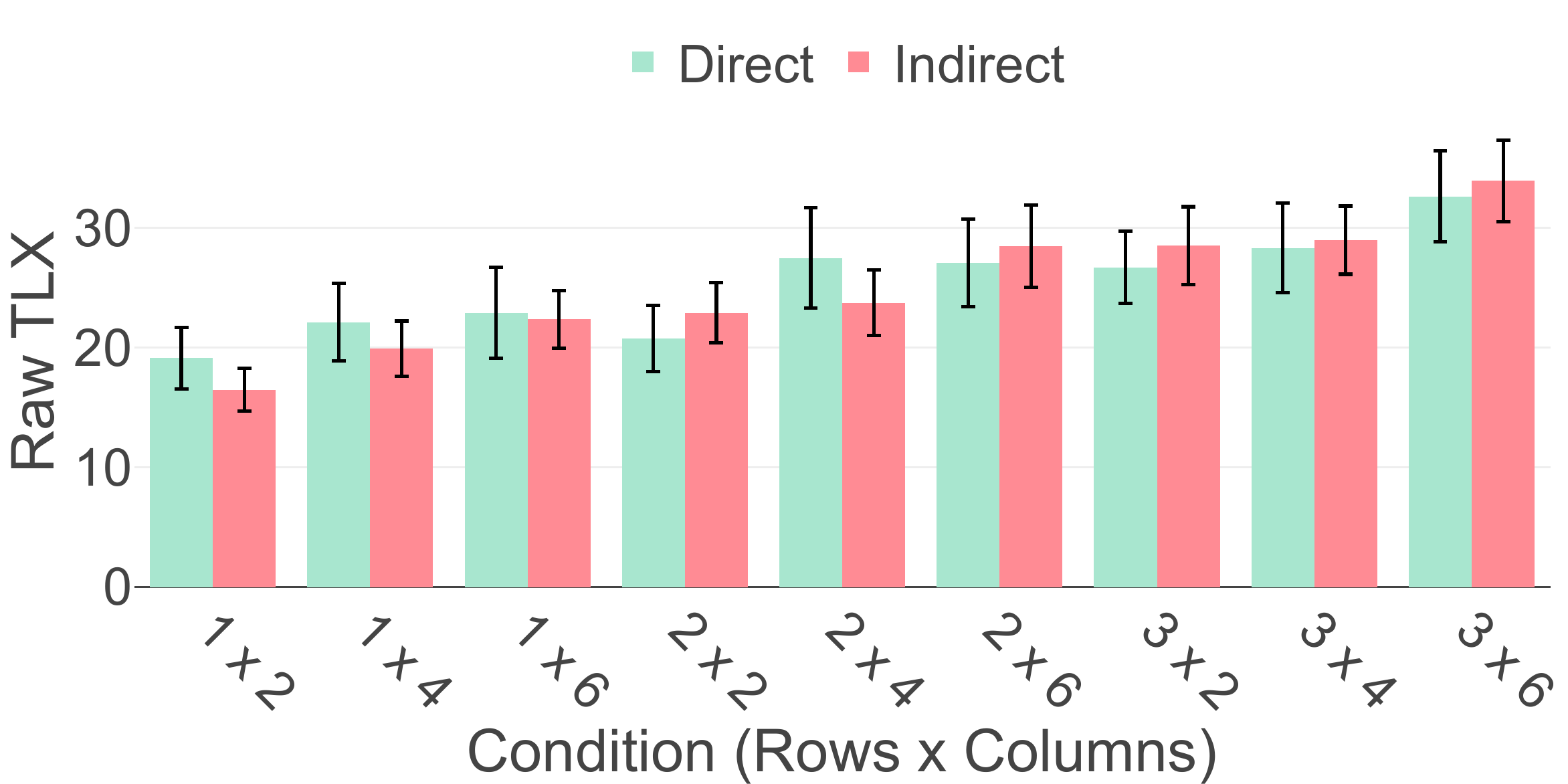}{The measured Raw-TLX rates in both experiments. All error bars depict the standard error.}

The \ac{RTLX} questionaire showed a significant influence of the number of rows (\anova{2}{34}{16.82}{<.001}{.047}) with a small effect size. Post-hoc tests confirmed a significant effect for the number of rows between the \row{1} (\emm{21.4}{3.28}{14.5}{28.3}) and \row{2} (\emm{25.1}{3.28}{18.3}{32.0}) conditions ($p<.05$), the \row{1} and \row{3} (\emm{29.2}{3.28}{22.4}{36.1}) conditions ($p<.001$) as well as between the \row{2} and \row{3} conditions ($p<.05$).

The analysis further showed a significant influence of the number of columns (\anovaCor{1.34}{22.85}{6.83}{<.01}{.672}{.023}) with a small effect size. Post-hoc tests showed significant differences between the \col{2} (\emm{22.2}{3.3}{15.3}{29.1}) and \col{4} (\emm{26.0}{3.3}{19.1}{32.9}) conditions as well as between the \col{2} and \col{6} (\emm{27.5}{3.3}{20.6}{34.4}) conditions. The analysis could not find any interaction effects between the factors (\anova{4}{68}{2.28}{>.05}{.005}). Figure \ref{fig:cheesyfoot/tlx} (\studyOneColor{}) depicts the measured values for all conditions.

\subsubsection{Confidence}

\textfig{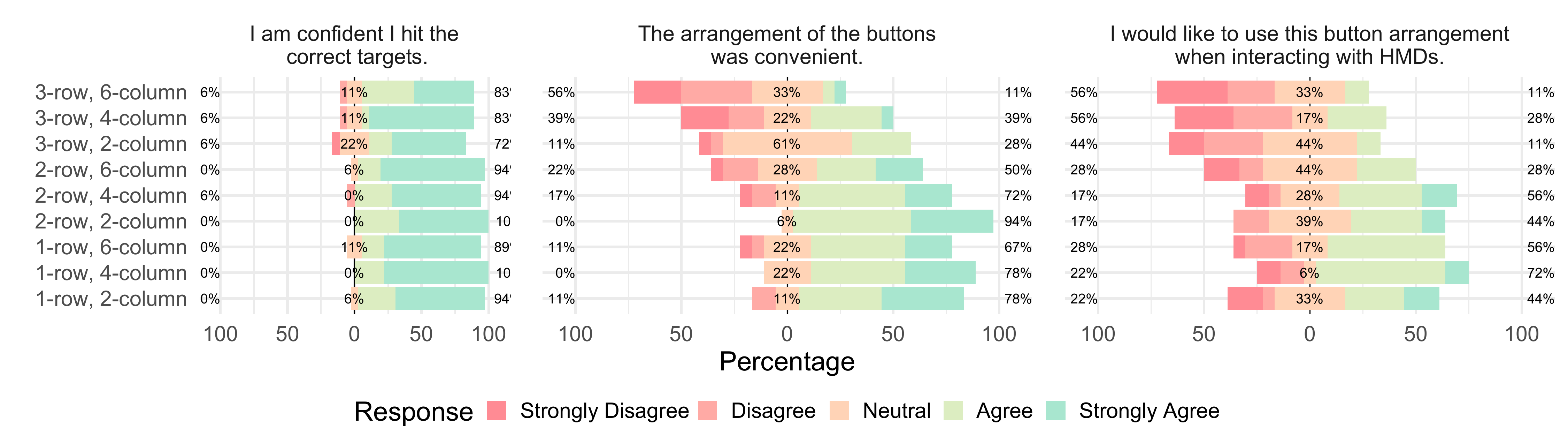}{The participant’s answers to our questions for direct interfaces on a 5-point Likert-scale.}

Matching the quantitative results, the participants felt very confident that they hit the correct targets across all conditions (\reffig{fig:cheesyfoot/likert-study1}). The analysis showed a significant effect for the number of columns (\anovaWithoutEffect{2}{34}{5.259}{<.05}). Post-hoc tests confirmed a significantly higher confidence for \col{4} conditions compared to \col{2} and \col{6} conditions (both $p<.05$). The analysis could not find effects for the number of rows (\anovaWithoutEffect{2}{34}{.831}{>.05}) but interaction effects between the two factors (\anovaWithoutEffect{4}{68}{5.057}{<.01}).

\subsubsection{Convenience}

The questionnaire asked the participants how convenient they felt with the layout to interact with information considering the number of input options and the physical and mental effort required to use them.. The analysis showed significant effects for both, the number of rows (\anovaWithoutEffect{2}{34}{23.984}{<.001}) as well as the number of columns (\anovaWithoutEffect{2}{34}{7.891}{<.01}). Post-hoc tests confirmed significantly lower ratings for the \row{3} conditions compared to the other levels (both $p<.001$). Regarding the number of columns, the analysis found a significant difference between the \col{2} and \col{6} conditions ($p<.01$). The analysis could not find interaction effects (\anovaWithoutEffect{4}{68}{1.065}{>.05}).

A closer look at the answers supports the statistical results and, thus, the strong influence of the number of rows: All but the \row{3} conditions are rated predominantly positively. Figure \ref{fig:cheesyfoot/likert-study1} depicts all answers from the participants.

\subsubsection{Willingness to Use}

Further, the questionnaire asked the participants if they would like to use this arrangement for interacting with HMDs. The analysis showed a significant effect for both, the number of rows (\anovaWithoutEffect{2}{34}{8.938}{<.001}) as well as the number of columns (\anovaWithoutEffect{2}{34}{6.087}{<.01}). The analysis could not find interaction effects (\anovaWithoutEffect{2}{68}{1.370}{>.05}). Post-hoc tests confirmed significantly lower ratings for the \row{3} conditions compared to \row{1} ($p<.001$) and \row{2} ($p<.05$) conditions. For the number of columns, the analysis found a significant higher rating for the \col{4} conditions compared to the \col{6} conditions ($p<.01$).

Again, the participants' ratings for all but the \row{3} conditions were predominantly positive (\reffig{fig:cheesyfoot/likert-study1} for all results).

\subsection{Qualitative Feedback}
\label{sec:cheesyfoot:experiment1:qualitative}

In general, all participants appreciated the idea of foot-based interactions with \acp{HMD} because it is \mpquote{easy to use}{P6, P8, P11, P12}, and \mpquote{not tiring [compared to the standard air-tap interface of the Hololens]}{P8, P17}.

Participants commented that the limitations of the used hardware - \mpquote{weight}{P1, P3, P4, P9}, \mpquote{field of view}{P5, P6, P7, P8, P15} - had a strong influence on their comfort because it forced them into an \pquote{unnatural}{14} posture during the study. P17 summarized: \enquote{Looking down all the time is a bit exhausting for the neck. So I wouldn’t use it for longer-lasting [interactions], but I would love this for quick and short [interactions]}.

\section{Experiment II: Indirect Interaction}
\label{sec:cheesyfoot:evaluation2_results}

This section reports the results of the second experiment focusing on RQ2 and, thus, on \indirect{} interfaces using the visualization in front of the participant as described in section \ref{sec:methodology:heads-up}. For this, 18 participants (5 female), aged between 21 and 31 years (\val{23.3}{2.8}), 3 left-footed, were recruited using our University’s mailing list. None of them had prior experience with \ac{AR}. During the analysis, 16 out of 8748 trials were excluded as outliers due to technical problems during recording. The analysis of the data was performed as described in section \ref{sec:rw:methodology}. 

Section \ref{sec:cheesyfoot:comparison} compares the results of this experiment (focusing on \indirect{} interfaces) to the results for \direct{} interfaces. Section \ref{sec:cheesyfoot:discussion} discusses the results of both experiments with regards to the research questions.

\subsection{Classification}

In the first experiment, the analysis used the physical dimensions of the targets (visible as \direct{} feedback on the floor) to calculate the accuracy rates. However, this approach could not be transferred directly to the second experiment, as the participants interacted with an \indirect{} visualization. There was, therefore, no direct definition of the accuracy of the participants’ hits and misses. As a result, the analysis was started with the construction of suitable classifiers.

The data was classified using \acp{SVM} and trained nine \ac{SVM} classifiers according to our nine conditions. For this, each corresponding partial data set was divided into an 80\% training set and a 20\% test set. The training sets were used to train per-condition \acp{SVM} with radial kernels. To avoid over-fitting to the data, the process used 10-fold cross-validation with 3 repetitions and predictions on the 20\% test sets to assess the quality of the \acp{SVM}. Furthermore, per-participant \acp{SVM} were trained and compared to the results of the models trained with the data of all participants. However, as there were only minor differences in the accuracy rates (+/- 2\%, depending on the condition), this section uses the generalized models for further analysis.

\subsection{Accuracy}

\begin{table}
	\centering
	
	\begin{tabularx}{\linewidth}{ccYYYY}
		& &   &   & \multicolumn{2}{c}{\textbf{95\% CI\textsuperscript{1}}}\\
		\cmidrule(lr){5-6}
		\textbf{Rows} & \textbf{Columns} & $\pmb{\mu}$ & $\pmb{\sigma}$ & \textbf{Lower} & \textbf{Upper}\\
		\midrule

1 & 2 & .995 & .008 & .931 & 1.0 \\
& 4 & .977 & .037 & .912 & 1.0 \\
& 6 & .846 & .122 & .782 & .911 \\
2 & 2 & .821 & .141 & .756 & .885 \\
& 4 & .746 & .148 & .682 & .810 \\
& 6 & .675 & .161 & .611 & .739 \\
3 & 2 & .675 & .172 & .611 & .740 \\
& 4 & .641 & .117 & .577 & .705 \\
& 6 & .585 & .148 & .521 & .650 \\

		\bottomrule
	\end{tabularx}
	
	\caption{The accuracy rates of indirect interfaces as measured in the second experiment. The table reports the recorded mean values $\mu$ together with the standard deviation $\sigma$. \textsuperscript{1} The confidence interval CI is based on the fitted \ac{EMM} model.}
	\label{tab:cheesyfoot/tables/accuracy_indirect}
\end{table}

The analysis showed that both independent variables, the number of rows (\anova{2}{30}{60.87}{<.001}{.460}) and the number of columns (\anova{2}{30}{11.61}{<.001}{.082}) had a significant influence on the accuracy with large and small effect size, respectively. Post-hoc tests confirmed significantly lower accuracy rates for a higher number of rows between all groups (all $p <.001$) and between 2 and 6 ($p<.001$) as well as 4 and 6 columns ($p<.05$). The analysis could not find any interaction effects between the variables (\anova{4}{60}{2.37}{>.05}{.007}).

Interestingly, a closer look revealed that, as the number of rows and columns increases, the falling accuracy is not directly dependent on the number of resulting targets: In both, the \row{1}, \col{4} condition as well as the \row{2}, \col{2} condition, the participants had to hit 4 different targets. However, we found a significantly higher accuracy rate for the \row{1}, \col{4} (\val{98 \%}{1.1 \%}) condition compared to the \row{2}, \col{2} (\val{83.6 \%}{13.4 \%}) condition ($p<.01$). The analysis found the same effect for the \row{1}, \col{6} (\val{85.1 \%}{12.3 \%}) condition compared to the \row{3}, \col{2} (\val{69.7 \%}{17.2 \%}) condition ($p<.01$). This indicates that the number of rows has a greater influence on the accuracy than the number of columns.

Considering the \ac{EMM} for the individual conditions, the analysis found a overall high accuracy rate for the \row{1} (\emmSiCI{94.3}{3}{89.0}{99.0}{\%}) conditions. Therefore, the accuracy for the \row{1} conditions of \indirect{} interfaces proved to be comparable to the accuracy found for \direct{} interfaces (see \ref{sec:cheesyfoot:experiment1:accuracy}). Section \ref{sec:cheesyfoot:comparison} presents a more detailed comparison of both interface styles. Figure \ref{fig:cheesyfoot/accuracy} (\studyTwoColor{}) depicts the measured accuracy rates for all conditions, table \ref{tab:cheesyfoot/tables/accuracy_indirect} lists the \acp{EMM} for the individual factors.

\subsection{Task Completion Time}
\label{sec:cheesyfoot:experiment2:tct}

\begin{table}
	\centering
	
	\begin{tabularx}{\linewidth}{ccYYYY}
		& &   &   & \multicolumn{2}{c}{\textbf{95\% CI\textsuperscript{1}}}\\
		\cmidrule(lr){5-6}
		\textbf{Rows} & \textbf{Columns} & $\pmb{\mu}$ & $\pmb{\sigma}$ & \textbf{Lower} & \textbf{Upper}\\
		\midrule
		1 & 2 & 1.44s & .17s & 1.34s & 1.55s  \\
		& 4   & 1.53s & .19s & 1.43s & 1.64s  \\
		& 6   & 1.58s & .17s & 1.48s & 1.69s  \\
		2 & 2 & 1.53s & .21s & 1.42s & 1.63s  \\
		& 4   & 1.50s & .20s & 1.40s & 1.61s  \\
		& 6   & 1.60s & .27s & 1.50s & 1.71s  \\
		3 & 2 & 1.51s & .19s & 1.41s & 1.62s  \\
		& 4   & 1.55s & .26s & 1.44s & 1.66s  \\
		& 6   & 1.57s & .17s & 1.46s & 1.68s  \\
		\bottomrule
	\end{tabularx}
	
	\caption{The task-completion times of indirect interfaces as measured in the second experiment (in seconds). The table reports the recorded mean values $\mu$ together with the standard deviation $\sigma$. \textsuperscript{1} The confidence interval CI is based on the fitted \ac{EMM} model.}
	\label{tab:cheesyfoot/tables/tct_indirect}
\end{table}

The analysis unveiled that the number columns of the condition had a significant (\anova{2}{30}{7.698}{<.01}{.032}) effect on the \acl{TCT} with a small effect size. Post-hoc tests confirmed a significantly lower \ac{TCT} for the \col{2} conditions (\emmSiCI{1.495}{.044}{1.402}{1.587}{s}) compared to the \col{6} conditions (\emmSiCI{1.585}{.044}{1.492}{1.677}{s}), $p<.001$. With regard to the number of rows (\ac{EMM} $\mu$ between $\SI{1.52}{s}$ and $\SI{1.54}{s}$), the analysis could not find any significant influence (\anova{2}{30}{.307}{>.05}{.003}). Also, the analysis could not find any interaction effects between the factors (\anova{4}{60}{1.314}{>.05}{.282}). Figure \ref{fig:cheesyfoot/tct} (\studyTwoColor{}) depicts the \acp{TCT} for all conditions, table \ref{tab:cheesyfoot/tables/tct_indirect} lists the \acp{EMM} for the individual factors.

\subsection{Footedness and Foot Used for the Interaction}

The analysis could not find any influence of the footedness of the participants on the accuracy (\anova{1}{14}{.145}{>.05}{.002}) nor on the TCT (\anova{1}{14}{2.08}{>.05}{arg5.08}).

As in the first experiment, almost all targets to the left of the participants’ line of sight were performed with the left foot and vice versa ($\mu > .97$ for all conditions). Again, the analysis found no significant influences of the number of rows (\anovaCor{1.27}{17.74}{.044}{>.05}{.633}{.001}), the number of columns (\anovaCor{1.33}{18.61}{.344}{>.05}{.665}{.003}) or the footedness of the participant (\anova{1}{14}{.048}{>.05}{.001}) on the foot used for interaction.

\subsection{Size of the Target Areas}
\label{sec:study2:size}

\textfig{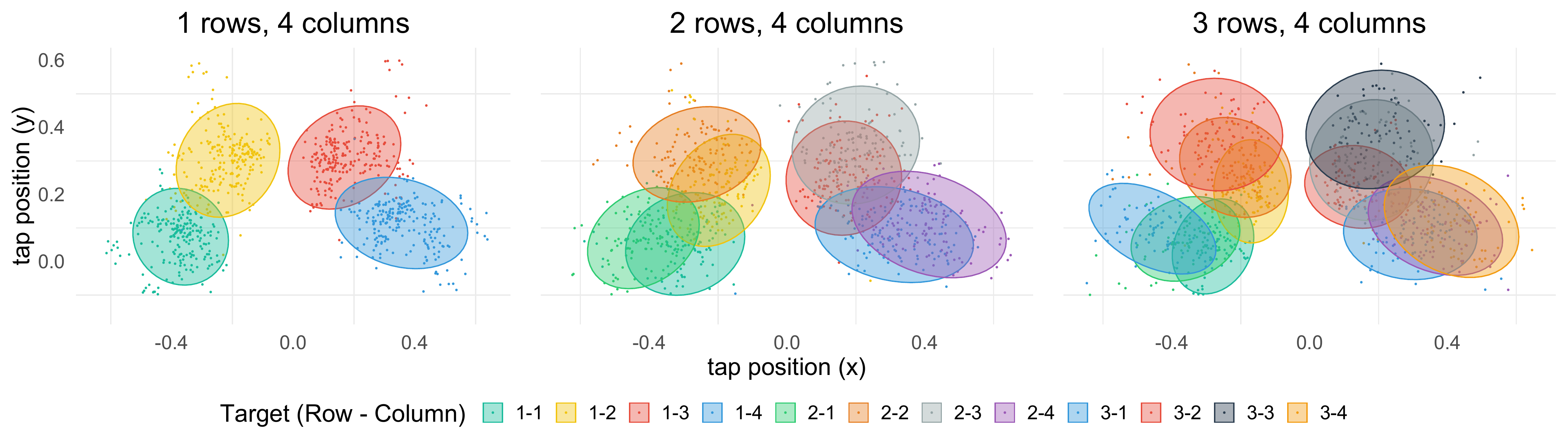}{Scatter plots with 95\% data probability ellipses for the \col{4} conditions with \indirect{} interfaces in the second experiment. While the data points for four columns can be separated, this is not possible for more than one row.}

Again, the analysis showed a siginificant influence of the target row on the area of the targets (\anova{2}{32}{8.90}{<.001}{.027}) with a small effect size. Post-hoc tests confirmed significantly smaller areas if the target was in \row{1} (\emmSiCI{.042}{.009}{.024}{.060}{m^2}) compared to \row{3} (\emmSiCI{.074}{.009}{.056}{.092}{m^2}), $p<.001$. The analysis could not find a significant influence of the target column (\anova{11}{176}{1.58}{>.05}{.029}) nor interaction effects (\anova{22}{352}{1.02}{>.05}{.028}).

\subsection{Overlap}
\label{sec:cheesyfoot:experiment2:overlap}

For conditions with multiple rows, there were noticeable overlaps in the distribution of the tapping points (\reffig{fig:cheesyfoot/scatters} for the 4 column conditions). As a measure for these overlaps, the analysis compared the number of points from adjacent targets in the row direction and in the column direction that fell into the 95\% data ellipse of each target.

The analysis showed a significant difference between the overlap in row and column direction (\anovaCor{1}{17}{324}{<.001}{.890}) with a large effect size. Post-hoc tests confirmed a significantly lower overlap in row direction (\valSi{4.0}{3.7}{\%}) compared to the column direction (\valSi{55.0}{12.7}{\%}), $p<.001$.

\subsection{TLX and Questionnaire}

The analysis showed a significant influence of the number of rows (\anova{2}{34}{31.02}{<.001}{.125}) with a medium effect size. Post-hoc tests confirmed a significantly higher perceived cognitive load for higher numbers of rows ($p<.001$ comparing \row{1} and \row{3}, $p<.01$ otherwise) from \emm{19.6}{2.55}{14.3}{24.9} (\row{1}) over \emm{25.0}{2.55}{19.7}{30.4} (\row{2}) to \emm{30.5}{2.55}{25.2}{35.8} (\row{3}).

The analysis further found a significant influence of the number of columns (\anova{2}{34}{10.481}{<.001}{.035}) with a small effect size. The post-hoc analysis showed rising estimated marginal means (\col{2}: \emm{22.6}{2.53}{17.4}{27.9}, \col{4}: \emm{24.2}{2.53}{18.9}{29.5}, \col{6}: \emm{28.3}{2.53}{23.0}{33.5}) with significant differences between 2 and 6 columns ($p<.001$) as well as between 4 and 6 columns ($p<.05$). We could not observe interaction effects between the number of rows and the number of columns (\anova{4}{68}{.447}{>.05}{.002}). Figure \ref{fig:cheesyfoot/tlx} (\studyTwoColor{}) depicts the measured values.

\subsubsection{Confidence}

\textfig{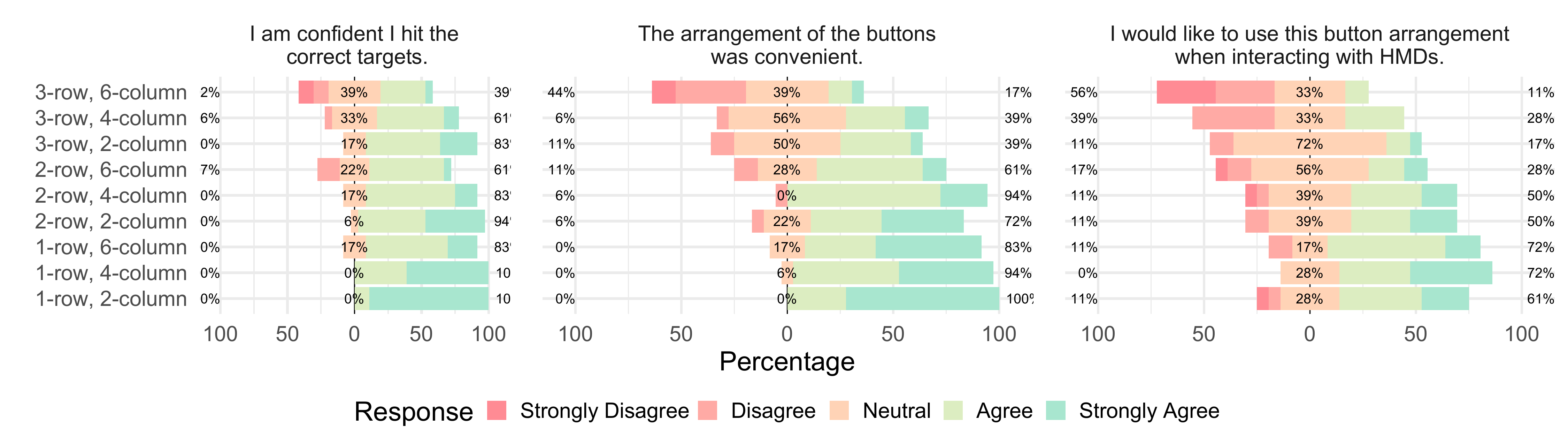}{The participant’s answers to our questions for indirect interfaces on a 5-point Likert-scale.}

The questionnaire asked the participants how confident they felt to have hit the correct targets. The analysis found significant effects for both, the number of rows (\anovaWithoutEffect{2}{34}{22.711}{<.001}) as well as the number of columns (\anovaWithoutEffect{2}{34}{35.345}{<.001}). Post-hoc tests confirmed significantly higher confidence ratings for \row{1} conditions compared to \row{2} and \row{3} conditions (both $p<.001$). For the number of columns, the analysis found significantly rising ratings between all levels (all $p<.001$). The analysis could not find interaction effects (\anovaWithoutEffect{4}{68}{.185}{>.05}).

The absolute numbers (\reffig{fig:cheesyfoot/likert-study2}) show a high agreement for all \row{1} conditions with decreasing confidence for higher numbers. Interestingly, the majority of the participants were convinced that they could keep the targets apart for all conditions (except \row{3}, \col{6}).

\subsubsection{Convenience}

The questionnaire further asked the participants how convenient the layout felt to interact with information. The analysis showed significant effects for both, the number of rows (\anovaWithoutEffect{2}{34}{56.462}{<.001}) and the number of columns (\anovaWithoutEffect{2}{34}{8.203}{<.01}). Post-hoc tests confirmed significantly falling ratings for higher numbers of rows between all levels (all $p<.001$). Regarding the number of columns, we found significantly lower ratings for the \col{6} conditions compared to the \col{2} ($p<.01$) and \col{4} ($p<.05$) conditions. The analysis could not find interaction effects (\anovaWithoutEffect{4}{68}{1.947}{>.05}).

All but the \row{3}, \col{6} condition were rated predominantly positive.

\subsubsection{Willingness to Use}

As the last question, the questionnaire asked the participants if they would like to use this arrangement for interacting with HMDs. The analysis showed a significant effect for the number of rows (\anovaWithoutEffect{2}{34}{26.849}{<.001}) and the number of columns (\anovaWithoutEffect{2}{34}{3.600}{<.05}) as well as interaction effects between the factors (\anovaWithoutEffect{4}{68}{3.286}{<.05}). Post-hoc tests confirmed significantly lower ratings for the \row{3} conditions compared to the \row{1} and \row{2} conditions (both $p<.001$). For the number of columns, post-hoc tests did not confirm significant differences.

\subsection{Qualitative Feedback}
\label{sec:cheesyfoot:experiment2:qualitative}

The participant's feedback proved to be more enthusiastic compared to the \direct{} interfaces as presented in section \ref{sec:cheesyfoot:experiment1:qualitative}. See section \ref{sec:cheesyfoot:comparison} and \ref{sec:cheesyfoot:discussion} for more detailed comparisons and discussion.

In general, all participants appreciated the idea of being able to interact with \acp{HMD} using their feet without looking at the floor. When asked for the reasons, participants told us that this interaction modality felt \pquote{novel}{1}, \pquote{fun to use}{12} and \pquote{very easy to perform in addition to other tasks}{15} as the \pquote{hands are not needed}{18} and \pquote{it's a low effort extension [\ldots] to interact}{9}. Participants found the \pquote{radial placement}{11} of the targets \pquote{nice}{11} and had the feeling that different columns were \pquote{relatively easy to discern}{7}. Four of the participants felt \mpquote{unsure}{P2, P5, P11, P12} about their performance with multiple rows. P11 even perceived more than one row as \enquote{\emph{inconvenient}}. P18 summarized: This \enquote{\emph{feels quite naturally in comparison to the strange in-air gestures that are used for the Hololens}}.

\section{Comparison of Interaction Techniques}
\label{sec:cheesyfoot:comparison}

This section compares the two techniques using two-way RM ANOVA with the interaction method as a between-subjects factor.

\subsection{Accuracy}
\label{sec:cheesyfoot:comparison:accuracy}
\textfig{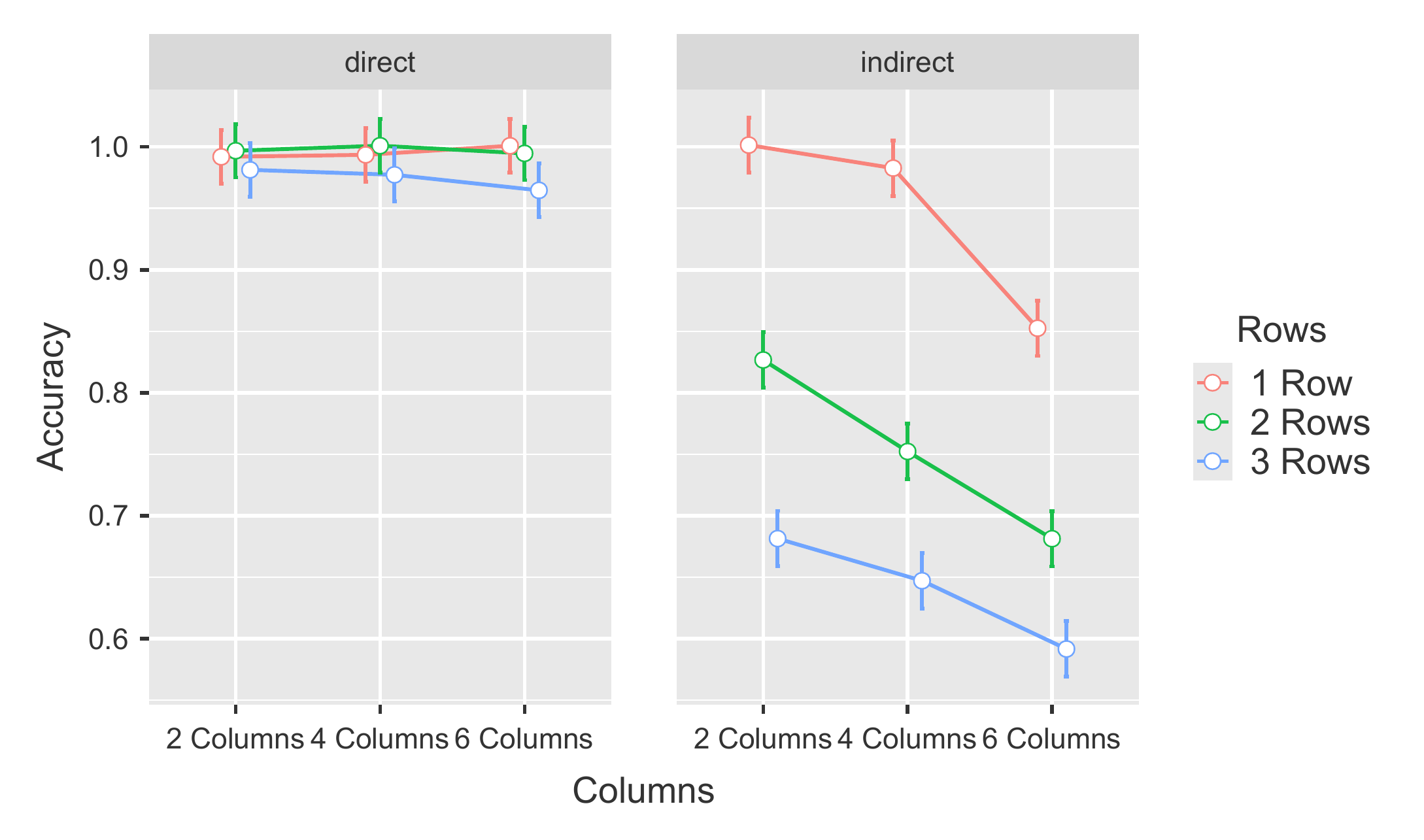}{Comparison of the \acp{EMM} of the accuracy rates for direct and indirect interfaces.}

As expected, the analysis found a significant effect of the interaction method on the accuracy with a large effect size (\anova{1}{32}{133.00}{<.001}{.386}). Post-hoc tests confirmed significantly higher accuracy rates for \direct{} (\emmSiCI{98.9}{1.4}{96.1}{100}{\%}) compared to \indirect{} (\emmSiCI{78.0}{1.4}{75.1}{80.8}{\%}) ($p<.001$). Figure \ref{fig:cheesyfoot/tables/accuracy_compared} depicts the \acp{EMM} for the individual factors.

\subsection{Task Completion Time}

\textfig{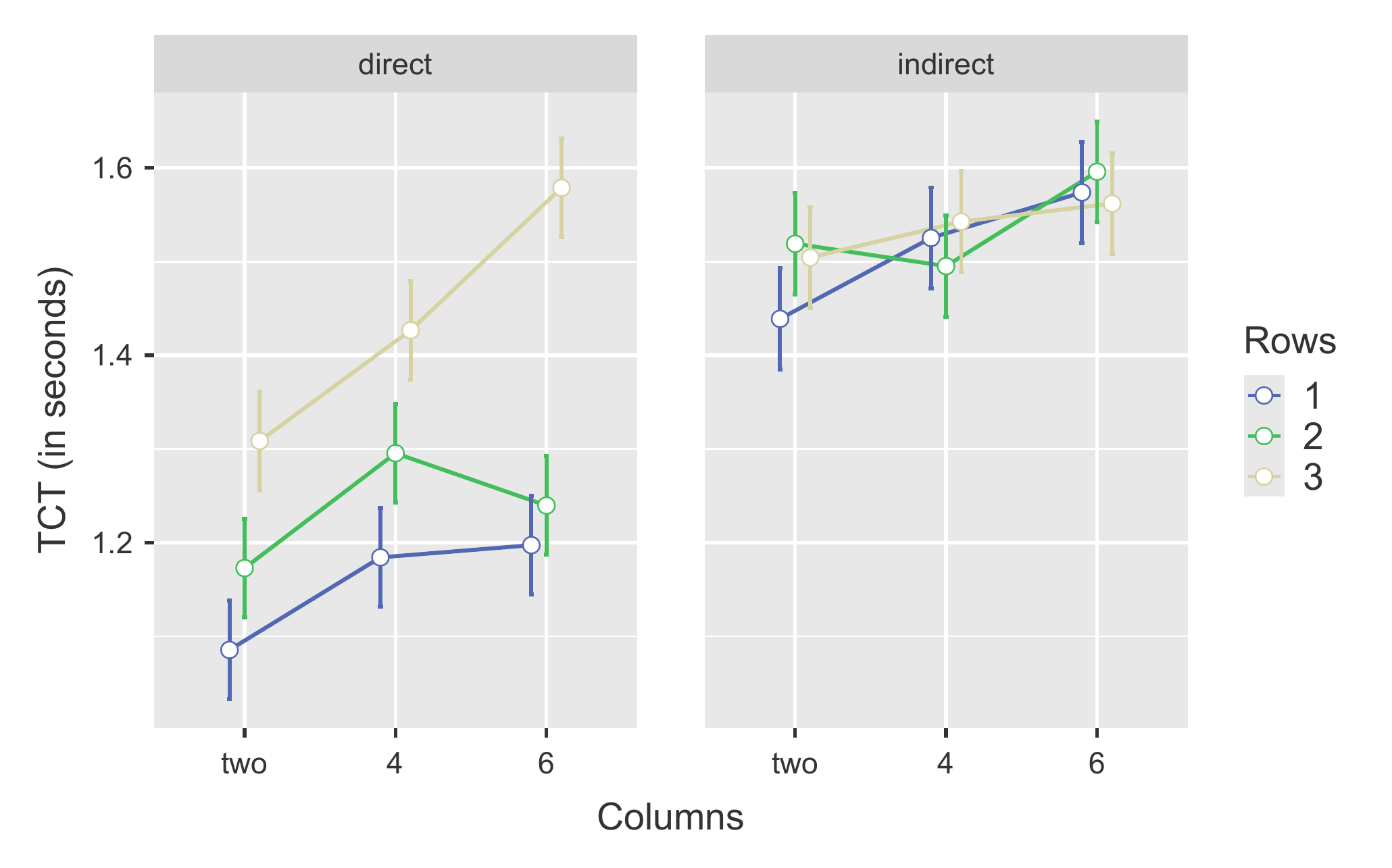}{Comparison of the \acp{EMM} of the \acp{TCT} of direct and indirect interfaces.}

The analysis unveiled a significant effect of the interaction method on the \ac{TCT} with a large effect size (\anova{1}{32}{17.8}{<.001}{.220}). Post-hoc tests confirmed significantly lower \acp{TCT} for \direct{} (\emmSiCI{1.277}{.042}{.119}{.136}{s}) compared to \indirect{} (\emmSiCI{1.529}{.042}{1.44}{1.61}{s}) ($p<.001$). Figure \ref{fig:cheesyfoot/tables/tct_compared} depicts the \acp{EMM} for the individual factors.

\subsection{Size of Target Areas}

The analysis found a significant effect of the interaction method on the size of the target areas with a small effect size (\anova{1}{33}{19.7}{<.001}{.042}). Post-hoc tests confirmed significantly smaller areas for \direct{} (\emmSiCI{.019}{.006}{.007}{.031}{m^2}) compared to \indirect{} (\emmSiCI{.055}{.006}{.045}{.076}{m^2}) interactions ($p<.001$).

\subsection{TLX and Questionaire}

The analysis did not show significant effect of the interaction method on the raw TLX (\anova{1}{34}{.002}{>.05}{.00}). The analysis found a significant influence of the interaction method on the confidence (\anovaWithoutEffect{1}{34}{14.05}{<.001}). Post-hoc tests confirmed significantly higher ratings for \direct{} compared to \indirect{} ($p<.001$). The analysis could not find significant effects of the interaction method on the convenience (\anovaWithoutEffect{1}{34}{1.83}{>.05}) or on the willingness to use (\anovaWithoutEffect{1}{34}{1.53}{>.05}).

\section{Discussion and Guidelines}
\label{sec:cheesyfoot:discussion}

The results of our controlled experiments suggest that foot-taps provide a viable interaction technique for \acp{HMD}. In both experiments, the evaluation showed \acp{TCT} suitable for fast interactions. While the analysis found significantly increasing \acp{TCT} for finer subdivisions of \direct{} interfaces (see section \ref{sec:cheesyfoot:experiment1:tct}), the \acp{TCT} of \indirect{} interfaces were stable across all conditions with only slight differences (see section \ref{sec:cheesyfoot:experiment2:tct}). Interestingly, for higher subdivisions, the \ac{TCT} seem to converge between both styles.

Based on the analysis of the two interaction styles, this section presents a set of guidelines.

\subsection{Favour the Division into Columns over Rows}

Our results suggest that more granular subdivisions through higher numbers of rows have a larger impact on the accuracy than finer subdivisions through the addition of columns. This impression was further supported for \indirect{} interfaces by investigating the overlap of the individual target areas: The analysis found a significantly larger overlap within a column (i.e., between several rows) compared with the overlap within a row (i.e., between several columns, see section \ref{sec:cheesyfoot:experiment2:overlap}). Also, in both experiments, the analysis showed a significantly growing spread of the tapping points for targets in more distant target rows (\reffig{fig:cheesyfoot/scatters}).

Therefore, the division into columns over rows should be favored when designing such interfaces.

\subsection{Use indirect interfaces for longer-term interactions that require less accuracy}

As expected, the accuracy rates for \indirect{} interactions were significantly lower compared to \direct{} interactions (see section \ref{sec:cheesyfoot:comparison:accuracy}). However, the difference was very low for the \row{1} conditions, in particular for 2 and 4 targets (see figure \ref{fig:cheesyfoot/accuracy}). Together with the differing overlaps in the row and column directions discussed above, this leads us to the conclusion that the participants - despite different self-perception - had great difficulties in distinguishing between different rows and, thus, the use of multiple rows for \indirect{} interfaces is not feasible. Regarding the Likert-questionnaires and the qualitative feedback, the analysis found greater popularity of the \indirect{} interfaces (see section \ref{sec:cheesyfoot:experiment1:qualitative} and \ref{sec:cheesyfoot:experiment2:qualitative}).

Taken together the greater enthusiasm, as well as the lower TLX scores (for \row{1} subdivisions), the use of \indirect{} interfaces is preferable in most situations. In particular, this applies to situations where 1) a lower number of options is sufficient and 2) a restricted view (as in the \direct{} interfaces, where the head is directed to the floor) could be problematic. Based on the analysis, a \row{1}, \col{4} layout for \indirect{} interfaces is feasible.

\subsection{Use direct interfaces for short-term and fine-grained interactions}

Direct interfaces delivered significantly higher accuracy rates compared to indirect interfaces (see sections \ref{sec:cheesyfoot:comparison:accuracy}). However, the analysis of qualitative feedback and answers in the Likert questionnaires showed a clear preference of participants for \indirect{} interfaces. The limitations of the hardware used in the experiment (e.g., weight, field of view) might have exerted a considerable influence on the opinion of the participants. However, in particular, the downward head posture seems to be rejected by the participants for longer-term interactions in general (see section \ref{sec:cheesyfoot:experiment1:qualitative}).

Therefore, for the tested design, \direct{} interfaces proved to be best suited for short-term interactions requiring high accuracy and a large number of input options. For such interfaces, a high degree of accuracy is still achieved with \row{3}, \col{6} layouts.

\section{Limitations and Future Work}
\label{sec:cheesyfoot:limitations}

The design and results of our experiments impose some limitations and directions for future work.

\subsection{Layout of the Targets}

The experiment used a fixed semicircular grid of targets. This layout was chosen because of the natural reachability of targets from a fixed standing position. However, other shapes (e.g., rectangular, oval) and arrangements (e.g., not equally sized targets) or adoptions to the footedness of the user could also be considered for future work. This is of particular interest as our experiment showed a larger spread for targets further away from the participant. These extensions enlargen the design space of such interfaces and, despite being outside of the scope of this work, provide interesting direction for future work. 

\subsection{Feedback for Indirect Interaction}

The goal of the presented experiment was to investigate the ability of users to use \indirect{} interfaces without visual feedback and, thus, create a baseline for future work. Therefore, the participants received no feedback about the position of their feet during the \indirect{} experiment as such additional feedback could strongly influence the performance of the participants.

Future Work is necessary to understand the implications of different forms of more direct feedback for users. Such direct feedback could be indicated by highlighting the currently selected option or by displaying a cursor moved by the foot.

\subsection{Other Styles of Interaction}

The experiment concentrated on interfaces, which, as an analogy to the traditional point-and-click interfaces, are operated with foot-taps. Other interaction styles, such as gestures for fine-granular control or taps with different parts of the foot (e.g., heel) may be beneficial for the future use of \acp{HMD}. 

Additionally, the presented experiment focused on one-time interactions. As a possible addition, cascading menus could help to keep the grid size small  and reduce necessary foot movements while maintaining a large set of options.

\subsection{The Midas Tap Problem}
\label{sec:cheesyfoot:limitations:midas}

Similar to the Midas Touch Problem~\ncite{Jacob1995} in eye gaze tracking, it is challenging to separate intentional input from natural motion when using foot-based input. A possible solution could be a special foot input mode, activated using a secondary input modality such as a toggle on the HMD or gaze interaction in the user interface. For \direct{} interfaces, just looking at the ground may be sufficient to activate this mode, as actions are only triggered after a subsequent tap. Further, sensor-based gait detection~\ncite{Jacob1995, Derawi2010} allows to only enable foot input while standing and, thus, help to prevent erroneous activation. Further work in this field is necessary to conclude on the Midas Tap problem.

\section{Conclusion}

This chapter explored foot-taps as an input modality for \acp{HMD}. More precisely, the chapter investigated two different interaction styles: 1)~\emph{direct interaction} with interfaces that are displayed on the floor and require the user to look down to interact and 2)~\emph{indirect interaction} with interfaces that, although operated by the user's feet, are displayed as a floating window in front of the user. The results confirmed the viability of foot-taps for accurate and pleasant interaction with \acp{HMD}.

To conclude, this chapter added to the body of research on interacting with \acp{HMD} in multiple areas:

\begin{enumerate}
	\item This chapter contributed interaction techniques for \acp{HMD} \emph{on the move}. These techniques allow for hands-free interaction and, thus, enable novel use cases for \acp{HMD} when the user's hands are encumbered.
	\item Second, this chapter contributed two controlled experiments, proving the viability of the presented concepts for accurate, efficient, and joyful interactions. Therefore, the chapter presented a first evaluation of hands-free and mobile interfaces for interacting with \acp{HMD}, opening up a new research field for future developments.
	\item Third, based on the results of the two experiments, this chapter presented a set of guidelines that can inform future development of interfaces for interacting with \acp{HMD} \emph{on the go}. 
\end{enumerate}

\subsection{Integration}

\textfigH{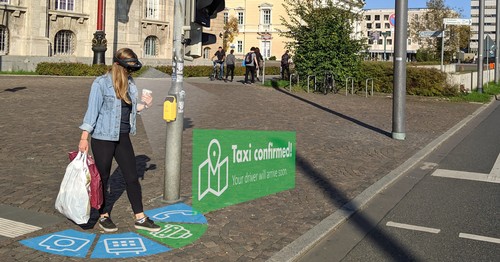}{Alice calls a taxi using foot-based interaction with her hands occupied.}

\interactionbox{alice_mindthetaps}{At the Traffic Light}{Alice is on her way home. She has taken a coffee with her and is still carrying her shopping, so both her hands are occupied. She reaches the exit of the mall and stands at a traffic light.  She wants to call an autonomous taxi for a ride home. She looks at the floor and chooses the taxi app with her foot (see figure \ref{fig:cheesyfoot/alice}). With a second foot tap, she confirms the pick-up location.
	
	The interaction takes place quickly and easily and purely by foot taps, without having to free the hands.}

The interaction technique presented in this chapter allows for fast interaction with with \emph{body-stabilized} interfaces leveraging our \emph{lower limbs} for input. This style of interaction supports \singleuserLower{}s with \discreteLower{} in situations, where the user's hands are encumbered (e.g., while carrying something during \mobilityLower{}). The two interaction styles presented thus cover a large number of situations, which could not be supported by the previous contributions of this thesis due to their focus on the \emph{upper limbs} (see chapter \ref{ch:proximity:merged} and chapter \ref{ch:cloudbits}) and, thus, contributes to the vision of ubiquitous \aroundbodyinteraction{} (see section \ref{sec:introduction:aroundbodyinteraction}).

As discussed in section \ref{sec:cheesyfoot:limitations}, there are multiple yet unexplored areas in the design space of such mobile and foot-based interactions that go beyond the scope of this work. Further modifications of the concept could potentially widen the applicability of foot-based interaction to more situations: By supporting continuous interaction (e.g., to adjust a slider precisely) and distinguishing the part of the foot that performed the interaction, the expressiveness could be considerably increased. The results presented in this chapter can provide a baseline for future work in these areas, providing the first step towards more comfortable and safer interaction with \acp{HMD} \emph{on the go}. 

\subsection{Outlook}

As a result of the Midas Tap problem identified in section \ref{sec:cheesyfoot:limitations:midas}, foot gestures, as presented in this chapter, cannot effectively be used for interaction while walking (and, thus, during an important part of \mobilityLower{}). Chapter \ref{ch:walktheline} discusses how this limitation can be mitigated by the use of \emph{world-stabilized} operated using the \emph{lower limbs}.

}

\chapter[Walk The Line]{Walk The Line: Leveraging Lateral Shifts of the Walking Path as an Input Modality for Head-Mounted Displays}\label{ch:walktheline}
{
	
\newcommand{\lane}[1]{#1-lane}

\newcommand{\seltime}[1]{\nicefrac{#1}{3}~s}

\newcommand{\interactionStyleBased}{walking-based}

\newcommand{\factorLanes}{\emph{number of lanes}}
\newcommand{\factorTime}{\emph{selection time}}

\newcommand{\prototypeCamera}{Camera}

\cptteaser{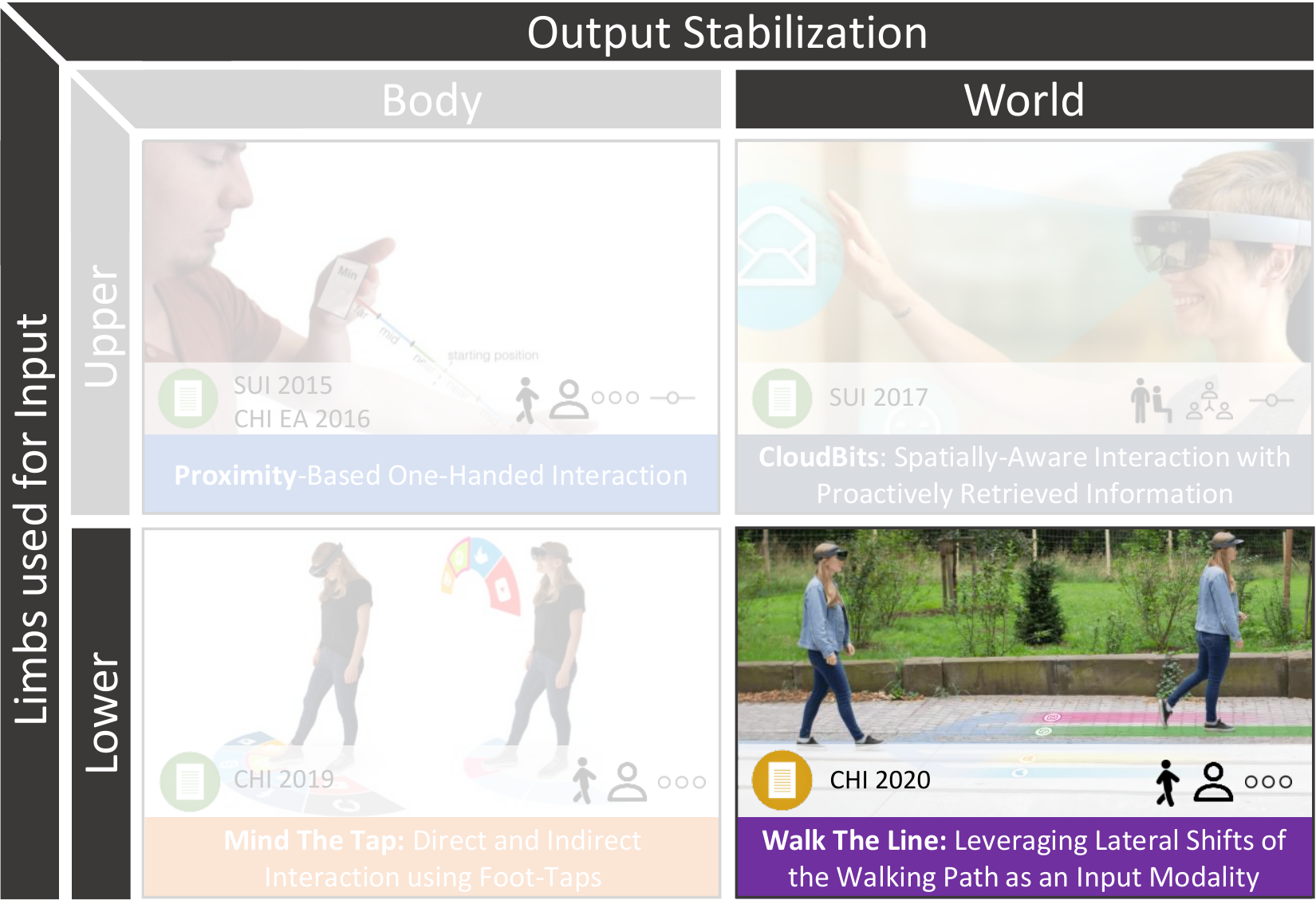}{This chapter presents an interaction technique for world-stabilized interfaces leveraging the lower limbs for input.}

The last chapter of this thesis explored foot-taps as an input modality for \acp{HMD}, leveraging the \emph{lower limbs} to interact with \emph{body-stabilized} interfaces. In addition to the presented advantages, the chapter also highlighted limitations of the presented interaction style (see section \ref{sec:cheesyfoot:limitations}). Parts of these limitations are rooted in the inherent limitations of foot-based operation: The user has to stand in a fixed position in order to operate such an interface to overcome the \emph{Midas Tap Problem} (see section \ref{sec:cheesyfoot:limitations:midas}).

While this limitation might be bearable for work, e.g., in an industrial context, it becomes a major challenge in genuinely mobile interaction situations: Mobility is strongly related to locomotion, rendering traditional foot-based interaction techniques difficult to use in such situations, as the user must stop the process of locomotion to interact with the system. If, however, the vision of ubiquitous interaction with information in a digitally augmented physical world is to become reality, a substantial part of the interaction with such devices will happen \emph{on the go}. This highlights the necessity for truly mobile interaction techniques for \acp{HMD} that not only support interaction while being at different places but also during the process of getting there - while walking.

As a possible solution, this chapter proposes the use of a \emph{world-stabilized} interface (see section \ref{sec:relatedwork:hmds:implementation}) that does not move together with the user (see figure \ref{fig:walktheline/overview_walktheline}). This fixation in the world can be leveraged to control an interface by changing the position relative to the interface, providing a solution for interaction with \acp{HMD} in \mobilityLower{} situations for \singleuserLower{}. A multitude of such shifts of the relative position to an interface as input dimension is conceivable. As a first step towards such a concept, this chapter focuses on shifts that occur orthogonally to the user's walking path, thus not modifying their original walking direction and, thereby, not interfering with the process of locomotion. While the contribution of this chapter focuses on \discreteLower{}, we are confident that it can also be transferred to \continuousLower{}, as outlined in the future work section of this chapter (see section \ref{sec:walktheline:limitations}).

The contribution of this chapter is two-fold: First, the chapter contributes the methodology and results of a controlled experiment assessing the accuracy and efficiency of such an interfaces. Second, based on the results of the controlled experiment, the chapter presents a prototype implementation of a \interactionStyleBased{} input modality for \acp{HMD} together with three example applications.

The remainder of this chapter is structured as follows: After reviewing the related works (section \ref{sec:walktheline:rw}), section \ref{sec:walktheline:concept} proposes a concept to interact with \acp{HMD} using lateral shifts of the walking path. Afterward, section \ref{sec:walktheline:methodology} describes the methodology and research questions of the controlled experiment. After that, section \ref{sec:walktheline:results} and \ref{sec:walktheline:discussion} present and discuss the results of the controlled experiment. Based on these results, section \ref{sec:walktheline:prototype} describes a prototype implementation of a walking-based input modality for interacting with \acp{HMD} together with three example applications. The chapter concludes with limitations and proposes directions for future work (section \ref{sec:walktheline:limitations}). 

\cptIntroBox{Muller2020}{The student \emph{Daniel Schmitt} implemented the study client application. \emph{Sebastian Günther}, \emph{Martin Schmitz}, \emph{Markus Funk} and \emph{Max Mühlhäuser} supported the conceptual design and contributed to the writing process.}
\section{Related Work}
\label{sec:walktheline:rw}

Chapter \ref{ch:relatedwork} discusses the related works on interaction techniques with \acp{HMD} and section \ref{sec:cheesyfoot:rw} discusses foot-based interaction techniques. In addition to this, the following section presents a set of requirements for hands-free interaction with \acp{HMD} while walking and, afterward, categorizes relevant research with regards to the requirements. 

\subsection{Requirements}

The following section presents a set of requirements for interactions with \acp{HMD} using the \emph{lower limbs} while walking derived from the related works. The requirements are then used to compare the most relevant related work (see table \ref{tab:req:walktheline}).

\begin{description}
	\item[R6.1: Adapted Input] The input methods of the system must compensate for the situationally-induced impairments of walking.
	\item[R6.2: Adapted Output] The output methods of the system must compensate for the situationally-induced impairments of walking.
	\item[R6.3: Generalizability of the Interaction Technique] The interaction technology must be suitable for general interaction with \acp{HMD}, not only for a specific use case.
	\item[R6.4: No Interruption of the Locomotion] The interaction with the system must occur completely during walking and may not require to interrupt the process of locomotion.
	\item[R6.5: Eyes-Free Operation] To avoid danger, a system must not capture the entire (visual) attention of the user. The user must retain awareness of his environment while interacting with the system.
\end{description}

\subsection{Interfaces for Use While Walking}
\label{sec:rw:whilewalking}

The proliferation of smartphones and the increasing usage during walking~\ncite{Yoshiki2017} led to a stream of research to mitigate the situationally-induced impairments~\ncite{sears03} that are introduced through walking~\ncite{Sarsenbayeva2018, Saulynas2018} and additional encumbrances such as carrying objects~\ncite{Ng2013, Ng2015} or ambient noises~\ncite{Sarsenbayeva2018a}.

\typcite{Kane2008} introduced the term \acp{WUI} for interfaces that are explicitly designed \enquote{to compensate for the effects of walking on mobile device usability.} The authors proposed the usage of increased button and text sizes to compensate for the reduced input performance. Following this stream of research, \typcite{Rahmati2009} used content stabilization to compensate for the shaking introduced from walking. Further examples to help users to overcome the situational impairments include the usage of other keyboard layouts~\ncite{Clawson2014} or text input modalities beyond touch-typing~\ncite{Fitton2013}.

Focusing on the safety aspects of usage, \typcite{Beuck2017} found that mobile applications actively interrupting smartphone usage when entering a potentially dangerous situation can help to prevent such situations. \typcite{Shikishima2018} showed how texting while walking can be detected.
Following the same path, \typcite{Hincapie-Ramos2013} presented an integrated alarm system that warns users of dangerous situations while being engaged with their smartphone. Further examples include other warning systems~\ncite{Wen2015,Tang2016a,Vinayaga-Sureshkanth2018a}, obstacle detection~\ncite{Wang2017}, or specialized support for texting~\ncite{Kong2017} or video watching~\ncite{Ahn2013}.

While most of the research on interfaces for use while walking focused on smartphones, this paper argues that \acp{HMD} are a better fit for the requirements of a truly mobile user interface to be operated while walking: Such devices do not require the user to look down to operate. Further, the user’s visual attention is not captured on an opaque screen, keeping the connection to the real world~\ncite{Lucero2014}.

Highly related, \typcite{Lages2019} explored how different adaptation strategies of user interfaces can support the user in interacting with \acp{HMD} during walking. However, this very inspirational work focused on adapting the \emph{output} to accommodate for the effects of walking. The authors did not address the changing requirements for \emph{input} while walking.

\subsection{Locomotion as Input}
\label{sec:rw:movement}

Research showed how the movement of the user’s body during locomotion, as well as the changing spatial relationships between users and objects, could be used as an input dimension for both, implicit and explicit interactions. 

Popular examples of implicit interaction using body motion as an input dimension can be found in the area of context-aware computing systems~\ncite{Chen2000}, e.g., for mobile navigation. Such systems use the (global) spatial position of the user as input and present navigation instructions through a variety of output modalities such as screen-based~\ncite{Kruger2004}, augmented~\ncite{Narzt2006}, vibrotactile~\ncite{Erp2005, Tsukada2004} or audio~\ncite{Holland2002}, or a combination of these. As another example, \typcite{Dow2005} showed how the spatial location of a user could be used to start the playback of location-specific content, allowing users to interactively explore content through walking. Further, \typcite{Vogel2004} showed how to use the spatial position of users relative to a public display to switch between different modes of interaction implicitly. However, the use of implicit interaction techniques is limited to specific application areas and, therefore, cannot be generalized to general interaction.

In recent years, research proposed more explicit methods for interacting with \acp{HMD} through walking. As the most prominent example, it is a widespread interaction paradigm in \ac{VR} and \ac{AR} to approach virtual objects in order to interact with them in place \ncite{Argelaguet2013}. In such systems, the user’s spatial movement acts as means for \emph{selection} of virtual objects. However, to the best of our knowledge, prior work did not explore how the locomotion of users during walking itself can be leveraged as a generic \emph{input} dimension for interaction with \acp{HMD} yet.

\renewcommand*\theadfont{\bfseries}
\settowidth\rotheadsize{\theadfont Interaction Without Full}
\renewcommand\theadgape{}
\renewcommand\theadalign{lc}
\renewcommand\rotheadgape{}
\begin{table}[h]
	\centering
	\begin{tabular}{m{5cm}ccccc}
		& \rothead{R6.1: Adapted Input} & \rothead{R6.2: Adapted Output} & \rothead{R6.3: Generalizability} & \rothead{R6.4: No Interruption of the Locomotion} & \rothead{R6.5: Eyes-Free Operation} \\
		\midrule
		
		\multicolumn{6}{c}{\emph{Walking User Interfaces}} \\
		\cite{Kane2008} & \reqPartially & \reqYes & \reqPartially & \reqYes & \reqNo \\
		\cite{Rahmati2009} & \reqNo & \reqYes & \reqNo & \reqYes & \reqNo \\
		\cite{Clawson2014} & \reqYes & \reqNo & \reqNo & \reqYes & \reqNo \\
		\cite{Fitton2013} & \reqYes & \reqPartially & \reqPartially & \reqYes & \reqNo \\
		\cite{Lages2019} & \reqNo & \reqYes & \reqPartially & \reqYes & \reqYes \\
		
		\multicolumn{6}{c}{\emph{Locomotion as Input}} \\
		
		\cite{Narzt2006} & \reqPartially & \reqYes & \reqNo & \reqYes & \reqYes \\
		\cite{Dow2005} & \reqPartially & \reqYes & \reqPartially & \reqNo & \reqYes \\
		\cite{Vogel2004} & \reqYes & \reqPartially & \reqYes & \reqNo & \reqNo \\
		\cite{Argelaguet2013} & \reqYes & \reqNo & \reqPartially & \reqNo & \reqYes \\
		\bottomrule
	\end{tabular}
	\caption{Fulfillment of requirements of the related works. \reqYes~ indicates that a requirement is fulfilled, \reqPartially~ indicates partial fulfillment.}
	\label{tab:req:walktheline}
\end{table}
\section{Concept}
\label{sec:walktheline:concept}
\label{sec:walktheline:intro-short}

\cptteaser{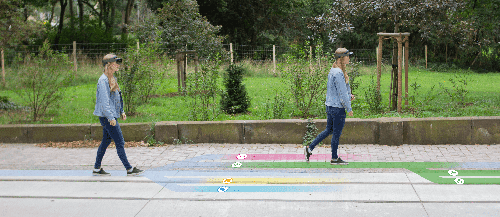}{Walk the Line leverages lateral shifts of the walking path as an input modality for \acp{HMD}. Options are visualized as \emph{lanes} on the floor parallel to the user's walking path. Users select options by shifting the walking path sideways. Following a selection, sub-options of a cascading menu appear as new \emph{lanes}.}

While walking, we routinely respond to changes in the environment by adapting the trajectory of our walking path to avoid obstacles, such as oncoming pedestrians or pavement damages. These trajectory changes occur quickly and accurately and without changing the original direction of travel, but by laterally shifting the walking path. This chapter argues that such lateral shifts of the user can be leveraged as a novel input modality for interaction \emph{on the go}.

Today, a large number of pedestrians interact with their smartphones as they walk, losing touch with the world around them~\ncite{Lin2017a}. Similar to distracted driving, distracted walking also leads to potentially dangerous situations: The lack of (visual) attention causes pedestrians to walk into obstacles, to collide with other persons or otherwise endanger themselves~\ncite{Schabrun2014a, Thompson2013}. As another approach to interaction while walking, related work proposed voice-based interfaces. However, such systems may perform badly depending on background noise and have social implications~\ncite{Koelle2017, Starner2002}. In addition, voice-based interfaces interfere with the communication between people, whether it is a local conversation or a phone call. To overcome the limitations of interaction while walking, research proposed ways to mitigate for the situational hindrances~\ncite{sears03} induced by walking, leveraging increased button sizes~\ncite{Kane2008} or content stabilization~\ncite{Rahmati2009}. 

The contribution of this chapter goes beyond the state-of-the-art by not only compensating for such situational hindrances but by actively exploiting the process of locomotion as an input modality: Mobile \ac{AR} interfaces potentially enable more comfortable and safer interaction while walking, as the visual attention is no longer captured purely by a display~\ncite{Lucero2014}. This chapter proposes to use \acp{HMD} to visualize different input options as augmented lanes on the ground parallel to the walking path of the user. By laterally shifting the path onto a lane and, subsequently, walking on the lane, users can select an option (see figure~\ref{fig:walktheline/teaser_big}).

For this, this chapter considers a system that displays multiple lanes parallel to the walking path of the user. Each lane represents an option the user can select. The lanes can be arranged on both sides of the user's walking path (see figure \ref{fig:walktheline/teaser_big}). The specific visualization of the lanes can be tailored to the application and adapted to the current situation of the user. For example, it can contain icons or text or can be connected to bubbles floating in the air which describe the information to be selected.

To interact with the system, users shift their path sideways until they walk on the desired option lane without a need to change their walking speed, just as they would when avoiding an obstacle on the sidewalk. The system highlights the lane the user is currently walking on by changing the visualization (e.g., the color) of the respective lane. This change affects the entire lane, which is also visible in front of the user. Therefore, users do not have to look to the ground to interact with the system, but can keep their head up. By walking along one of the lanes for a certain period of time, the respective option can be selected, analogously to the concept of selection by dwell time in eye-gaze interaction~\ncite{Qian2017}. This section refers to the time-to-select that users need to walk on a lane as the \emph{selection time}. In our concept, this \emph{selection time} is visualized to the user by changing the opacity of the lanes: While walking on a lane, all other lanes are gradually faded out.

In addition to the option lanes, the concept proposes a non-active \emph{null lane} that covers the path directly in front of the user, which remains free. Therefore, if users continue walking straight ahead without adjusting their path, the system does not interpret this as an interaction and does not trigger any actions.

\section{Methodology}
\label{sec:walktheline:methodology}

This section presents the methodology of an evaluation of the proposed concept of a \interactionStyleBased~input modality for \acp{HMD} in a controlled experiment. The controlled experiment investigated the following research questions:

\begin{description}

\item[RQ1] How does the width of the \emph{lanes} affect the accuracy, efficiency, and user experience of the system?

\item[RQ2] How does the \emph{selection time} affect the accuracy, efficiency, and user experience of the system?

\item[RQ3] Are there interaction effects between the width of the lanes and the selection time on the accuracy, efficiency, and user experience?

\end{description}

\subsection{Design and Task}

\begin{figure*}[ht!]
\subfloat[8+1 Lanes]
  {\includegraphics[width=.3\linewidth]{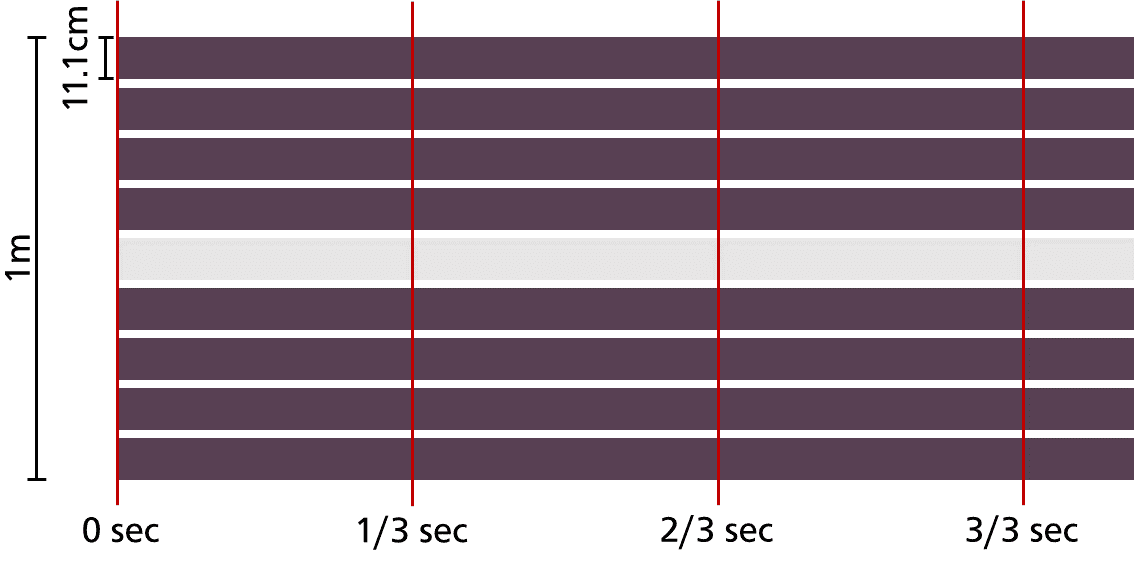}}\hfill
\subfloat[12+1 Lanes]
{\includegraphics[width=.3\linewidth]{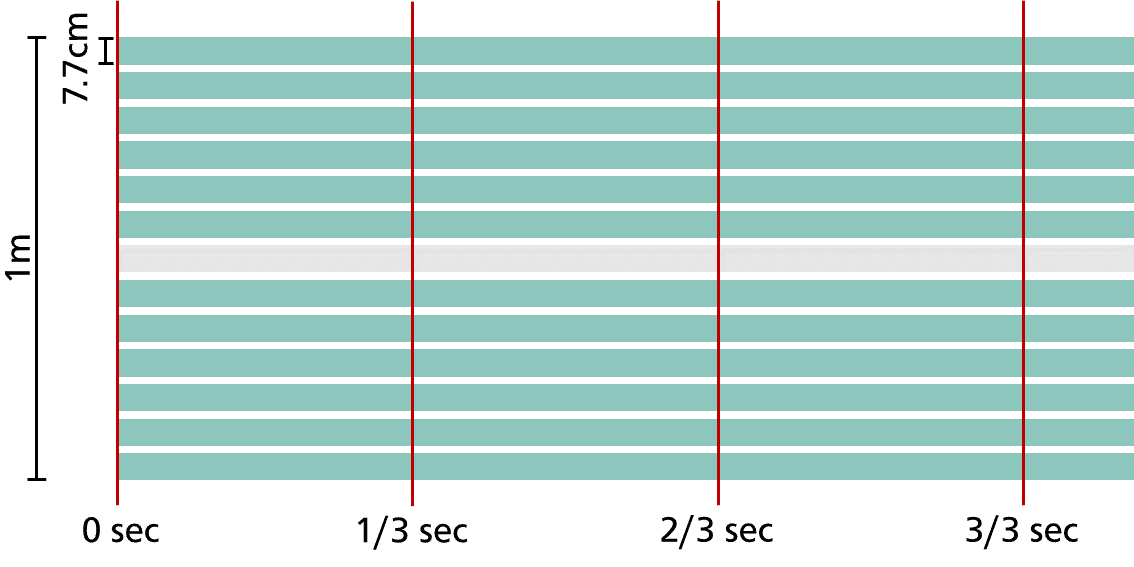}}\hfill
\subfloat[16+1 Lanes]
  {\includegraphics[width=.3\linewidth]{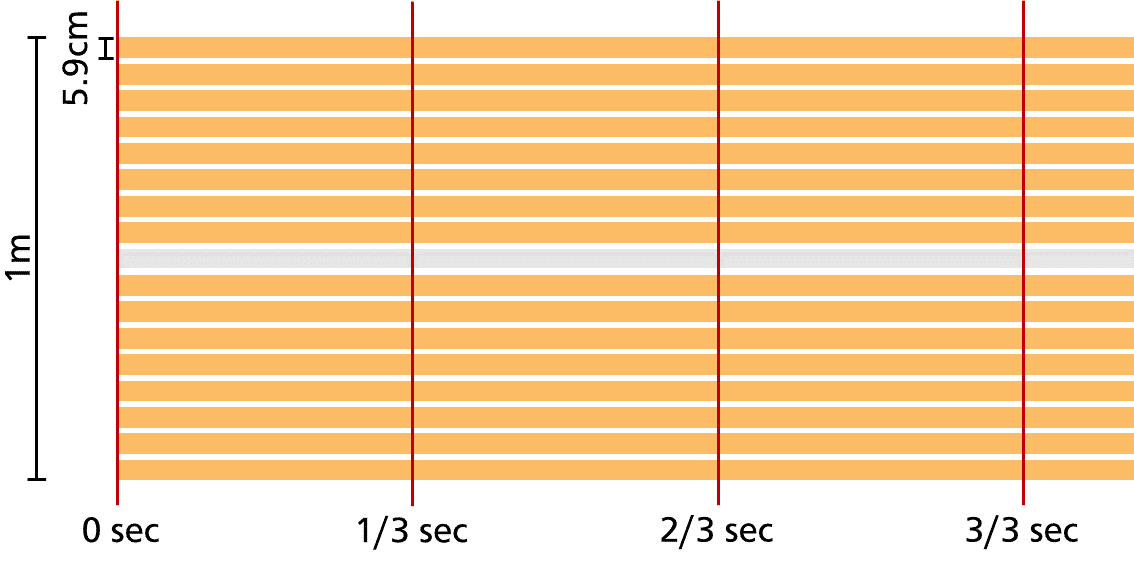}}
\caption{The levels of the two independent variables, \factorLanes~and \factorTime, as tested in the experiment.}
\label{fig:walktheline/conditions}
\end{figure*}

To answer the presented research questions, a controlled experiment was designed, in which users interacted with a system as described in section \ref{sec:walktheline:concept}. The participants’ task was to laterally shift their walking path to the highlighted target lane and to stay within its bounds for a certain period of time while keeping average walking speed.

As the first independent variable, the experiment varied the \factorLanes{} on a fixed-width interaction area. The experiment varied the equal-sized width of the lanes in each condition to fill the available interaction space, thus also varying the width of the individual lanes. This design also allowed us to conclude on the influence of the width of lanes.%

As the second independent variable, the experiment varied the \factorTime{} as the time participants had to walk on a lane to select it. Shorter dwell times on a lane did not select the respective lane and, thus, could be used to cross lanes to reach targets further to the side.

The experiment varied both independent variables in a repeated measures design with three levels each (\factorLanes{}: \lane{8}, \lane{12}, and \lane{16}, \factorTime{}: \seltime{1}, \seltime{2}, and \seltime{3}), resulting in a 2-factorial study design with a total of $3x3=9$ conditions (see Figure \ref{fig:walktheline/conditions}). The levels were chosen based on the results of pre-tests, which suggested that wider lanes or longer selection times did not yield higher accuracy or efficiency rates. Further, the design included two repetitions per target lane in each condition, resulting in a total of $3 \cdot (8+12+16) \cdot 2 = 216$ trials per participant. The order of the conditions was counterbalanced using a Balanced Latin Square design. For each condition, the system randomized the series of targets while assuring that each target was repeated two times.

To specify the dimensions of the longitudinal area used in the study, the calculation started from the typical width of a sidewalk of \SI{2.5}{m}~\ncite{Kim2011a}. The available width was halved to take into account oncoming traffic from other pedestrians and decreased by a safety distance of \SI{.25}{m}, resulting in \SI{1}{m} of interaction width. Since the experiment varied the \factorLanes{} on a fixed-width area, the experiment also varied the \emph{width} of the individual lanes. Therefore, the width of the individual lanes in each condition was $\SI{1}{m} / ($ \factorLanes{} $+1)$, resulting in an absolute lane width of $\sim$~\SI{11}{cm} (for the \lane{8} conditions) to $\sim$~\SI{6}{cm} (for the \lane{16} conditions). For the length of the area, the setup used \SI{20}{m} as informal pre-tests showed that this distance allowed the participants to perform the interaction in all conditions without reaching the end of the area.

\subsection{Experiment Setup and Apparatus}
\label{sec:walktheline:methodology:setup}

\textfig{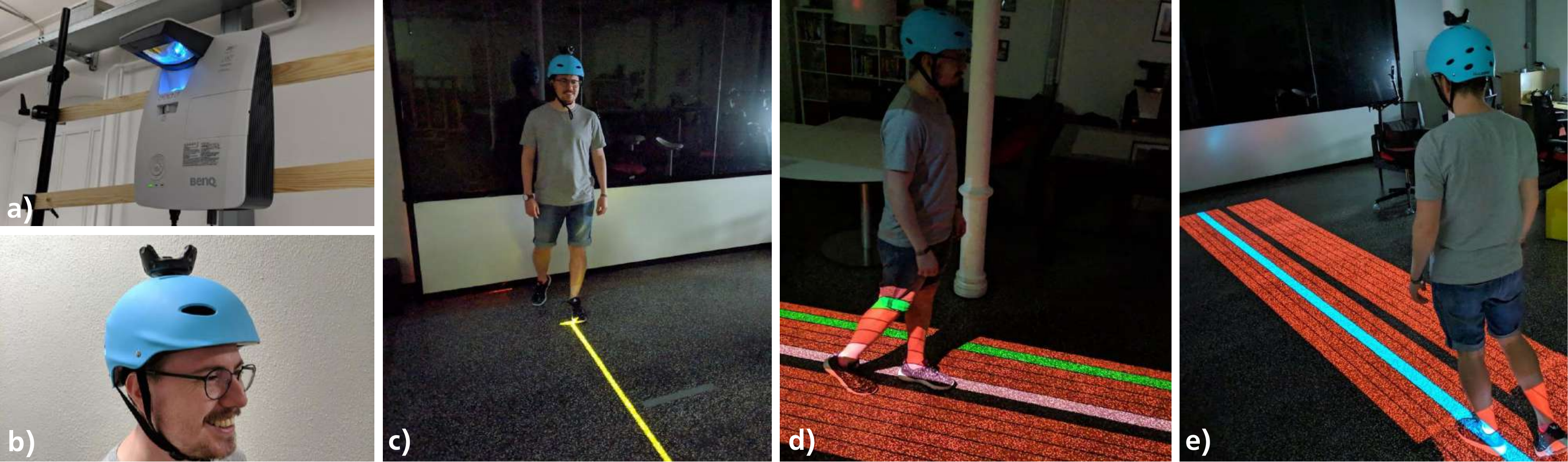}{Setup and procedure of the experiment: The setup used two projectors (a) for visual output and tracked the participants using an HTV VIVE Tracker (b). Participants initiated each trial by starting to walk (c). Thereafter, the system projected the lanes with a highlighted target lane (d). The participants' task was to shift their path to the target lane (e) and stay within the bounds for the respective \factorTime{} of the condition.}

The setup of the experiment did not use AR glasses, such as the Microsoft Hololens, because the current generations of such devices still suffer from technical limitations (e.g., weight, the field of view) that could influence the measurements, rendering the results unusable for future developments. Further, pre-tests indicated that the visual SLAM approach for inside-out tracking used by the current generation of such devices is not accurate and robust enough for the presented experiment. 

Therefore, the setup of the experiment consisted of two short-throw 1080p projectors (BenQ MH856UST) to simulate the visual output of an \ac{HMD}. For this, two projectors were mounted at a distance of \SI{7}{m} to wooden slats, which, in turn, were attached to two tripods at the height of \SI{3.5}{m} (see Figure \ref{fig:walktheline/study_setup_procedure}, a). This setup allowed covering a longitudinal range of \SI{20}{m} with visual output on the floor. The setup combined this visual output with the robust and accurate tracking of the participants’ position and orientation using an HTC VIVE Tracker (position tracking error < \SI{0.02}{cm} \ncite{Niehorster2017}). Therefore, the setup also included two VIVE lighthouses at the far edges of the area covered by visual output, to allow the same physical space to be tracked by the system. The implementation used OpenCV to calibrate\footnote{\url{https://docs.opencv.org/2.4/doc/tutorials/calib3d/camera_calibration/camera_calibration.html}} the projected image with the tracking of the VIVE system by displaying calibration points and positioning a VIVE Tracker on the displayed positions, achieving a 3x20m interaction space with combined input and output.

To capture the position of the participants' heads in space, the setup used a modified bicycle helmet equipped with a VIVE Tracker (see Figure \ref{fig:walktheline/study_setup_procedure}, b). The setup used the head position of the participants as input for the system (in contrast to, for example, the position of the two feet) to simulate the type of tracking available in today’s \acp{HMD}. A desktop PC located next to the study area orchestrated the VIVE tracking as well as the two projectors. The PC was further used to render the visual output as well as for data logging. Figure \ref{fig:walktheline/study_setup_procedure} depicts the complete setup and apparatus of the study.

Further, the desktop PC hosted a study operator application that allowed the investigator to set the task. For each trial, the study client logged the following dependent variables:

\begin{description}
	\item[Trajectory] as the trace of the participants’ \emph{walking path} (i.e., the path of the participants’ head movements),
	\item[Task Completion Time (TCT)] as the time between displaying the task and entering the lane which was subsequently selected (i.e., the time until the activation of the lane minus the \factorTime{}),
	\item[Accuracy Rate] as the rate of successfully selecting the target lane of the trial,
	\item[Stabilizing Error Rate] as the rate of participants walking past the boundaries of the target lane after initially reaching it. This includes overshooting errors (i.e., leaving the target lane while maintaining the initial direction of the lateral shift) as well as swing-back errors (i.e., leaving the target lane in the opposite direction to the initial shifting direction).
\end{description}

The experiment was conducted in a room of the institute's building, where there was a sufficiently large area available. For the duration of the study, the area was closed to regular public access in order to exclude external influences. 

\subsection{Procedure}

After welcoming the participants, the investigator introduced them to the concept and goals of the study and measured their body height as it was expected to influence the participants' performance. In the following, the investigator asked the participants to fill a consent form and an introductory questionnaire asking for demographic data. After starting and calibrating the system, the investigator asked the participant to put on the modified bicycle helmet. To avoid learning effects, the participants began the study by freely testing the system to get used to the hardware and interfaces.

To start the first condition, the investigator asked the participants to go to the starting position. Participants were free to start each trial whenever they wanted by starting to walk (see Figure \ref{fig:walktheline/study_setup_procedure}, c). After a few steps (i.e., after reaching an average walking speed of around 1-1.5 m/s~\ncite{bohannon1997}), the system showed the task to the participants (see Figure \ref{fig:walktheline/study_setup_procedure}, d). The system randomly selected the exact starting point (2 +/- .5m) of each trial in order to avoid influencing the participants by learned positions. The study interface consisted of red lanes, which indicated the \factorLanes{} of the respective condition. The system highlighted the target with green and the currently active lane with a lighter gradation of red (for the regular lanes, see Figure \ref{fig:walktheline/study_setup_procedure}, d) or blue (for the target lane, see Figure \ref{fig:walktheline/study_setup_procedure}, e). As soon as the participant shifted their position to the side (leaving the \emph{null lane} in the middle), the visualization showed the \factorTime{} by fading out the other lanes through animating their opacity. The selection timer was reset once the participant left the lane and restarted for the newly active lane. When the participant walked on a lane for the \factorTime{} of the particular condition, the system logged the result and signaled the end of the trial to the participant with a sound. If the participant had not made a selection by the end of the available interaction space or selected a wrong lane, the system logged this as a failed attempt. After finishing a trial, the participant walked back to the starting position and proceeded to the next trial.

The investigator instructed the participants to maintain their average walking speed over the entire course. After each condition, the investigator asked the participants to fill a questionnaire regarding their experiences on a 5-point Likert-scale (1: strongly disagree, 5: strongly agree). Additionally, the participants filled a NASA TLX~\ncite{Hart2006} questionnaire. The investigator further enforced a 5-minute break between the conditions so that the participants could rest. During this break, the participants gave qualitative feedback regarding their experiences in a semi-structured interview. Each experiment took about 80 minutes per participant. 

\subsection{Participants}

For the experiment, 18 participants (8 male, 8 female, 2 identified as gender variant/non-conforming) aged between 16 and 55 (\val{30.83}{9.6}) were recruited from the University's mailing list. All participants voluntarily took part in the study and no compensation was paid.

\textfigStudybox{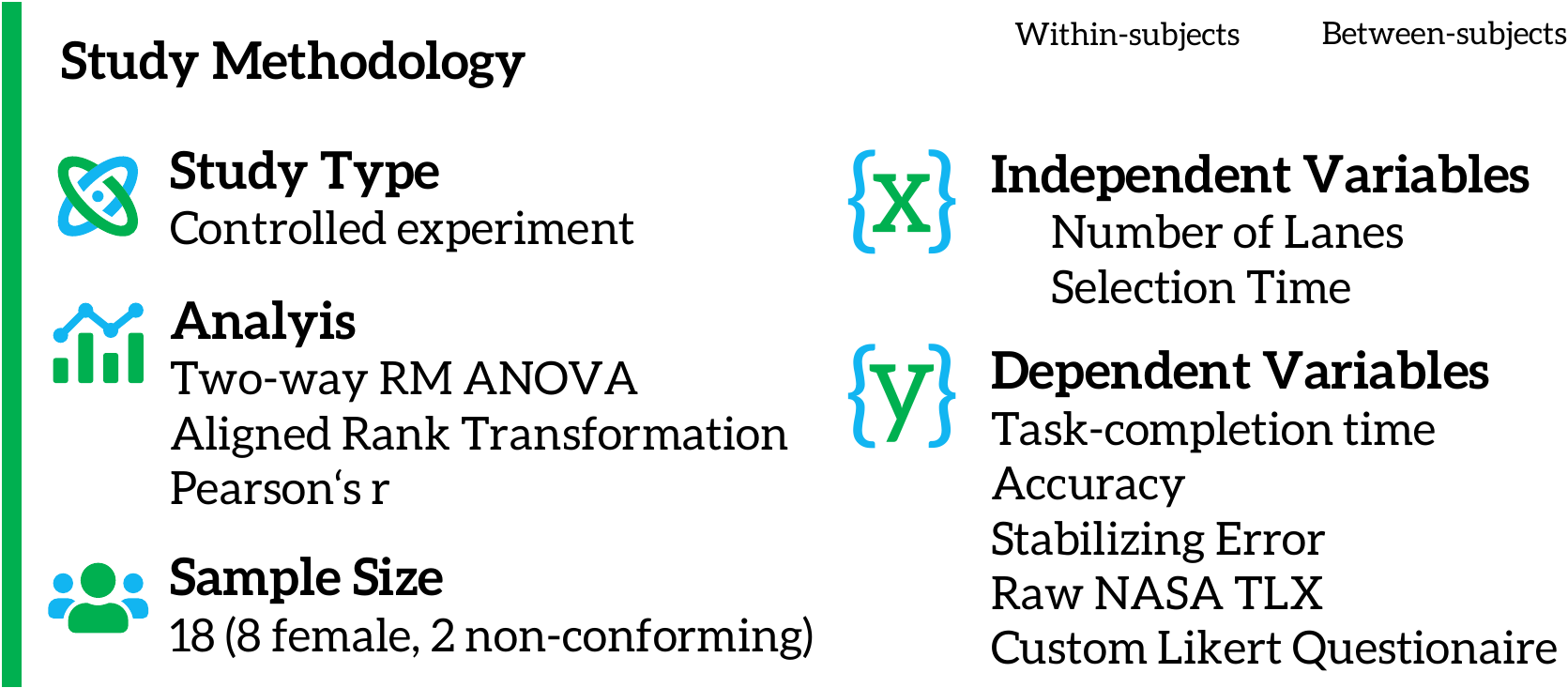}
\section{Results}
\label{sec:walktheline:results}

The following section reports the results of the controlled experiment investigating the research questions RQ1 - RQ3. The analysis of the data was performed as described in section \ref{sec:rw:methodology}. 

\newpage
\subsection{Accuracy}

\colfig{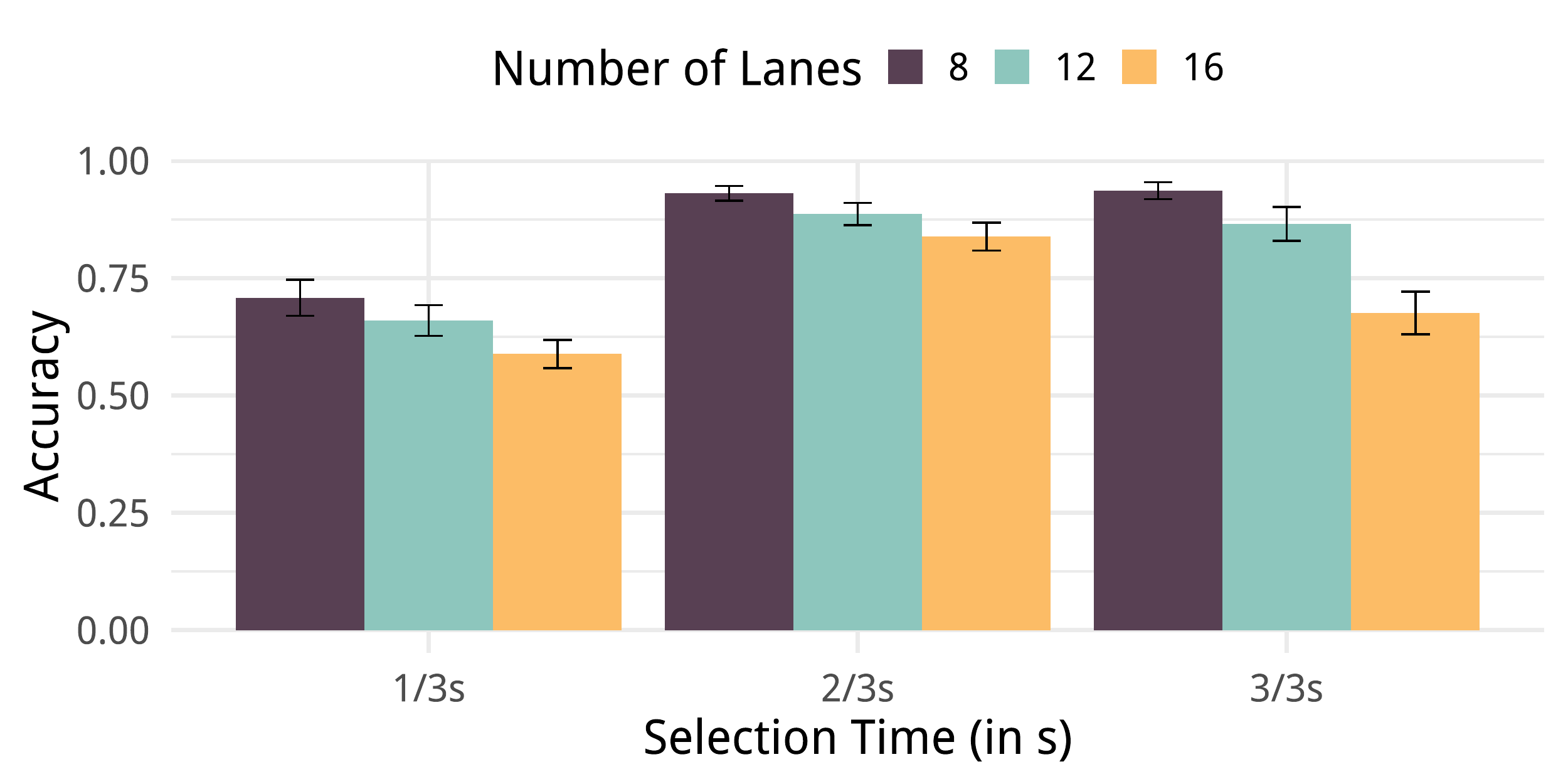}{The measured accuracy rates in the controlled experiment. All error bars depict the standard error.}

The accuracy of participants was analyzed as the rate of successful trials. The analysis revealed that the \emph{number of lanes} had a significant (\anova{2}{34}{27.05}{<.001}{.134}) influence on the participants' accuracy with a medium effect size. Post-hoc tests confirmed significant differences between the \lane{8} (\emmP{85.8}{2.2}) and \lane{16} (\emmP{70.1}{2.2}) conditions as well as between the \lane{12} (\ac{EMM} \emmP{80.4}{2.2}) and \lane{16} conditions (both $p<.001$).

Further, the analysis showed a significant (\anovaCor{1.44}{24.45}{37.57}{<.001}{.719}{.307}) effect for the \emph{selection time} on the participants' accuracy with a large effect size. Post-hoc tests confirmed significant differences between the \seltime{1} (\emmP{65.2}{2.4}) and both, the \seltime{2} (\emmP{88.5}{2.4}) and the \seltime{3} (\emmP{82.6}{2.4}) conditions (both $p<.001$).

Additionally, the analysis showed significant (\anovaCor{2.57}{43.74}{4.28}{<.05}{.643}{.033}) interaction effects between both factors with a small effect size.

In the experiment, the measurements showed accuracy rates ranging from \emmP{93.6}{3.1} (\lane{8}, \seltime{3}) to \emmP{58.9}{3.1} (\lane{16}, \seltime{1}). Table \ref{tab:walktheline/accuracy_table} lists the measured accuracy rates for the tested conditions, figure \ref{fig:walktheline/accuracy_small} depicts the mean values.

\begin{table}
	\centering
	
	\begin{tabularx}{\linewidth}{ccYYYY}
		& &   &   & \multicolumn{2}{c}{\textbf{95\% CI\textsuperscript{1}}}\\
		\cmidrule(lr){5-6}
		\textbf{Number of Lanes} & \textbf{Selection Time} & $\pmb{\mu}$ & $\pmb{\sigma}$ & \textbf{Lower} & \textbf{Upper}\\
		\midrule
		8 & \seltime{1} & .708 & .163 & .646 & .771 \\
		  & \seltime{2} & .931 & .067 & .868 & .993 \\
		  & \seltime{3} & .936 & .077 & .874 & .998 \\
		12& \seltime{1} & .660 & .139 & .597 & .722 \\
		  & \seltime{2} & .887 & .100 & .824 & .949 \\
		  & \seltime{3} & .866 & .154 & .804 & .928 \\
		16& \seltime{1} & .589 & .127 & .526 & .651 \\
		  & \seltime{2} & .839 & .127 & .776 & .901 \\
		  & \seltime{3} & .675 & .193 & .613 & .738 \\
		
		\bottomrule
	\end{tabularx}
	
	\caption{The accuracy rates per combination of \emph{selection time} and \emph{number of lanes} as measured in the experiment. The table reports the recorded mean values $\mu$ together with the standard deviation $\sigma$. \textsuperscript{1} The confidence interval CI is based on the fitted \ac{EMM} model.}
	\label{tab:walktheline/accuracy_table}
\end{table}

\subsection{Stabilizing Error}

\colfigH{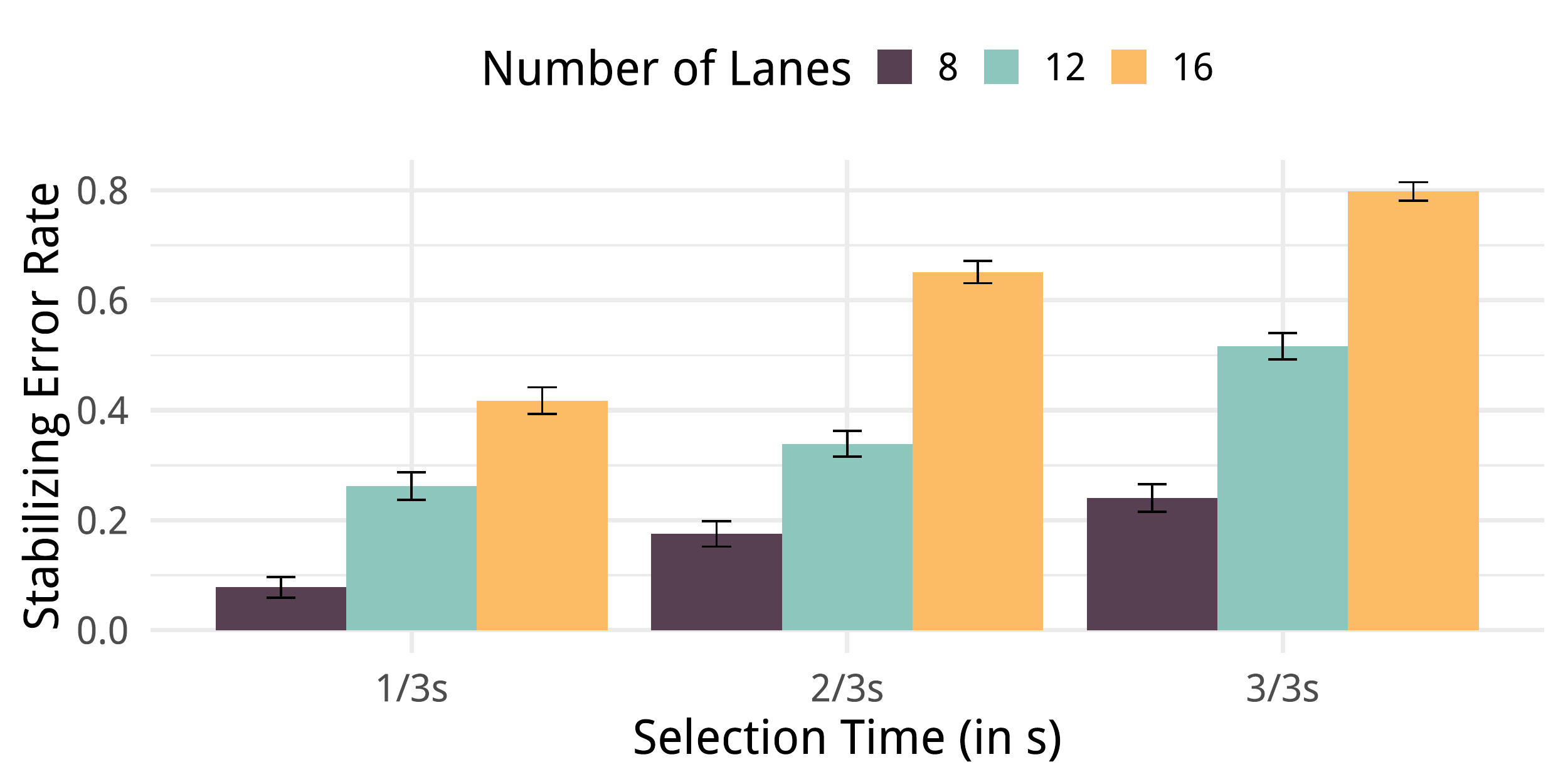}{The measured stabilizing error rates in the controlled experiment. All error bars depict the standard error.}

The stabilizing error rate was calculated by counting the number of trials when participants left the target lane after initially reaching it. The analysis showed a significant (\anova{2}{34}{127.3}{<.001}{.45}) influence of the \emph{number of lanes} on the stabilizing error rate with a large effect size. Post-hoc tests confirmed significantly higher stabilizing error rates for higher \emph{numbers of lanes} (and thus smaller lanes) between all levels (\lane{8}: \emmP{16.2}{3.5}, \lane{12}: \emmP{36.8}{3.5}, \lane{16}: \emmP{61.7}{3.5}, all $p<.001$). 

Further, the \emph{selection time} also proved to have an significant (\anova{2}{34}{67.07}{<.001}{.164}) influence on the stabilizing error rate in the experiment with a large effect size. Post-hoc tests confirmed significantly higher stabilizing error rates for longer selection times between all levels (\seltime{1}: \emmP{24.3}{3.3}, \seltime{2}: \emmP{38.7}{3.3}, \seltime{3}: \emmP{51.8}{3.3}, all $p<.001$).

Lastly, the analysis also showed significant (\anova{4}{68}{6.73}{<.001}{.023}) interaction effects between both factors with a small effect size. 

The analysis found stabilizing error rates ranging from \emmP{7.8}{4.1} (\lane{8}, \seltime{1}) to \emmP{79.8}{4.1} (\lane{16}, \seltime{3}). Table \ref{tab:walktheline/overshoot_table} lists the measured accuracy rates for the tested conditions, figure \ref{fig:walktheline/overshoot_small} depicts the mean values.

\begin{table}
	\centering
	
	\begin{tabularx}{\linewidth}{ccYYYY}
		& &   &   & \multicolumn{2}{c}{\textbf{95\% CI\textsuperscript{1}}}\\
		\cmidrule(lr){5-6}
		\textbf{Number of Lanes} & \textbf{Selection Time} & $\pmb{\mu}$ & $\pmb{\sigma}$ & \textbf{Lower} & \textbf{Upper}\\
		\midrule
		8 & \seltime{1} & .074 & .116 & .0   & .155 \\
		  & \seltime{2} & .172 & .137 & .091 & .254 \\
		  & \seltime{3} & .240 & .159 & .158 & .321 \\
		12& \seltime{1} & .250 & .141 & .169 & .332 \\
		  & \seltime{2} & .338 & .218 & .256 & .419 \\
		  & \seltime{3} & .516 & .219 & .435 & .597 \\
		16& \seltime{1} & .404 & .187 & .323 & .486 \\
		  & \seltime{2} & .649 & .191 & .568 & .731 \\
		  & \seltime{3} & .798 & .145 & .716 & .879 \\
		
		\bottomrule
	\end{tabularx}
	
	\caption{The stabilizing error rates per combination of \emph{selection time} and \emph{number of lanes} as measured in the experiment. The table reports the recorded mean values $\mu$ together with the standard deviation $\sigma$. \textsuperscript{1} The confidence interval CI is based on the fitted \ac{EMM} model.}
	\label{tab:walktheline/overshoot_table}
\end{table}

\subsection{Task Completion Time}
\label{sec:walktheline:results:tct}

\colfig{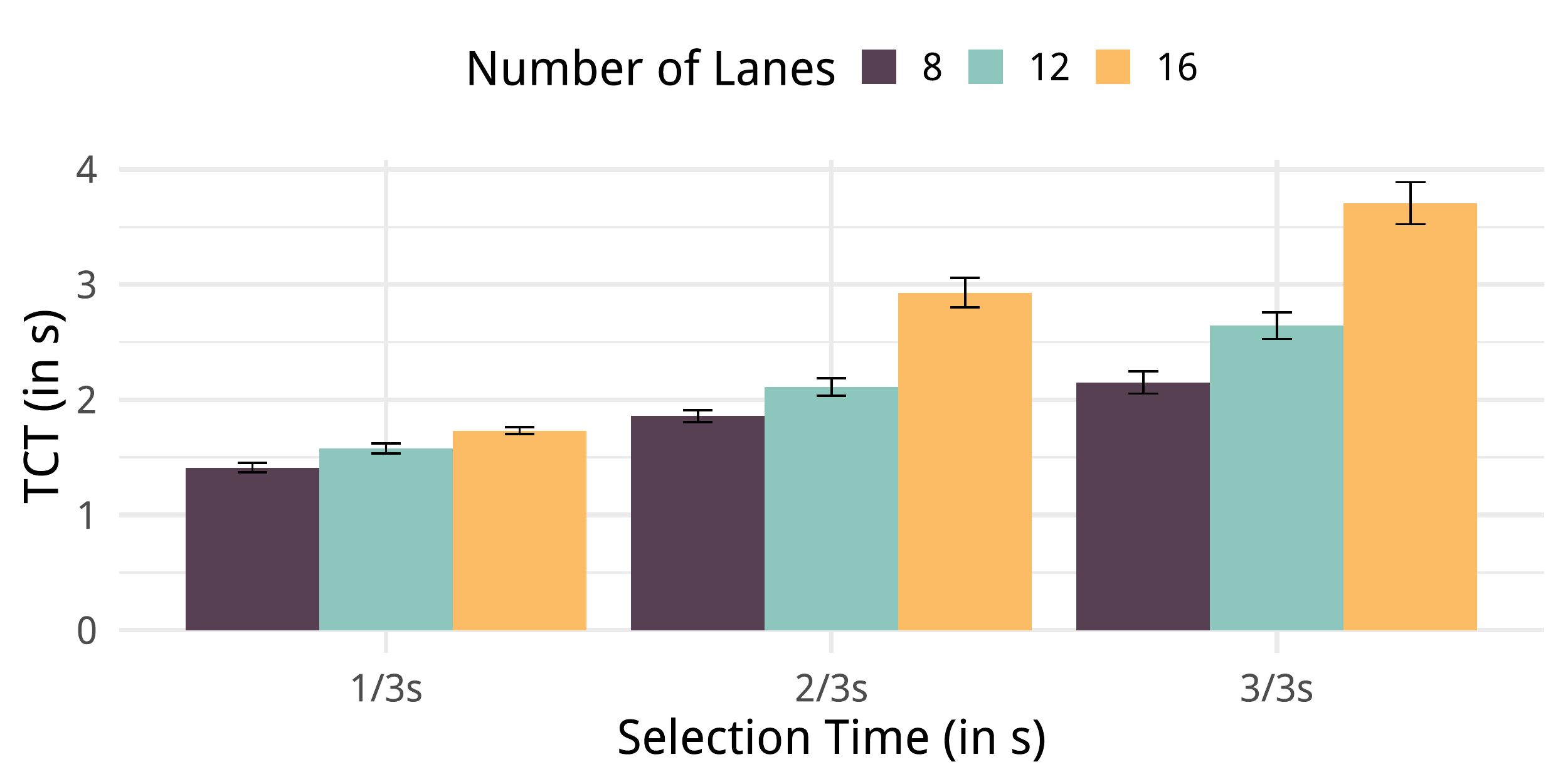}{The measured task-completion times (in seconds) in the controlled experiment. All error bars depict the standard error.}

The \ac{TCT} was measured as the time to successful activation of a lane after subtracting the respective selection time to keep the \acp{TCT} comparable. The time was measured from the moment the target was displayed to the participant. The analysis only considered the \acp{TCT} of the successful trails, as the different accuracy rates would otherwise influence the results. 

The analysis showed a significant (\anova{2}{34}{117.8}{<.001}{.262}) influence of the \emph{number of lanes} on the \ac{TCT} with a large effect size. Post-hoc tests confirmed rising \acp{TCT} for higher numbers of lanes between all levels (\lane{8}: \emmSi{1.81}{.07}{s}, \lane{12}: \emmSi{2.11}{.07}{s}, \lane{16}: \emmSi{2.79}{.07}{s} all $p<.001$). 

Interestingly, despite subtracting of the selection time from the \ac{TCT}, the analysis also showed a significant (\anova{2}{34}{123.3}{<.001}{.413}) effect of the \emph{selection time} on the \ac{TCT} with a large effect size. Post-hoc tests confirmed significantly higher \acp{TCT} for higher selection times between all levels (\seltime{1}: \emmSi{1.57}{.08}{s}, \seltime{2}: \emmSi{2.30}{.08}{s}, \seltime{3}: \emmSi{2.83}{.08}{s}, all $p<.001$). As depicted in figure \ref{fig:walktheline/tct_small}, the \acp{TCT} for the different selection times are close together for the \lane{8} conditions. For higher numbers of lanes, the \acp{TCT} grow faster for longer selection times.

Further, the analysis again showed significant (\anovaCor{2.71}{46.06}{25.3}{<.001}{.677}{.073}) interaction effects between the factors with a medium effect size.%

The graphical analysis of the \acp{TCT} showed strong visual correlations with the stabilizing error rates as presented above (see Figure \ref{fig:walktheline/overshoot_small} and \ref{fig:walktheline/tct_small}). Calculating Pearson's $r$ supported the visual impression by confirming a very strong~\ncite{Evans1996} correlation between \emph{stabilizing error rate} and \emph{\ac{TCT}} (\pearson{.925}{<.001}).

The analysis found \acp{TCT} ranging from \emmSi{1.41}{.10}{s} (\lane{8}, \seltime{1}) to \emmSi{3.71}{.08}{s} (\lane{16}, \seltime{3}). Table \ref{tab:walktheline/tct_table} and figure \ref{fig:walktheline/tct_small} depict the measured \acp{TCT} for all conditions.

\newcommand{\myunit}[1]{#1~s}

\begin{table}
	\centering
	
	\begin{tabularx}{\linewidth}{ccYYYY}
		& &   &   & \multicolumn{2}{c}{\textbf{95\% CI\textsuperscript{1}}}\\
		\cmidrule(lr){5-6}
		\textbf{Number of Lanes} & \textbf{Selection Time} & $\pmb{\mu}$ & $\pmb{\sigma}$ & \textbf{Lower} & \textbf{Upper}\\
		\midrule
		8 & \seltime{1} & \myunit{1.410} & \myunit{.170} & \myunit{1.216} & \myunit{1.605} \\
		  & \seltime{2} & \myunit{1.858} & \myunit{.226} & \myunit{1.663} & \myunit{2.052} \\
	      & \seltime{3} & \myunit{2.149} & \myunit{.409} & \myunit{1.955} & \myunit{2.344} \\
		12& \seltime{1} & \myunit{1.578} & \myunit{.187} & \myunit{1.384} & \myunit{1.773} \\
		  & \seltime{2} & \myunit{2.111} & \myunit{.327} & \myunit{1.916} & \myunit{2.305} \\
		  & \seltime{3} & \myunit{2.643} & \myunit{.491} & \myunit{2.448} & \myunit{2.837} \\
		16& \seltime{1} & \myunit{1.732} & \myunit{.123} & \myunit{1.537} & \myunit{1.926} \\
		  & \seltime{2} & \myunit{2.929} & \myunit{.548} & \myunit{2.735} & \myunit{3.124} \\
		  & \seltime{3} & \myunit{3.705} & \myunit{.773} & \myunit{3.511} & \myunit{3.900} \\
		
		\bottomrule
	\end{tabularx}
	
	\caption{The task-completion times (in seconds) per combination of \emph{selection time} and \emph{number of lanes} as measured in the experiment. The table reports the recorded mean values $\mu$ together with the standard deviation $\sigma$. \textsuperscript{1} The confidence interval CI is based on the fitted \ac{EMM} model.}
	\label{tab:walktheline/tct_table}
\end{table}

\subsection{Walked Distance}

\colfig{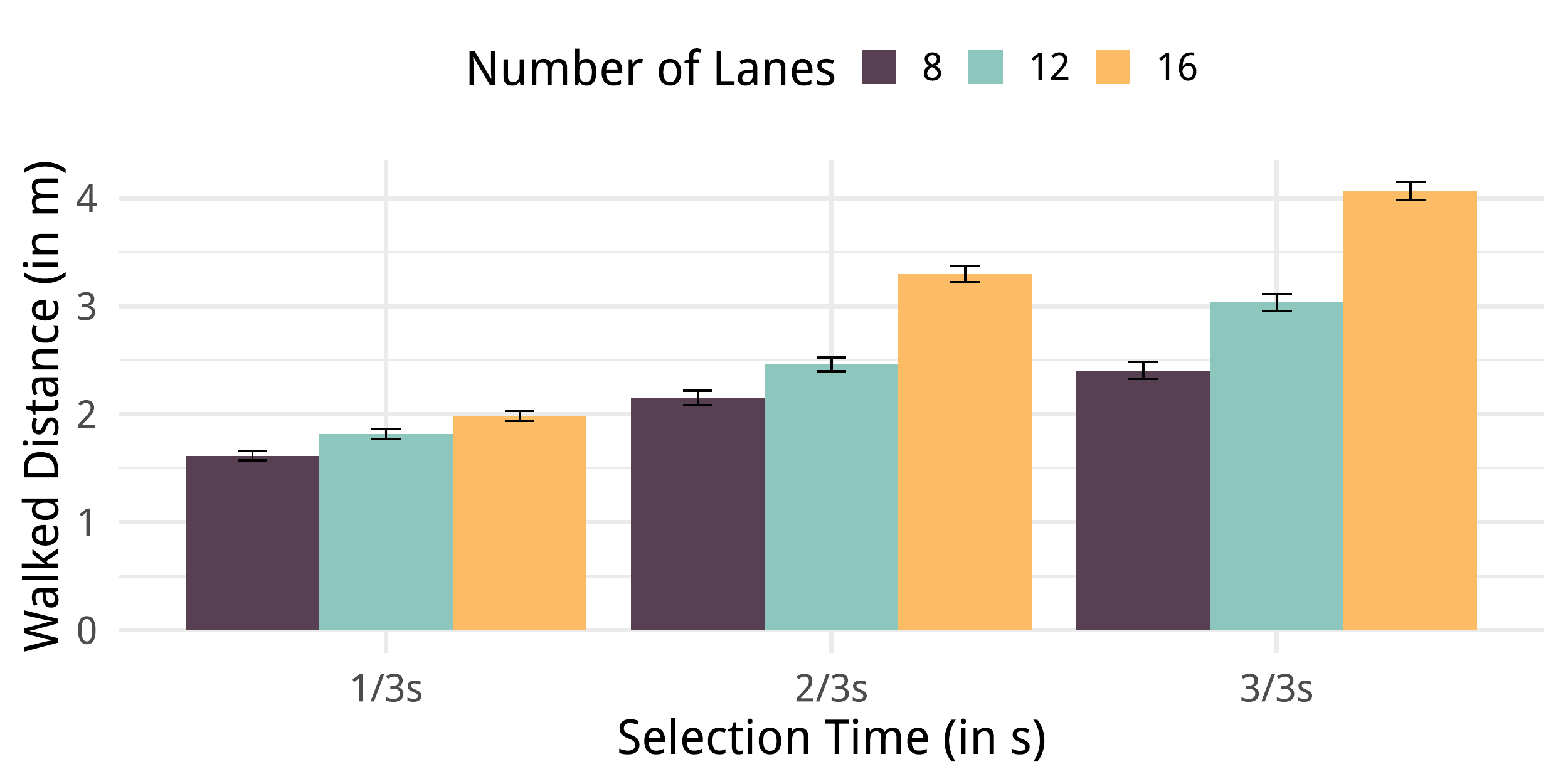}{The measured distances to selection in the controlled experiment. All error bars depict the standard error.}

To take into account different walking speeds of the participants, the analysis assessed the walking distance necessary to activate a target. Similar to the \ac{TCT}, the analysis only considered the distance that was necessary to select a lane without the distance walked during the \factorTime~of the respective condition. Therefore, the system measured the distance the participants walked from the beginning of the task within the \ac{TCT} as defined above. %

The analysis showed a significant (\anovaCor{1.45}{24.65}{84.0}{<.001}{.725}{.159}) influence of the \emph{number of lanes} on the distance with a large effect size. Post-hoc tests confirmed significantly higher distances for higher \emph{numbers of layers} between all levels (\lane{8}: \emmSi{2.06}{.18}{m}, \lane{12}: \emmSi{2.43}{.18}{m}, \lane{16}: \emmSi{3.11}{.18}{m} , all $p<.001$).

As for the \ac{TCT}, the analysis also showed a significant (\anovaCor{1.41}{23.89}{102.5}{<.001}{.703}{.263}) effect for the \emph{selection time} on the distance with a large effect size. Again, post-hoc tests confirmed significantly higher distances for higher selection times between all levels (\seltime{1}: \emmSi{1.80}{.18}{m}, \seltime{2}: \emmSi{2.63}{.18}{m}, \seltime{3}: \emmSi{3.17}{.18}{m}, all $p<.001$).

Further, the analysis showed significant (\anovaCor{2.96}{50.32}{21.9}{<.001}{.74}{.042}) interaction effects between both factors with a small effect size. 

As for the \ac{TCT}, the visual analysis of the measured distances showed correlations with the stabilizing error rates (see Figure \ref{fig:walktheline/overshoot_small} and \ref{fig:walktheline/distance_small}). Again, calculating Pearson's $r$ supported the visual impression, confirming a strong correlation between the \emph{stabilizing error rate} and the needed \emph{distance} to walk (\pearson{.924}{<.001}).

The analysis found distances ranging from \emmSi{1.62}{.19}{m} (\lane{8}, \seltime{1}) to \emmSi{4.07}{.19}{m} (\lane{16}, \seltime{3}). Table \ref{tab:walktheline/distances_table} and figure \ref{fig:walktheline/distance_small} depicts the measured walking distances for all conditions in the experiment.

The results presented in this section can only provide an approximation to a potential lower limit of interaction distances. In an urban environment, environmental influences such as obstacles or road conditions can influence these results.

\newcommand{\meterunit}[1]{#1~m}

\begin{table}
	\centering
	
	\begin{tabularx}{\linewidth}{ccYYYY}
		& &   &   & \multicolumn{2}{c}{\textbf{95\% CI\textsuperscript{1}}}\\
		\cmidrule(lr){5-6}
		\textbf{Number of Lanes} & \textbf{Selection Time} & $\pmb{\mu}$ & $\pmb{\sigma}$ & \textbf{Lower} & \textbf{Upper}\\
		\midrule
		8 & \seltime{1} & \meterunit{1.616} & \meterunit{.505} & \meterunit{1.217} & \meterunit{2.015} \\
	      & \seltime{2} & \meterunit{2.153} & \meterunit{.648} & \meterunit{1.753} & \meterunit{2.552} \\
		  & \seltime{3} & \meterunit{2.402} & \meterunit{.707} & \meterunit{2.002} & \meterunit{2.801} \\
		12& \seltime{1} & \meterunit{1.811} & \meterunit{.549} & \meterunit{1.411} & \meterunit{2.210} \\
		  & \seltime{2} & \meterunit{2.457} & \meterunit{.748} & \meterunit{2.057} & \meterunit{2.856} \\
		  & \seltime{3} & \meterunit{3.034} & \meterunit{1.02} & \meterunit{2.634} & \meterunit{3.433} \\
		16& \seltime{1} & \meterunit{1.981} & \meterunit{.580} & \meterunit{1.582} & \meterunit{2.381} \\
		  & \seltime{2} & \meterunit{3.291} & \meterunit{1.08} & \meterunit{2.892} & \meterunit{3.690} \\
		  & \seltime{3} & \meterunit{4.066} & \meterunit{1.25} & \meterunit{3.667} & \meterunit{4.466} \\
		
		\bottomrule
	\end{tabularx}
	
	\caption{The walked distances to complete a task (in meters) per combination of \emph{selection time} and \emph{number of lanes} as measured in the experiment. The table reports the recorded mean values $\mu$ together with the standard deviation $\sigma$. \textsuperscript{1} The confidence interval CI is based on the fitted \ac{EMM} model.}
	\label{tab:walktheline/distances_table}
\end{table}

\subsection{TLX}

\colfig{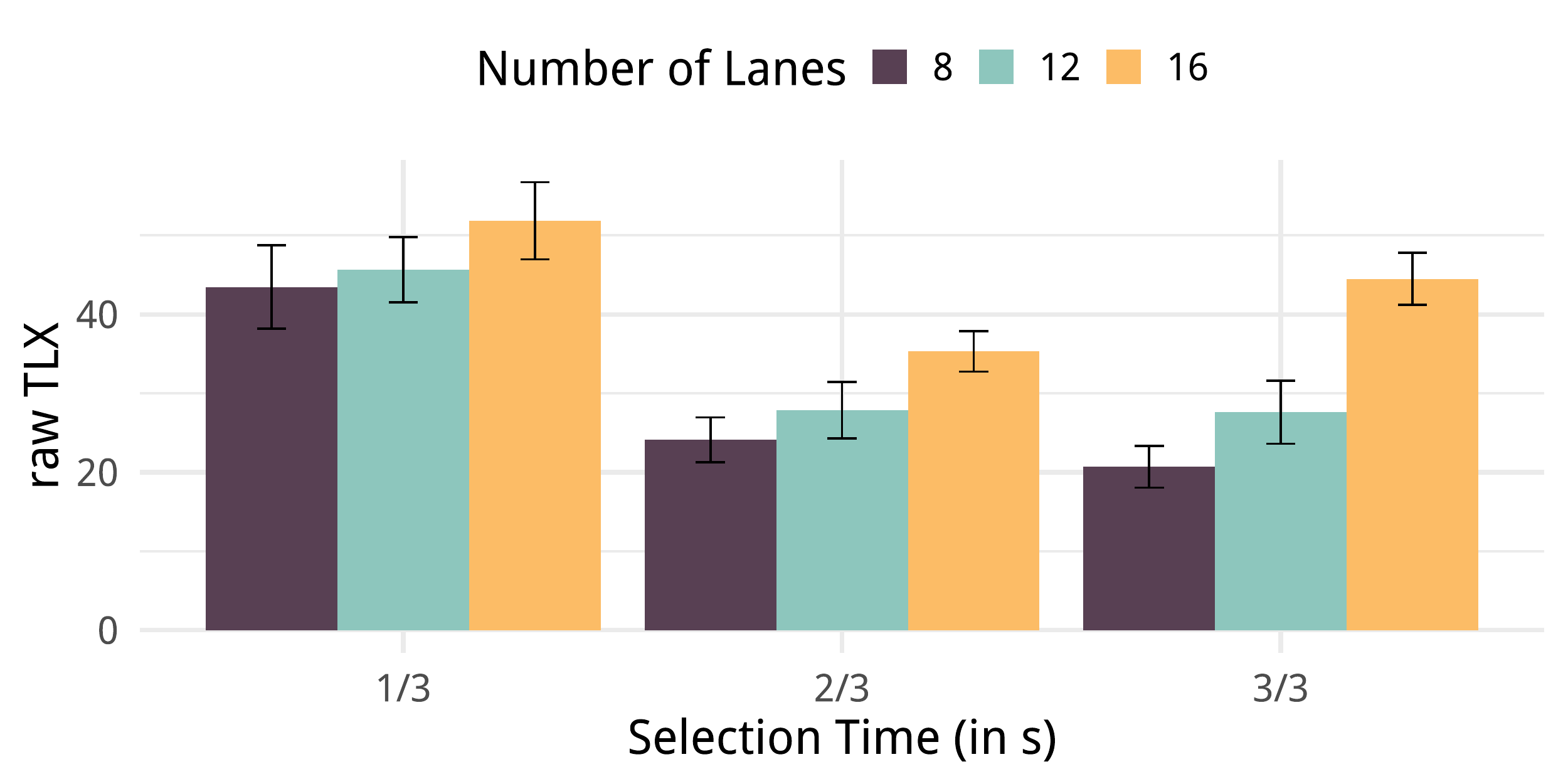}{The \acl{RTLX} measured in the controlled experiment. All error bars depict the standard error.}

To assess the differences in the mental load induced by the two factors, the analysis considered the influence of the factors on the \acf{RTLX}. The analysis showed a significant (\anovaCor{1.43}{24.36}{18.96}{<.001}{.716}{.104}) influence of the \emph{number of lanes} with a medium effect size. Post-hoc tests confirmed significantly higher values for the \lane{16} (\emm{43.9}{2.99}) conditions compared to both, the \lane{8} (\emm{29.4}{2.99}) and \lane{12} (\emm{33.7}{2.99}) conditions (both $p<.001$).

Further, the analysis showed a significant (\anovaCor{1.18}{20.01}{21.7}{<.001}{.588}{.182}) effect for the \emph{selection time} on the \ac{RTLX} with a large effect size. Post-hoc tests confirmed significantly higher values for the \seltime{1} (\emm{47.0}{3.16}) conditions compared to the \seltime{2} (\emm{29.1}{3.16}) and \seltime{3} (\emm{30.9}{3.16}) conditions (both $p<.001$).

Lastly, the analysis showed significant (\anova{4}{68}{3.12}{<.05}{.022}) interaction effects between the factors with a small effect size. 

The analysis found \ac{RTLX} values ranging from \emm{20.7}{3.8} (\lane{8}, \seltime{3}) to \emm{51.9}{3.8} (\lane{16}, \seltime{1}). Table \ref{tab:walktheline/tlx_table} and figure \ref{fig:walktheline/tlx_small} show the mean raw TLX values for the tested conditions.

\begin{table}
	\centering
	
	\begin{tabularx}{\linewidth}{ccYYYY}
		& &   &   & \multicolumn{2}{c}{\textbf{95\% CI\textsuperscript{1}}}\\
		\cmidrule(lr){5-6}
		\textbf{Number of Lanes} & \textbf{Selection Time} & $\pmb{\mu}$ & $\pmb{\sigma}$ & \textbf{Lower} & \textbf{Upper}\\
		\midrule
		8 & \seltime{1} & 43.47 & 22.50 & 35.87 & 51.07 \\
		  & \seltime{2} & 24.12 & 12.03 & 16.52 & 31.72 \\
		  & \seltime{3} & 20.69 & 11.17 & 13.09 & 28.30 \\
		12& \seltime{1} & 45.65 & 17.54 & 38.05 & 53.25 \\
		  & \seltime{2} & 27.87 & 15.20 & 20.27 & 35.47 \\
		  & \seltime{3} & 27.59 & 16.88 & 19.99 & 35.19 \\
		16& \seltime{1} & 51.85 & 20.73 & 44.25 & 59.45 \\
		  & \seltime{2} & 35.32 & 10.89 & 27.72 & 42.93 \\
		  & \seltime{3} & 44.49 & 13.98 & 36.89 & 52.09 \\
		
		\bottomrule
	\end{tabularx}
	
	\caption{The raw \ac{RTLX} values per combination of \emph{selection time} and \emph{number of lanes} as measured in the experiment. The table reports the recorded mean values $\mu$ together with the standard deviation $\sigma$. \textsuperscript{1} The confidence interval CI is based on the fitted \ac{EMM} model.}
	\label{tab:walktheline/tlx_table}
\end{table}

\subsection{Height}

To assess the influence of the participants height and, thus, differences in step sizes, the analysis assessed the between subject effects of the \emph{height} on the independent variables. However, the analysis did not show any influence on the \emph{accuracy} (\anova{1}{16}{1.78}{>.05}{.031}), the stabilizing error (\anova{1}{16}{1.31}{>.05}{.041}), the \ac{TCT} (\anova{1}{16}{.09}{>.05}{.002}) nor on the distance (\anova{1}{16}{1.15}{>.05}{.053}).

\subsection{Location of the Target Lane}

The analysis assessed the effect of the location of the target lane by comparing the measurements grouped by \emph{outer} (i.e., lanes on the far left and right as well as the lanes next to the central \emph{zero-lane}) and \emph{inner} (i.e., all other lanes) target lanes.

The analysis showed a significant influence of the target location on the accuracy (\anova{1}{17}{35.95}{<.001}{.058}) with a small effect size. Post-hoc tests confirmed significantly higher accuracy rates for outer (\emmP{85.0}{2.01}) compared to inner (\emmP{75.4}{2.01}) target lanes ($p<.001$).

Besides the accuracy, the analysis did not show any significant effects for the stabilizing error rate (\anova{1}{17}{3.49}{>.05}{.003}), the \ac{TCT} (\anova{1}{17}{2.43}{>.05}{.001}) nor the walked distance (\anova{1}{17}{1.92}{>.001}{.001}).

\subsection{Questionnaire}

After each condition, the participants filled a questionnaire asking questions regarding their experiences on a 5-point Likert-scale (1: strongly disagree, 5: strongly agree). The following section analyses the participants' answers.

\subsubsection{Confidence}

\textfigH{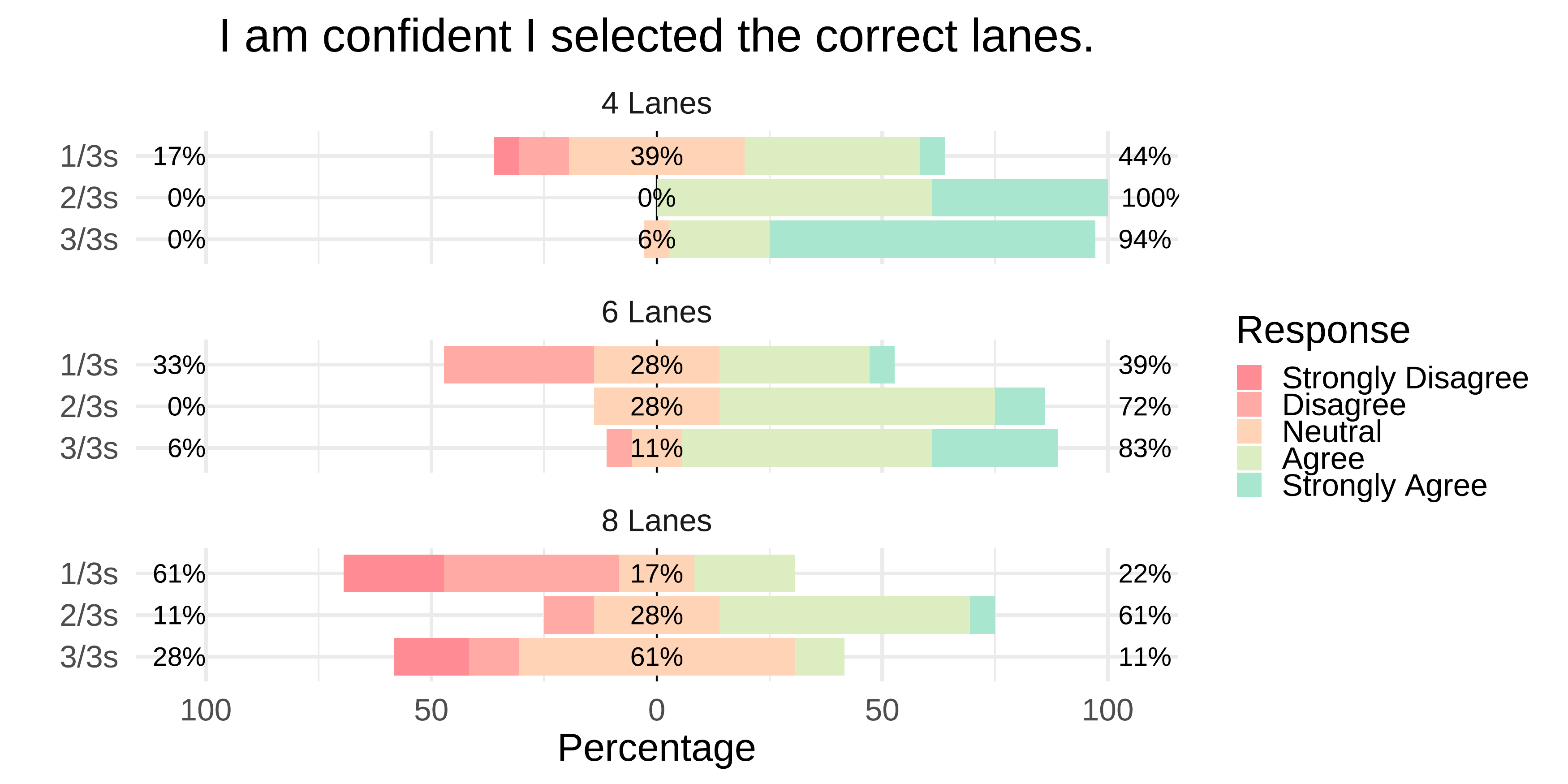}{The answers of the participants the \emph{Confidence} question in the questionnaire.}

The Likert questionnaire asked the participants about their \emph{confidence} to have successfully hit the target lanes in the condition. The analysis showed a significant (\anovaWithoutEffect{2}{34}{61.92}{<.001}) effect of the \emph{number of lanes} on the participants' confidence. Post-hoc tests confirmed significantly lower approval for the \lane{16} conditions compared to the \lane{8} and \lane{12} conditions (both $p<.001$).

Additionally, the analysis showed a significant (\anovaWithoutEffect{2}{34}{16.67}{<.001}) effect for the \emph{selection time} on the participants' confidence. Post-hoc tests revealed significantly lower approval rates for the \seltime{1} conditions compared to the \seltime{2} and \seltime{3} conditions (both $p<.001$).

The analysis found significant (\anovaWithoutEffect{4}{68}{6.11}{<.01}) interaction effects. Figure \ref{fig:walktheline/likert_q1_diss} depicts all the answers of the participants.

\subsubsection{Convenience}

\textfigH{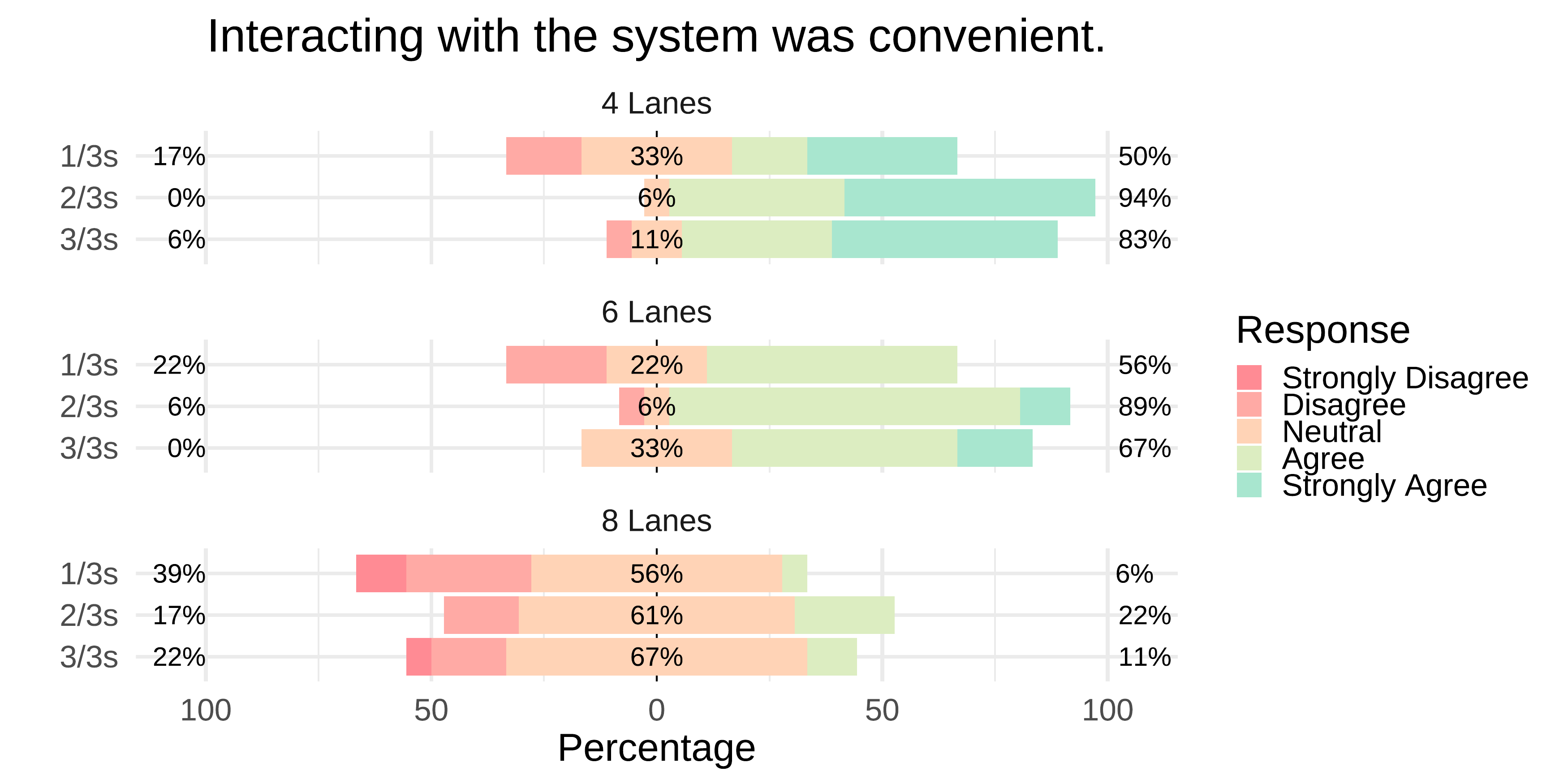}{The answers of the participants the \emph{convenience} question in the questionnaire.}

Further, the Likert questionnaire asked the participants if the combination of \emph{number of lanes} and \emph{selection time} was convenient to use. The analysis showed a significant (\anovaWithoutEffect{2}{34}{48.53}{<.001}) effect for the \emph{number of lanes} on the participants' ratings of the convenience. Post-hoc tests revealed significantly higher convenience ratings for \lane{8} and \lane{12} conditions compared to \lane{16} conditions (both $p<.001$).

Further, the analysis found a significant (\anovaWithoutEffect{2}{34}{11.47}{<.001}) influence of the \emph{selection time} on the ratings. Post-hoc tests confirmed significantly higher approval ratings for \seltime{2} ($p<.001$) and \seltime{3} ($p<01$) compared to \seltime{1} conditions.

The analysis found no interaction effects (\anovaWithoutEffect{4}{68}{0.16}{>.05}). Figure \ref{fig:walktheline/likert_q2_diss} depicts all the answers of the participants.

\subsubsection{Willingness to Use}

\textfigH{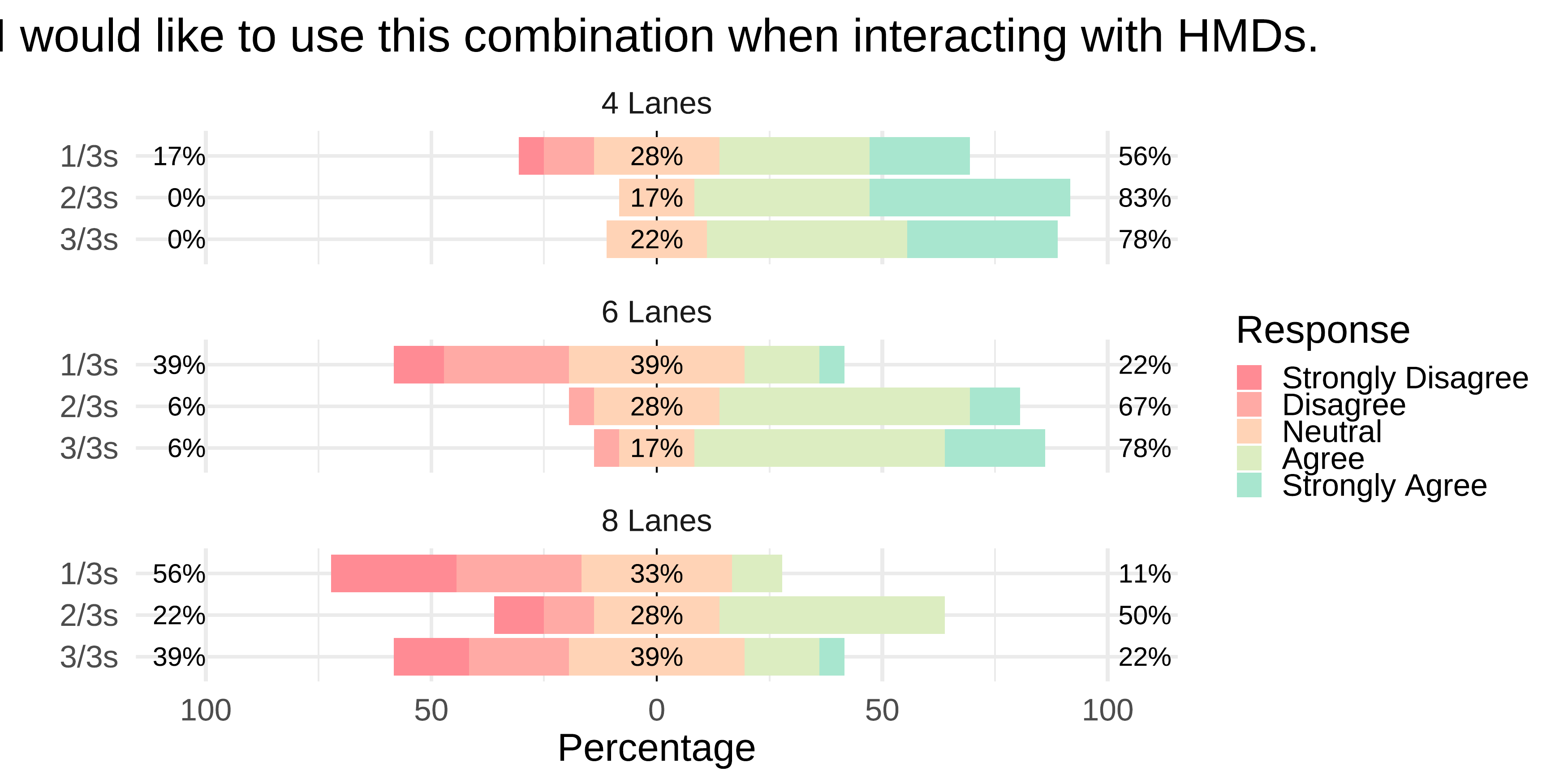}{The answers of the participants the \emph{Willingness to Use} question in the questionnaire.}

As a last question, the questionnaire asked the participants if they would like to use this combination of the \emph{number of lanes} and the \emph{selection time} for interacting with \acp{HMD}. The analysis showed a significant (\anovaWithoutEffect{2}{34}{28.13}{<.001}) influence of the \emph{number of lanes} on the participants' ratings. Post-hoc tests revealed significantly rising approval ratings for lower numbers of lanes between all levels ($p<.01$ comparing \lane{8} and \lane{12}, otherwise $p<.001$).

Further, the analysis unveiled a significant (\anovaWithoutEffect{2}{34}{17.86}{<.001}) influence of the \emph{selection time} on the ratings. Post-hoc tests showed significantly lower approval rates for \seltime{1} conditions compared to \seltime{2} and \seltime{3} conditions (both $p<.001$).

The analysis did not indicate any significant (\anovaWithoutEffect{4}{68}{1.59}{>.05}) interaction effects between the two factors. Figure \ref{fig:walktheline/likert_q2_diss} depicts all the answers of the participants.

\subsection{Qualitative Results}

In general, all participants showed strong approval for the idea of hands-free interaction with \acp{HMD} through walking. Asked for the reasons, participants told that it felt \pquote{fun}{8}, \pquote{novel}{15}, \pquote{fast}{12} and \pquote{convenient}{1,8}, and would be especially \pquote{helpful [...] while doing other things}{8}.

The participants noted that the \factorLanes{} had a strong influence on their experience. P14 summarized: \enquote{With many lanes it is frustrating. I have to concentrate a lot to accomplish that.} P8 added: \enquote{With the small lanes, it almost feels like I have to walk on a balance beam.} 

Concerning the \factorTime{}, the opinions of the participants diverged. While almost all participants agreed that \seltime{1} is \mpquote{too short}{P1, P2, P8, P13, P17}, both other selection times were equally popular. P7 explained the problem of identifying the \enquote{best} selection time: \enquote{It's complicated. With the thin lanes, I'm annoyed [...] by too much [selection] time because balancing is difficult. With the wide lanes, on the other hand, I find longer [selection] times easier.} %
\section{Discussion}
\label{sec:walktheline:discussion}

The results of the controlled experiment suggest that the usage of lateral shifts of the walking path of users provides a viable interaction technique for \acp{HMD}. The analysis showed the highest accuracy rates ($\approx 94\%$) for 8 interaction lanes (with an additional inactive \emph{zero lane} in the middle, resulting in a lane width of ~\SI{11}{cm}) with a selection time of \seltime{2}. The following section discusses the results of the experiment with respect to the research questions as presented above.

\subsection{RQ1: Influence of the Number of Lanes}

The analysis revealed a strong dependence of both, the accuracy and the efficiency, on the \emph{number} and - since the experiment varied the number of lanes on a fixed-width area - the \emph{width} of lanes. Higher numbers of lanes reduced the accuracy across all conditions. Further, higher numbers of lanes also led to higher \acp{TCT} and also increased the required walking distance to activate a target, decreasing efficiency.

The reduced accuracy and efficiency for higher numbers of lanes can be attributed to the higher stabilizing error rates through overshooting and swing-back errors caused by thinner lanes. This effect was further amplified by the natural lateral oscillation of the head that occurs during walking: The steps cause the head to constantly move slightly to the left and right of the actual path while walking, causing participants to oscillate out of the target lane. At comfortable walking speeds of around 1-1.5 m/s~\ncite{bohannon1997}, this effect occurs at the stride frequency of approximately \SI{1}{Hz}, and is responsible for a lateral translation of \SIrange{10}{15}{mm} in each direction~\ncite{MOORE2006}. This equates to 9-14\% (for the \lane{8} conditions) and up to 17-25\% (for the \lane{16} conditions) of the lane widths tested in the experiment. 

Further, the analysis showed significantly higher \ac{RTLX} values indicating a higher mental load for higher numbers of lanes. The results of the Likert-questionnaires supported the general discomfort of the participants with higher numbers of lanes. The participants answered all three questions - regarding confidence, convenience, and willingness to use - with significantly lower scores for \lane{16} compared to \lane{12} and \lane{8} conditions. The qualitative feedback of the participants further supported these findings, most of whom were in favor of lower lane numbers. 

\subsection{RQ2: Influence of the Selection Time}

Concerning the \factorTime, the analysis found a more complicated relationship to the accuracy than with the \factorLanes. The analysis showed that different selection times had a strong influence on the accuracy. Surprisingly, the middle selection time (\seltime{2}) was the one that achieved the highest accuracy rates. 

On the one hand, too short selection times led to participants accidentally selecting wrong lanes, as they spent too much time over an intermediate lane when changing lanes, thus decreasing accuracy. On the other hand, too long selection times increased the chance of participants accidentally leaving the target lane before the selection, as indicated by increased stabilizing error rates for higher selection times measured in the experiment. The observations during the experiment showed that - after such an incident - participants very carefully re-approached the target lane in order not to overshoot again, spending long periods of time on the adjacent lane. This behavior increased the chance of accidentally selecting a wrong target lane and, thereby, again reduced the accuracy.

Interestingly, the analysis also showed an influence of the \factorTime{} on the efficiency of participants, even though the respective \factorTime{} was explicitly subtracted from the \acl{TCT}. This effect can be attributed to the extended periods of time participants had to stay on a lane, increasing the chance of accidentally oscillating out of the lane and, thus, restarting the \factorTime. The restarted timer increased the \ac{TCT} and, thereby, the necessary distance. The analysis of the data provided further support for the assumption that the increased \ac{TCT} for higher numbers of lanes and higher selection times are related to oscillating out of the target lane: The analysis showed a strong correlation between stabilizing error rates, measured as the rate of trials in which the target lane was left, and the \ac{TCT}.

While the analysis of the \ac{RTLX} and the Likert questionnaires showed no differences between \seltime{2} and \seltime{3} conditions, both were rated significantly better than the \seltime{1} conditions. The qualitative feedback of participants supported this result: While participants' opinions were mixed for the \enquote{best} condition between \seltime{2} and \seltime{3}, there was a clear agreement that \seltime{1} was too short.

\subsection{RQ3: Interaction Effects between Number of Lanes and Selection Time}

The analysis of the experiment showed interaction effects between the \factorLanes{} and \factorTime{} for both, the accuracy and the efficiency measurements. This effect can be attributed to a mutual reinforcement of the influences of the individual factors described above: Lower numbers and, thus, wider lanes led to an increased width of the intermediate lanes between the participant and the target lane. With shorter selection times, this resulted in lower accuracy rates, because a larger lateral distance had to be crossed, resulting in more false selections in between. The reverse effect applies for higher numbers of and, thus, thinner lanes: Due to the thinner lanes, it generally became more difficult for the participants to select a lane by walking on it, resulting in lower efficiency and accuracy rates. This effect is further intensified by forcing the user to walk longer on the lane through higher selection times, increasing the chance of accidentally walking out of the lane.

Taken together, this explains the interaction effects found in the experiment on accuracy and efficiency: Wider lanes require longer selection times, thinner lanes require shorter selection times to attain high accuracy and effectiveness. %
\section{Implementation and Example Applications}
\label{sec:walktheline:prototype}

\begin{figure*}[ht!]
\subfloat[Assistant\label{fig:walktheline:uc:assistant}]
  {\includegraphics[width=.33\linewidth]{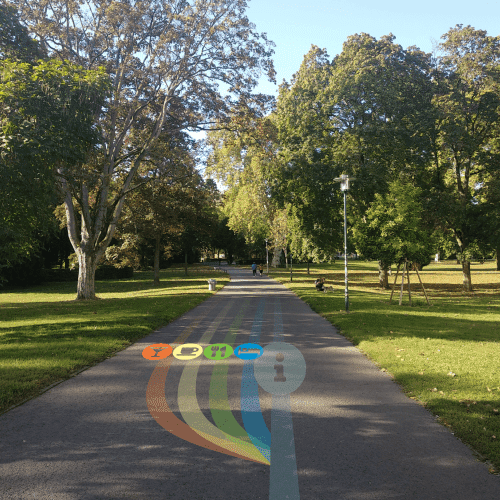}}\hfill
\subfloat[\prototypeCamera\label{fig:walktheline:uc:camera}]
  {\includegraphics[width=.33\linewidth]{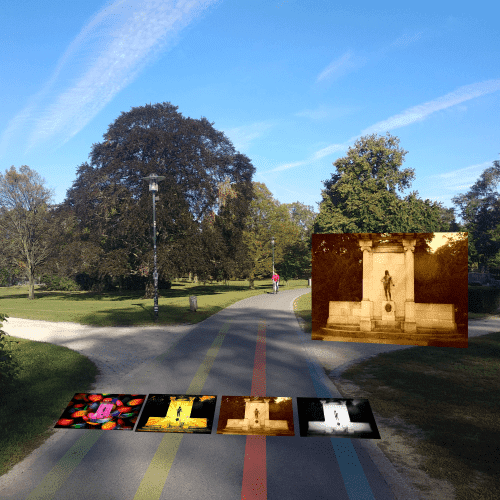}}\hfill
\subfloat[Music Walker\label{fig:walktheline:uc:music}]
  {\includegraphics[width=.33\linewidth]{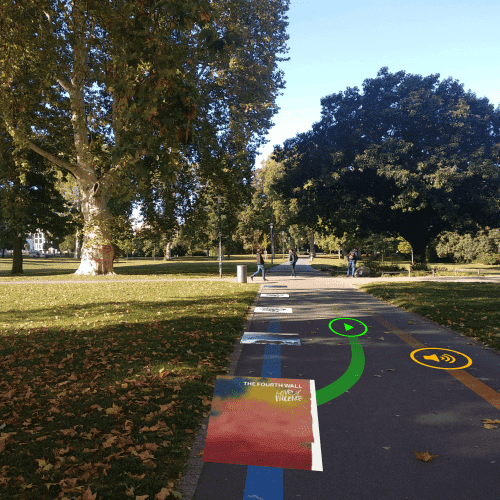}}
\caption{Three example applications to illustrate the presented concepts. \emph{Assistant} (a) allows users to explore nearby services (e.g., coffee shops, hotels, restaurants) after walking on the assistant lane. \emph{\prototypeCamera} (b) allows users to take pictures, apply filters and share the results to social media platforms. \emph{Music Walker} (c) allows users to walk through the playlist and select songs by leaving the lane. Further, users can continuously change the volume by walking on the respective lane.}
\end{figure*}

To eliminate influences of the restrictions of the current hardware generations of \acp{HMD}, the experiment used an artificial setup consisting of projectors and external tracking of the participants (see section \ref{sec:walktheline:methodology:setup}). However, the results of our experiment indicated that the accuracy of the inside-out tracking used by the Microsoft Hololens (average deviation of \SI{1.25}{cm}~\ncite{liu2018technical}) would be sufficient for a real-world implementation of the concepts. Therefore, based on the results of the experiment, this section presents the implementation of a \interactionStyleBased~input modality for the Microsoft Hololens. 

The software augments lanes onto the ground in front of the user, parallel to their walking path. Using the internal inside-out tracking of the Hololens, the system calculates the intersection between the orthogonal projection of the user's head position and the augmented lanes to identify the currently selected lane. The implementation works as a standalone application without modifications to the Hololens or additional external tracking. 

\subsection{Example Applications}

To show the practical applicability of the concept, this section presents three example applications: \nameref{sec:walktheline:prototype:assistant}, \nameref{sec:walktheline:prototype:camera} and \nameref{sec:walktheline:prototype:music}. The presented applications are not restricted to the general interaction concept presented in this chapter, but extend it by additional elements to make interaction possible in an urban context.

\subsubsection{Assistant}
\label{sec:walktheline:prototype:assistant}

Assistant is a personal assistance service, which - similar to commercial assistance solutions like Google Assistant, Alexa, or Cortana - can offer personalized recommendations. For this, an unobtrusively visualized assistant lane is displayed at the right edge of the user's field of view. This lane serves as a minimalist means of starting to interact with the system. If users want to access the service, they shift their walking path to the lane. This movement opens various options that allow the user to access personalized local services such as recommended restaurants or shops (see Figure \ref{fig:walktheline:uc:assistant}). By selecting an element through walking on it, the user can walk further down the options tree of a cascading menu.

\subsubsection{\prototypeCamera}
\label{sec:walktheline:prototype:camera}

By entering the photo lane (i.e., a specific lane within \emph{Assistant} or a standalone lane), the user can activate the camera. Exiting this lane to the \enquote{take a picture} side starts a countdown to take a picture of the current view of the participant without the augmented content. In the following, the user can apply various filters to the image, which are displayed as new lanes. The effect of each filter is previewed as soon as the user enters the corresponding lane (see Figure \ref{fig:walktheline:uc:camera}). By walking on a lane for a longer time, the user can select one of the filters and, in the next step, share the edited photo to different social media platforms, again visualized as newly appearing lanes.

\subsubsection{Music Walker}
\label{sec:walktheline:prototype:music}

Besides discrete inputs investigated in the experiment, walking-based interfaces can also be used to control continuous interfaces (see section \ref{sec:walktheline:limitations}). This section presents \emph{Music Walker} as an example of such a type of interaction.

Music Walker is a music player application. The user can continuously change the volume by walking on the \emph{volume up} or \emph{volume down} lane. The longer the user stays on a lane, the further the volume is increased or decreased, respectively. Further, the \emph{playlist} lane allows the user to walk up a list of the upcoming tracks. Leaving the lane allows the user to select a new song. Figure \ref{fig:walktheline:uc:music} depicts the interaction. %
\section{Limitations and Future Work}
\label{sec:walktheline:limitations}

The presented results provide valuable insights to the applicability of \interactionStyleBased{} input for the interaction with \acp{HMD}. However, the study design, as well as the results of the experiment, impose some limitations and directions for future work.

\subsection{Continuous Interaction}

The experiment focused on discrete interaction steps, that is, the sequential calling of options. The approach was chosen to define the basic requirements for the design of such interfaces in terms of minimum width and time needed to interact. However, such \interactionStyleBased{} interfaces can also be of great use for continuous interaction as suggested in the \emph{Music Walker} example (see section \ref{sec:walktheline:prototype:music}): For such interfaces, a) the deviation of the user from the direct path or b) the time spent on a lane could be mapped directly to a cursor or other interface elements. Future work in this area is necessary to asses the accuracy and efficiency of such interfaces.

\subsection{Shapes beyond Straight Lines}

This chapter investigated the deviation from a straight line in front of the user as an input modality. In many real-world scenarios, however, a straight line may not be a suitable baseline for interaction (because of e.g., obstacles, directional changes of the user). Therefore, further work in this field is necessary to conclude on these challenges.

\subsection{Other modes of Locomotion}

Beyond walking investigated in this chapter, this type of interaction can also be of great use for other modes of locomotion such as jogging, cycling, riding e-scooters, or when using wheelchairs. In particular, a study of the influence of the real world (obstacles, oncoming traffic) on the interaction with such a system is of interest. Future work is necessary to assess the influence of other modes of locomotion and, thus, also speeds on the feasibility, accuracy, efficiency, and safety of such interfaces. In addition, when using driving equipment, users often have their hands on a steering mechanism (e.g., steering wheel) that can be included in the interaction. Further work is necessary to evaluate the possibilities of such an extended design space. %
\section{Conclusion}

This chapter explored a way to leverage minimal changes in the user's walking path as an additional input modality for \acp{HMD}. The proposed concept augments lanes parallel to the user's walking direction on the floor in front of the user, representing individual options. The user can select one of these options by shifting the walking path sideways. The results of the controlled experiment confirmed the viability of such interfaces for fast, accurate, and fun interactions. Based on the results, the chapter presented a prototype implementation for Microsoft Hololens together with three example applications. 

To conclude, this chapter added to the body of research on interacting with \acp{HMD} in multiple areas:

\begin{enumerate}
	\item This chapter contributed an interaction technique for \acp{HMD} \emph{on the go}. The interaction is hands-free and can be performed while walking without interfering with locomotion. Thus, this work contributes to advancing the vision of ubiquitous interaction with information in a digitally augmented physical world.
	\item This chapter investigated the accuracy and efficiency of the envisioned interface in a controlled experiment, focusing on the influence of \factorLanes{} and the \factorTime{}. Furthermore, the chapter identified interdependencies between these factors that influence the future design of such interfaces. Therefore, this chapter contributed an initial evaluation of an interface designed specifically for interaction while walking, opening up a new field of research. For future research, the quantitative results contributed in this chapter provide a valuable baseline.
	\item Finally, this chapter presented three example applications illustrating the interaction with \acp{HMD} while walking in an urban context. Thus, this chapter demonstrated a safer and easier way to interact with information on the go compared to today's usage of smartphones.
\end{enumerate}

\subsection{Integration}
\textfigH{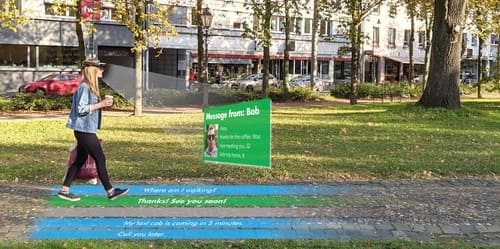}{Alice answers a text message by selecting a predefined message through walking on the corresponding lane.}

\interactionbox{alice_walktheline}{On the Go}{Alice is on her way to the pick-up point of the autonomous taxi. Meanwhile, she gets a message from Bob who thanks her for the great time they spent together. Automatically, four predefined responses appear as lanes in front of Alice, suggested as possible responses to Bob's message from her \ac{HMD}. Alice shifts her walkway to the side and, therby, chooses one of the options (see figure \ref{fig:walktheline/alice}).
	
Alice's interaction takes place as she walks without having to stop or free her hands.}

The interaction technique presented in this chapter allows for interaction with \emph{world-stabilized} interfaces leveraging our \emph{lower limbs} for input. The interaction is hands-free and takes place while walking, closing the gap left by previous interaction techniques for the lower limbs (see chapter \ref{ch:cheesyfoot}). This interaction technique thus supports users by providing \discreteLower{} in \mobilityLower{} situations, contributing to the vision of ubiquitous \aroundbodyinteraction{} (see section \ref{sec:introduction:aroundbodyinteraction}).

As described in section \ref{sec:walktheline:limitations}, the design space of such \interactionStyleBased{} interaction modalities for \acp{HMD} offers many more degrees of freedom that go beyond the scope of this work. The possible improvements of the concept listed in the section could further enable a broader applicability of the concept, e.g. by adapting the lanes to the real world or by integrating continuous interaction. However, the concepts as well as quantitative and qualitative results presented in this chapter provide a first step towards more comfortable and safe interaction with \acp{HMD} \emph{on the go} and, thus, can inform for future work in this area.

\subsection{Outlook}

This chapter presented the last main contribution of this work. Chapter \ref{ch:conclusion} concludes this work by integrating the individual contributions and presenting directions for future work.

}

\setcounter{ptc}{4}
\ctparttext{\parttoc}
\part{Conclusions}\label{pt:conclusion}
\chapter[Conclusion, Integration and Future Work]{A Vision for Around-Body Interaction with Head-Mounted Displays}\label{ch:conclusion}
{
This thesis argued for \aroundbodyinteraction{} (see section \ref{sec:introduction:aroundbodyinteraction}) leveraging 1) the \emph{upper} and \emph{lower} limbs to interact with 2) \emph{body}- and \emph{world-stabilized} interfaces. For each combination in this design space, this thesis presented a suitable interaction technique based on the individual requirements of the combinations and evaluated them in terms of efficiency and accuracy.

In this work, a concrete interaction technique was derived from the respective general requirements for each quadrant of the presented design space. These interaction techniques do not represent the only possible interaction techniques, and other approaches may yield different results depending on the situation. However, we are convinced that - while each of the presented interaction techniques is tailored towards a specific setting - the combination of these interaction techniques already supports a wide range of situations encountered during everyday usage of such devices.

This chapter shortly summarizes the main contributions and, based on the presented interaction techniques, outlines a vision for an integrated concept for joyful and efficient mobile interaction with \acp{HMD} using \aroundbodyinteraction{} (section \ref{sec:conclusion:summary}). Further, section \ref{sec:conclusion:futurework} highlights open questions and gives directions for future research.

\section{Contributions}
\label{sec:conclusion:summary}

\subsection{Upper Limbs}

The first two main contributions of this thesis focused on interaction techniques for the \emph{upper limbs} to operate \emph{body-stabilized} and \emph{world-stabilized} interfaces, respectively. 

\subsubsection{Body-Stabilized Interfaces}

First, \projProximity explored a one-handed interaction technique for \emph{body-stabilized} interfaces to be operated using the \emph{upper limbs}. The chapter presented an interaction technique that leverages the degree of freedom of the elbow joint to allow users to explore a one-dimensional interaction space along the line of sight. By flexing or extending the arm, the user can browse through successive layers. The visual content of the layers is anchored to the user's hand, enabling quick and immediate access to information, anytime and anywhere. The approach was evaluated in a controlled experiment, which showed short interaction times and high accuracy rates. 

As illustrated in the chapter, the support for both, \discreteLower{} and \continuousLower{}, enables the usage of this technique for a wide range of \singleuserLower{} applications. The contribution focuses entirely on one-handed interaction and, thus, considerably increases the \mobilityLower{} of fine-grained mobile interaction with digitally augmented information, as the second hand is still available for interaction with the real world. Therefore, the presented interaction technique contributes to the vision of \aroundbodyinteraction{} by providing fast and accurate one-handed interactions for \acp{HMD}.

\subsubsection{World-Stabilized Interfaces}

Second, \projCloudbits shifted the focus from body-stabilized to \emph{world-stabilized} interfaces, where information is not anchored to the user, but the environment. The chapter presented interaction techniques that leverage this world-stabilization of interfaces to support collaborative use cases where multiple users can exploit the spatial layout of information for collaborative access and manipulation.

The interaction techniques support \continuousLower{} in \inplaceLower{} and \multiuserLower{} scenarios. As shown in the evaluation, the contributed techniques alleviate the burdens connected to the retrieval and ease the interaction with information in conversation scenarios. Therefore, the presented interaction techniques contribute to the vision of \aroundbodyinteraction{} by providing by allowing users to interact with, sort, and share information easily and intuitively using the degrees of freedom offered by the joints of their body.

\subsection{Lower Limbs}

The third and the fourth main contributions of this thesis focused on interaction techniques for the \emph{lower limbs} to operate \emph{body-stabilized} and \emph{world-stabilized} interfaces, respectively. 

\subsubsection{Body-Stabilized Interfaces}

Third, \projCheesyfoot assessed the viability of foot-tapping as an interaction technique for direct and indirect interaction with \acp{HMD}. This technique visualizes a semicircular grid on the ground (directly) or in the air in front of the user (indirectly). In both cases, the user can select input options using foot taps. The analysis showed promising results that indicated foot-tapping as a viable interaction technique for hands-free interaction with \acp{HMD}.

Therefore, the presented interaction technique allows \singleuserLower{} to perform \discreteLower{} in situations, where the hands are not available, supporting \mobilityLower{}. Depending on the requirements of the application, the two interaction styles presented support both highly accurate (direct) as well as fast and casual interactions (indirect), where the user does not need to lower his head. Therefore, the presented interaction technique contributes to the vision of \aroundbodyinteraction{} by providing an alternative input modality for situations where the users' hands are busy interacting with the real world.

\subsubsection{World-Stabilized Interfaces}

Forth, \projCheesyfootToGo focused on foot-based interaction while walking and presented an approach to leverage locomotion as an additional input modality. The approach augments world-anchored lanes parallel to the walking path of the user to the ground. Each of the lanes represents an input option. Through lateral displacements of the user's walking path, the user can walk on a line to select it. The analysis proved this to be a promising approach for hands-free interaction while walking without detracting the user's visual attention.

Therefore, the presented interaction technique supports \singleuserLower{}s during \mobilityLower{} situations and provides \discreteLower{}. This interaction technique is of particular importance for situations in which the other techniques presented in this paper are not available to the user due to the situational impairments (e.g., walking with things carried in hand). Therefore, the presented interaction technique contributes to the vision of \aroundbodyinteraction{} by providing a highly mobile interaction technique that allows for hands-free interaction during locomotion.
\section{Integration and Future Work}
\label{sec:conclusion:futurework}

As outlined in the introduction of this thesis (chapter \ref{ch:introduction}), a single body-based interaction technique cannot support all situations a user might encounter during the day. Instead, an integrated set of interaction techniques is necessary that allows interaction while particular body-parts are encumbered. 

As illustrated throughout the thesis, the interaction techniques presented already support a variety of situations which users may encounter during a day. However, to allow for ubiquitous interaction with such devices in all everyday situations, it must be possible to perform all types of interaction using all limbs. Therefore, it will be necessary to first identify a set of interaction patterns which are necessary for interaction with this device class. In a second step, this must be translated into interaction techniques to support the respective situations in each of the presented quadrants of the design space. This is a major challenge that is outside of the scope of this thesis. Therefore, further work in this area is necessary to conclude on these challenges.

Approaches for the continuation of the work in the individual areas are given at the end of the respective chapters. This section, in addition, focuses on open questions and challenges for future research that arise from the vision of an integrated set of \aroundbodyinteraction{} techniques for interacting with \acp{HMD}.

\textfigH{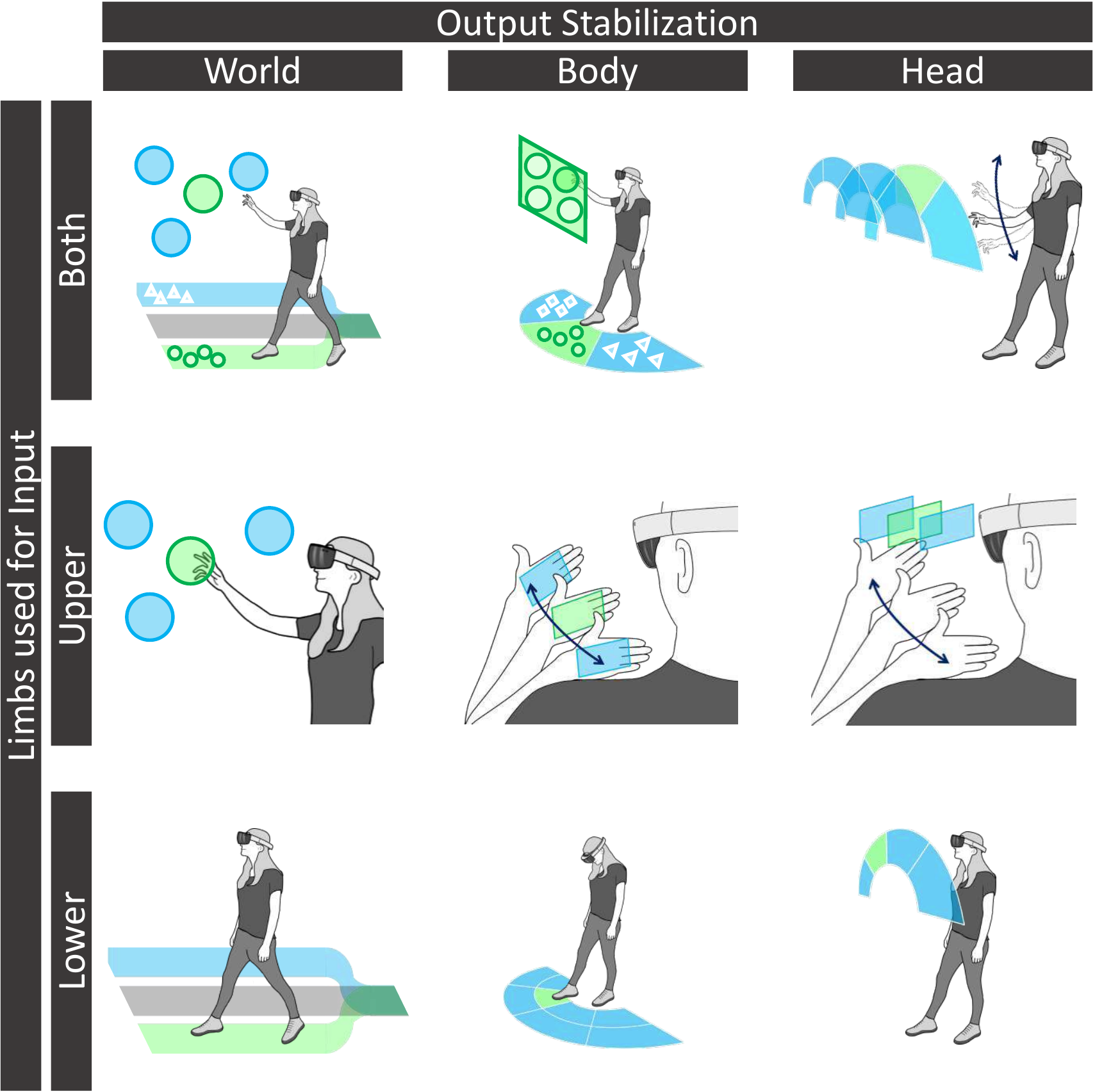}{The contributions of this thesis and possible directions for future work in the design space of \aroundbodyinteraction{}.}

\subsection{Head-Stabilized Interfaces}

Even though the rapid development in \ac{HMD} hardware and software led to a decreased usage of \emph{head-stabilized} interfaces, they are still useful in specific situations and, thus, interaction techniques should support such interfaces.

Despite the focus on \emph{body-stabilized} interfaces, we want to argue that the presented interaction techniques in the contributions \projProximityS (for the \emph{lower limbs}) and \projCheesyfootS (for the \emph{upper limbs}) can be used as a foundation to inspire future \emph{head-stabilized} interfaces. 

\paragraph{Upper Limbs} Instead of anchoring the visual output to the user's palm, the visual output can also be displayed as a \ac{HUD} visualization. Through the sense of proprioception, users can continuously scroll through or discretely select layers without looking at their hand. 

\paragraph{Lower Limbs} The second experiment focusing on indirect interfaces used a floating visualization in front of the user. While the interface was still anchored to the body of the user as it stayed in position when the user rotated the head, we still consider the results to be transferable to \emph{head-stabilized} interfaces.

\subsection{Interfaces for Simultaneous Use of Upper and Lower Limbs}

The simultaneous use of multiple limbs limits the possibility of use in many situations, as limbs used to interact with the system are no longer available to interact with the real world. Therefore, this work focused on interaction using individual limbs.

However, inspired by the field of whole-body interaction, interacting using multiple limbs simultaneously can increase the expressiveness of interactions and is, therefore, an important field for future work. In the following, this section presents ideas for the integration of upper and lower limb interaction, grouped by the stabilization of the output. Figure \ref{fig:conclusion/designspace} depicts these examples of the integration of interaction techniques..

\paragraph{Head-Stabilized Interfaces}

For head-stabilized interfaces, the contributions \projProximityS and \projCheesyfootS could be combined to a system that allows the selection of content (displayed as a \acp{HUD} interface) using movements of the user's arm. Foot-taps of the user can further provide input to the selected layer. As shown in chapter \ref{ch:cheesyfoot}, the sense of proprioception allows users to split the area in front of their feet into multiple interactive zones that can be tapped with the feet without visual guidance. 

\paragraph{Body-Stabilized Interfaces}

Similar to the head-stabilized interface, body-stabilized interfaces can combine the input using the upper and lower limbs. Again, foot-tapping can act as a second input modality to provide input for individual layers selected using the hand of the user.

\paragraph{World-Stabilized Interfaces}

For world-stabilized interfaces, the interaction techniques in \projCheesyfootToGoS could be used for the selection of applications which, once selected, pop out of the floor. As a result, the interaction techniques from \projCloudbitsS and \projProximityS could be used for further interaction and spatial classification of these applications and information.

\subsection{Selection of a Suitable Interaction Technique Depending on the Context}

As discussed before, interaction techniques can be rendered unsuitable by the context of use. When we carry things in our hands, we cannot use our hands for interaction. When we walk, we cannot use foot-tapping for interacting and, vice versa, when we stand, we cannot use a locomotion-based interaction technique. This raises the problem of how the user can select a suitable interaction technique. 

Simultaneous availability of all interaction techniques as the most straightforward solution leads to many problems similar to the \emph{Midas Touch Problem} in gaze-based interaction: How can natural movement be distinguished from the interaction? The constant availability of all interaction techniques would, thus, lead to a large number of false actions by the system.

Another simple solution could be the explicit selection of an interaction type by the user. However, this selection must also be carried out somehow - possibly by operating a switch, which again requires the use of the hands and is therefore not possible in all situations.

In this field, therefore, further work is still necessary to reach a truly integrated system, which - based on the context and the situation of the user - can decide which interaction techniques are appropriate for the situation.

\section{Concluding Remarks}

The recent history of technological progress teaches us that we are on the way to ever smaller and more powerful \acp{HMD}, that could one day take on the role that smartphones play in our lives today. Thus, the interaction with such devices could - 50 years after Sutherland's initial steps - finally move out of the laboratory and into the real world; to private homes, remote rural areas and lively urban spaces. 

Certainly, this work cannot provide a conclusive picture of how we will interact with these devices in the future. However, it has made a substantial contribution towards a vision for future interaction with such devices by harnessing the degrees of freedom offered by our bodies for more natural, pleasant and fun interactions.
}

\addtocontents{toc}{\protect\partbegin}

\setcounter{ptc}{5}
\appendix
\cleardoublepage
\bookmarksetup{startatroot}%
\bookmarksetupnext{level=0}%
\cleardoublepage
\pagestyle{scrheadings}

\defbibheading{bibintoc}[\bibname]{%
	\phantomsection
	\manualmark
	\markboth{\spacedlowsmallcaps{#1}}{\spacedlowsmallcaps{#1}}%
	\addtocontents{toc}{\protect\vspace{\beforebibskip}}%
	\addcontentsline{toc}{chapter}{\tocEntry{#1}}%
	\chapter*{#1}%
}

\clearpage
\begingroup
\let\clearpage\relax
\let\cleardoublepage\relax

\phantomsection
\chapter*{List of Publications}
\manumarkboth{List of Publications}
\addtocontents{toc}{\protect\vspace{\beforebibskip}}%
\addcontentsline{toc}{chapter}{\tocEntry{List of Publications}}

Parts of this work have already been published in the proceedings of international peer-reviewed conferences. This thesis uses parts of the content of the respective publications verbatim.

\section*{Publications Directly Related to this Thesis}

\begin{refsection}[relevantpubs]
	\small
	\nocite{*} %
	\printbibliography[heading=none]
\end{refsection}

\section*{Other Publications}

In addition to the contributions listed above, which form the foundation of this work, the author has also (co-)authored several other publications which are situated in related research areas.

\begin{refsection}[otherownpubs]
	\small
	\nocite{*} %
	\printbibliography[heading=none]
\end{refsection}

\newpage

\phantomsection
\chapter*{List of Third-Party Materials}
\manumarkboth{List of Third-Party Materials}
\addcontentsline{toc}{chapter}{\tocEntry{List of Third-Party Materials}}

Some graphics used in this work are not copyrighted by the author but have been created by third parties. This section lists these works with the names of their authors.

\attributegraphic{Content/Figures/General/mobility}{Freepik from \url{flaticon.com}}

\attributegraphic{Content/Figures/General/in-place}{Freepik from \url{flaticon.com}}

\attributegraphic{Content/Figures/General/multi-user}{Good Ware from \url{flaticon.com}}

\attributegraphic{Content/Figures/General/single-user}{Good Ware from \url{flaticon.com}}

\attributegraphic{Content/Figures/General/continuous}{Smashicons from \url{flaticon.com}}

\attributegraphic{Content/Figures/General/discrete}{Smashicons from \url{flaticon.com}}

\attributegraphic{Content/Figures/General/published}{conquine from \url{pngtree.com}}

\attributegraphic{Content/Figures/General/submitted}{conquine from \url{pngtree.com}}

\newpage

\phantomsection
\manumarkboth{\listfigurename}
\addcontentsline{toc}{chapter}{\tocEntry{\listfigurename}}
\listoffigures

\newpage

\phantomsection
\manumarkboth{\listtablename}
\addcontentsline{toc}{chapter}{\tocEntry{\listtablename}}
\listoftables

\newpage

\phantomsection
\manumarkboth{List of Acronyms}
\addcontentsline{toc}{chapter}{\tocEntry{List of Acronyms}}
\markboth{\spacedlowsmallcaps{Acronyms}}{\spacedlowsmallcaps{Acronyms}}
\chapter*{Acronyms}
\begin{acronym}[UMLX]
  \acro{HMD}{head-mounted display}
  \acro{GUI}{graphical user interface}
  \acro{HUD}{head-up display}
  \acro{TCT}{task-completion time}
  \acro{SVM}{support vector machine}
  \acro{EMM}{estimated marginal mean}
  \acro{AR}{Augmented Reality}
  \acro{VR}{Virtual Reality}
  \acro{RTLX}{Raw Nasa-TLX}
  \acro{WUI}{Walking User Interface}
  \acro{CSCW}{Computer-Supported Cooperative Work}
  \acro{HCI}{Human-Computer Interaction}
  \acro{ABI}{Around-Body Interaction}
  \acro{SLAM}{Simultaneous Location and Mapping}
\end{acronym}

\endgroup
\cleardoublepage\defbibheading{bibintoc}[\bibname]{%
  \phantomsection
  \manualmark
  \markboth{\spacedlowsmallcaps{#1}}{\spacedlowsmallcaps{#1}}%
  \addtocontents{toc}{\protect\vspace{\beforebibskip}}%
  \addcontentsline{toc}{chapter}{\tocEntry{#1}}%
  \chapter*{#1}%
}

\printbibliography[heading=bibintoc, notcategory=ignore]

\cleardoublepage\pdfbookmark[0]{Declaration}{declaration}
\chapter*{Erklärung}
\thispagestyle{empty}

Hiermit erkläre ich, die vorgelegte Arbeit zur Erlangung des akademischen Grades Doktor rerum naturalium (Dr. rer. nat.) mit dem Titel \medskip

\begin{center}
\emph{\myTitle{}: \mySubtitle{}}\medskip
\end{center}

selbständig und ausschließlich unter Verwendung der angegebenen Hilfsmittel erstellt zu haben. Ich habe bisher noch keinen Promotionsversuch unternommen.

\bigskip

\noindent\textit{\myLocation, \myTime}

\smallskip

\begin{flushright}
    \begin{tabular}{m{6cm}}
        \\ \hline
        \centering\myName \\
    \end{tabular}
\end{flushright}

\end{document}